\documentclass[a4paper,10pt,5p,preprint]{elsarticle}

\usepackage[mathlines,switch]{lineno}
\usepackage{amsmath, amssymb, amsfonts, physics, mathtools}
\usepackage{rotating}
\usepackage{graphicx,color}
\usepackage{bm}
\usepackage{color}
\usepackage{multirow}
\usepackage{enumitem}
\usepackage{tikz}
\usepackage{breakcites}
\usepackage{epstopdf}
\usepackage{morefloats}
\usepackage[pdfpagelabels, hypertexnames=true, plainpages=false, naturalnames=false, breaklinks=true]{hyperref}
\usepackage{subcaption}
\usepackage[capitalise,noabbrev]{cleveref}
\usepackage{pgfplots}
\usetikzlibrary{shapes.geometric, arrows}
\usepackage{algorithm} 
\usepackage{algorithmic}  
\usepackage[algo2e]{algorithm2e}

\newcommand*\patchEnvForLineno[1]{%
	\expandafter\let\csname old#1\expandafter\endcsname\csname #1\endcsname
	\expandafter\let\csname oldend#1\expandafter\endcsname\csname end#1\endcsname
	\renewenvironment{#1}%
	{\linenomath\csname old#1\endcsname}%
	{\csname oldend#1\endcsname\endlinenomath}}%
\newcommand*\patchBothEnvForLineno[1]{%
	\patchEnvForLineno{#1}%
	\patchEnvForLineno{#1*}}%
\AtBeginDocument{%
	\patchBothEnvForLineno{equation}%
}

\renewcommand*{\today}

\journal{Advances in Water Resources}

\begin{document}

\begin{frontmatter}

\title{Projection-based embedded discrete fracture model (pEDFM) for flow and heat transfer in real-field geological formations with corner-point grid geometries}

\author[tud-1]{Mousa HosseiniMehr}
\ead{S.Hosseinimehr@tudelft.nl}

\author[tud-1]{Janio Piguave Tomala}
\ead{janio.pigguave@gmail.com}

\author[tud-1]{Cornelis Vuik}
\ead{C.Vuik@tudelft.nl}

\author[pi-address]{Mohammed Al Kobaisi}
\ead{mohammed.alkobaisi@ku.ac.ae}

\author[tud-1]{Hadi Hajibeygi\corref{correspondingauthor}}
\ead{h.hajibeygi@tudelft.nl}

\cortext[correspondingauthor]{Corresponding author}

\address[tud-1]{Delft University of Technology, P.O. Box 5048, 2600 GA Delft, the Netherlands.}
\address[pi-address]{Khalifa University of Science and Technology, P.O. Box 2533, Abu Dhabi, United Arab Emirates.}

\begin{abstract}
In this work, the projection-based embedded discrete fracture model (pEDFM) for corner-point grid (CPG) geometry is developed for simulation of flow and heat transfer in fractured porous media. Unlike classical embedded discrete fracture approaches, this method allows for using any geologically-relevant model with a complex geometry and generic conductivity contrasts between the rock matrix and the fractures (or faults). The mass and energy conservation equations are coupled using the fully-implicit (FIM) scheme, allowing for stable simulation. Independent corner-point grids are imposed on the rock matrix and all fractures. The connectivities between the non-neighboring grid cells are described such that a consistent discrete representation of the embedded fractures occurs within the corner-point grid geometry. Various numerical tests including geologically-relevant and real-field models are conducted to demonstrate the performance of the developed method. It is shown that pEDFM can accurately capture the physical influence of both highly conductive fractures and flow barriers on the flow and heat transfer fields in complex reservoir geometries. This development casts a promising approach for flow simulations of real-field geo-models, increasing the discretization flexibility and enhancing the computational performance for capturing explicit fractures accurately.
\end{abstract}


\end{frontmatter}


\section{Introduction}\label{Sec:Introduction}
In a variety of geo-engineering fields, regardless of their applications in energy production (e.g., hydrocarbons and geothermal energy) or storage (e.g., CO2 storage and hydrogen storage), a detailed understanding of fluids and heat transport, their physical and chemical interactions together with rock, and their impact on the geological formation is greatly necessary. While achieving field development plans successfully, accurate, efficient and scalable modeling of mass and heat transfer in the subsurface porous media plays a crucial role in the fulfillment of the scientific, economical and societal expectations. Such computer models and their resulting predictions, contribute to efficient and safe operations on the production or storage facilities with regards to any of the geo-engineering applications. These predictions provide valuable insights on the optimization of hydrocarbon extractions \cite{Jansen2005}, the energy production outlines and the life-time of geothermal systems \cite{OSullivan2001,Axelsson2003,Burnell2012,Burnell2015}, the practical capacities that can be offered by the underground formations to store $\text{CO}_2$ or hydrogen, and many more.

However, while modeling subsurface flow, the geo- engineering community faces a number of key challenges. The geological formations are often in large scales. While they are located few kilometers deep in the subsurface (crust) and have a thickness of tens (if not hundreds) of meters, their areal extents can easily be in orders of kilometers. In order to reflect the geological and geometrical properties of the subsurface accurately, high-resolution computational grids are often imposed on the domain. This results in significant computational complexity, which makes it impossible to run the computer models with such large domains using conventional methods. Moreover, strong spatial heterogeneity contrasts are observed between various physical and chemical properties in the formations. These heterogeneities affect the flow and transport properties of the rock (i.e., storage capacity and conductivity) in several orders of magnitude. The discretization of the governing partial differential equations, or PDEs, results in ill-conditioned linear systems of equations creating challenges for numerical solution schemes to solve such heterogeneous systems. In addition, the measurement of the heterogeneous properties several kilometers beneath the subsurface involves a great deal of uncertainty. In order to minimize the impact of such uncertainties, instead of one realization, hundreds (if not thousands) of realizations are created to obtain uncertainty quantification (UQ) and a large number of simulations have to be run. Thus, the complexity of the system can have a huge impact on providing predictions in a reasonable time scale.

Furthermore, geological formations are often defined with complex geometry and stratigraphy. Using Cartesian grid geometry, even though it allows for simpler conceptual modeling analyses, can result in oversimplified and inaccurate predictions. In addition, the presence of faults and fractures has significant effect on fluid and heat flow patterns through the subsurface formations. The heterogeneity contrasts in the length scales and conductivities caused by these complex networks of fractures and faults can cause extreme challenges in solving the linear systems using numerical methods \cite{gan2016production,gholizadeh2016study,salimzadeh2019novel}. Moreover, the strong coupling of mass and heat transport results in severe non-linearity which negatively impacts the stability and convergence in the system. In case of multi-phase flow (e.g., high-enthalpy geothermal systems) these issues become more drastic \cite{WONG2018236}. The elastic and plastic deformations in the geo-mechanical interactions  \cite{Rossi2018,Garipov2016,gholizadeh2016evaluation}, reactive transport (e.g., geo-chemical interaction between the substances) \cite{geochem_1,Saar_2,salimzadeh2019coupled} and compositional alterations in the fluid and rock are among the list of other noteworthy challenges. Therefore, there is a high demand for developing advanced simulation methods that are computationally efficient and scalable, yet accurate at the desired level. Therefore, the development of a reliable computer model for simulation of subsurface flow and transport in fractured porous media is critical to address the challenges in practical applications. To address these challenges, many advanced numerical methods have been developed and offered.\\

To represent the real-field geological formations accurately, instead of using Cartesian grids, more complex and flexible gridding structure are needed as these formations are more conveniently represented by flexible grids \cite{Lie_unst,Rainer2006}. The grid geometry should create a set of discrete cell volumes that approximate the reservoir volume, yet fit the transport process physics, and avoid over complications as much as possible \cite{Ahmed2015_DFM}. Unstructured grids allow for many flexibilities, which need to be carefully applied to a computational domain so that the discrete systems do not become over-complex \cite{KarimiFard2004,rami_edfmDFM}. Without introducing the full flexibility (and at the same time complexity) of the fully unstructured grids, corner-point grid (CPG) geometry allows for many possibilities in better representation of the geological structures. This has made CPG attractive in the geoscience industry-grade simulations \cite{ponting1989corner,ding1995use,geoquest2014eclipse,lie2019introduction}.

Fractures have often small apertures (size of millimeters) but pose a serious impact on flow patterns due to large contrast of permeability between fractures and their neighboring rock matrix \cite{berkowitz2002characterizing,KUMAR2020109138}. Therefore, consistent representation of these geological features is important in predicting the flow behavior using numerical simulations \cite{berkowitz2002characterizing}. Different approaches have been proposed to model the effect of fractures on flow patterns. To avoid direct numerical simulation (DNS) and posing extremely high resolution grids in the length scales of fracture apertures, it is possible to upscale fractures by obtaining averaged and effective properties (e.g., permeability) between fractures (or faults) and the hosting rock (also known as the rock matrix). This introduces a porous media representation without fractures but with approximated conductivities. However, such models raise concerns about the inaccuracy of the simulation results due to the employed excessively upscaled parameters, especially in presence of high conductivity contrasts between the matrix and fractures. Therefore, two distinct methods have been introduced in fracture modeling approaches; the so-called dual continuum models (also known as dual porosity or dual porosity-dual permeability) \cite{Warren1963,Barenblatt1983,Kazemi1996} and the discrete fracture model (DFM) \cite{RainerBook2005}. In the dual porosity method, the matrix plays the role of fluid storage and the fluid only flows inside the fractures as it is assumed that there is no direct connection between the matrix cells. In the dual porosity-dual permeability method, both matrix and fracture have connections. Both dual porosity and dual porosity-dual permeability models homogenize the fracture domain in a computing block and neglect specific fracture features such as orientation and size. DFM, on the other hand, considers fractures as a separate system in a lower dimensional domain than that of the rock matrix, and couples them through a flux transfer function. In $2$D domains the fractures are represented by $1$D line-segments and in $3$D domains each fracture is modeled by a $2$D plane-segment. DFM provides more accurate results. Thus, it has been developed and evolved quite significantly during the past several years (See, e.g., \cite{KarimiFard2004,Lee_Jensen_1999,Lee_Lough_2001,Hajibeygi2012_JCP,Li2008,Ahmed2015_DFM,Dimitrios2013,Rainer2006,Moinfar2014}, and the references therein). Two different DFM approaches have been presented in the literature: the Embedded DFM (EDFM) and the Conforming DFM (CDFM) \cite{FLEMISCH2018239,LI2021108657,Ali_EDFM01}. The main difference between these two techniques resides in the flexibility to the grid geometry \cite{SHAH201636}. In CDFM, the fracture elements are located at the interfaces between the unstructured matrix grid-cells \cite{SANDVE20123784}. The effect of the fractures is represented by modifying transmissibilities at those interfaces. Therefore, there is an accurate consideration of flux transfer between the matrix and the fractures \cite{KarimiFard2004,Rainer2006,Ahmed2015_DFM}. However for highly dense fracture networks the number of matrix grid cells should be very high with very fine grid cells close to the fracture intersections, to account for the fractures. In addition, in case of fracture generation and propagation, the matrix grid has to be redefined at various steps of the simulation which reduces the efficiency of such approach. All of these complexities can limit the application of CDFM in real-field applications. In EDFM, fractures are discretized separately and independently from the matrix on a lower dimensional domain by using non-conforming grids \cite{Lee_Jensen_1999,Li2008}. Once the grid cells are created and the discretization is done, the fractures and matrix are coupled together using conservative flux transfer terms that calculate the flow between each fracture element and its overlapping neighbors (non-neighboring connectivities) \cite{Hajibeygi2011c_Hierarchical,Hosseinimehr2018,Ali_EDFM02}. Having two independent grids allows for modeling of complex fracture networks with simpler grids for the matrix.

While EDFM can provide acceptable results for highly conductive fractures, it cannot accurately represent flow barriers (such as non-conductive fractures and sealing faults). To overcome this restriction, projection-based EDFM (pEDFM) was introduced, for the first time, by Tene et al. \cite{Tene2017} and extended to multilevel multiscale framework in a fully 3D Cartesian geometry \cite{Hosseinimehr2020_Geothermal}. pEDFM provides consistent connectivity values between the rock matrix and the fractures, thus can be applied to fractured porous media with a any range of conductivity contrasts between the rock and the fractures (either highly conductive or impermeable). The original pEDFM concept \cite{Tene2017} has been applied to more geoscientific applications (e.g. in \citep{Rami_pedfm_copyCatz}).\\

In this work, the projection-based embedded discrete fracture model (pEDFM) on corner-point grid (CPG) geometry is presented. To cover a more general application criteria, different flow environments are considered, i.e., multiphase fluid flow model (isothermal) in fractured porous media and single-phase coupled mass-heat flow in low-enthalpy fractured geothermal reservoirs. The finite volume method (FVM) is used for discretization of the continuum domain. To represent realistic and geologically relevant domains, corner-point grid geometry is used. The sets of nonlinear equations are coupled using fully-implicit (FIM) coupling strategy. The flux terms in the mass and energy conservation equations are discretized with an upwind two-point flux approximation (TPFA) in space and with a backward (implicit) Euler scheme in time. The pEDFM is employed in order to explicitly and consistently represent fractures and to provide computational grids for the rock matrix and the fractures independently regardless of complex geometrical shapes of domains. Here, the applicability of the pEDFM implementation \cite{Tene2017,Hosseinimehr2020_Geothermal} has been extended to include fractures with generic conductivity contrasts (either highly conductive or impermeable) with any positioning and orientation on the corner-point grid geometry. This matter is paramount for practical field-scale applications. In addition to geometrical flexibility of EDFM, the matrix-matrix and fracture-matrix connectivities are altered to consider the projection of fractures on the interfaces of matrix grid cells. Using various synthetic and geologically-relevant real-field models the performance of the pEDFM on corner-point grid geometry is shown.

This article is structured as follows: First, the governing equations are described in section \ref{Sec:Governing_Equations}. The discretization and simulation strategy are explained in section \ref{Sec:Simulation_Strategy}. In section \ref{Sec:CornerPointGrid}, the corner-point grid geometry and calculation of the transmissibilities are briefly covered. The pEDFM approach for corner-point grid geometry is presented in section \ref{Sec:pEDFM_CPG}. The test cases and the numerical results are shown in section \ref{Sec:Simulation_Results}. At last, the paper is concluded in section \ref{Sec:Conclusion}.

\section{Governing Equations}\label{Sec:Governing_Equations}
Two different flow environments are considered in this article and are covered separately. 

\subsection{Multiphase flow in fractured porous media (isothermal)}
\label{Sec:GoverningEquations_IsothermalMultiphaseFlow}
Mass conservation for phase $\alpha$ in the absence of mass-exchange between phases, capillary, and gravitational effects, in porous media with $n_{\text{frac}}$ explicit fractures is given as
\begin{multline}\label{Eq:IsothermalMassBalance_m}
	\frac{\partial}{\partial t}\left(\phi \rho_{\alpha} S_{\alpha}\right)^{m} - \nabla \cdot \left(\rho_{\alpha} {\bm{\lambda_\alpha}} \cdot \nabla p \right)^{m} = \\
	\rho_\alpha q_\alpha^{m,w} + \sum\limits_{i=1}^{n_{\text{frac}}} \rho_\alpha\mathcal{Q}_{\alpha}^{m,f_i}  \quad \text{on} \quad \Omega_{m} \subseteq \Re^{n}
\end{multline}
for the rock matrix $m$ and
\begin{multline}\label{Eq:IsothermalMassBalance_f}
	\frac{\partial}{\partial t}\left(\phi \rho_{\alpha} S_{\alpha}\right)^{f_i} - \nabla \cdot \left(\rho_{\alpha} {\bm{\lambda_\alpha}} \cdot \nabla p \right)^{f_i} = \\
	\rho_\alpha q_\alpha^{f_i,w} + \rho_\alpha\mathcal{Q}_{\alpha}^{f_i,m} + \sum\limits_{j=1}^{n_{\text{frac}}}\left(\rho_\alpha\mathcal{Q}_{\alpha}^{f_i,f_j}\right)_{j\not=i} \\
	\quad \text{on} \quad  \Omega_{f_i} \subseteq \Re^{n-1} \quad \forall \, i \in \{1,...,n_{\text{frac}}\}
\end{multline}
for the lower-dimensional fracture $f_i$. There exist $n_\alpha$ phases. Moreover, the superscripts $m$, $f$ and $w$ in Eqs. \eqref{Eq:IsothermalMassBalance_m}-\eqref{Eq:IsothermalMassBalance_f} indicate, respectively, the rock matrix, the fractures and the wells. Here, $\phi$ is the porosity of the medium, $\rho_\alpha$, $S_\alpha$, $\bm{\lambda_\alpha}$ are, respectively, the density, saturation, and mobility of phase $\alpha$. In addition, $\bm{\lambda} = \frac{k_{r_\alpha}}{\mu_\alpha} \mathbf{K}$ holds, where $k_r$, $\mu$ and $\mathbf{K}$ are phase relative permeability, viscosity and rock absolute permeability tensor, respectively. Also, $q_\alpha$ is the phase source term (i.e., wells). Finally, $\mathcal{Q}_{\alpha}^{m,f_i}$ and $\mathcal{Q}_{\alpha}^{f_i,m}$ are the phase flux exchanges between matrix and the $i$-th fracture, whereas $\mathcal{Q}_{\alpha}^{f_i,f_j}$ represents the influx of phase $\alpha$ from $j$-th fracture to the $i$-th fracture. Note that the mass conservation law enforces $\iiint\limits_{V} \mathcal{Q}_{\alpha}^{m,f_i} dV =   - \iint\limits_{A_{f_i}} \mathcal{Q}_{\alpha}^{f_im} dA$ and $\iint\limits_{A_{f_i}} \mathcal{Q}_{\alpha}^{f_i,f_j} dA =   - \iint\limits_{A_{f_j}} \mathcal{Q}_{\alpha}^{f_j,f_i} dA$.

The Peaceman well model \cite{Peaceman1978} is used to obtain the well source terms of each phase for the rock matrix as
\begin{equation}\label{Eq:WellSourceTerm_MultiPhase_m}
	q^{m,w}_\alpha = \frac{WI \cdot \lambda^*_\alpha \cdot (p^w - p^m)}{\Delta V}
\end{equation}
and for the fractures as
\begin{equation}\label{Eq:WellSourceTerm_MultiPhase_f}
	q^{f_i,w}_\alpha = \frac{WI \cdot \lambda^*_\alpha \cdot (p^w - p^{f_i})}{\Delta A}.
\end{equation}

Here, $WI$ denotes the well productivity index and $\lambda^*_\alpha$ is the effective mobility of each phase ($\lambda = \frac{k_{r_\alpha}}{\mu_\alpha} K$) between the well and the grid cell penetrated by the well in each medium. In the discrete system for the rock matrix, the control volume is defined as $\Delta V$ and in the discrete system for the fracture, the control area is written as $\Delta A$.

The flux exchange terms $\mathcal{Q}_{\alpha}^{m,f_i}$, $\mathcal{Q}_{\alpha}^{f_i,m}$ (matrix-fracture connectivities) and $\mathcal{Q}_{\alpha}^{f_i,f_j}$ (fracture-fracture connectivities) are written as:

\begin{align}\label{Eq:MassFluxExchange_MultiPhase_Matrix_Fractures}
	\mathcal{Q}_{\alpha}^{m,f_i} = CI^{m,f_i} \cdot \lambda_{\alpha}^* \cdot (p^{f_i} - p^m)\nonumber
	\\
	\mathcal{Q}_{\alpha}^{m,f_i} = CI^{f_i,m}  \cdot \lambda_{\alpha}^* \cdot (p^m - p^{f_i})
	\\
	\mathcal{Q}_{\alpha}^{f_i,f_j} = CI^{f_i,f_j}  \cdot \lambda_{\alpha}^* \cdot (p^{f_j} - p^{f_i})\nonumber,
\end{align}
where $CI$ indicates the connectivity index between each two non-neighboring elements (see equation \eqref{Eq:ConductivityIndex}).

Equations \eqref{Eq:IsothermalMassBalance_m}-\eqref{Eq:IsothermalMassBalance_f}, subject to proper initial and boundary conditions, form a well-posed system for $n_\alpha$ unknowns, once the $\sum\limits_{\alpha=1}^{n_{ph}} S_\alpha = 1$ constraint is employed to eliminate one of the phase saturation unknowns. Here, this system of equations is solved for a two phase flow fluid model with the primary unknowns of $p$ and $S_1$ (from now on indicated as $S$).

\subsection{Mass and heat flow in low-enthalpy fractured porous media}
\label{Sec:GoverningEquations_GeothermalSinglePhaseFlow}
In low-enthalpy geothermal systems it is commonly assumed that the phase exchange (i.e., evaporation of liquid phase into vapor phase and vice versa) is neglected \cite{WANG2020114693}. Therefore, this system is assumed to be single-phase flow. Two sets of equations are described for this system, i.e., the mass balance and the energy balance equations. The effects of capillarity and gravity are neglected in all the equations here.
\subsubsection{Mass Balance Equations}
Mass balance equation for thermal single-phase fluid flow in porous media with $n_{\text{frac}}$ explicit fractures is given as

\begin{multline}\label{Eq:GeothermalSinglePhaseMassBalance_m}
	\frac{\partial}{\partial t}\left(\phi \rho_{(fl)} \right)^{m} - \nabla \cdot \left(\rho_{(fl)} {\bm{\lambda}} \cdot \nabla p \right)^{m} = \\
	\rho_{(fl)} q^{m,w} + \sum\limits_{i=1}^{n_{\text{frac}}} \rho^*_{(fl)} \ \mathcal{Q}^{m,f_i}, \quad \text{on} \quad \Omega_{m} \subseteq \Re^{n},
\end{multline}
for the rock matrix ($m$) and

\begin{multline}\label{Eq:GeothermalSinglePhaseMassBalance_f}
	\frac{\partial}{\partial t}\left(\phi \rho_{(fl)} \right)^{f_i} - \nabla \cdot \left(\rho_{(fl)} {\bm{\lambda}} \cdot \nabla p \right)^{f_i} = \\
	\rho_{(fl)} q^{f_i,w} + \rho^*_{(fl)} \ \mathcal{Q}^{f_i,m} + \sum\limits_{j=1}^{n_{\text{frac}}}\left(\rho^*_{(fl)} \mathcal{Q}^{f_i,f_j}\right)_{j\not=i},\\
	\quad \text{on} \quad  \Omega_{f_i} \subseteq \Re^{n-1} \quad \forall \, i \in \{1,...,n_{\text{frac}}\},
\end{multline}
for the lower dimensional fracture ($f_i$).
In the equations above, pressure $p$ is the primary unknown. $\phi$ is the porosity. In addition, $\bm{\lambda} = \frac{\mathbf{K}}{\mu_{(fl)}}$ is the mobility calculated for the fluid in which $\mu_{(fl)}$ is the fluid viscosity and $\mathbf{K}$ is the rock absolute permeability. Here, $\mathbf{K}$ is written as a tensor in case of generic anisotropy. The superscripts $m$, $f_i$ and $w$ denote the rock matrix, the $i$-th fracture and the wells, respectively. The subscripts $fl$ and $r$ refer to the fluid and the rock. $\rho_{(fl)}$ indicates the density of the fluid. Moreover, $q^{m,w}$ and $q^{f_i,w}$ are the source terms (i.e., wells) on the matrix $m$ and the fracture $f_i$. In addition, $\mathcal{Q}^{m,f_i}$ and $\mathcal{Q}^{f_i,m}$ are the flux exchanges between the rock matrix $m$ and the overlapping fracture $f_i$ for the grid cells that they overlap. $\mathcal{Q}^{f_i,f_j}$ is the flux exchange from $j$-th fracture to the $i$-th fracture where the elements intersect. Due to mass conservation, one can conclude that $\iiint\limits_{V} \mathcal{Q}^{m,f_i} dV =   - \iint\limits_{A_{f_i}} \mathcal{Q}^{f_i,m} dA$, and $\iint\limits_{A_{f_i}} \mathcal{Q}^{f_i,f_j} dA =   - \iint\limits_{A_{f_j}} \mathcal{Q}^{f_j,f_i} dA$.

Using the Peaceman well model the well source terms are calculated as
\begin{equation}\label{Eq:WellSourceTerm_SinglePhase_m}
	q^{m,w} = \frac{WI \cdot \lambda^* \cdot (p^w - p^m)}{\Delta V}
\end{equation}
for the rock matrix and
\begin{equation}\label{Eq:WellSourceTerm_SinglePhase_f}
	q^{f_i,w} = \frac{WI \cdot \lambda^* \cdot (p^w - p^{f_i})}{\Delta A}.
\end{equation}
for the fractures. In these equations, $WI$ is the well productivity index and $\lambda^*$ denotes the effective mobility ($\lambda = K / \mu$) between the well and the penetrated grid cell each medium. The flux exchange terms $\mathcal{Q}^{m,f_i}$, $\mathcal{Q}^{f_i,m}$ (matrix-fracture connectivities) and  $\mathcal{Q}^{f_i,f_j}$ (fracture-fracture connectivities) read:

\begin{align}\label{Eq:MassFluxExchange_SinglePhase_Matrix_Fractures}
	\mathcal{Q}^{m,f_i} = CI^{m,f_i} \cdot \lambda^* \cdot (p^{f_i} - p^m)\nonumber\\
	\mathcal{Q}^{m,f_i} = CI^{f_i,m}  \cdot \lambda^* \cdot (p^m - p^{f_i})\\
	\mathcal{Q}^{f_i,f_j} = CI^{f_i,f_j}  \cdot \lambda^* \cdot (p^{f_j} - p^{f_i})\nonumber,
\end{align}
where $CI$ is the connectivity index and explained in the next section.

\subsubsection{Energy Balance Equations}
The energy balance with assumption of local equilibrium is given as

\begin{multline}\label{Eq:GeothermalSinglePhaseEnergyBalance_m}
	\frac{\partial}{\partial t}\left( (\rho \, U)_{eff} \right)^{m} -
	\nabla \cdot \left( \rho_{(fl)} H_{(fl)} {\bm{\lambda}} \cdot \nabla p \right)^{m} - \\
	\nabla \cdot \left( \Lambda_{eff} \cdot \, \nabla T \right)^{m} = \\
	\rho_{(fl)} H_{(fl)} q^{m,w} + \sum\limits_{i=1}^{n_{\text{frac}}} \rho^*_{(fl)} H^*_{(fl)} \, \mathcal{Q}^{m,f_i} + \sum\limits_{i=1}^{n_{\text{frac}}} \, \mathcal{R}^{m,f_i}, \\
	\quad \text{on} \quad \Omega_{m} \subseteq \Re^{n},
\end{multline}
for the rock matrix ($m$) and
\begin{multline}\label{Eq:GeothermalSinglePhaseEnergyBalance_f}
	\frac{\partial}{\partial t}\left( (\rho \, U)_{eff} \right)^{f_i} - 
	\nabla \cdot \left(\rho_{(fl)} H_{(fl)} {\bm{\lambda}} \cdot \nabla p \right)^{f_i} - \\
	\nabla \cdot \left( \Lambda_{eff} \cdot \nabla T \right)^{f_i} = \\
	\rho_{(fl)} H_{(fl)} q^{f_i,w} + \rho^*_{(fl)} H^*_{(fl)} \, \mathcal{Q}^{f_i,m} + \\ \sum\limits_{j=1}^{n_{\text{frac}}}\left(\rho^*_{(fl)} H^*_{(fl)} \mathcal{Q}^{f_i,f_j}\right)_{j\not=i} + 
	\mathcal{R}^{f_i,m} + \sum\limits_{j=1}^{n_{\text{frac}}}\left(\mathcal{R}^{f_i,f_j}\right)_{j\not=i},\\
	\quad \text{on} \quad  \Omega_{f_i} \subseteq \Re^{n-1} \quad \forall \, i \in \{1,...,n_{\text{frac}}\},
\end{multline}
for the explicit fracture ($f_i$). Here, the temperature $T$ is the secondary unknown and is assumed identical in both the fluid and the solid rock. $H_{(fl)}$ denotes the specific fluid enthalpy. $(\rho \, U)_{eff}$ is the effective property defined as

\begin{equation}\label{Eq:SpecificInternalEnergy}
	(\rho \, U)_{eff} = \phi \rho_{(fl)} U_{(fl)} + (1-\phi) \rho_r \, U_r,
\end{equation}

with $U_f$ and $U_r$ indicating the specific internal energy in the fluid and the rock respectively. The dependent terms are non-linear functions of the pressure and the temperature and calculated via a few correlations \cite{Hosseinimehr2020_Geothermal}. Moreover, $\Lambda_{eff}$ is the effective thermal conductivity in the medium and is given as

\begin{equation}\label{Eq:ThermalConductivity}
	\Lambda_{eff} = \phi \, \Lambda_{(fl)} + (1-\phi) \Lambda_r,
\end{equation}

where, $\Lambda_{(fl)}$ and $\Lambda_r$ are the thermal conductivities in fluid and rock, respectively. Moreover, $\mathcal{R}^{m,f_i}$ and $\mathcal{R}^{f_im,}$ denote the conductive heat flux exchange between the rock matrix $m$ and the overlapping fracture $f_i$. $\mathcal{R}^{f_i,f_j}$ is the conductive heat flux exchange from $j$-th fracture to the $i$-th fracture at the location of the intersections. Please note the mass and heat flux exchanges are non-zero only for the existing overlaps or intersections. To honor energy conservation law, $\iiint\limits_{V} \mathcal{R}^{m,f_i} dV =   - \iint\limits_{A_{f_i}} \mathcal{R}^{f_i,m} dA$, and $\iint\limits_{A_{f_i}} \mathcal{R}^{f_i,f_j} dA =   - \iint\limits_{A_{f_j}} \mathcal{Q}^{f_j,f_i} dA$ hold.

The conductive heat flux exchanges, i.e., $\mathcal{R}^{m,f_i}$, $\mathcal{R}^{f_i,m}$ (matrix-fracture connectivities) and $\mathcal{R}^{f_i,f_j}$ (fracture-fracture connectivities), are obtained using the embedded discrete scheme as
\begin{align}\label{Eq:ConductiveHeatFluxExchangeMatrixFractures}
	\mathcal{R}^{m,f_i} = CI^{m,f_i}  \cdot \Lambda_{eff}^* \cdot (T^{f_i} - T^m)\nonumber\\
	\mathcal{R}^{m,f_i} = CI^{f_i,m}  \cdot \Lambda_{eff}^* \cdot (T^m - T^{f_i})\\
	\mathcal{R}^{f_i,f_j} = CI^{f_i,f_j}  \cdot \Lambda_{eff}^* \cdot (T^{f_j} - T^{f_i})\nonumber,
\end{align}
with $\Lambda_{eff}^*$ being a harmonically-averaged property between the two non-neighboring elements.

\section{Discretization of the equations and simulation strategy}\label{Sec:Simulation_Strategy}
The discretization of the nonlinear equations is done using the finite volume method (FVM). The equations are discretized with a two-point-flux-approximation (TPFA) finite-volume scheme in space and a backward (implicit) Euler scheme in time. In this section, for the sake of shortness, only the discretization of the isothermal multiphase flow equations is explained here. More details on discretization of both equations set can be found in the literature (\cite{Hosseinimehr2018,Hosseinimehr2020_Geothermal}).
Independent structured grids are generated for a three-dimensional (3D) porous medium and 2D fracture plates. The discretization is done for each medium. For a corner-point grid geometry, an illustration is presented in Fig. \ref{Fig:Matrix_Fractures_Grids_Norne}. The coupled system of non-linear equations \eqref{Eq:IsothermalMassBalance_m}-\eqref{Eq:IsothermalMassBalance_f} is discretized by calculating the fluxes. The advective TPFA flux of phase $\alpha$ between control volumes $i$ and $j$ reads

\begin{equation}\label{Eq:IsothermalMassFlux_mm}
	F_{\alpha, ij} = \rho_\alpha^{*} \frac{k_{r_\alpha}^{*}}{\mu^{*}_\alpha} T_{ij}(p_i - p_j).
\end{equation}

Here, $T_{ij} = \frac{A_{ij}}{d_{ij}}K^H_{ij}$ denotes the transmissibility between the neighboring cells $i$ and $j$. $A_{ij}$ and $d_{ij}$ are the interface area and the distance between these two cells centers respectively. The term $K^H_{ij}$ is the harmonic average of the two permeabilities. The superscript $*$ indicates that the corresponding terms are evaluated using a phase potential upwind scheme. Following the EDFM and pEDFM paradigms \cite{Hajibeygi2011c_Hierarchical,Tene2017,Hosseinimehr2018}, the fluxes between a matrix cell $i$ and a fracture cell $j$ are modeled as
\begin{equation}\label{Eq:IsothermalMassFlux_mf}
	\mathcal{F}^{m,f}_{\alpha, ij} = - \mathcal{F}^{f,m}_{\alpha, ij} = - \rho_\alpha^{*} \frac{k_{r_\alpha}^{*}}{\mu^{*}_\alpha} T^{m,f}_{ij}(p_i^{m} - p_j^{f}),
\end{equation}

In this equation, $T^{m,f}_{ij}$ is the geometrical transmissibility in the mass flux between cell $i$ belonging to the rock matrix $m$ and the element $j$ belonging to the fracture $f_i$ and it reads:

\begin{equation}\label{MassFluxTransmissibility_mf}
	T^{m,f}_{ij} = K^{H}_{ij} \times CI_{ij}.
\end{equation}

\begin{figure}[!htb]
	\centering
	\includegraphics[trim={0cm 0cm 0cm 0cm}, clip, width = 0.24\textwidth]{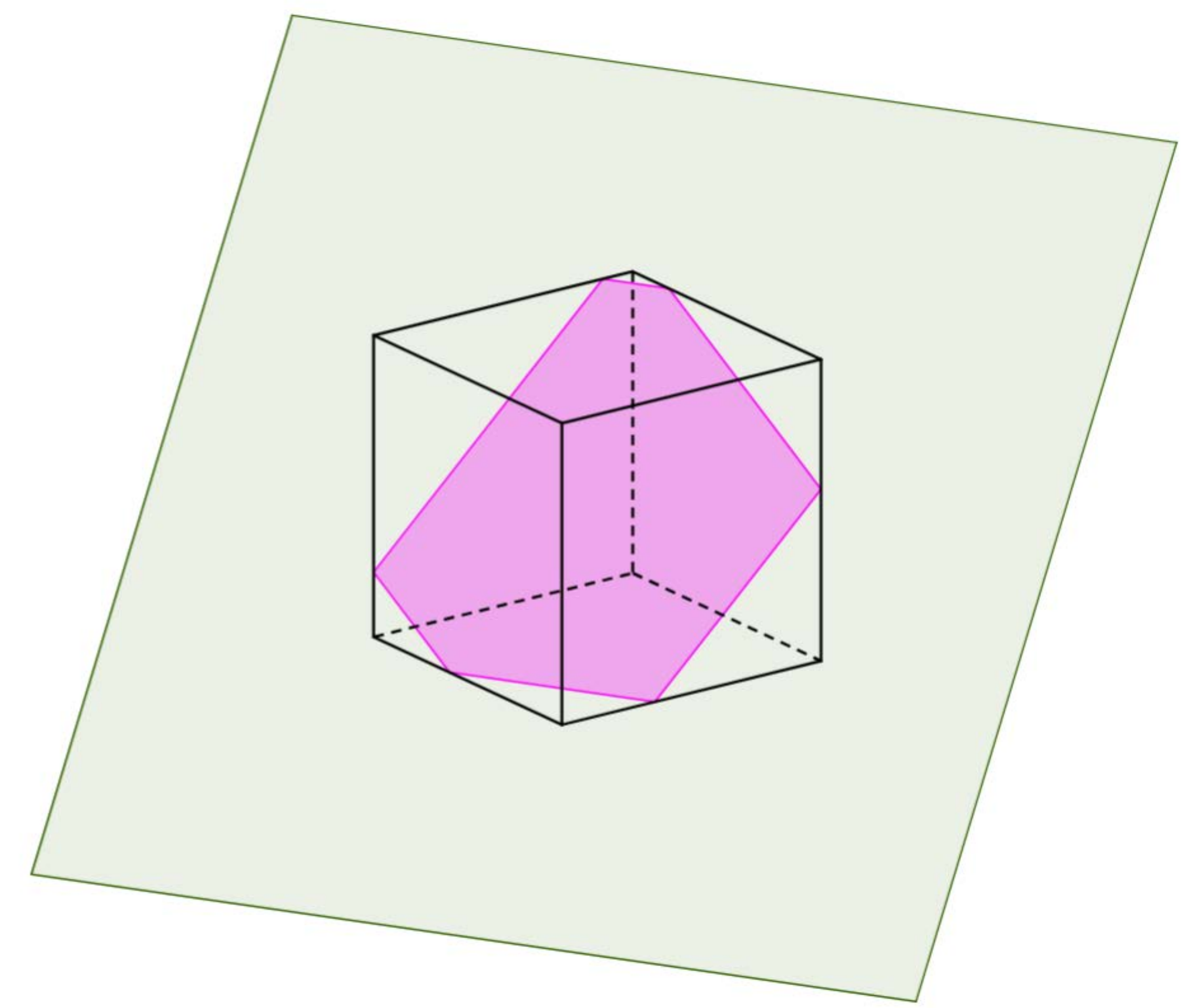}
	\includegraphics[trim={0cm 0cm 0cm 0cm}, clip, width = 0.22\textwidth]{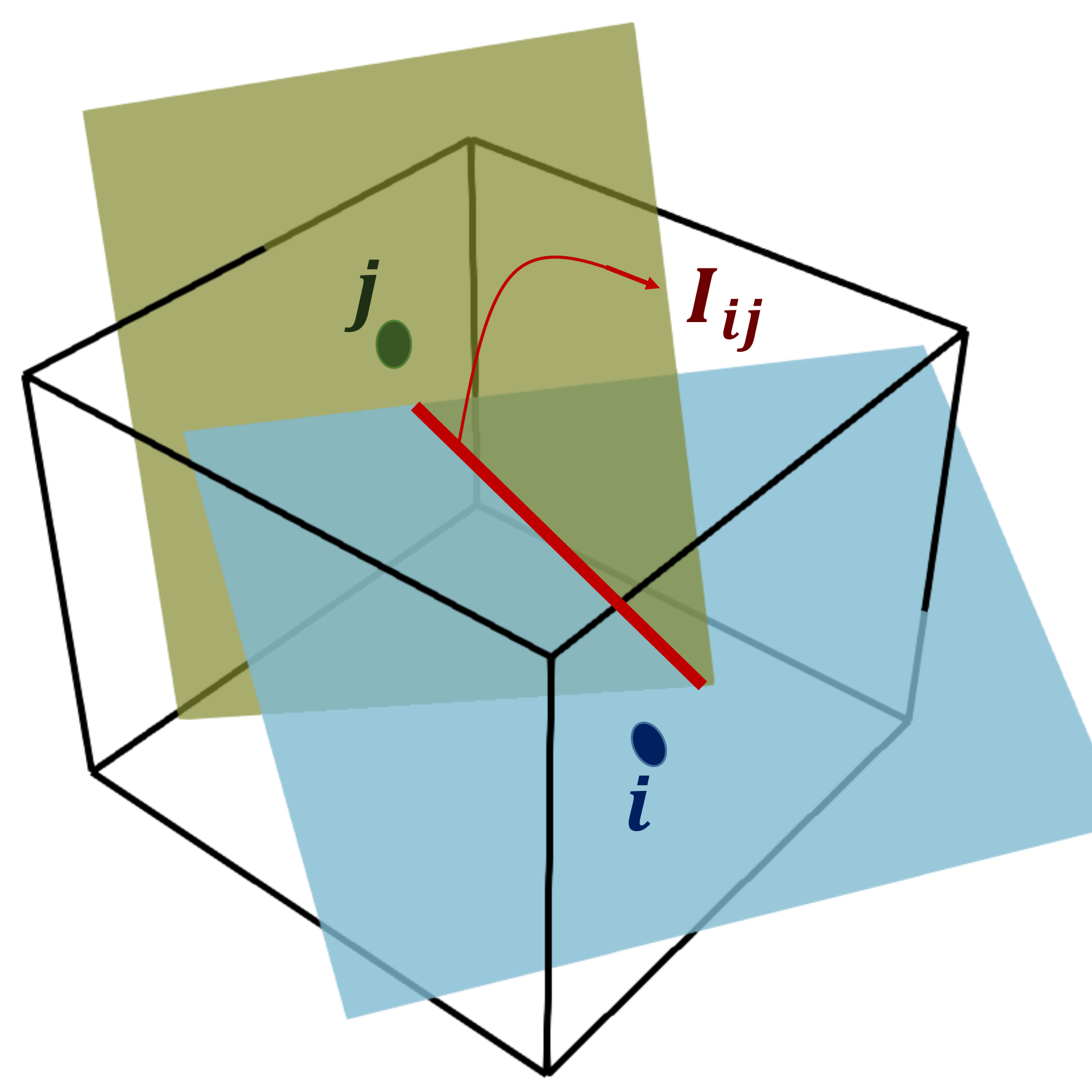}
	\caption{Visualization of a matrix-fracture overlap and a fracture-fracture intersection. The figure on the left shows a fracture element overlapping with a matrix grid cell. The overlapping section forms an irregular polygon. The figure on the right illustrates intersecting of two elements from two fracture plates inside a matrix grid cell with the intersection line colored in red.}
	\label{Fig:matrix_frac_overlap}
\end{figure}

In the equation above, $K^{H}_{ij}$ denotes the harmonically averaged permeability between the rock matrix and the overlapping fracture elements. Moreover, $CI^{m,f}_{ij}$ is the connectivity index between the two overlapping elements. The EDFM and pEDFM model the matrix-fracture connectivity index as:

\begin{equation}\label{Eq:ConductivityIndex}
	CI^{m,f}_{ij} = \frac{A^{m,f}_{ij}}{\langle d \rangle_{ij}},
\end{equation}

with $A^{m,f}_{ij}$ being the area fraction of fracture cell $j$ overlapping with matrix cell $i$ (see figure \ref{Fig:matrix_frac_overlap}, on the left) and $\langle d \rangle_{ij}$ being the average distance between these two cells \cite{Hajibeygi2011c_Hierarchical}.

Similarly, the flux exchange between intersecting fracture elements $i$ (belonging to fracture $f_i$) and $j$ (belonging to fracture $f_j$) is modeled as
\begin{equation}\label{Eq:IsothermalMassFlux_ff}
	\mathcal{F}^{f_i,f_j}_{\alpha, ij} = - \mathcal{F}^{f_j,f_i}_{\alpha, ij} = - \rho_\alpha^{*} \frac{k_{r_\alpha}^{*}}{\mu^{*}_\alpha} T^{f_i,f_j}_{ij}(p_i^{f_i} - p_j^{f_j}).
\end{equation}

Here, $T^{f_i,f_j}_{ij}$ is the geometrical transmissibility in the mass flux between element $i$ in the fracture $f_i$ and the element $j$ in the fracture $f_j$, which reads:

\begin{equation}\label{MassFluxTransmissibility_ff}
	T^{f_i,f_j}_{ij} = K^{H}_{ij} \,\, \frac{CI^{f_i}_{iI_{ij}} \times CI^{f_j}_{jI_{ij}} }{CI^{f_i}_{iI_{ij}} + CI^{f_j}_{jI_{ij}} }.
\end{equation}

Please note that the geometrical transmissibility $T^{f_i,f_j}_{ij}$ between the two non-neighboring (intersecting) fracture cells is obtained on a lower dimensional formulation. This is needed due to the fact that the intersection between two 2D fracture plates forms a line-segment and the intersection between two 1D fracture line-segments results in a point. Figure \ref{Fig:matrix_frac_overlap} (on the right) visualizes an example of an intersection between two non-neighboring 2D fracture elements. The result of the intersection is a line segment $I_{ij}$ (colored in red) with the average distances from the intersection segment written as $\langle d \rangle^{f_i}_{iI_{ij}} \neq \langle d \rangle^{f_j}_{jI_{ij}}$. This is the reason why these transmissibilities are computed using a harmonic-average formulation as shown above.

Thus, at each time-step the following system of equations is solved
\begin{multline}\label{Eq:IsothermalMassBalanceDiscrete_m}
	\left( \frac{\left( \phi \rho_\alpha S_\alpha \right)_i^{n+1} - \left( \phi \rho_\alpha S_\alpha \right)_i^{n}}{\Delta t} \right)^m + \left(\sum \limits_{j=1}^{N_n} F_{\alpha,ij}\right)^m + \\
	\sum\limits_{k=1}^{n_{\text{frac}}} \left( \sum\limits_{j=1}^{N_{f_k}} \mathcal{F}^{m,f_k}_{\alpha, ij} \right)  =  \rho_\alpha q^{m,w}_{\alpha, i}, \\
	\quad \forall \, i \in \{1,...,N_m\} 
\end{multline}
in the matrix and
\begin{multline}\label{Eq:IsothermalMassBalanceDiscrete_f}
	\left( \frac{\left( \phi \rho_\alpha S_\alpha \right)_i^{n+1} - \left( \phi \rho_\alpha S_\alpha \right)_i^{n}}{\Delta t} \right)^{f_h} + \left(\sum \limits_{j=1}^{N_n} F_{\alpha,ij}\right)^{f_h} + \\
	\sum\limits_{j=1}^{N_m} \mathcal{F}^{f_h,m}_{\alpha, ij} + \sum\limits_{k=1}^{n_{\text{frac}}} \left( \sum\limits_{j=1}^{N_{f_k}} \mathcal{F}^{f_h,f_k}_{\alpha, ij} \right)  =  \rho_\alpha q^{f_h,w}_{\alpha, i}, \\
	\quad \forall \, i \in \{1,...,N_{f_h}\}
\end{multline}
in each fracture $f_h$. Here, $N_m$ and $N_{f_{k}}$ are the number of cells in the matrix and number of the cells in fracture $f_k$, respectively. $N_n$ indicates the number of neighboring cells ($2$ in 1D, $4$ in 2D, $6$ in 3D).

Equations \eqref{Eq:IsothermalMassBalanceDiscrete_m}-\eqref{Eq:IsothermalMassBalanceDiscrete_f} can be written in residual form as

\begin{multline}\label{Eq:IsothermalMassBalanceDiscreteResidual_m}
	\left( r^{n+1}_{\alpha,i} \right)^m = \rho_\alpha q^{m,w}_{\alpha, i} - \left( \frac{\left( \phi \rho_\alpha S_\alpha \right)_i^{n+1} - \left( \phi \rho_\alpha S_\alpha \right)_i^{n}}{\Delta t} \right)^m - \\
	\left(\sum \limits_{j=1}^{N_n} F_{\alpha,ij}\right)^m - \sum\limits_{k=1}^{n_{\text{frac}}} \left( \sum\limits_{j=1}^{N_{f_k}} \mathcal{F}^{m,f_k}_{\alpha, ij} \right), \\
	\quad \forall \, i \in \{1,...,N_m\} 
\end{multline}
for the rock matrix, and
\begin{multline}\label{Eq:IsothermalMassBalanceDiscreteResidual_f}
	\left( r^{n+1}_{\alpha,i} \right)^{f_h} = \rho_\alpha q^{f_h,w}_{\alpha, i} - \left( \frac{\left( \phi \rho_\alpha S_\alpha \right)_i^{n+1} - \left( \phi \rho_\alpha S_\alpha \right)_i^{n}}{\Delta t} \right)^{f_h} - \\
	\left(\sum \limits_{j=1}^{N_n} F_{\alpha,ij}\right)^{f_h} - \sum\limits_{j=1}^{N_m} \mathcal{F}^{f_h,m}_{\alpha, ij} - \sum\limits_{k=1}^{n_{\text{frac}}} \left( \sum\limits_{j=1}^{N_{f_k}} \mathcal{F}^{f_h,f_k}_{\alpha, ij} \right), \\
	\quad \forall \, i \in \{1,...,N_{f_h}\}
\end{multline}

for fracture $f_h$. Let us define $r^{n} = [(r^{m})^{n}, (r^{f_1})^{n} ... \\ ... (r^{f_{n_{\text{frac}}}})^{n}]^{T}$ where $(r^{k})^{n}$ is the residual vector of medium $k$ at time-step $n$. Similarly, $p^{n}$ and $S^{n}$ indicate the vectors of pressure and saturation unknowns (of all media). The residual $r^{n+1}$ is a non-linear function of the primary unknowns $p^{n+1}$ and $S^{n+1}$. Thus, at each time-step a Newton-Raphson method is employed to solve the non-linear system iteratively, i.e.
\begin{equation}\label{Eq:IsothermalNewtonMethod}
	r^{\nu+1}_\alpha \approx r^{\nu}_\alpha + \frac{\partial r_\alpha}{\partial p}\bigg\lvert^{\nu} \, \delta p^{\nu+1} + \frac{\partial r_\alpha}{\partial S}\bigg\lvert^{\nu} \delta S^{\nu+1}, 
\end{equation}
where the superscript $\nu$ is the iteration index. Consequently, at each Newton's iteration the linearized system $\mathbf{J}^{\nu} \delta x^{\nu+1} = - r^{\nu}$ is solved. Here, $\mathbf{J}^{\nu}$ is the Jacobian matrix with $\delta x^{\nu + 1} = [\delta p, \, \delta S]^{T}$.
Therefore, assuming two phases (the indices $1$ and $2$ representing the equations of the first and the second phases respectively), the linear system of equations can be written as
\begin{equation}\label{Eq:Isothermal_FineScale_FIM_LinearSystem}
	\underbrace{
		\begin{pmatrix}
			\begin{pmatrix}
				J_{1,p}^{m,m} & J_{1,p}^{m,f}\\ \\
				J_{1,p}^{f,m} & J_{1,p}^{f,f}
			\end{pmatrix}
			\begin{pmatrix}
				J_{1,S}^{m,m} & J_{1,S}^{m,f}\\ \\
				J_{1,S}^{f,m} & J_{1,S}^{f,f}
			\end{pmatrix}\\ \\
			\begin{pmatrix}
				J_{2,p}^{m,m} & J_{2,p}^{m,f}\\ \\
				J_{2,p}^{f,m} & J_{2,p}^{f,f}
			\end{pmatrix}
			\begin{pmatrix}
				J_{2,S}^{m,m} & J_{2,S}^{m,f}\\ \\
				J_{2,S}^{f,m} & J_{2,S}^{f,f}
			\end{pmatrix}
		\end{pmatrix}^\nu}_{\mathbf{J}^{\nu}}
	\underbrace{
		\begin{pmatrix}
			\delta p^{m}\\ \\
			\delta p^{f}\\ \\
			\delta S^{m}\\ \\
			\delta S^{f}\\
		\end{pmatrix}^{\nu+1}}_{\delta x^{\nu+1}} \! \! \! \! \! \! \! \! \! \! \! \! = 
	- \underbrace{\begin{pmatrix}
			r_{1}^{m}\\ \\
			r_{1}^{f}\\ \\
			r_{2}^{m}\\ \\
			r_{2}^{f}\\
		\end{pmatrix}^{\nu}}_{r^{\nu}}
\end{equation}

In this formulation, non-linear convergence is reached when the following conditions are satisfied:
\begin{align}\label{Eq:Isothermal_Iteration_Criteria}
	\Bigg(
	\frac{||r^{\nu+1}_1||_2}{||r^0_1||_2} < \epsilon_{(r_1)}
	\enspace \vee \enspace
	\frac{||r^{\nu+1}_1||_2}{||rhs_1||_2} < \epsilon_{(r_1)}
	\Bigg)
	\enspace &\wedge \nonumber \\
	\Bigg(
	\frac{||r^{\nu+1}_2||_2}{||r^0_2||_2} < \epsilon_{(r_2)}
	\enspace \vee \enspace
	\frac{||r^{\nu+1}_2||_2}{||rhs_2||_2} < \epsilon_{(r_2)}
	\Bigg)
	\enspace &\wedge \\
	\Bigg(
	\frac{||\delta p||_2}{||p||_2} < \epsilon_{(p)}
	\enspace \wedge \enspace
	\frac{||\delta S||_2}{||S||_2} < \epsilon_{(S)}
	\Bigg) \nonumber
\end{align}
Here, $\epsilon_{(r_1)}$, $\epsilon_{(r_2)}$, $\epsilon_{(p)}$ and $\epsilon_{(S)}$, are the user-defined tolerances that are set initially as input at the beginning of the simulation. The notation $||x||_2$ is the second norm of the vector $x$. The superscript $0$ denotes the value of its corresponding vector at the initial state of the iteration step. Please note that in some systems the condition $\frac{||r^{\nu+1}||_2}{||rhs||_2} < \epsilon_{(r)}$ can result in a better convergence when compared to $\frac{||r^{\nu+1}||_2}{||r^0||_2} < \epsilon_{(r)}$ and vice versa. Therefore both conditions are checked and either of them can implicate the convergence signal.

\begin{figure}[!htbp]
	\centering
	\includegraphics[width=0.9\linewidth]{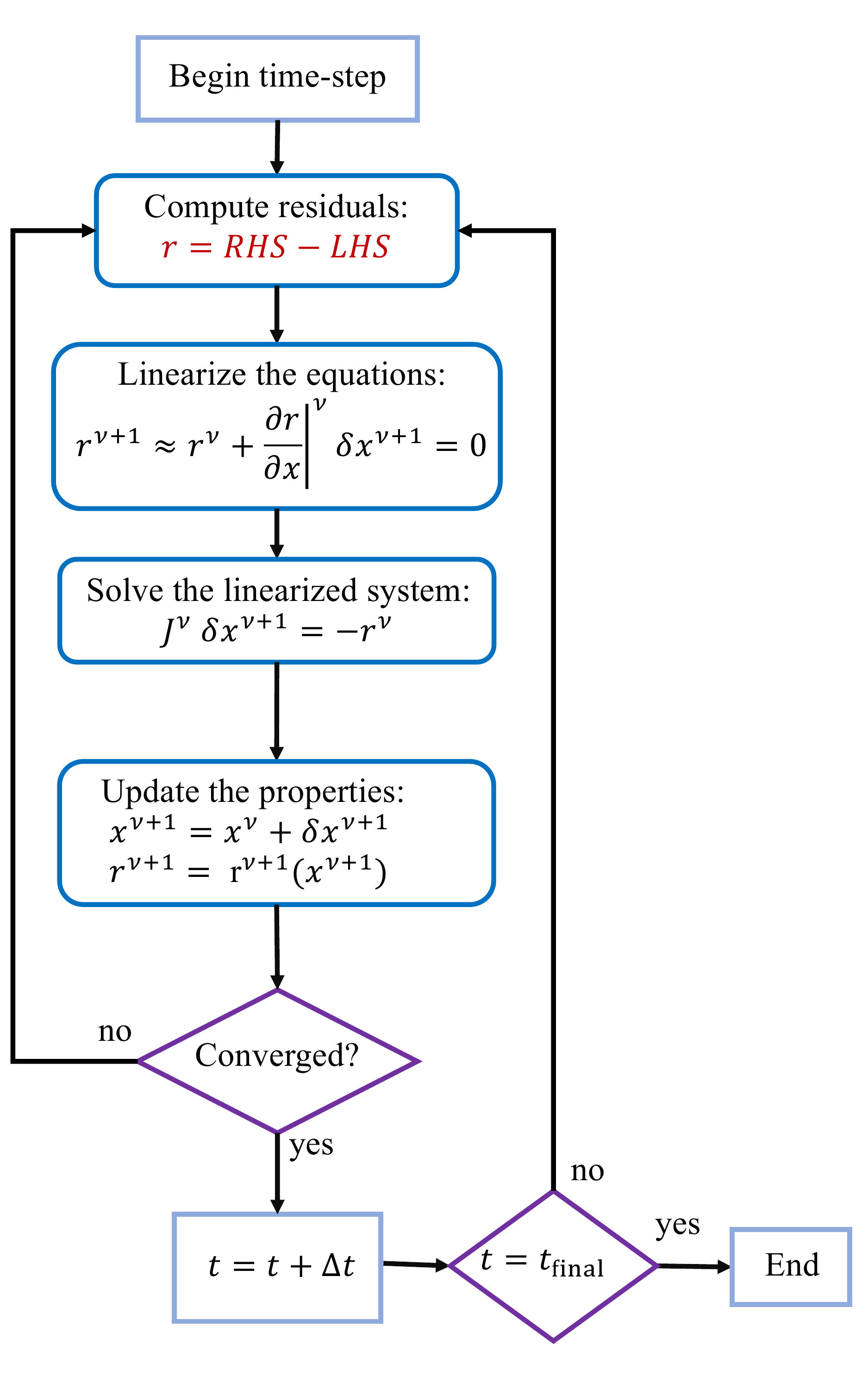}
	\caption{Schematic of fully-implicit simulation flowchart.}
	\label{Fig:SimFlowChart}
\end{figure}

Figure \ref{Fig:SimFlowChart} illustrates the schematic of the fully-implicit (FIM) simulation flowchart.

\section{Corner-point grid geometry}\label{Sec:CornerPointGrid}
A corner-point grid (CPG) is defined with a set of straight pillars outlined by their endpoints over a Cartesian mesh in the lateral direction \cite{lie2019introduction}. On every pillar, a constant number of nodes (corner-points) is set, and each cell in the grid is set between $4$ neighboring pillars and two neighboring points on each pillar. Every cell can be identified by integer coordinates ($i$,$j$,$k$); where the $k$ coordinate runs along the pillars, and $i$ and $j$ coordinates span along each layer. The cells are ordered naturally with the $i$-index ($x$-axis) cycling fastest, then the $j$-index ($y$-axis), and finally the $k$-index (negative of $z$-direction). 

For establishing vertical and inclined faulting more accurately, it is advantageous to define the position of the grid cell by its corner point locations and displace them along the pillars that have been aligned with faults surfaces. Similarly, for modeling erosion surfaces and pinch-outs of geological layers, the corner point format allows points to collapse along coordinate lines. The corner points can collapse along all four lines of a pillar so that a cell completely disappears in the presence of erosion surfaces. If the collapse is present in some pillars, the degenerate hexahedral cells may have less than six faces. This procedure creates non-matching geometries and non-neighboring connections in the underlying $i$-$j$-$k$ topology \cite{lie2019introduction}.

\begin{figure}[!htb]
	\centering
	\includegraphics[trim={10cm 4cm 10cm 4cm}, clip, width = 0.33\textwidth]{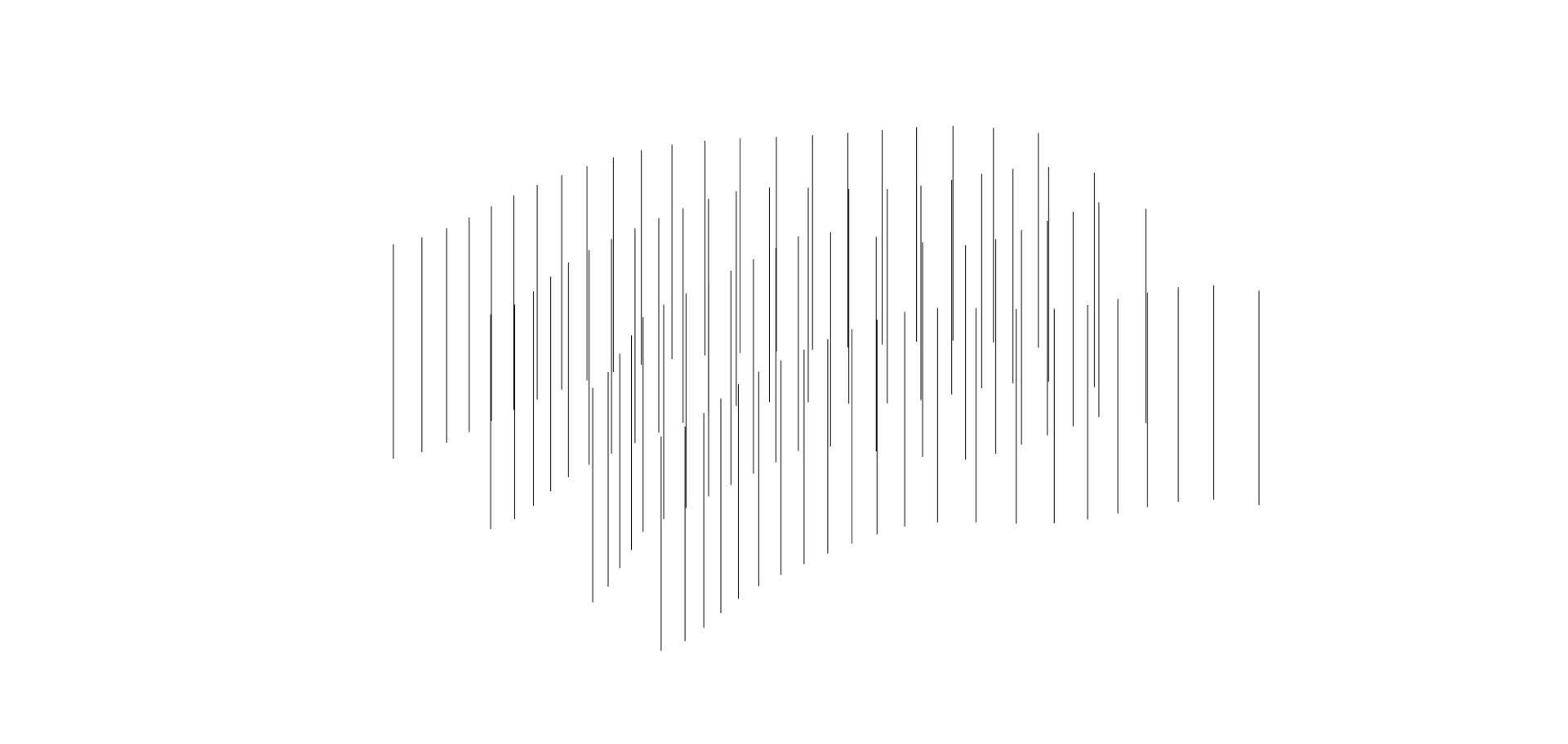}
	\\
	\includegraphics[trim={10cm 4cm 10cm 4cm}, clip, width = 0.33\textwidth]{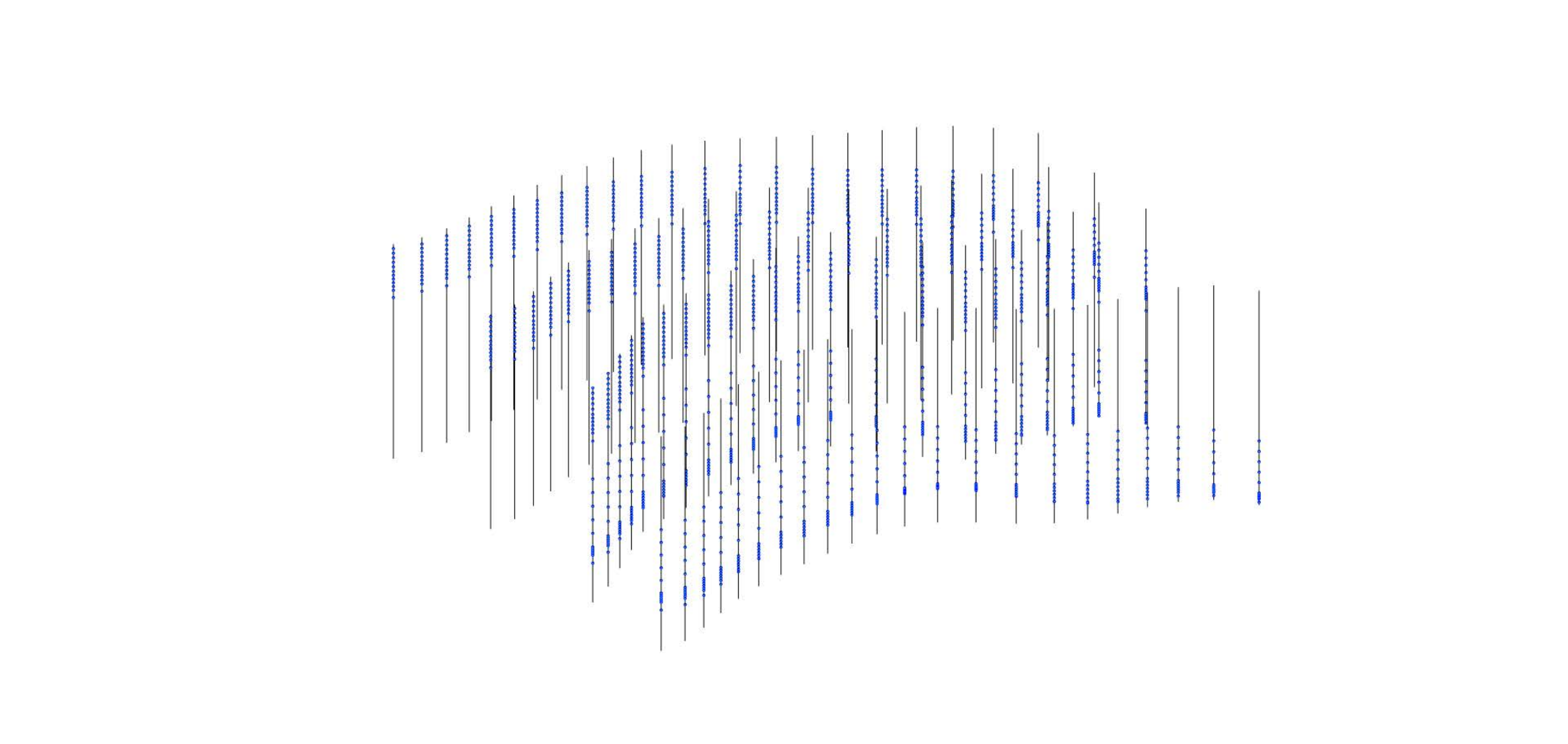}
	\\
	\includegraphics[trim={10cm 4cm 10cm 4cm}, clip, width = 0.33\textwidth]{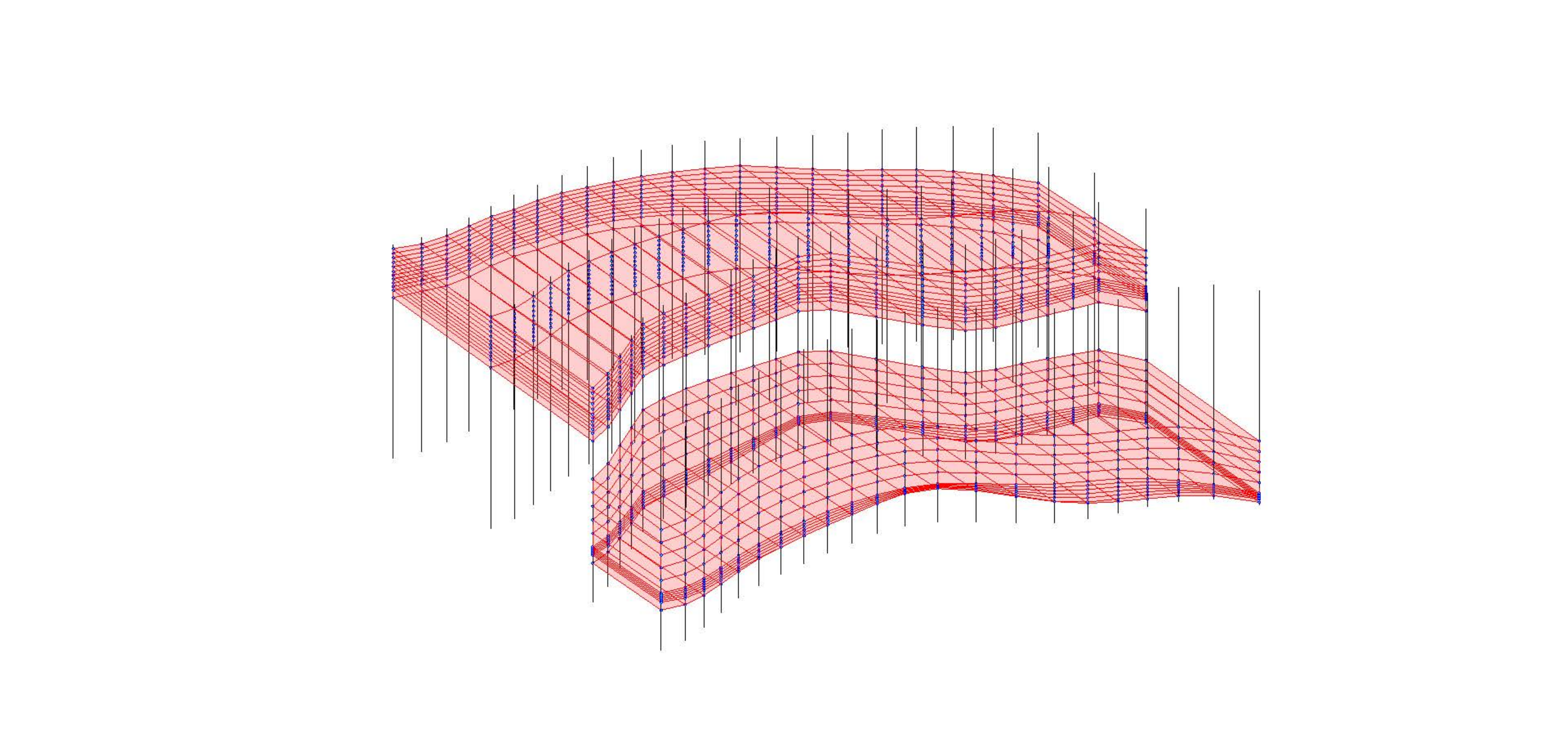}
	\caption{Construction of a corner-point grid: Starting from the coordinate lines defining pillars (top), the corner-points are added to them (middle). A stack of cells is created for each set of four lines defining a pillar and at last the full grid is obtained (bottom).}
	\label{Fig:CPG_Construction}
\end{figure}

\subsection{Two-point flux approximation in corner-point grid geometry}
In order to only highlight the calculation of the two-point flux approximation in corner-point grid geometry and avoid complexities in presenting fully detailed governing equations, a simplified linear elliptic equation is used which serves as a model pressure equation for incompressible fluids, i.e.,
\begin{equation}\label{Eq:EllipticPressure}
	\nabla \cdot u = f,
\end{equation}
where $f$ is the source/sink term (wells), and $u$ is the Darcy velocity, defined as
\begin{equation}\label{Eq:Velocity}
	u = - \mathbf{K} \, \nabla p.
\end{equation}

Finite volume discrete systems can be obtained by rewriting the equation in integral form, on discrete cell $\Omega_i$, as
\begin{equation}\label{Eq:Integral1}
	\int_{\partial \Omega_i} \vec{u} \cdot \vec{n} \,\, dS = - \int_{\Omega_i} q \,\, d\vec{x}.
\end{equation}

The flux between the two neighbouring cells i and k can be then written as 
\begin{equation}\label{Eq:Integral2}
	u_{i,k}= \int_{\Gamma_{i,k}} \vec{u} \cdot \vec{n} \,\, dS.
\end{equation}

The faces $\Gamma_{i,k}$ are denominated half face as they are linked with a grid cell $\Omega_i$ and a normal vector $\vec{n}_{i,k}$. It is assumed that the grid is matching to another one so that each interior half-face will have a twin half-face $\Gamma_{k,i}$ that has an identical area $A_{i,k} = A_{k,i}$ but an opposite normal vector $\vec{n}_{i,k} = - \vec{n}_{k,i}$.  The integral over the cell face is approximated by the midpoint rule, and Darcy's law, i.e.,
\begin{equation}\label{Eq:MidPointRule1}
	u_{i,k} \approx  A_{i,k} \,\, \big(\mathbf{K} \, \nabla p\big) \big( \vec{x}_{i,k} \big) \cdot \vec{n}_{i,k}
\end{equation}

where $\vec{x}_{i,k}$ indicates the centroid of $\Gamma_{i,k}$.

The one-sided finite difference is used to determine the pressure gradient as the difference between the pressure $\pi_{i,k}$ at the face centroid and the pressure at some point inside the cell. The reconstructed pressure value at the cell centre is equal to the average pressure $p_i$ inside the cell, thus,
\begin{equation}\label{Eq:MidPointRule2}
	u_{i,k} \approx A_{i,k} \mathbf{K_i} \frac{ (p_i - \pi_{i,k})\vec{c}_{k,i} }{ \mid \vec{c}_{k,i}\mid^2 } \cdot \vec{n}_{i,k} \,\, \Rightarrow u_{i,k} \approx  T_{i,k} (p_i - \pi_{i,k}).
\end{equation}

The vectors $\vec{c}_{k,i}$ are defined from cell centroids to face centroids. Face normals are assumed to have a length equal to the corresponding face areas $A_{i,k} \cdot \vec{n}_{i,k}$, i.e.,

\begin{equation}\label{Eq:CPG_Trans}
	T_{i,k} = A_{i,k} \mathbf{K_i} \frac{\vec{c}_{k,i} \cdot \vec{n}_{i,k}}{\mid \vec{c}_{k,i}\mid^2}
\end{equation}

\begin{figure}[!htb]
	\centering
	\includegraphics[trim={0cm 0cm 0cm 0cm}, clip, width = 0.3\textwidth]{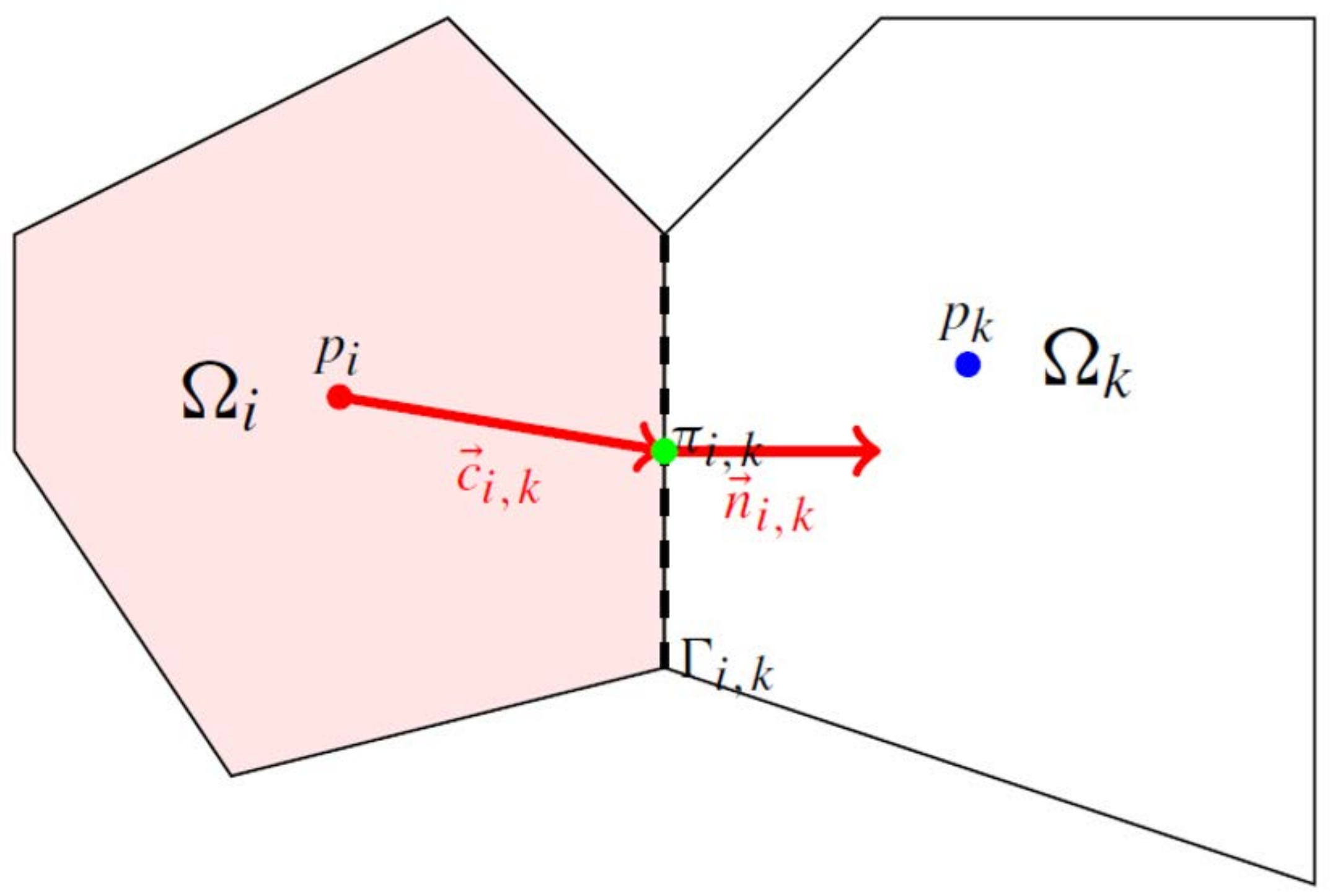}
	\\
	\includegraphics[trim={0cm 0cm 0cm 0cm}, clip, width = 0.4\textwidth]{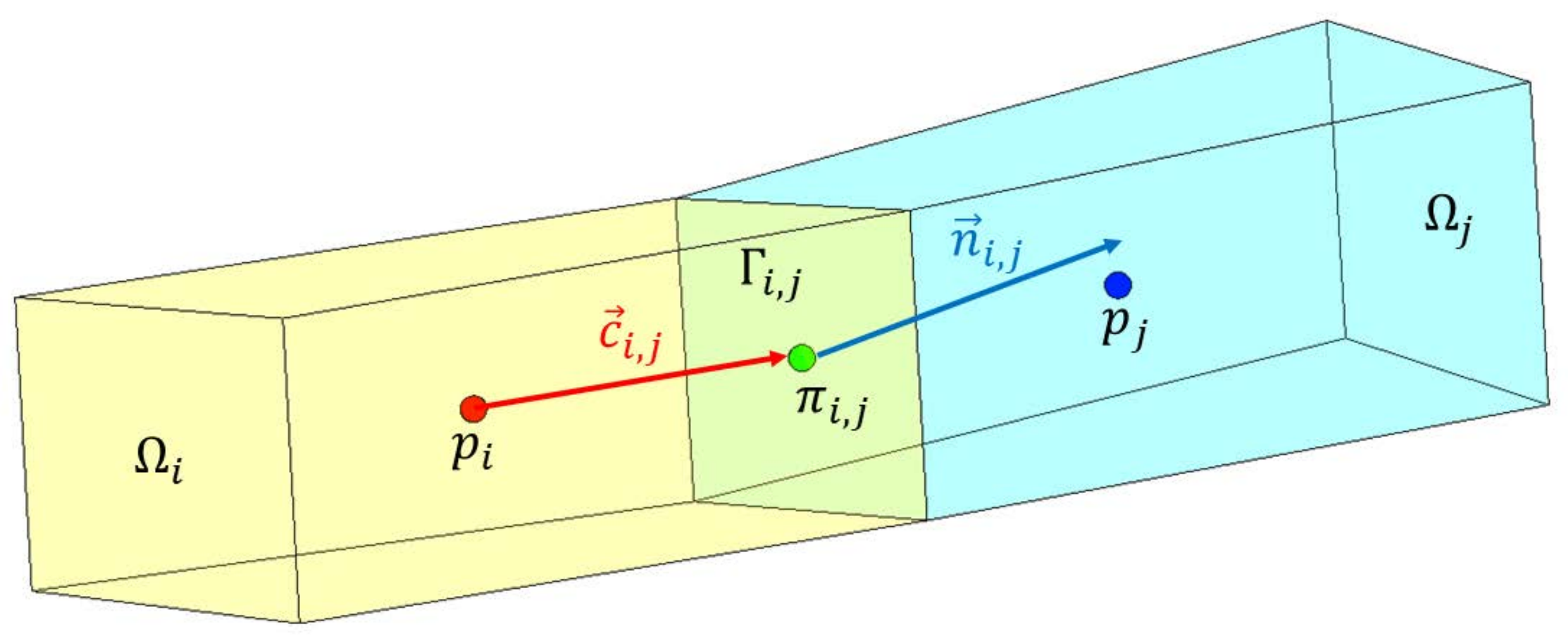}
	\caption{Two cells used to define the two-point discretization on general 2D polygon cells (see figure on top) and 3D polyhedral cells (see figure at the bottom).}
	\label{Fig:CPG_TPFA}
\end{figure}

The one-sided transmissibilities $T_{i,k}$ are related to a single cell and provide a two-point relation between the flux across a cell face and the pressure difference between the cell and face centroids. The proper name for these one-sided transmissibilities is half-transmissibilities as they are associated with a half-face \cite{KarimiFard2004,Bosma2017145}.

Finally, the continuity of fluxes across all faces, $u_{i,k} = - u_{k,i}$, as well as the continuity of face pressures $\pi_{i,k} = \pi_{k,i} = \pi_{ik}$ are set. This leads to 
\begin{equation}\label{Eq:CPG_Trans_half_1}
	T_{i,k}^{-1} \,\, u_{ik} = p_i - \pi_{ik}
\end{equation}
\begin{equation}\label{Eq:CPG_Trans_half_2}
	-T_{k,i}^{-1} \,\, u_{ik} = p_k - \pi_{ik}.
\end{equation}

The interface pressure $\pi_{ik}$ is then eliminated and the two-point flux approximation (TPFA) scheme is defined as
\begin{equation}\label{Eq:CPG_TFlux}
	u_{ik} = \big[ T_{i,k}^{-1} + T_{k,i}^{-1} \big]^{-1} \,\, (p_i - p_k) = T_{ik} \,\, (p_i - p_k).
\end{equation}

$T_{ik}$ is the transmissibility associated with the connection between the two cells. The TPFA scheme uses two ``points", the cell averages $p_i$ and $p_k$, to approximate the flux across the interface $\Gamma_{i,k}$ between cells $\Omega_i$ and $\Omega_k$. The TPFA scheme in a compact form obtains a set of cell averages that meet the following system of equations
\begin{equation}\label{Eq:CPG_TPFA}
	\sum\limits_{k} T_{ik} \,\, (p_i - p_k) = q_i \,, \quad \forall \Omega_i \subset \Omega.
\end{equation}

\section{pEDFM for corner-point grid geometry}\label{Sec:pEDFM_CPG}
\begin{figure}[!htb]
	\centering
	\subcaptionbox{matrix grid}{\includegraphics[width = 0.45\textwidth]{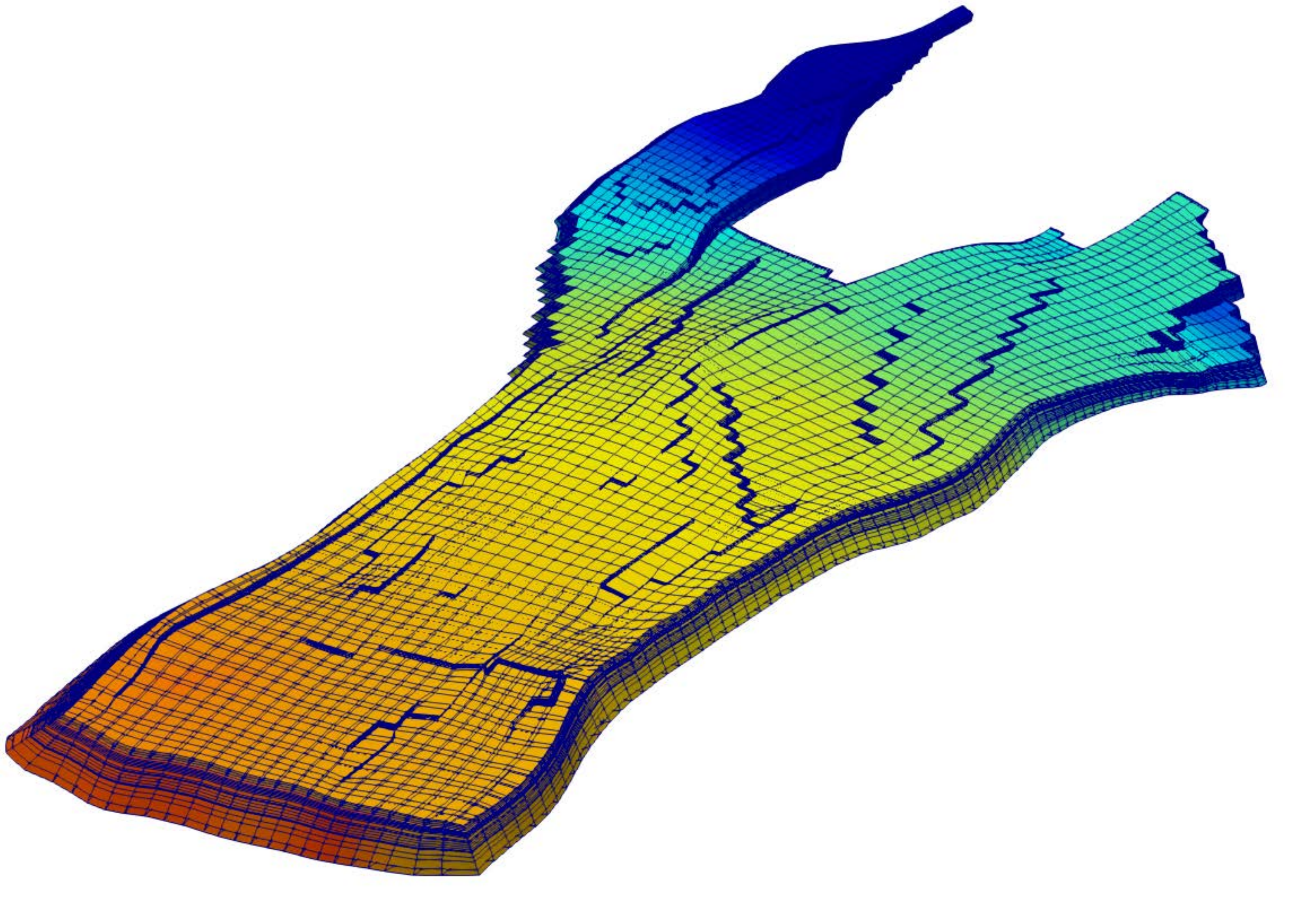}}
	\\
	\subcaptionbox{fracture grids}{\includegraphics[width = 0.45\textwidth]{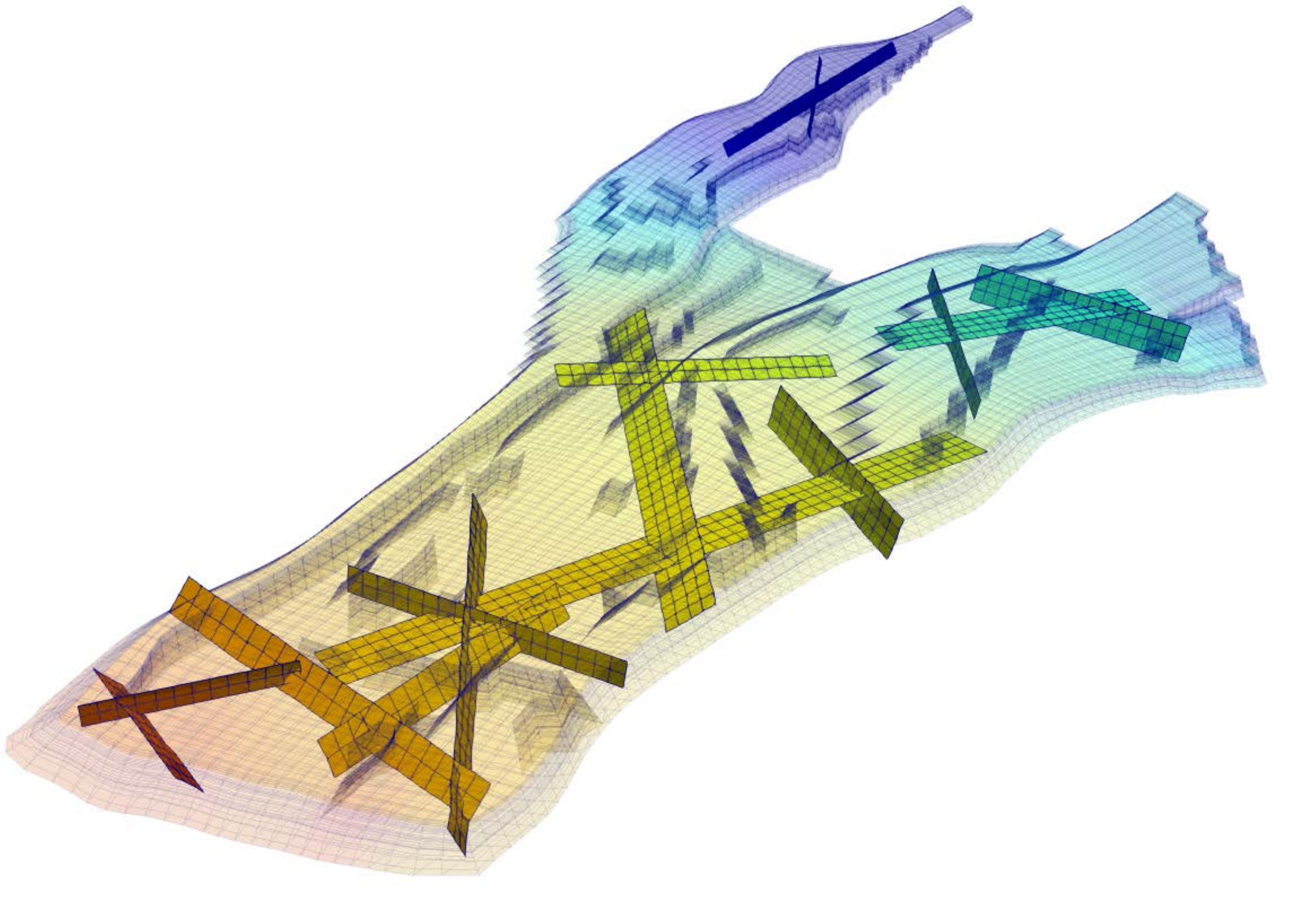}}
	\caption{An example of a fractured domain on corner-point grid geometry. The domain presented in the left image is the well-known Norne oil-field which is a true representative of the real-field geometry \cite{lie2019introduction}. The figure on the right is a realization of a fracture network inside the domain that was exclusively designed by the authors of this paper. Note that each sub domain (matrix, and individual fractures) entail independent grid resolutions, and can have independent complexities (e.g. 3D orientation).}
	\label{Fig:Matrix_Fractures_Grids_Norne}
\end{figure}

\begin{figure}[!htb]
	\centering
	\includegraphics[width = 0.23\textwidth]{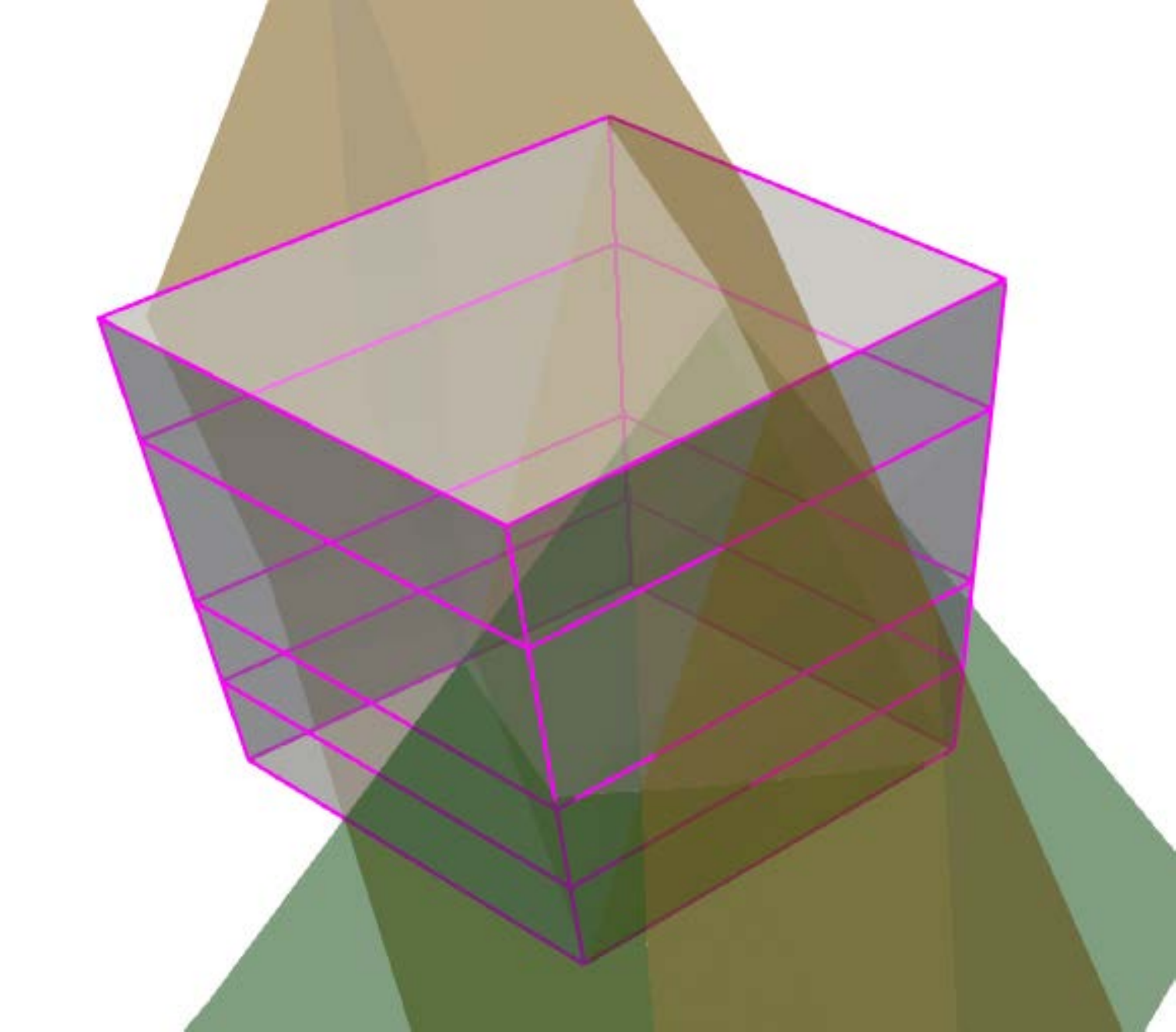}
	\includegraphics[width = 0.23\textwidth]{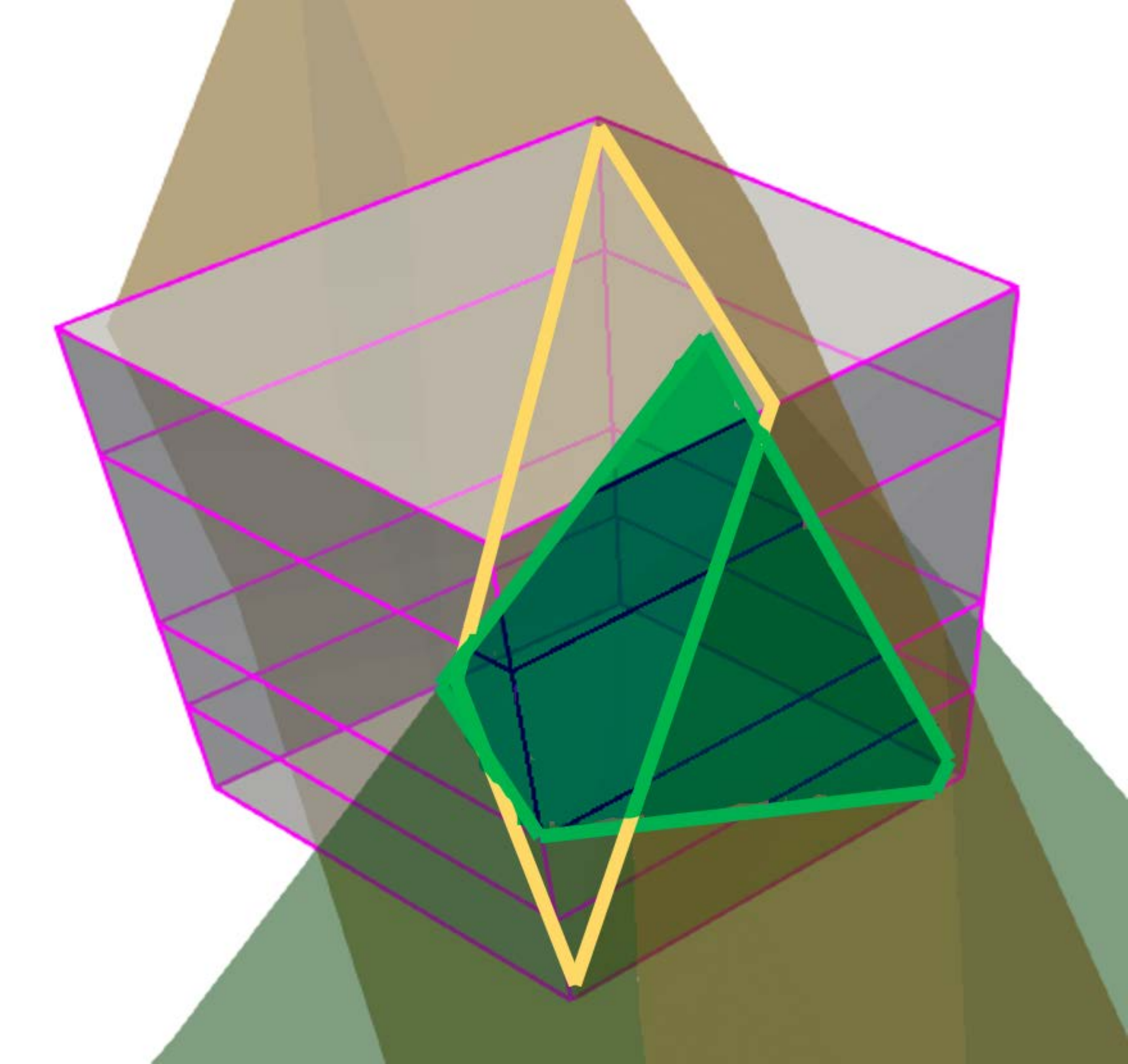}
	\caption{The intersection between to fractures and a hexahedron from corner-point grid geometry is illustrated here. The figure on the right highlights the area fraction of the two separate fracture plates inside the matrix grid cell. The overlapping segment of green fracture forms an irregular pentagon where the orange fracture has a tetragon overlapping segment.}
	\label{Fig:Matrix_Fracture_Overlap_CPG}
\end{figure}

As stated in the section for discretization of governing equations, sets of flux exchange terms are defined between the matrix and the explicit fractures. Inside each term, the connectivity index ($CI_{ij} = \frac{A_{ij}}{\langle d \rangle_{ij}}$) is considered. In corner-point grid geometry, to calculate the area fraction ($A_{ij}$) of each overlapping fracture element inside the corresponding matrix grid cell, various geometrical functions are defined which can obtain the intersection between a tetragon (the 2D planar fracture grid cell in 3D geometry) and a hexahedron (the matrix grid cell in corner-point grid geometry). Once the intersection is obtained and the area fraction is calculated, the average distance ($\langle d \rangle_{ij}$) between the two overlapping elements is calculated as well. Figures \ref{Fig:Matrix_Fractures_Grids_Norne} and \ref{Fig:Matrix_Fracture_Overlap_CPG} illustrate the geometry of CPG-based pEDFM grids. Note that the fractures can have any orientations in 3D, and arbitrary crossing lines with other fractures. 

\begin{figure}[!htbp]
	\centering
	\includegraphics[trim={0cm 0cm 0cm 0cm}, clip, width = 0.49\textwidth]{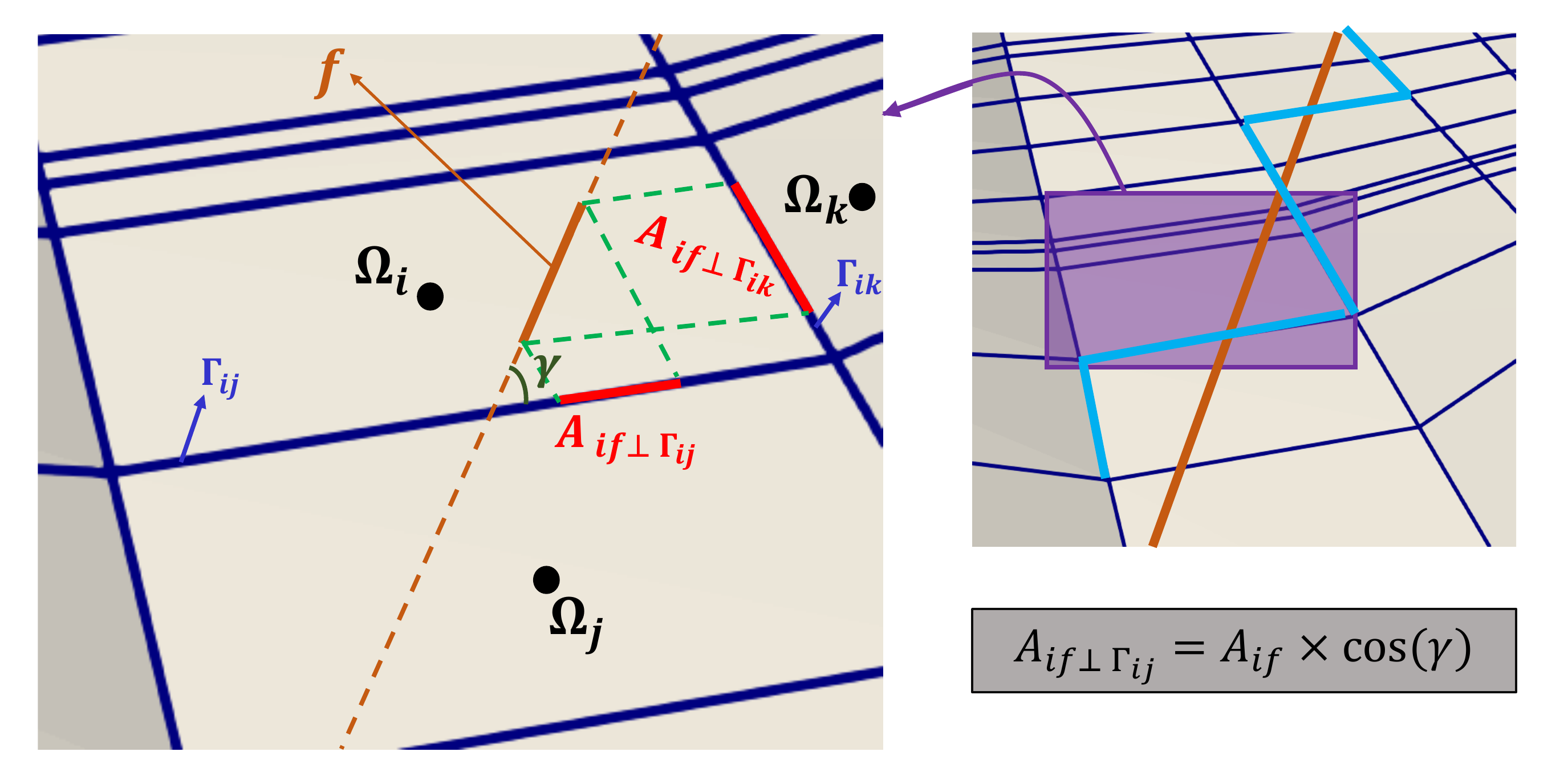}
	\caption{pEDFM Illustration for a rock matrix in corner-point grid geometry and an overlapping fracture. Due to pEDFM modifications, new non-neighboring connections between the fracture elements and the non-overlapped matrix cells are defined. The matrix-matrix connectivities are modified only if the fracture plate interrupts the line-segment that passes through the cell centers of each two neighboring matrix grid cells.}
	\label{Fig:pEDFM_Illustration_CPG}
\end{figure}

The first step in development of pEDFM for corner-point grid geometry is to flag all the interfaces between the matrix cells that a fracture plate interrupts the connection between their cell centers. Thereafter, a continuous projection path (shown in figure \ref{Fig:pEDFM_Illustration_CPG} as solid lines in light-blue color) is obtained on the interfaces. This projection path, disconnects the connections between the neighboring cells on both sides of this path, thus allowing a non-parallel and consistent flux exchange (i.e., through matrix-fracture-matrix). In figure \ref{Fig:pEDFM_Illustration_CPG}, the fracture element $f$ is assumed to overlap with the matrix grid cell $\Omega_i$ with an area fraction of $A_{if}$. A set of projections is defined on the interfaces between the overlapped matrix grid cell $\Omega_i$ and its neighboring grid cells that are affected by the crossing (i.e., $\Omega_j$ and $\Omega_k$). There will be two projections for 2D cases and three projection for 3D cases. For the interface between grid cells $\Omega_i$ and $\Omega_j$ (denoted as $\Gamma_{i,j}$) the projection area fraction $A_{if \perp \, \Gamma_{i,j}}$ is obtained via

\begin{equation}\label{Eq:pEDFM_Projection_Af_CPG}
	A_{if \perp \, \Gamma_{i,j}} = A_{if} \times \cos(\gamma).
\end{equation}

Here, $\gamma$ is the angle between the fracture element $f$ and the interface $\Gamma_{i,j}$ connecting the matrix grid cell $\Omega_i$ and the neighboring grid cell (in this example, $\Omega_j$). On the zoomed-in section of figure \ref{Fig:pEDFM_Illustration_CPG}, this projection area fraction is highlighted in red color. Similarly, the projection area fractions on the interfaces between all the neighboring matrix grid cells that are intersected by fracture elements are calculated based on the same formulation. A set of new transmissibilities are defined to provide connection between the fracture element $f$ and each non-neighboring matrix grid cell (i.e., $j$ and $k$ in the example shown in figure. \ref{Fig:pEDFM_Illustration_CPG}):

\begin{equation}\label{Eq:pEDFM_Projection_Transmissiblility1_CPG}
	T_{i_e f} = \frac{A_{if \perp \, \Gamma_{i,j}}}{\langle d \rangle_{i_e f}} \, \lambda_{i_e f},
\end{equation}

with $\langle d \rangle_{i_e f}$ defined as the average distance between the fracture element $f$ and matrix grid cell $i_e$, and $\lambda_{i_e f}$ being the effective fluid mobility between these two cells. Therefore, the transmissibility between the matrix grid cell $i$ and its corresponding neighboring cells is re-adjusted as

\begin{equation}\label{Eq:pEDFM_Projection_Transmissiblility2_CPG}
	T_{i i_e} = \frac{A_{i i_e} - A_{if \perp \, \Gamma_{i,j}}}{\Delta x_e} \, \lambda_{i_e f}, \quad \mathfrak{X} \in \{ x,y,z \}.
\end{equation}

These transmissibilities are modified via multiplication of a factor $\alpha$ defined as a fraction of the projection area, and the total area of that interface. One should consider that for all the overlapping fracture elements (except for the boundaries of the fractures), the projection will cover all the area of the interface. This means that $\alpha$ is $1.0$ for most of the cases, resulting in zero transmissibilities between the matrix grid cells (i.e., $T_{i i_e}=0$) affected by the projection, thus eliminating the parallel connectivities \cite{Tene2017}.

\section{Simulation Results}\label{Sec:Simulation_Results}
Numerical results of various test cases are presented in this section. The first two test cases compare the pEDFM model on Cartesian grid with pEDFM on corner-point grid geometry visually. For these two test cases, the fine-scale system is obtained on low-enthalpy geothermal fluid model from section \ref{Sec:GoverningEquations_GeothermalSinglePhaseFlow}. 
The third test case demonstrates the pEDFM result on a non-orthogonal grid model. The fluid model used for this test case is isothermal multiphase flow (see section \ref{Sec:GoverningEquations_IsothermalMultiphaseFlow}). Thereafter, we move towards a series of geologically relevant fields (all with isothermal multiphase fluid model). Using pEDFM on corner-point grid geometry, a number of synthetic (highly conductive) fractures and (impermeable) flow barriers are added to the geologically relevant models. The computational performance of this method will not be benchmarked as the purpose of these simulation results is to demonstrate pEDFM on corner-point grid geometry as a proof-of-concept.

Tables \ref{Tab:pEDFM_CPG_Input_Parameters_Isothermal} and \ref{Tab:pEDFM_CPG_Input_Parameters_Geothermal} show the mutual input parameters that are used for the test cases with isothermal multiphase and geothermal single-phase flow models respectively.

\begin{table}[!htb]
	\caption{Input parameters of fluid and rock properties of isothermal multiphase flow used in the geothermal test cases of pEDFM on corner-point geometry.}
	\label{Tab:pEDFM_CPG_Input_Parameters_Isothermal}
	\centering
	\begin{tabular}{l|c}
		Property & value \\
		\hline
		Matrix porosity ($\phi$)                  & $0.2 [-]$ \\
		Fractures permeability (min)              & $10^{-20} \ [\text{m}^2]$\\
		Fractures permeability (max)              & $10^{-8} \ [\text{m}^2]$\\
		Fractures aperture                        & $5 \times 10^{-3} \ [\text{m}]$\\
		Fluid viscosity (phase 1, $\mu_1$)        & $0.001 \ [\text{Pa.S}]$\\
		Fluid viscosity (phase 2, $\mu_2$)        & $0.003 \ [\text{Pa.S}]$\\
		Fluid density (phase 1, $\rho_1$)         & $1000 \ [\text{kg}/\text{m}^3]$\\
		Fluid density (phase 2, $\rho_2$)         & $850 \ [\text{kg}/\text{m}^3]$\\
		Initial pressure of the reservoir         & $2 \times 10^7 \ [\text{Pa}]$\\
		Initial saturation (phase 1, $S_1$)       & $0.0 [-]$\\
		Initial saturation (phase 1, $S_2$)       & $1.0 [-]$\\
		Injection Pressure                        & $5 \times 10^7 \ [\text{Pa}]$\\
		Production Pressure                       & $1 \times 10^7 \ [\text{Pa}]$
	\end{tabular}
\end{table}

\begin{table}[!htb]
	\caption{Input parameters of fluid and rock properties of geothermal single-phase flow used in the some test cases of pEDFM on corner-point geometry.}
	\label{Tab:pEDFM_CPG_Input_Parameters_Geothermal}
	\centering
	\begin{tabular}{l|c}
		Property & value \\
		\hline
		Rock thermal conductivity ($\Lambda_r$)  & $4 \ [{\text{W}}/{\text{m}. \text{K}}]$\\
		Fluid thermal conductivity ($\Lambda_f$) & $0.591 \ [{\text{W}}/{\text{m}. \text{K}}]$\\
		Rock density ($\rho_r$)                  & $2750 \ [{\text{kg}}/{\text{m}^3}]$\\
		Fluid specific heat ($C_{p_f}$)          & $4200 \ [{\text{J}}/{\text{kg}.\text{K}}]$\\
		Rock specific heat ($C_{p_r}$)           & $790 \ [{\text{J}}/{\text{kg}.\text{K}}]$\\
		Matrix porosity ($\phi$)                 & $0.2 [-]$ \\
		Fractures permeability (min)             & $10^{-20} \ [\text{m}^2]$\\
		Fractures permeability (max)             & $10^{-8} \ [\text{m}^2]$\\
		Fractures aperture                       & $5 \times 10^{-3} \ [\text{m}]$\\
		Initial pressure of the reservoir        & $1.5 \times 10^7 \ [\text{Pa}]$\\
		Initial temperature of the reservoir     & $400 \ [\text{K}]$\\
		Injection Pressure                       & $2 \times 10^7 \ [\text{Pa}]$\\
		Injection Temperature                    & $300 \ [\text{K}]$\\
		Production Pressure                      & $1 \times 10^7 \ [\text{Pa}]$
	\end{tabular}
\end{table}
\subsection{Test Case 1: 2D Heterogeneous fractured reservoir (square)}
\begin{figure}[!htbp]
	\centering
	\subcaptionbox{Permeability\label{Fig:pEDFM_CPG_TestCase1_K}}
	{\includegraphics[width=0.20\textwidth]{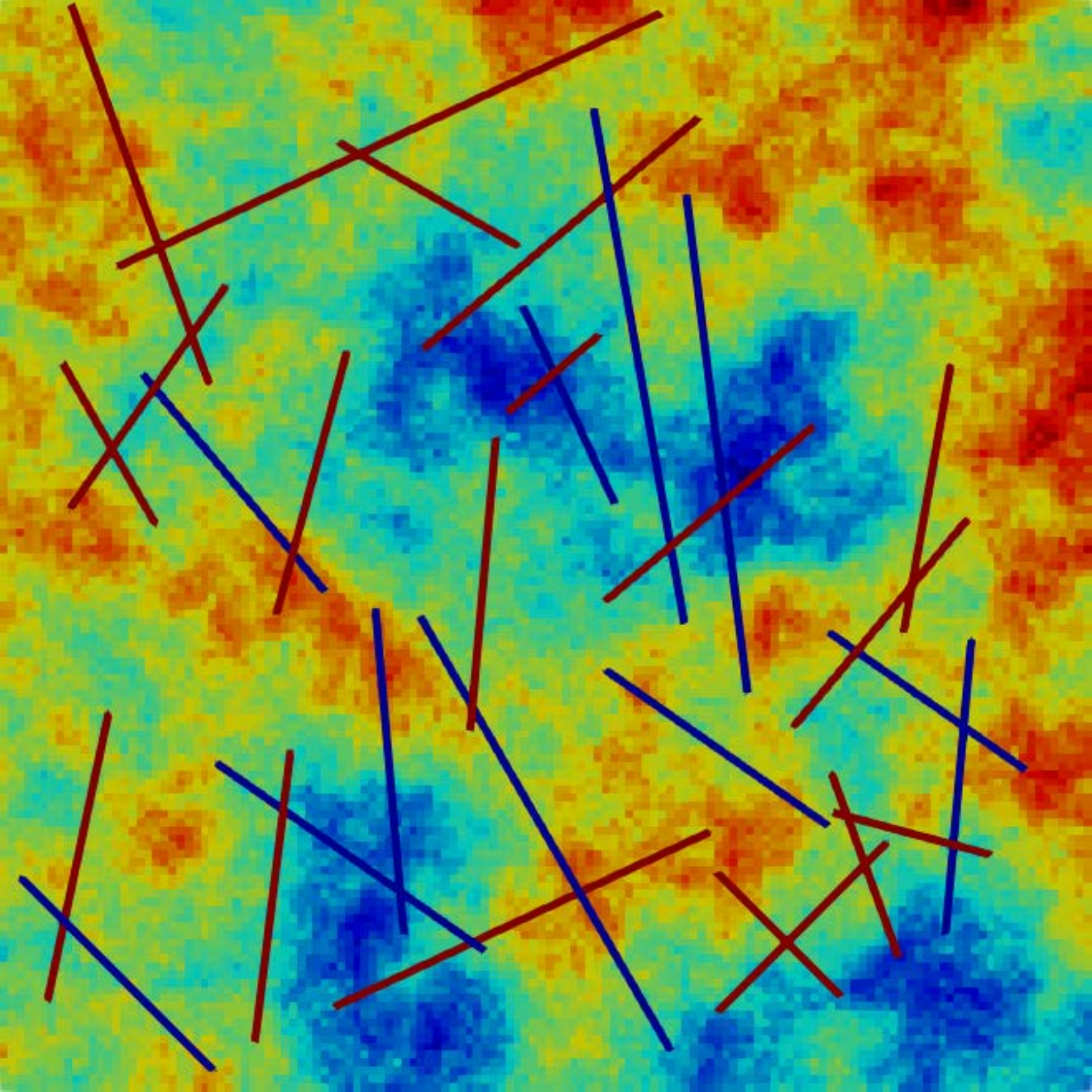}
	 \includegraphics[width=0.06\textwidth]{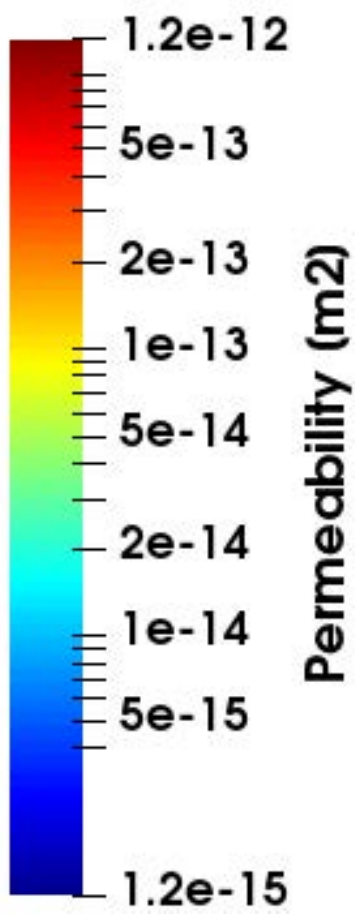}}
	\\
	\centering
	\subcaptionbox{{\footnotesize Pressure - Cartesian Grid}\label{Fig:pEDFM_CPG_TestCase1_P_Cart}}
	{\includegraphics[width=0.20\textwidth]{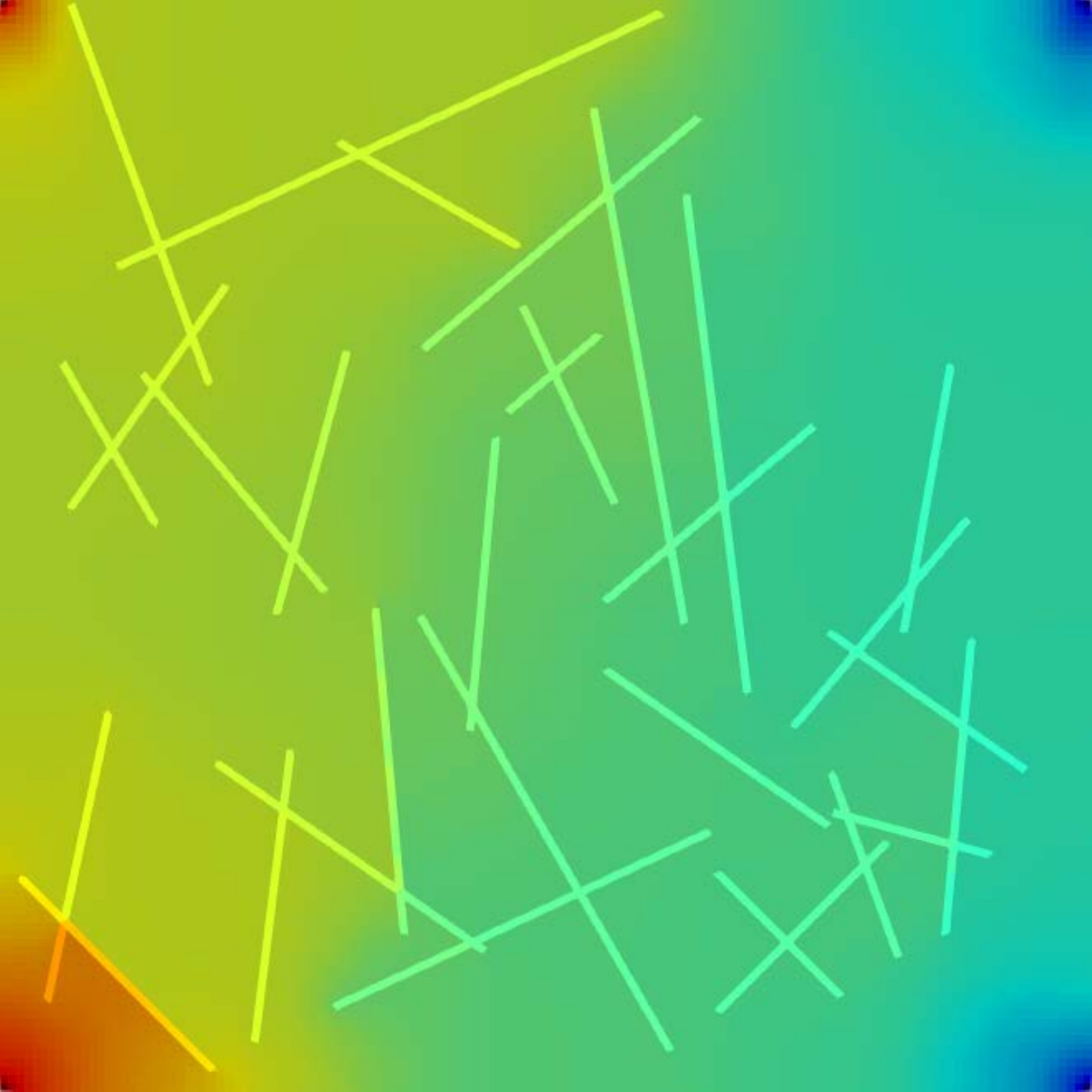}}
	\includegraphics[width=0.07\textwidth]{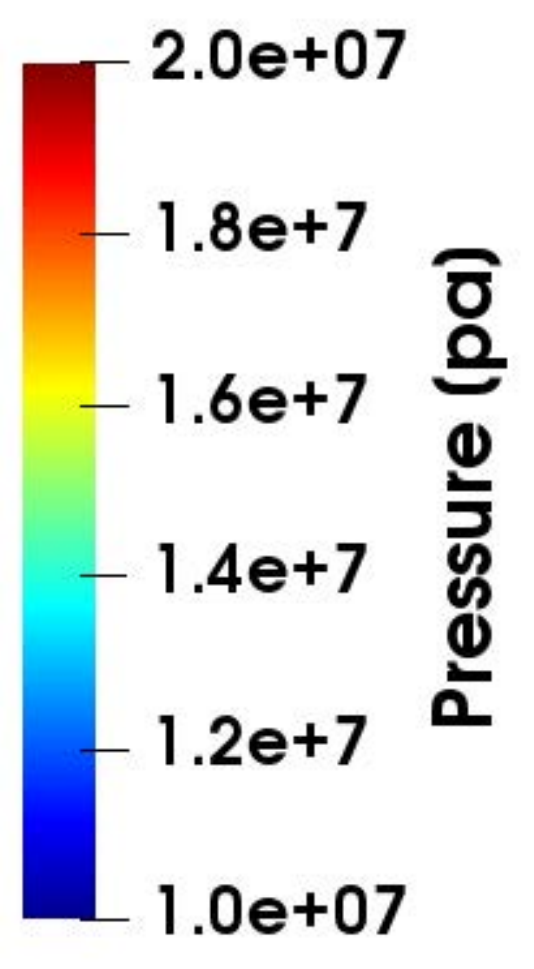}
	\subcaptionbox{{\footnotesize Pressure - Corner-point Grid}\label{Fig:pEDFM_CPG_TestCase1_P_CPG}}
	{\includegraphics[width=0.20\textwidth]{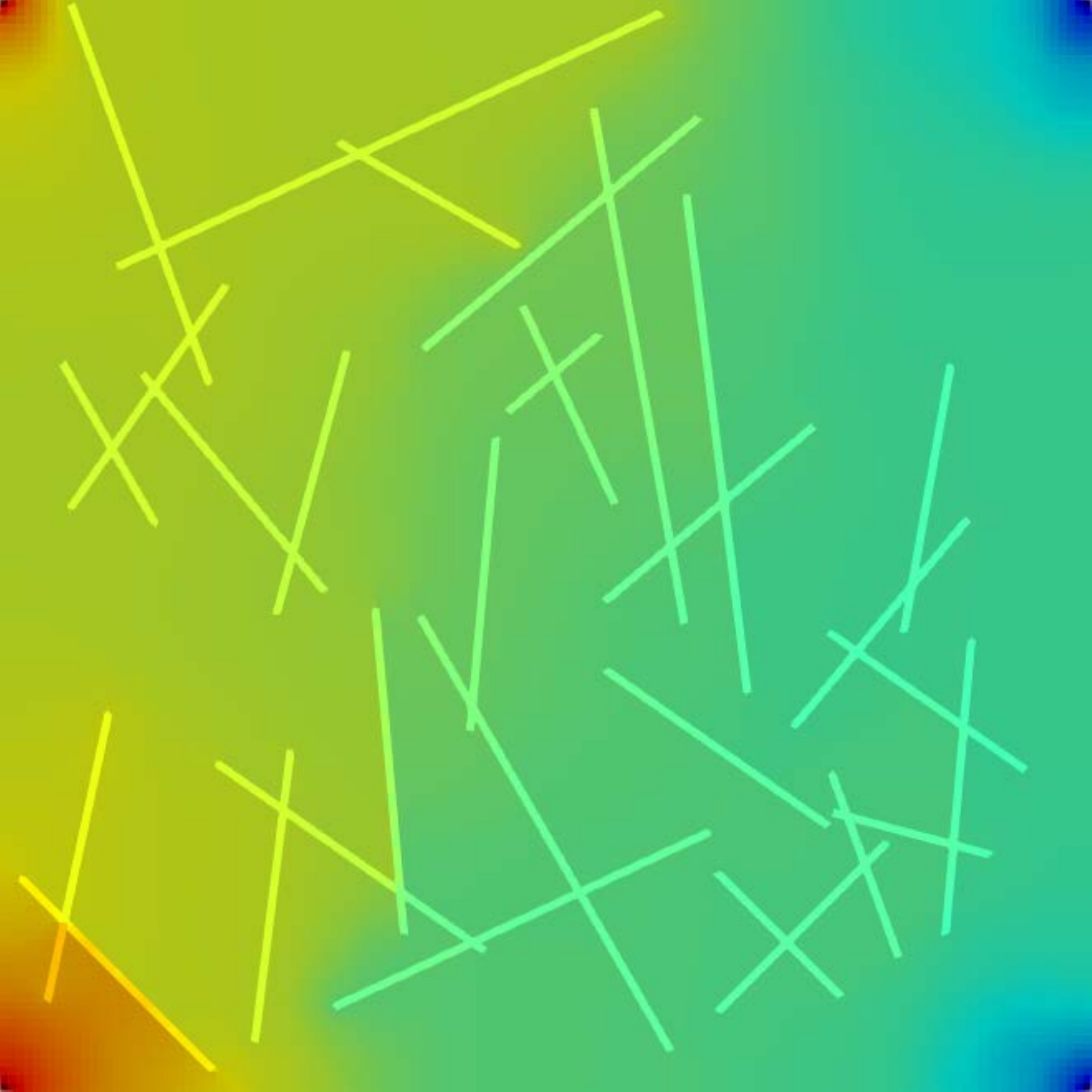}}
	\\
	\subcaptionbox{{\footnotesize Temperature - Cartesian Grid}\label{Fig:pEDFM_CPG_TestCase1_T_Cart}}
	{\includegraphics[width=0.20\textwidth]{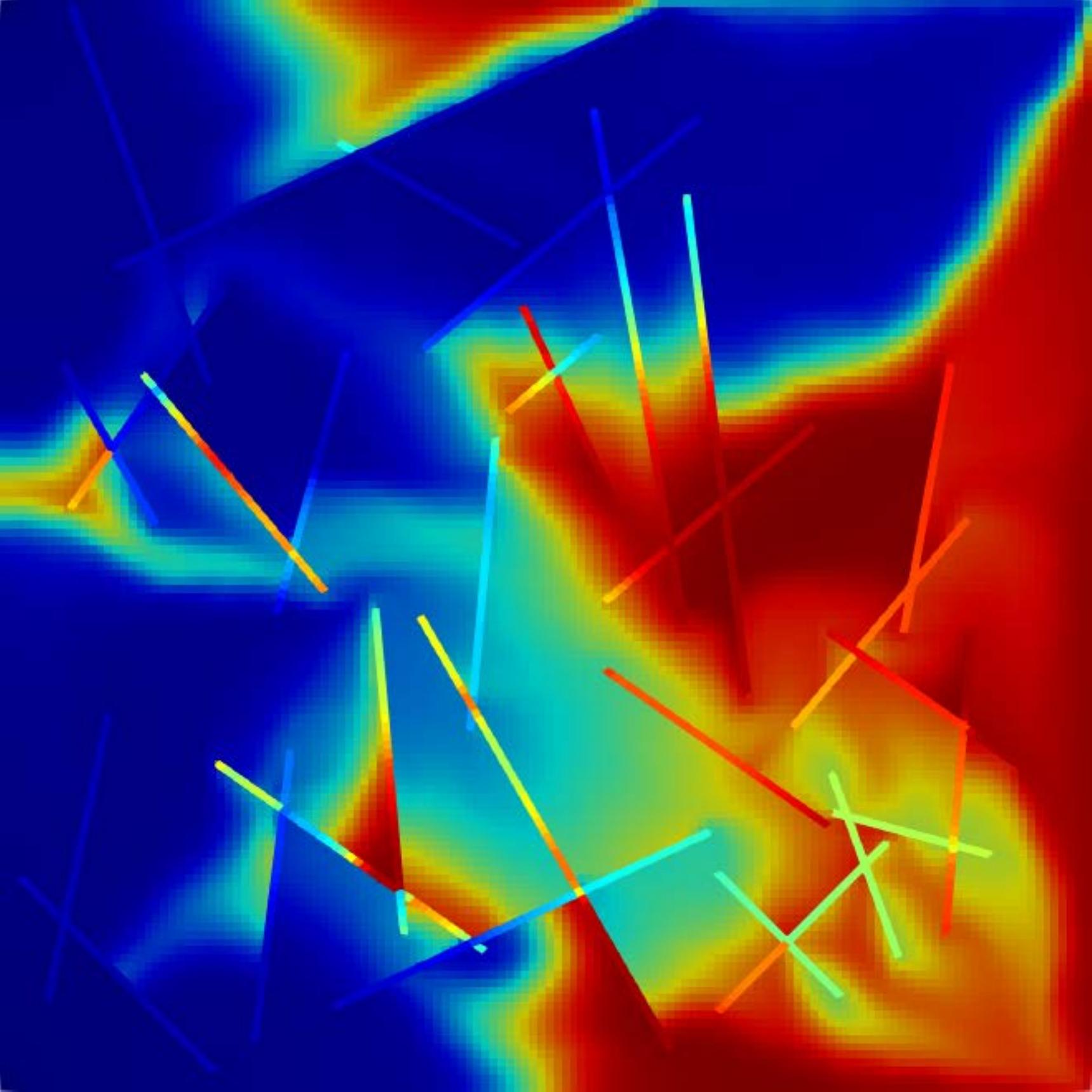}}
	\includegraphics[width=0.07\textwidth]{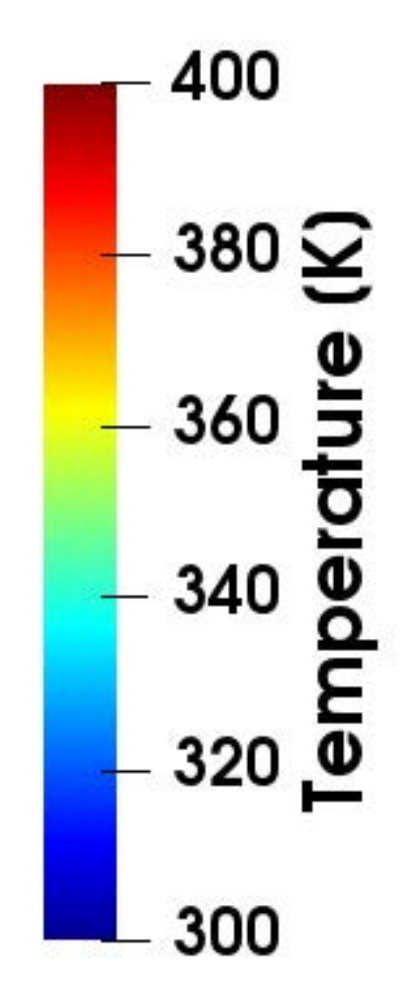}
	\subcaptionbox{{\footnotesize Temperature - Corner-point Grid}\label{Fig:pEDFM_CPG_TestCase1_T_CPG}}
	{\includegraphics[width=0.20\textwidth]{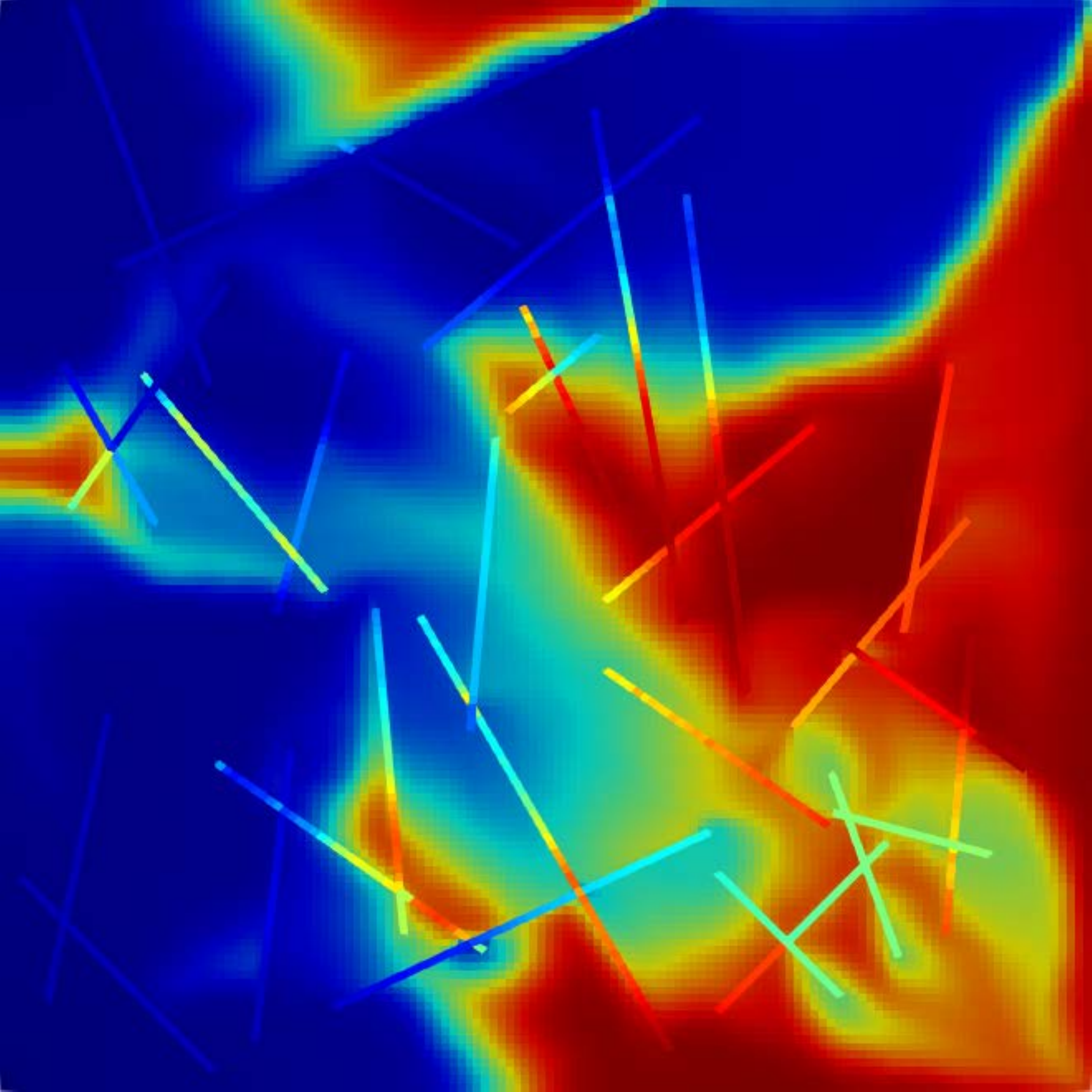}}
	\caption{Test case 1: 2D Heterogeneous. Fig. \ref{Fig:pEDFM_CPG_TestCase1_K} illustrates the permeability map of the system. The figures \ref{Fig:pEDFM_CPG_TestCase1_P_Cart} and \ref{Fig:pEDFM_CPG_TestCase1_P_CPG} show the pressure solution on a specific time-step for Cartesian grid and corner-point grid geometry respectively. The figures on the bottom row (\ref{Fig:pEDFM_CPG_TestCase1_T_Cart} and \ref{Fig:pEDFM_CPG_TestCase1_T_CPG}) visualize the temperature solutions on the same time-step.}
	\label{Fig:pEDFM_CPG_TestCase1}
\end{figure}

In this test case, pEDFM on Cartesian grid versus corner-point grid geometry is visually compared. For this reason, a box-shaped heterogeneous $100\,[\text{m}] \times 100 \, [\text{m}]$ domain containing $30$ fractures with mixed conductivities is considered. The length of each fracture is different but the size of their aperture is identical and set to  $a_f = 5 \cdot 10^{-3}\, [\text{m}]$. A $136 \times 136$ grid is imposed on the rock matrix and the fracture network consists of $1024$ grid cells (in total $19520$ cells). The permeability of the matrix ranges from $K_{m_{min}} = 1.2 \times 10^{-15}\, [\text{m}^2]$ to $K_{m_{max}} = 1.2 \times 10^{-12}\, [\text{m}^2]$, and the permeability of the fracture network has the range of $K_{f_{min}} = 10^{-20}\, [\text{m}^2]$ and $K_{f_{max}} = 10^{-8}\, [\text{m}^2]$. Two injection wells are located at the bottom left and top left corners with injection pressure of $p_{\text{inj}} = 2 \times 10^7 \, [\text{Pa}]$. Additionally, there are two production wells at the bottom right and the top right corners with pressure of $p_{\text{prod}} = 1 \times 10^7 \, [\text{Pa}]$. Table \ref{Tab:pEDFM_CPG_Input_Parameters_Geothermal} demonstrates the input parameters of this test case. Figure \ref{Fig:pEDFM_CPG_TestCase1} shows the results of the simulation using both Cartesian Grid and corner-point geometry.

Please note that in this test case (and the test case $2$), the $x,y,z$ coordinates of the grids of the Cartesian geometry and corner-point grid geometry are identical. However, the small differences in the simulation results arise from the treatment of the edges (boundaries) of the fractures. In Cartesian geometry, the pEDFM model computes the projection area of the tips of the fracture (where the fracture elements may not completely block the interface of the matrix grid cell it overlaps with). This results in an ``alpha factor” of between $0$ and $1$. In corner-point grid geometry however, the value of alpha factor is either $0$ or $1$ as the computation of the projection area in the mentioned boundary region is done differently. This is an approximation which is negligible due to the fact that the affecting length scale is below the length scale of the fine-scale grid structure. The differences is noticeable only if the computational grids imposed on the matrix are not at high resolution. At higher resolutions, this difference will be negligible.

\subsection{Test Case 2: 3D Homogeneous fractured reservoir (box)}
\begin{figure}[!htbp]
	\centering
	\subcaptionbox{{\footnotesize Pressure - Cartesian Grid}\label{Fig:pEDFM_CPG_TestCase2_P_Cart}}
	{\includegraphics[width=0.21\textwidth]{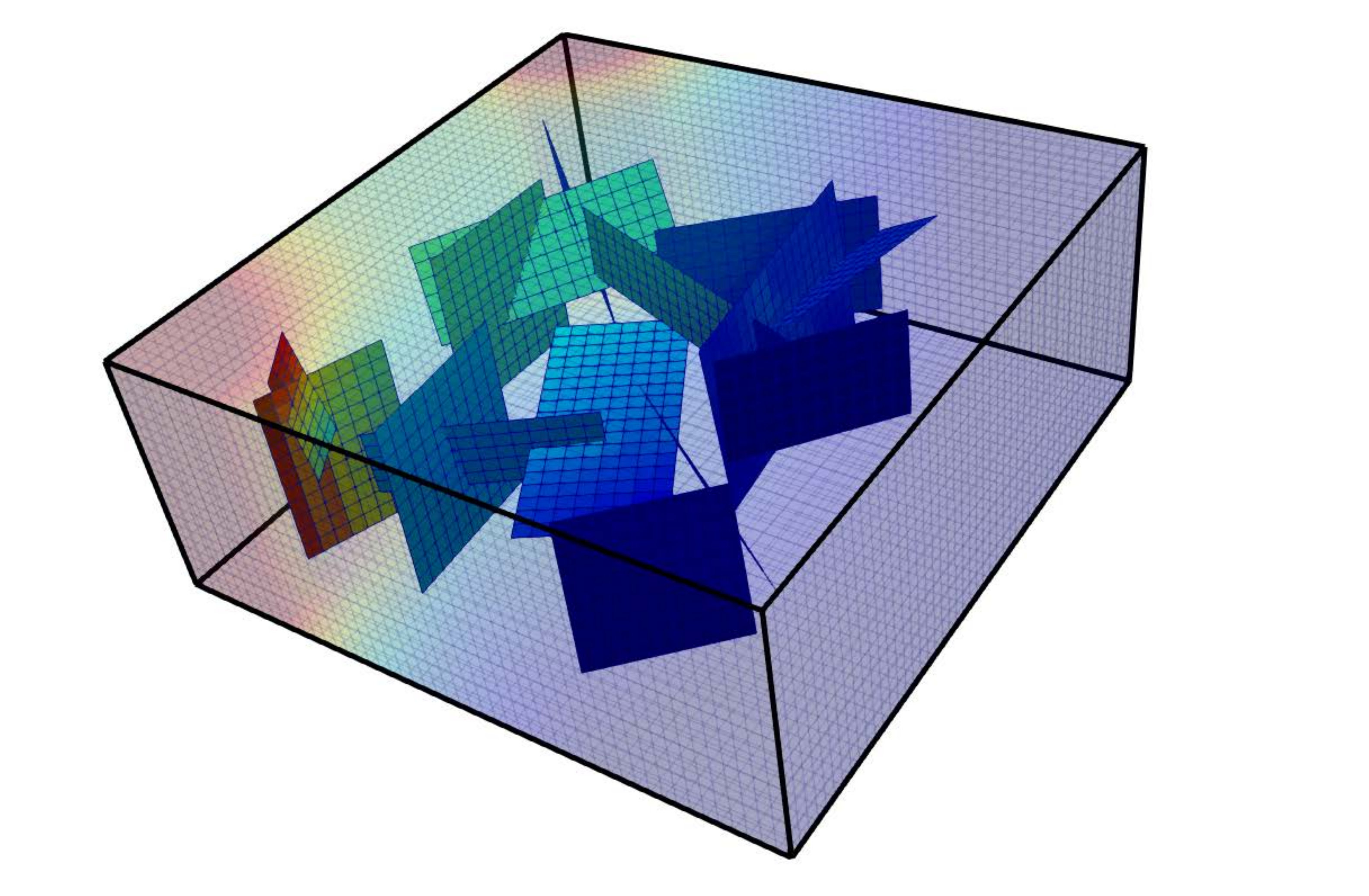}}
	\includegraphics[width=0.05\textwidth]{Figures/Colorbar_P_Jet.pdf}
	\subcaptionbox{{\footnotesize Pressure - Corner-point Grid}\label{Fig:pEDFM_CPG_TestCase2_P_CPG}}
	{\includegraphics[width=0.21\textwidth]{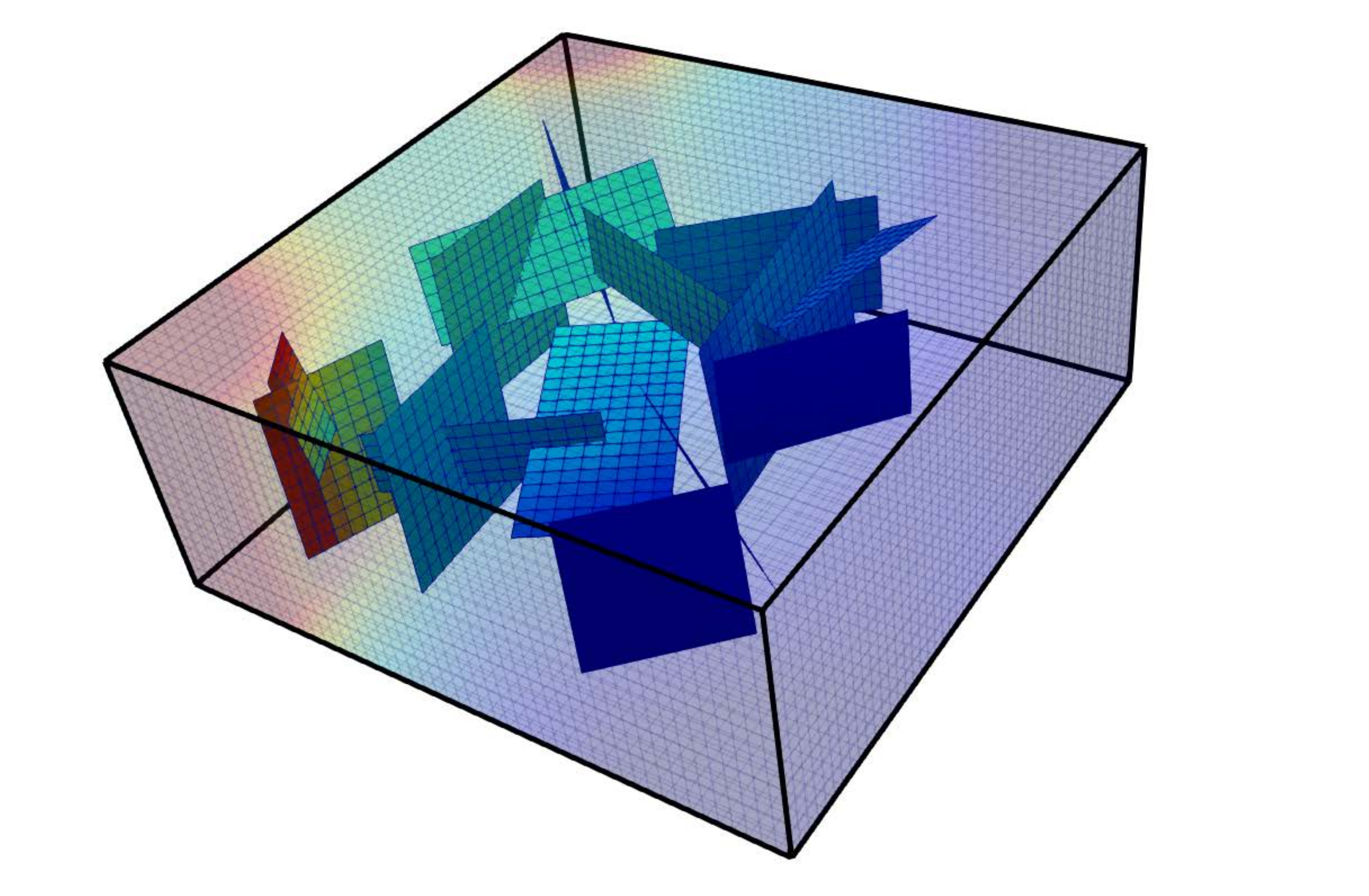}}
	\\
	\subcaptionbox{{\footnotesize Temperature - Cartesian Grid}\label{Fig:pEDFM_CPG_TestCase2_T_Cart}}
	{\includegraphics[width=0.21\textwidth]{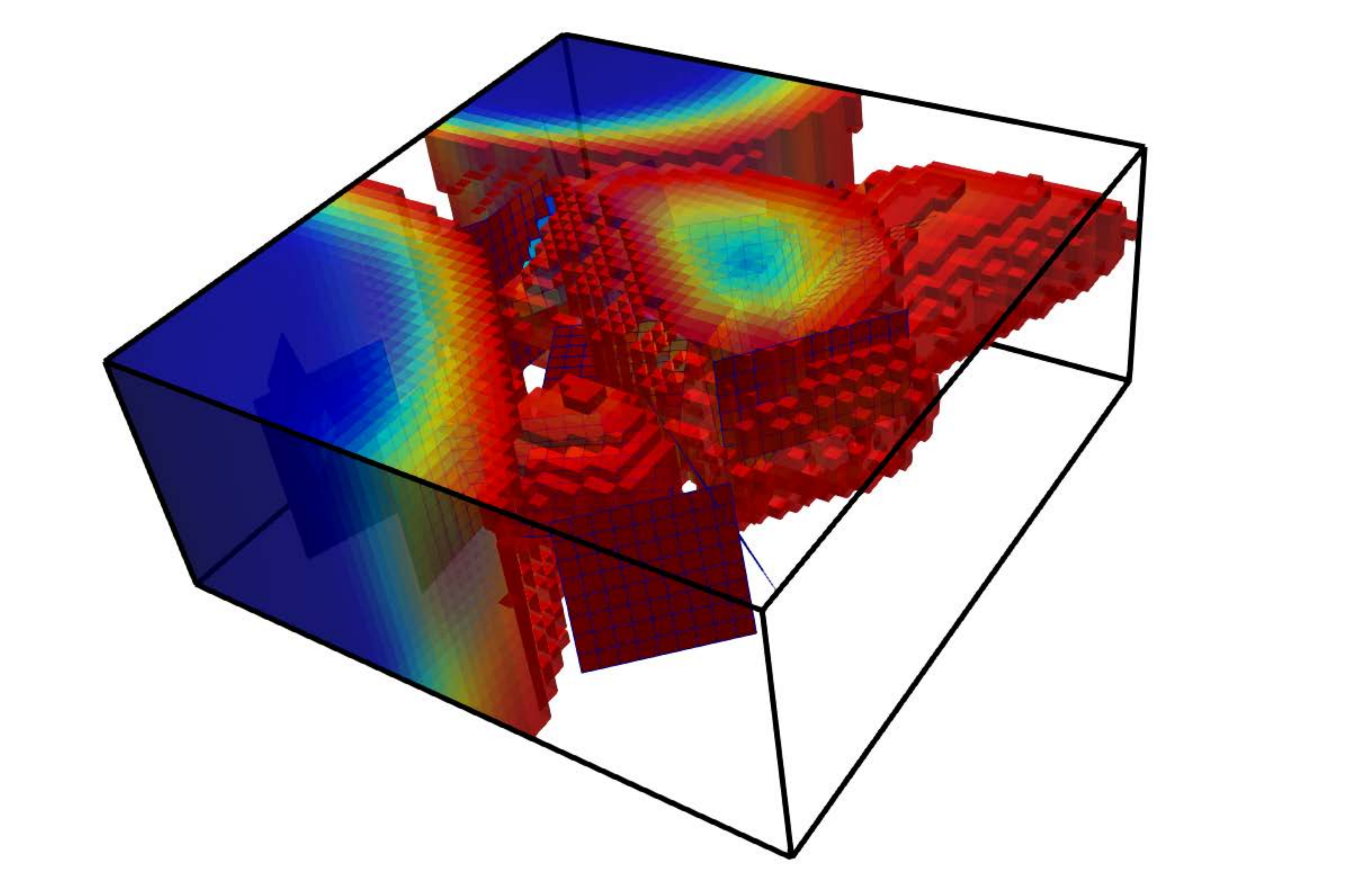}}
	\includegraphics[width=0.05\textwidth]{Figures/Colorbar_T_Jet.pdf}
	\subcaptionbox{{\footnotesize Temperature - Corner-point Grid}\label{Fig:pEDFM_CPG_TestCase2_T_CPG}}
	{\includegraphics[width=0.21\textwidth]{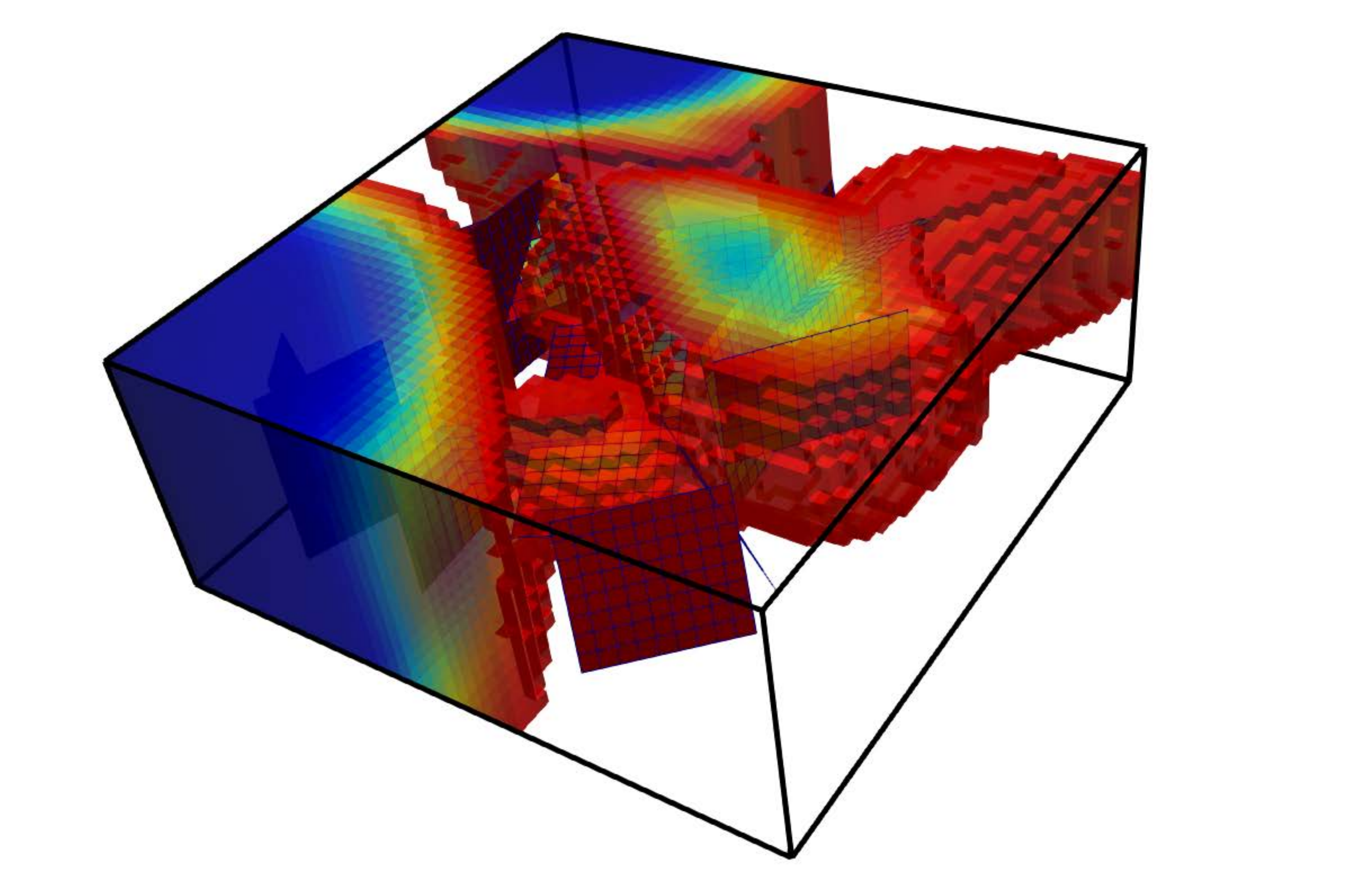}}
	\caption{Test case 2: 3D Homogeneous box. The figures \ref{Fig:pEDFM_CPG_TestCase2_P_Cart} and \ref{Fig:pEDFM_CPG_TestCase2_P_CPG} show the pressure solution on a specific time-step for Cartesian grid and corner-point grid geometry respectively. The figures on the bottom row (\ref{Fig:pEDFM_CPG_TestCase2_T_Cart} and \ref{Fig:pEDFM_CPG_TestCase2_T_CPG}) visualize the temperature solutions on the same time-step.}
	\label{Fig:pEDFM_CPG_TestCase2}
\end{figure}

This test cases, similar to the test case 1, shows a visual comparison for pEDFM on Cartesian drid versus corner-point grid geometry. A 3D $100 \, [\text{m}] \times 100 \, [\text{m}] \times 40 \, [\text{m}]$ domain containing $15$ lower dimensional fractures with different geometrical properties is considered. A $50 \times 50 \times 20$ grid is imposed on rock matrix. The fracture network contains $1414$ grid cells (total of $51414$ grid cells). The rock matrix has permeability of $K_m = 10^{-14}\, [\text{m}^2]$. Fracture network consists of both highly conductive fractures with permeability of $K_f = 10^{-8}\, [\text{m}^2]$ and flow barriers with permeability of $K_f = 10^{-20}\, [\text{m}^2]$. Two injection wells exist on the bottom left and top left boundaries with pressure of $p_{\text{inj}} = 2 \times 10^7 \, [\text{Pa}]$. Similarly, two production wells are located at the bottom right and top right boundaries with pressure of $p_{\text{prod}} = 1 \times 10^7 \, [\text{Pa}]$. All wells are vertical and perforate the entire thickness of the reservoir. Figure \ref{Fig:pEDFM_CPG_TestCase2} illustrates the results of the simulation using both Cartesian Grid and corner-point geometry.

\subsection{Test Case 3: 3D reservoir with non-orthogonal grids}
\begin{figure}[!htbp]
	\centering
	\subcaptionbox{{\footnotesize Fractures permeability (scenario 1)}\label{Fig:pEDFM_CPG_TestCase3_K_1}}
	{\includegraphics[width=0.23\textwidth]{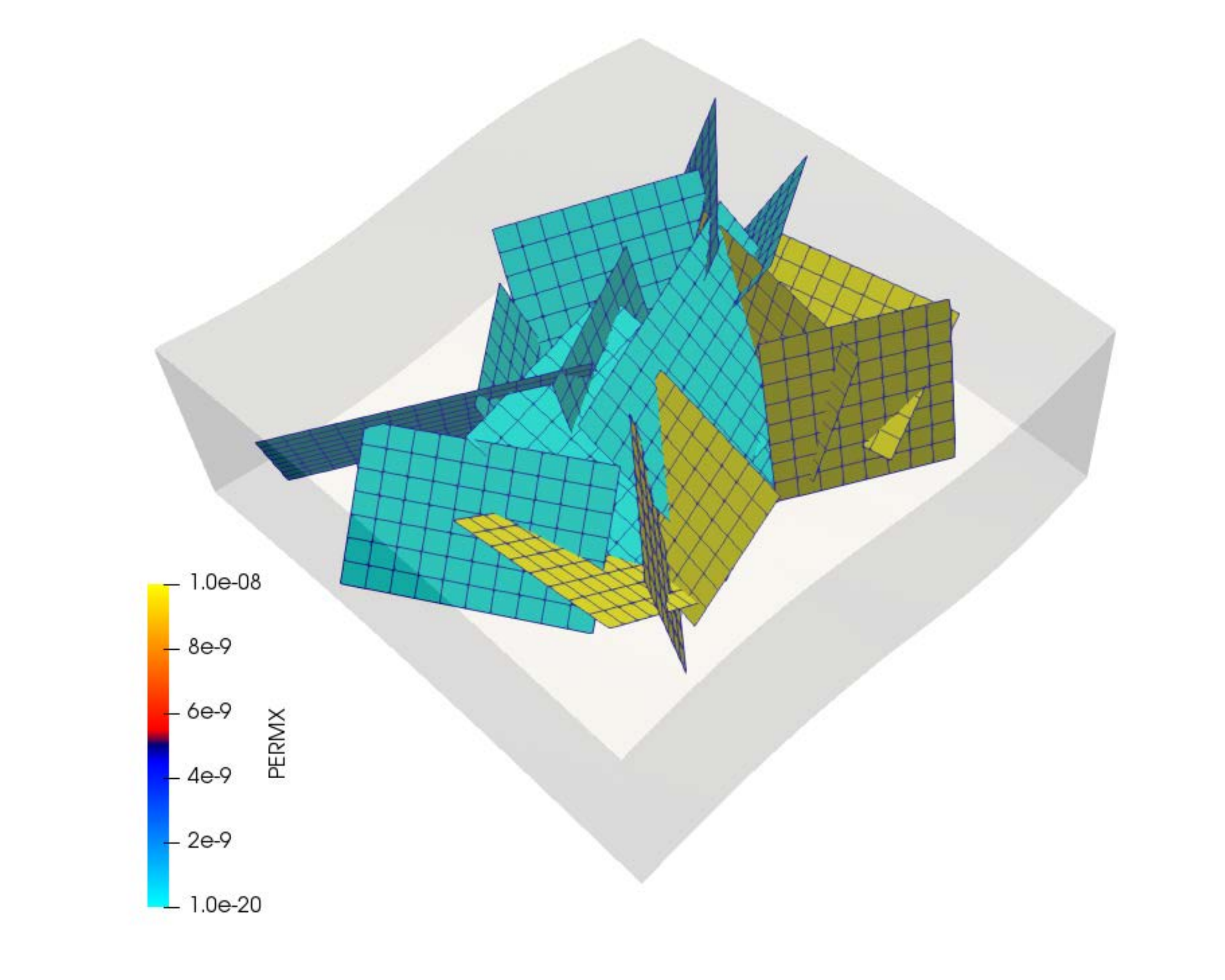}}
	\subcaptionbox{{\footnotesize Fractures permeability (scenario 2)}\label{Fig:pEDFM_CPG_TestCase3_K_2}}
	{\includegraphics[width=0.23\textwidth]{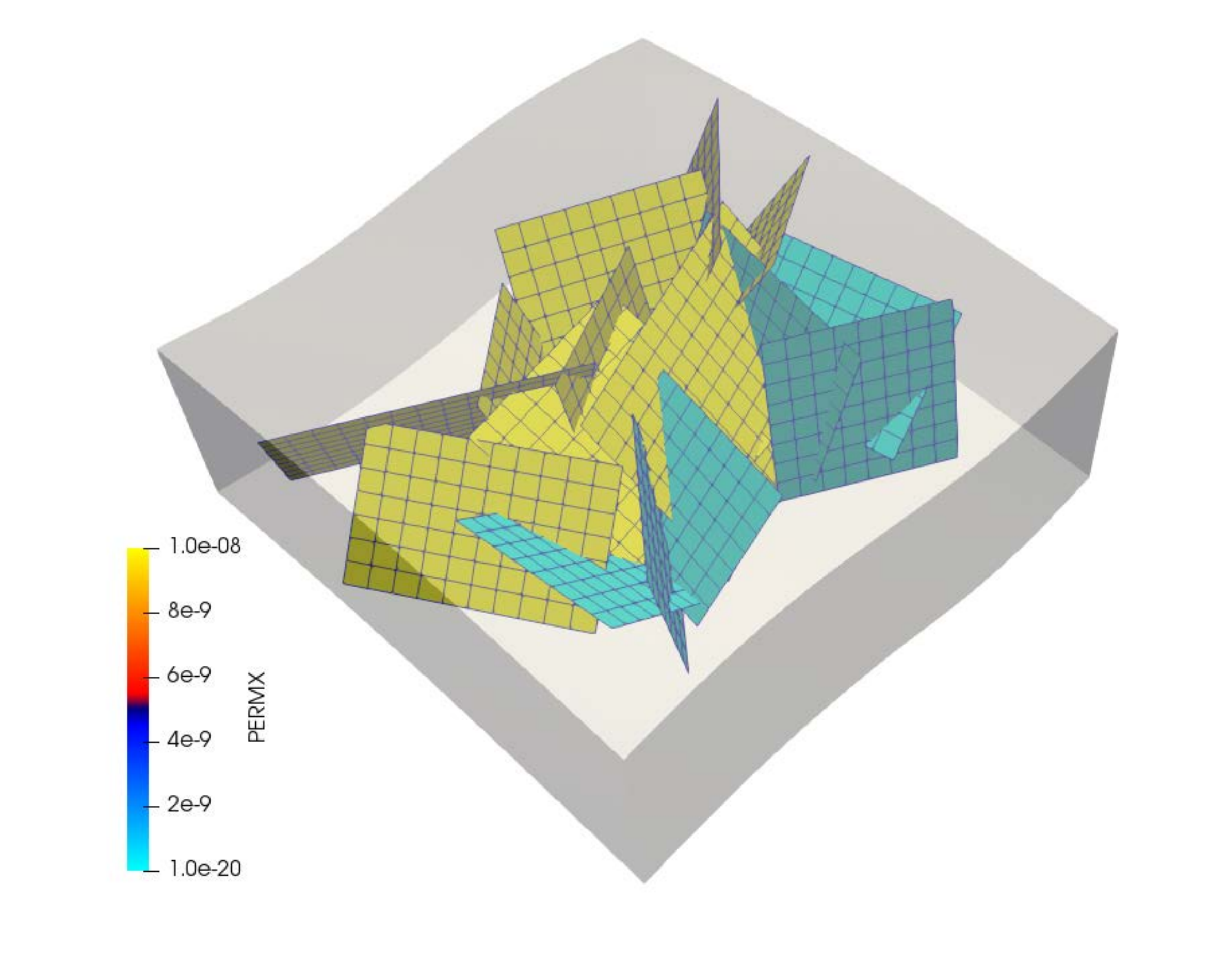}}
	\\
	\subcaptionbox{{\footnotesize Matrix cells overlapped by fractures}\label{Fig:pEDFM_CPG_TestCase3_Fractured_Cells}}
	{\includegraphics[width=0.23\textwidth]{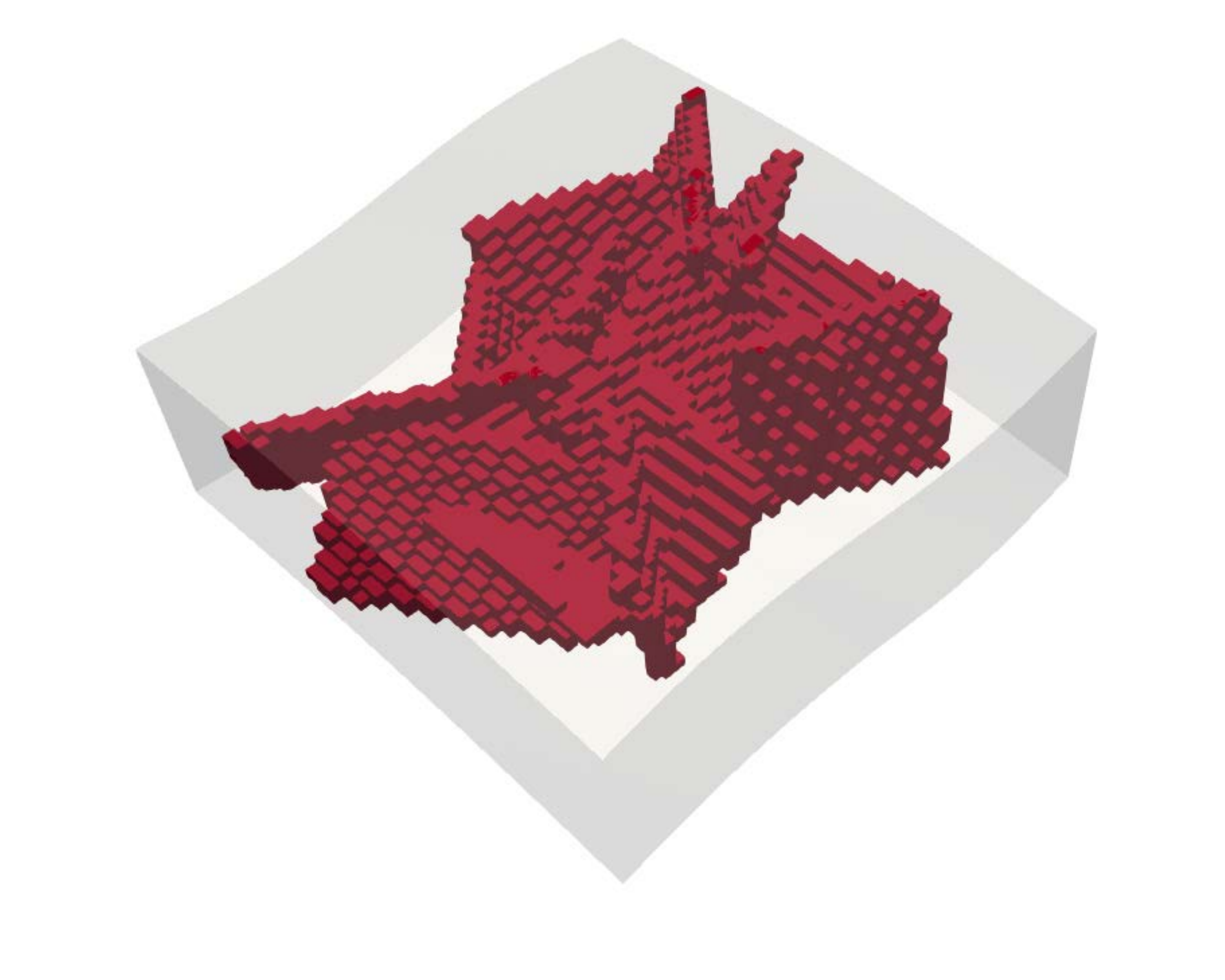}}
	\subcaptionbox{{\footnotesize The wells geometry}\label{Fig:pEDFM_CPG_TestCase3_Wells}}
	{\includegraphics[width=0.23\textwidth]{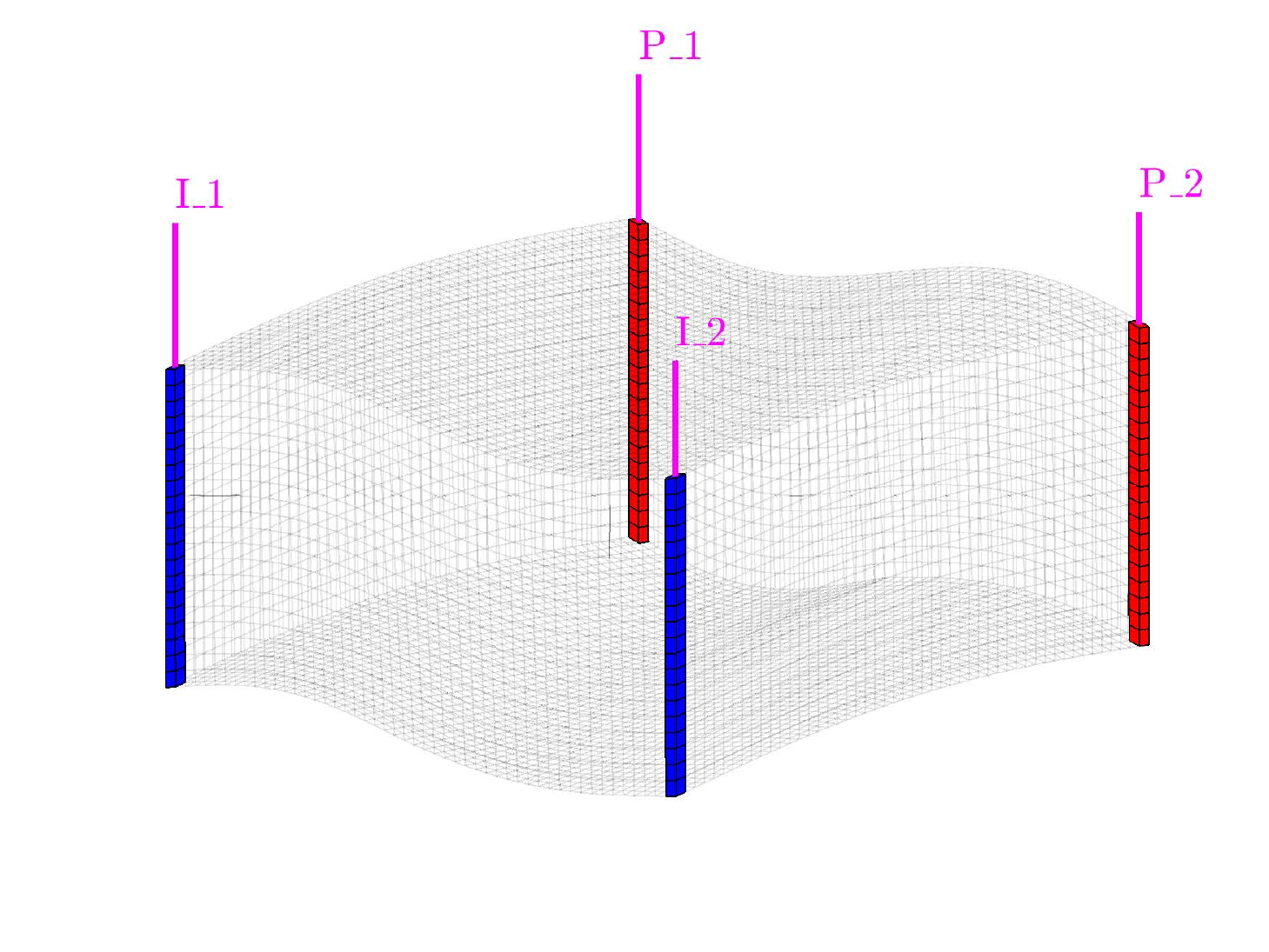}}
	\caption{Test case 3: A 3D fractured deformed box with non-orthogonal grid corner-point grid geometry. The figures \ref{Fig:pEDFM_CPG_TestCase3_K_1} and \ref{Fig:pEDFM_CPG_TestCase3_K_2} on top show the permeability of the fractures for the scenarios 1 and 2 respectively. The figure \ref{Fig:pEDFM_CPG_TestCase3_Fractured_Cells} on the bottom left illustrates the matrix grid cells that are overlapped by the fractures. Figure \ref{Fig:pEDFM_CPG_TestCase3_Wells} on the bottom right shows the geometry of injection and production wells.}
	\label{Fig:pEDFM_CPG_TestCase3_Geometry}
\end{figure}

The third test case (figure \ref{Fig:pEDFM_CPG_TestCase3_Geometry}) demonstrates the capability of pEDFM method on the reservoir model based on corner point grids. The grid cells in test case $2$ were deformed to create a distorted version of that model. The model allows to test the pEDFM implementation in a non-orthogonal grid system. The same dimensions and gridding from test case $2$ are used in this test case. The fracture network consisting of $15$ fractures is discretized in $876$ grids, and a total of $50876$ grid cells are imposed on the entire domain.

Two different scenarios are considered in this test case. In the first scenario some fractures are considered as highly conductive while the others are given a very low permeability and are considered to be flow barriers (shown in figure \ref{Fig:pEDFM_CPG_TestCase3_K_1} with yellow color for high permeability and blue color for low permeability). In the second scenario, the permeability of the fractures is chosen as in inverse of scenario 1 (\ref{Fig:pEDFM_CPG_TestCase3_K_2}), i.e., the low conductive fractures are now highly conductive and vice versa. The values of permeability for the matrix and (low and high conductive) fractures are identical to the ones in test case $2$. The well pattern and pressure restrictions are also the same as in the previous test case.

\begin{figure}[!htbp]
	\centering
	\subcaptionbox{{\footnotesize Pressure (scenario 1)}\label{Fig:pEDFM_CPG_TestCase3_P_1}}
	{\includegraphics[width=0.23\textwidth]{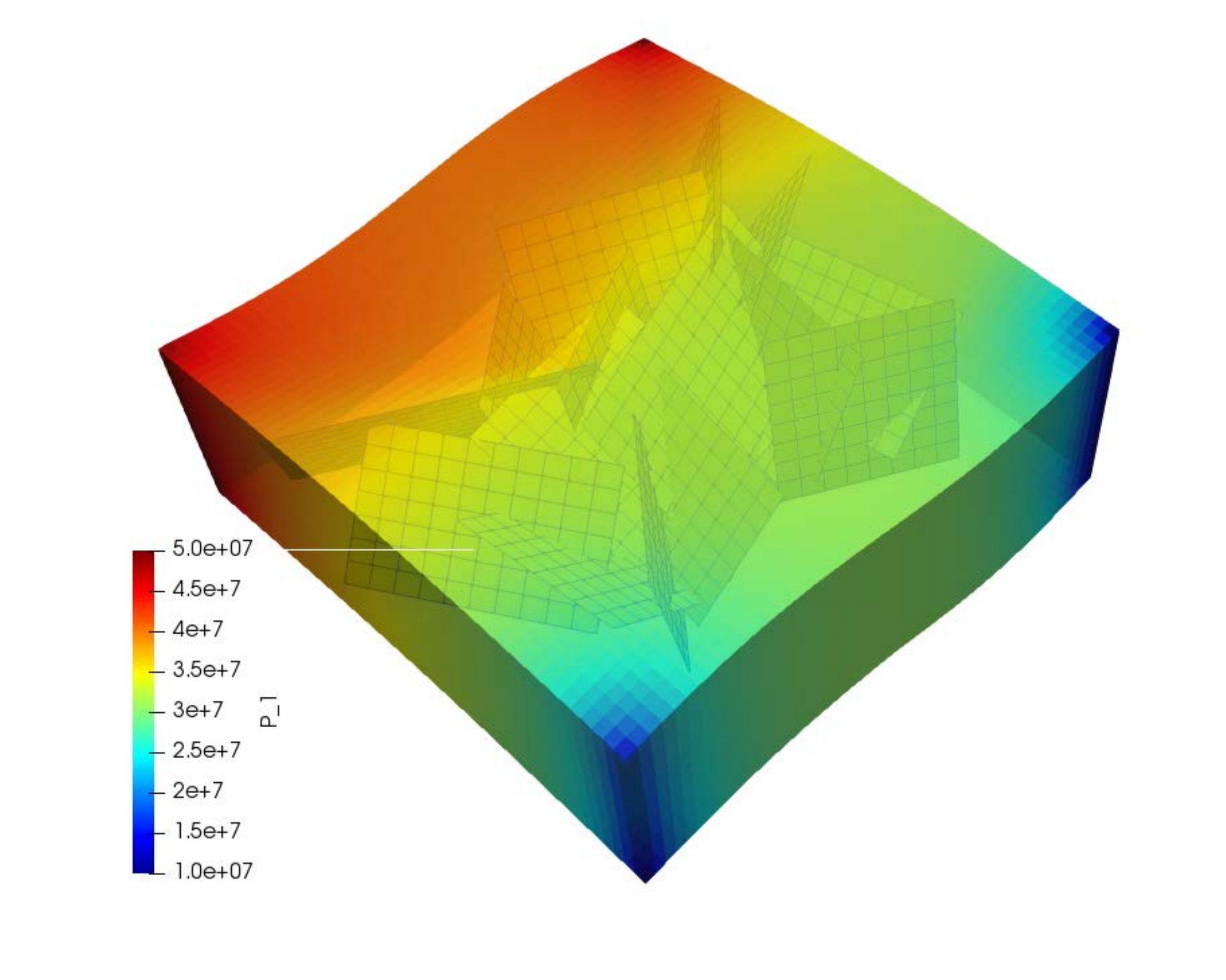}}
	\subcaptionbox{{\footnotesize Pressure (scenario 2)}\label{Fig:pEDFM_CPG_TestCase3_P_2}}
	{\includegraphics[width=0.23\textwidth]{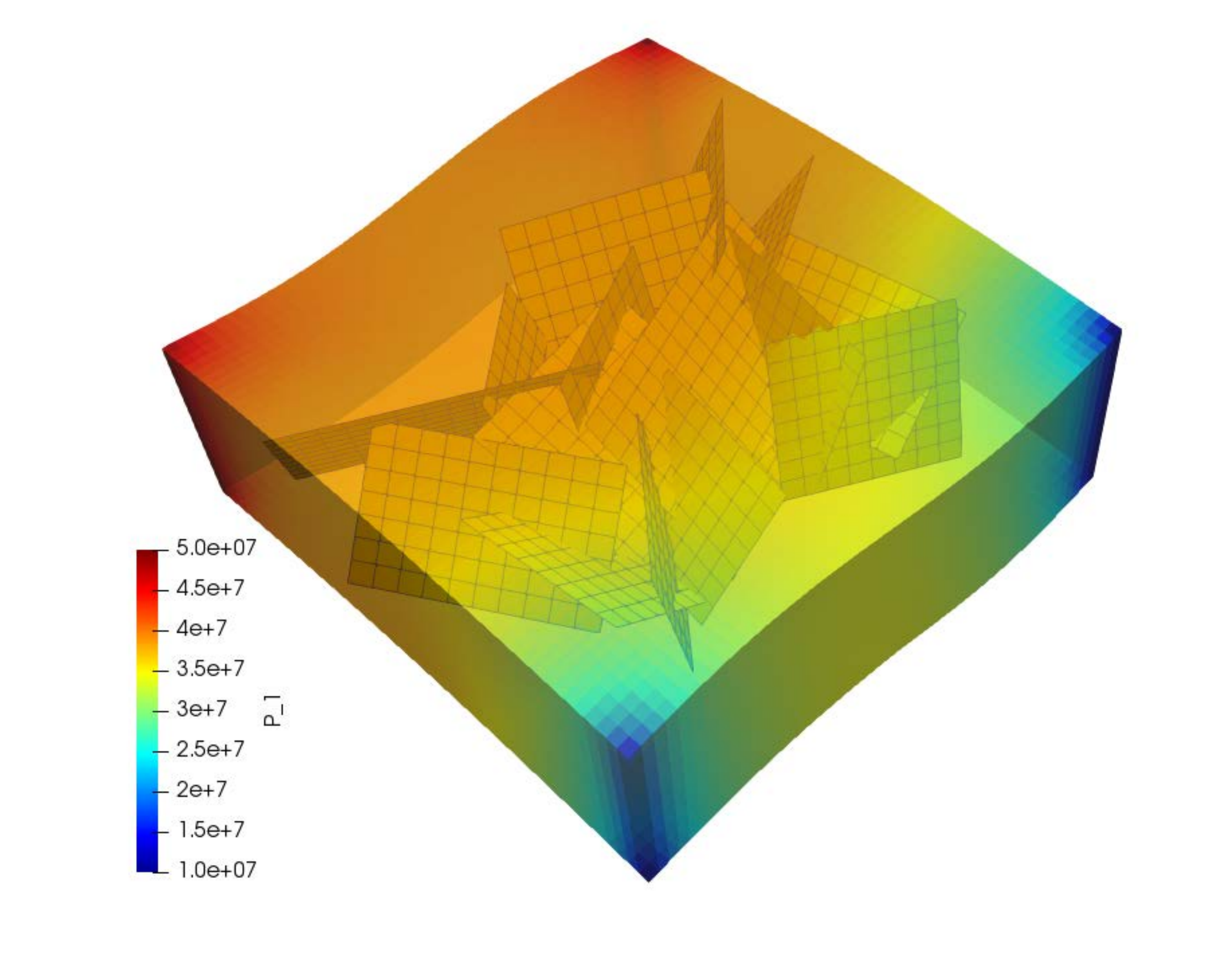}}
	\\
	\subcaptionbox{{\footnotesize Saturation (scenario 1)}\label{Fig:pEDFM_CPG_TestCase3_S_1}}
	{\includegraphics[width=0.23\textwidth]{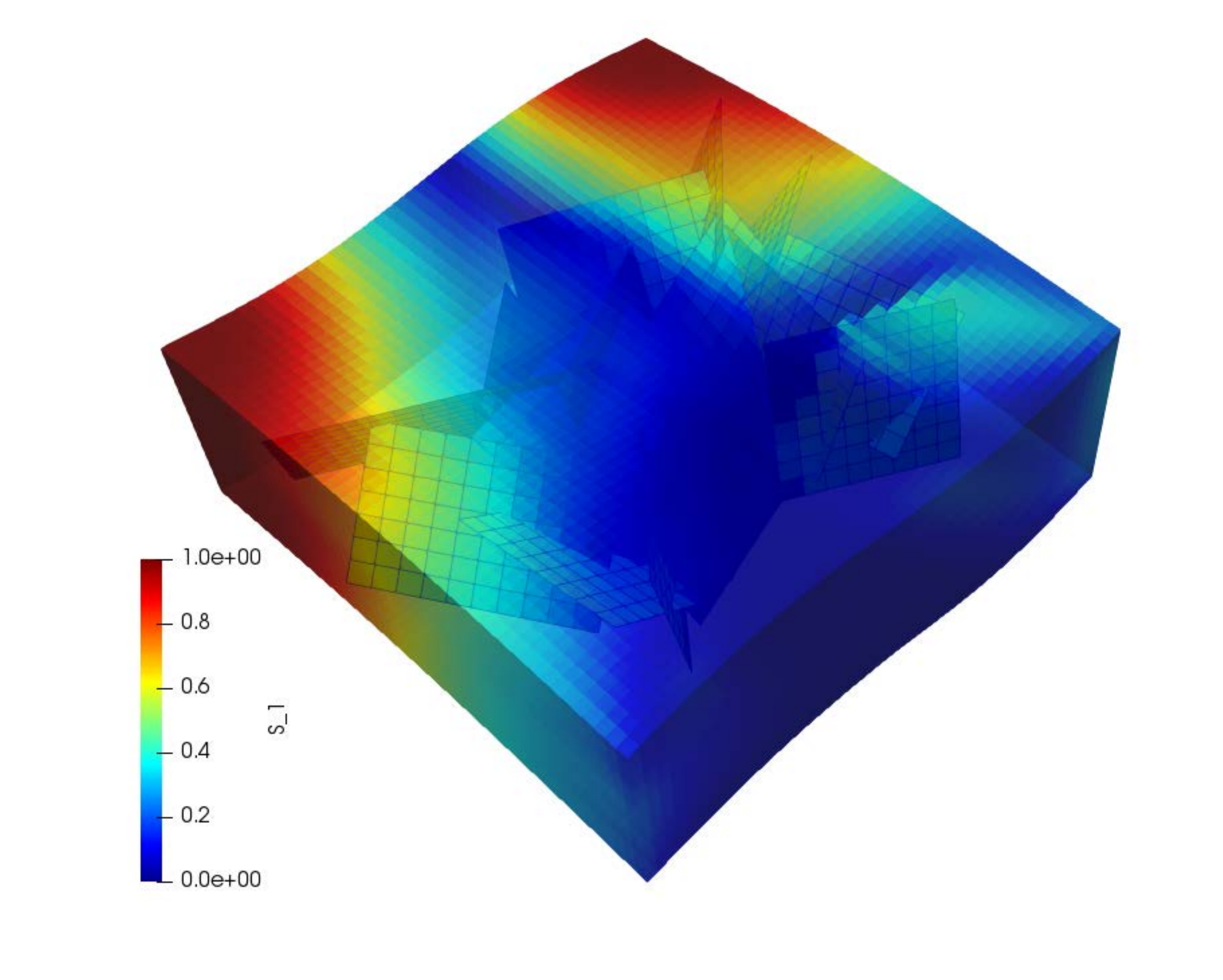}}
	\subcaptionbox{{\footnotesize Saturation (scenario 2)}\label{Fig:pEDFM_CPG_TestCase3_S_2}}
	{\includegraphics[width=0.23\textwidth]{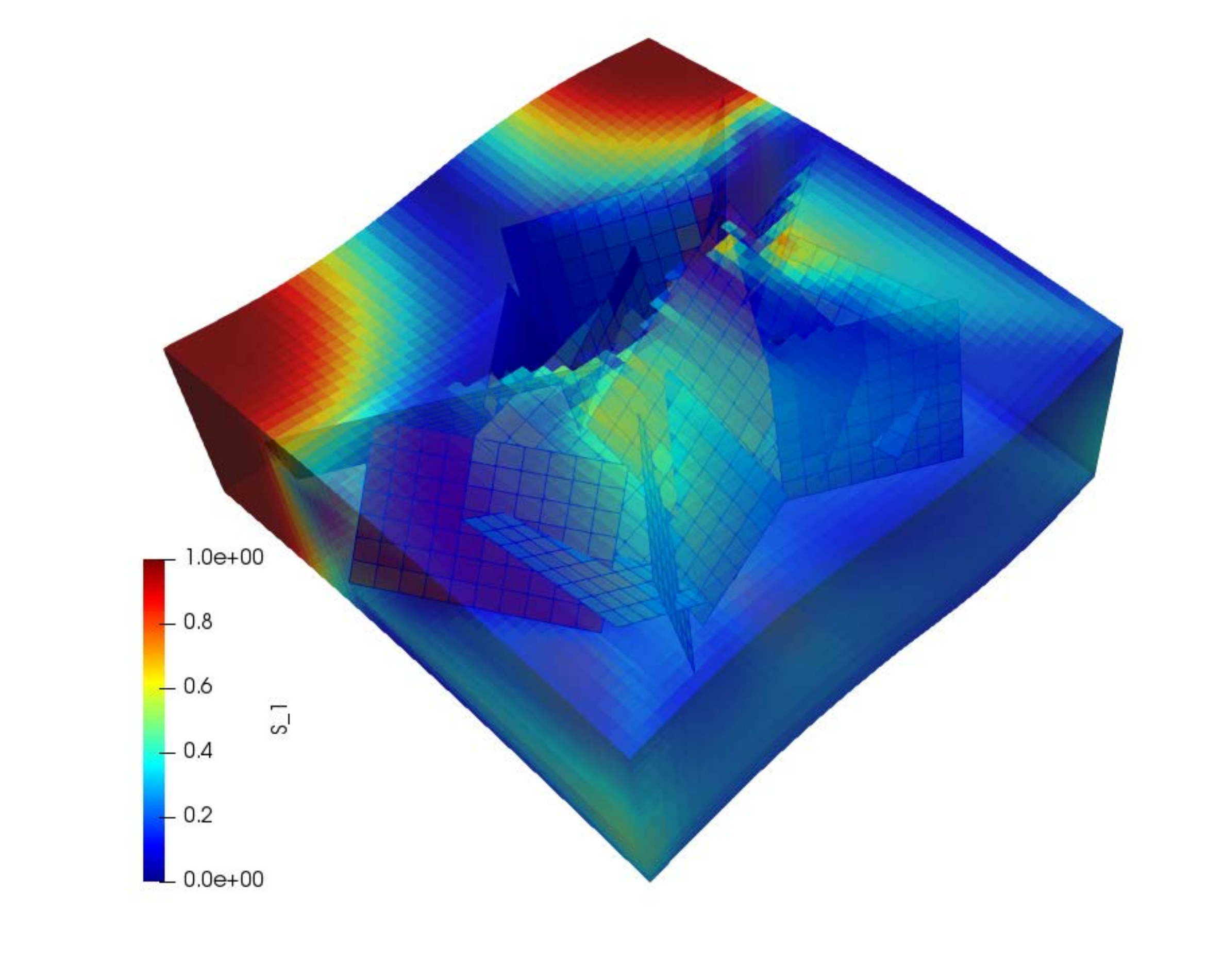}}
	\caption{Test case 3: A 3D fractured deformed box with non-orthogonal grid corner-point grid geometry. The figures \ref{Fig:pEDFM_CPG_TestCase3_K_1} and \ref{Fig:pEDFM_CPG_TestCase3_K_2} on top show the permeability of the fractures for the scenarios 1 and 2 respectively. The figure \ref{Fig:pEDFM_CPG_TestCase3_Fractured_Cells} on the bottom left illustrates the matrix grid cells that are overlapped by the fractures. Figure \ref{Fig:pEDFM_CPG_TestCase3_Wells} on the bottom right shows the geometry of injection and production wells.}
	\label{Fig:pEDFM_CPG_TestCase3_Results}
\end{figure}

The pressure and saturation results of scenario 1 are shown in figures \ref{Fig:pEDFM_CPG_TestCase3_P_1} and \ref{Fig:pEDFM_CPG_TestCase3_S_1} respectively (at the left side of figure \ref{Fig:pEDFM_CPG_TestCase3_Results}). As the grid geometry and the gridding system of this test case is not similar to the previous test case, it is not possible to compare the two test cases. The pressure and saturation distribution of the second scenario (at the same simulation time) can be observed at right side of the figure \ref{Fig:pEDFM_CPG_TestCase3_Results}.

In the first scenario, the flow barriers are close to the injection wells, thus restricting the displacement of the injecting phase towards the center of the domain (figure \ref{Fig:pEDFM_CPG_TestCase3_S_1}). Therefore, a high pressure gradient is visible near the injection wells (figure \ref{Fig:pEDFM_CPG_TestCase3_P_1}) as the low permeability fractures limit the flux through the domain. In the second scenario, the highly conductive fractures are near the injection wells, and the saturation profile increases through the whole thickness of the domain (figure \ref{Fig:pEDFM_CPG_TestCase3_S_2}). The pressure profile is uniformly distributed in the reservoir as there is no flux restriction near the wells (figure \ref{Fig:pEDFM_CPG_TestCase3_P_2}).

\subsection{Test Case 4: The Johansen formation}\label{Sec:TestCase4_Johansen}
The water-bearing Johansen formation was a potential candidate for CO\textsubscript{2} storage in a project promoted by the Norwegian government. The Norwegian continental margin has excellent potential for CO\textsubscript{2} storage options in saline aquifers.

\begin{figure}[!htbp]
	\centering
	\begin{subfigure}[]{0.17\textwidth}
		\includegraphics[width=\textwidth]{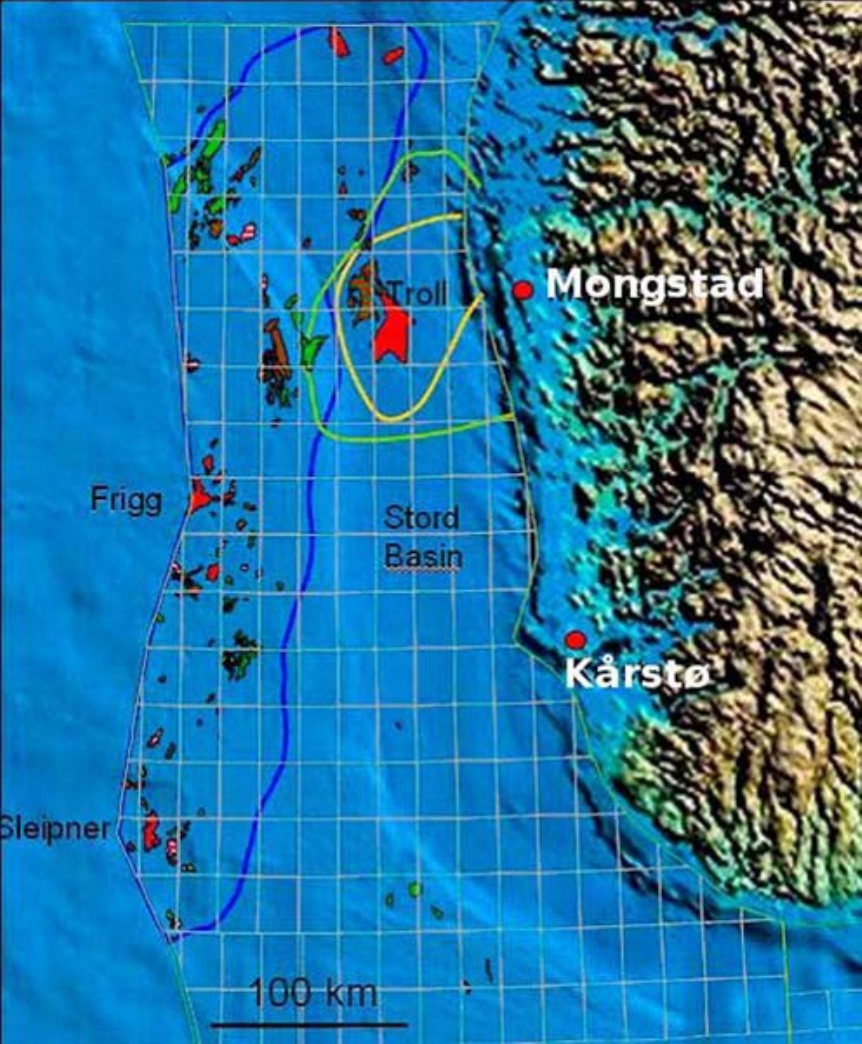}
	\end{subfigure}
	\begin{subfigure}[]{0.30\textwidth}
		\includegraphics[width=\textwidth]{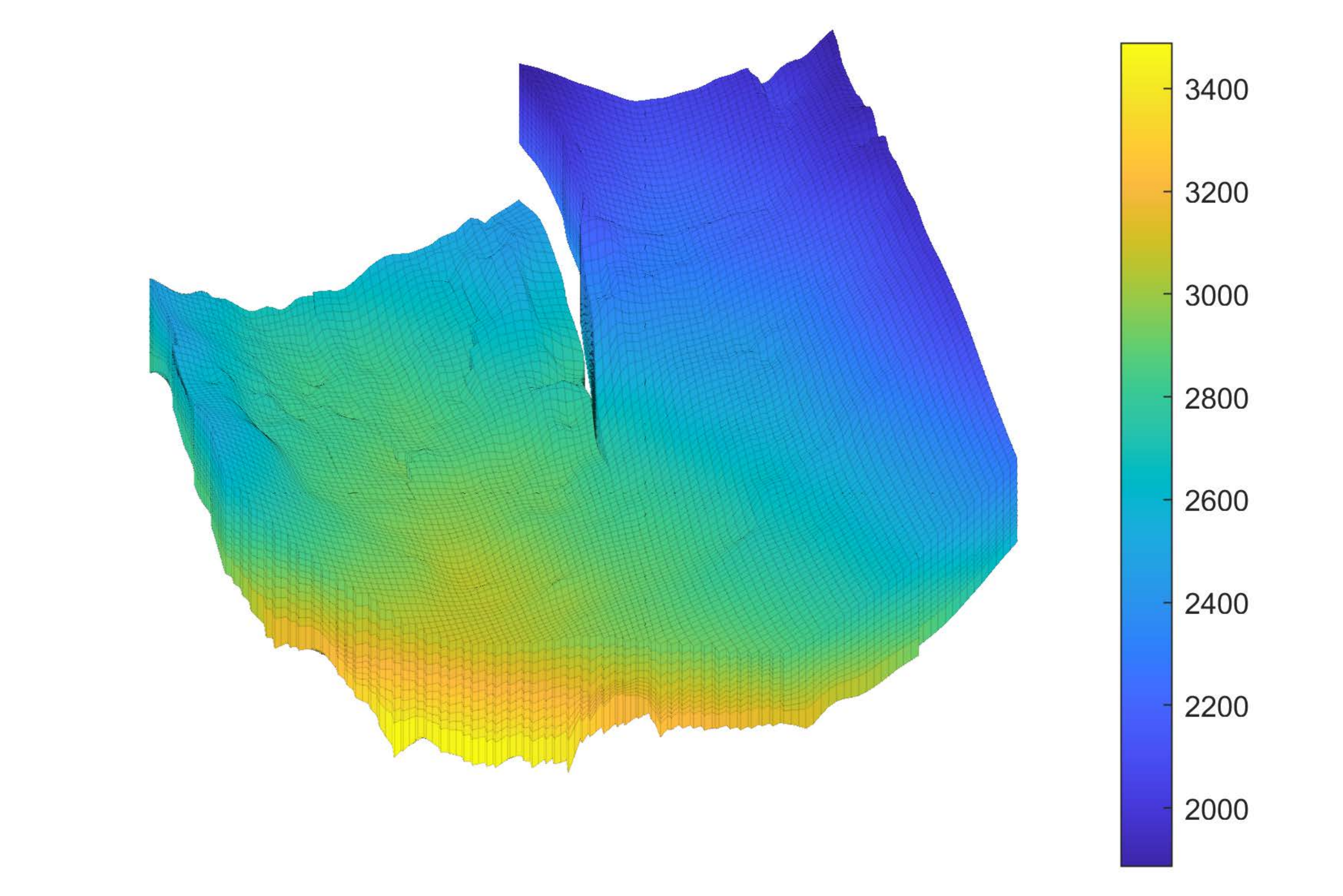}
	\end{subfigure}
	\caption{Test case 4: The location of the Johansen formation can be observed on the left figure. This formation is located within the green curve in the map, and the yellow curve represents areas where seismic data has been acquired (courtesy of Gassnova). The figure on the right shows the depth map of the Johansen model (NPD5 data set).}
	\label{Fig:pEDFM_CPG_TestCase4_Johansen_Location}
\end{figure}

The Johansen formation \cite{Eigestad2009} is located in the deeper part of the Sognefjord delta, $40–90 [\text{km}]$ offshore Mongstad on Norway's southwestern coast (show figure \ref{Fig:pEDFM_CPG_TestCase4_Johansen_Location}). It belongs to the Lower Jurassic Dunlin group and is interpreted as a laterally extensive sandstone, and it is overlaid by the Dunlin shale and below by the Amundsen shale. A saline aquifer exists in the depth levels ranging from $2200 [\text{m}]$ to $3100 [\text{m}]$ below the sea level. The depth range makes the formation ideal for CO\textsubscript{2} storage due to the pressure regimes existent in the field (providing a thermodynamical situation where CO\textsubscript{2} is in its supercritical phase). 

These formations have uniquely different permeabilities and perform very different roles in the CO\textsubscript{2} sequestration process. The Johansen sandstone has relatively high porosity and permeability, and it is suitable as a container to store CO\textsubscript{2}. The low-permeability overlaying Dunlin shale acts as a seal that avoids the CO\textsubscript{2} from leaking to the sea bottom layers.

The Johansen formation has an average thickness of nearly $100 [\text{m}]$, and the water-bearing region extends laterally up to $60 [\text{km}]$ in the east-west direction
and $100 [\text{km}]$ in the north-south direction. The aquifer has good sand quality with average porosities of roughly $25\%$. This implies that the Johansen formation's theoretical storage capacity can exceed one Gigaton of CO\textsubscript{2} providing the assumption of residual brine saturation of about $20\%$. The northwestern parts of the Johansen formation are located some $500 [\text{m}]$ below the operating Troll field, one of the North Sea's largest hydrocarbon fields.

\subsubsection{Data set}
\begin{figure}[!htbp]
	\centering
	\begin{subfigure}[]{0.23\textwidth}
		\includegraphics[width=\textwidth]{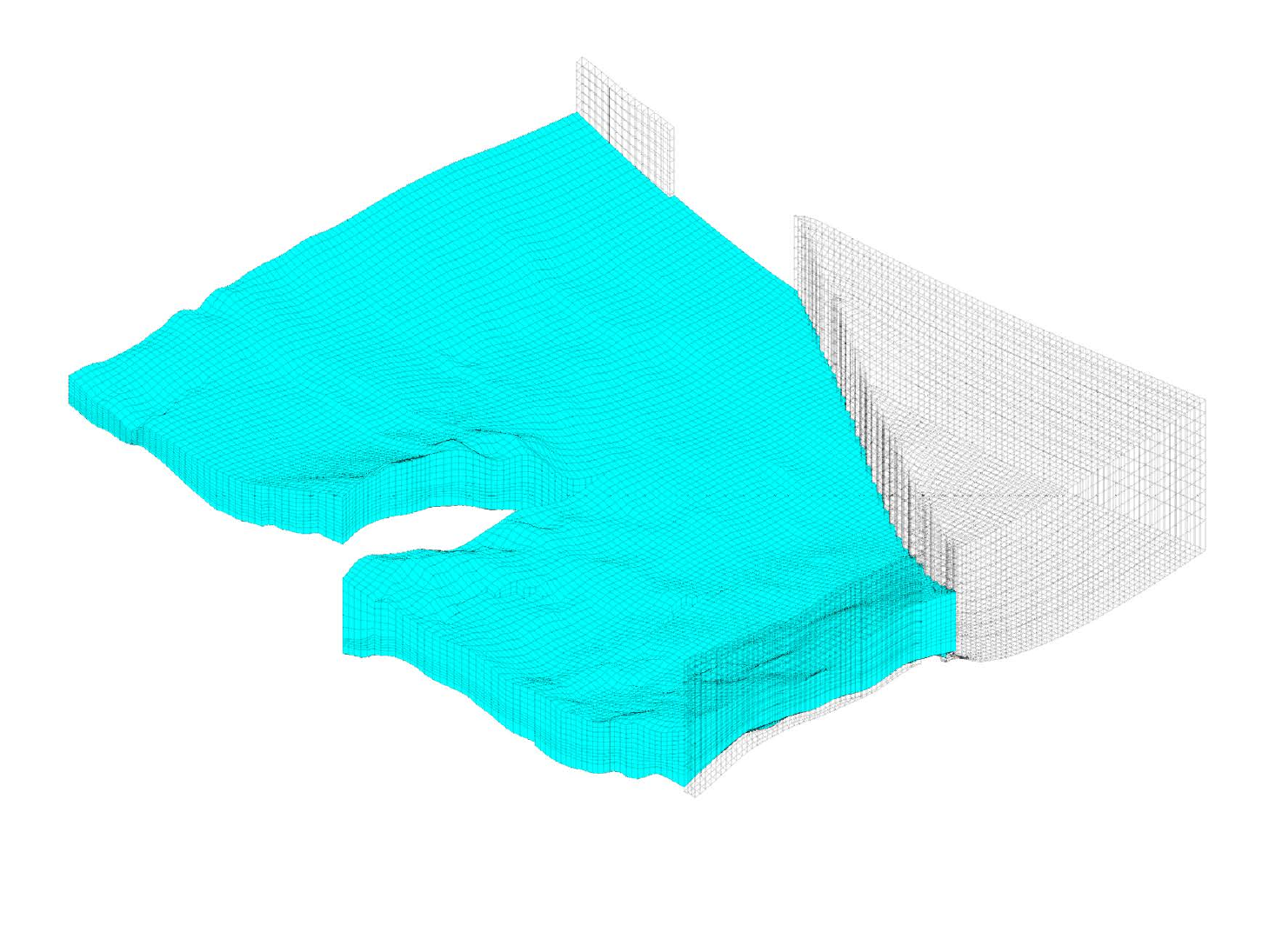}
	\end{subfigure}
	\begin{subfigure}[]{0.23\textwidth}
		\includegraphics[width=\textwidth]{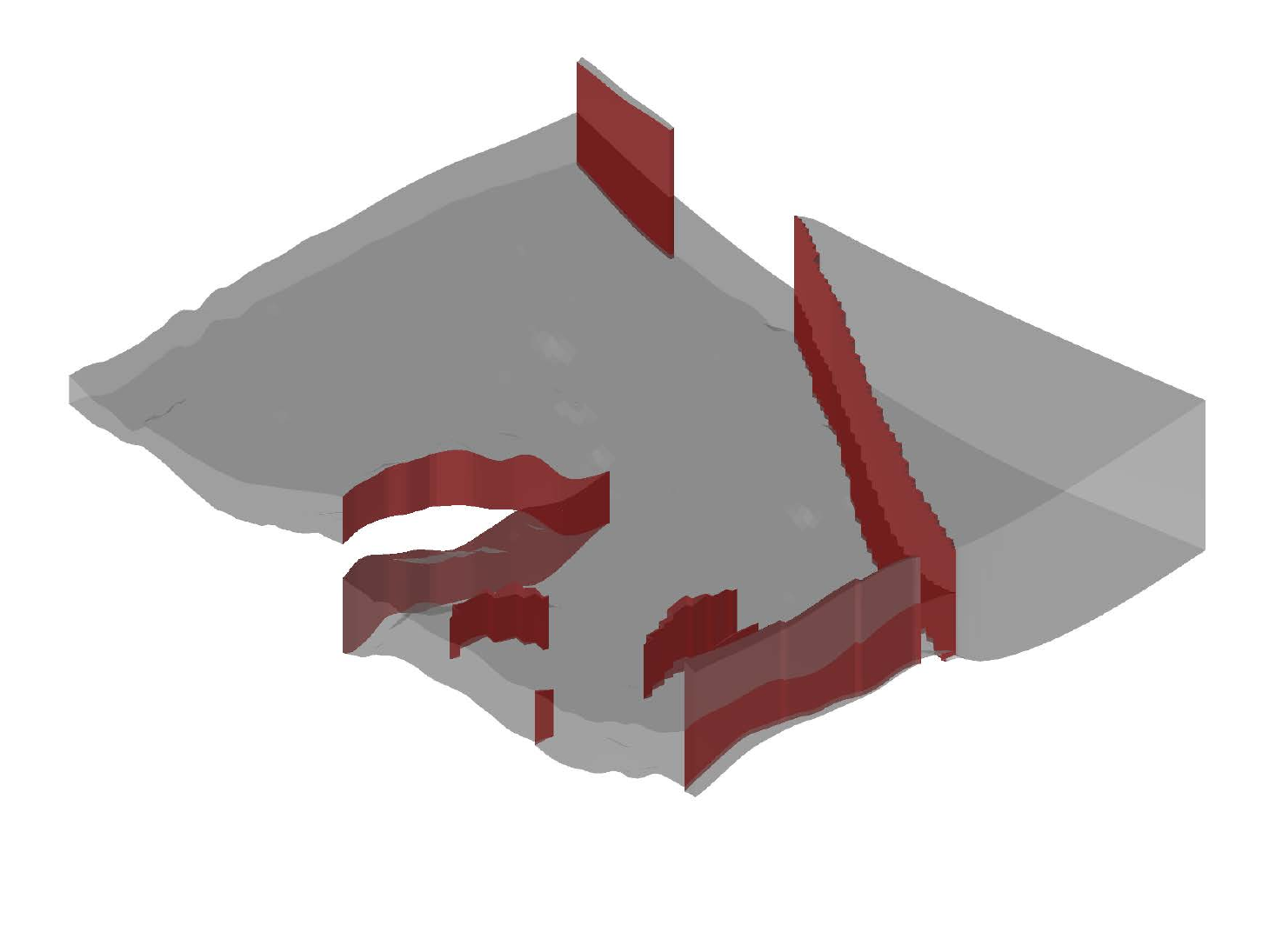}
	\end{subfigure}
	\caption{Test Case 4: Illustration of the Johansen model (NPD5 data set). The left figure represents the active section of the model or NPD5, highlighted with blue color, and the right figure shows the faults marked with red color.}
	\label{Fig:pEDFM_CPG_TestCase4_Johansen_Dataset}
\end{figure}

The MatMoRA project has created five models of the Johansen formation: one full-field model ($149 \times 189 \times 16$ grids), three homogeneous sector models ($100 \times 100 \times n$ for $n = 11,16,21$), and one heterogeneous sector model ($100 \times 100 \times 11$) also known as the NPD5 sector. In this work, the last data set (NPD5) has been used. The NPD5 sector can be seen in figure \ref{Fig:pEDFM_CPG_TestCase4_Johansen_Dataset}. In the left side of this figure, the NPD5 sector is highlighted with blue color.

In the discretized computational grids, the Johansen formation is represented by five layers of grid cells. The Amundsen shale below the Johansen formation and the low-permeable Dunlin shale above are characterized by one and five cell layers, respectively. The Johansen formation consists of approximately $80\%$ sandstone and $20\%$ claystone, whereas the Amundsen formation consists of siltstones and shales, and the Dunlin group has high clay and silt content.

\subsubsection{Rock properties}
\begin{figure}[!htbp]
	\centering
	\subcaptionbox{{\footnotesize Porosity of the entire formation}\label{Fig:pEDFM_CPG_TestCase4_Johansen_Porosity_Full}}
	{\includegraphics[width=0.23\textwidth]{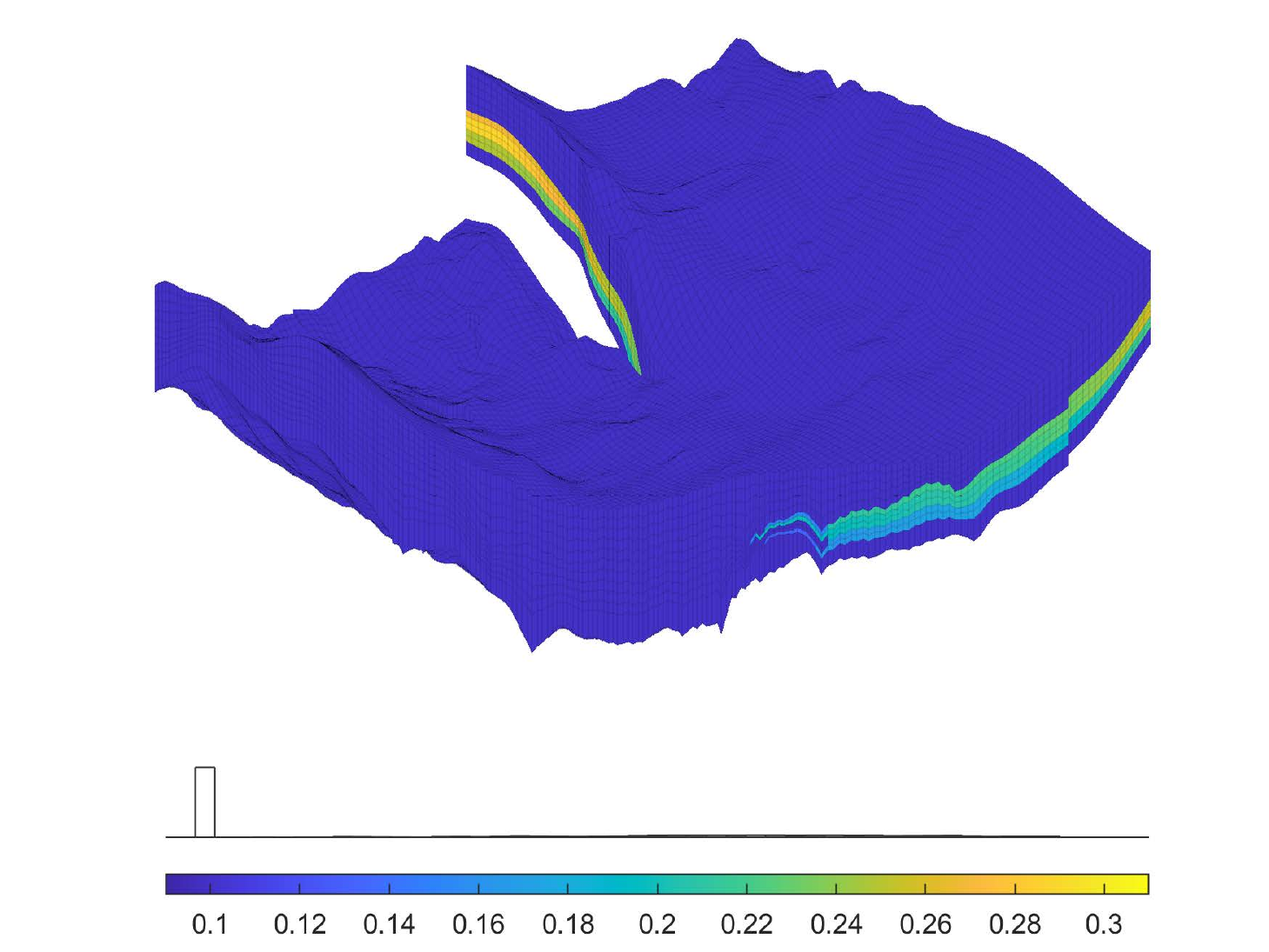}}
	\subcaptionbox{{\footnotesize Porosity of the NPD5 sector}\label{Fig:pEDFM_CPG_TestCase4_Johansen_Porosity_NPD5}}
	{\includegraphics[width=0.23\textwidth]{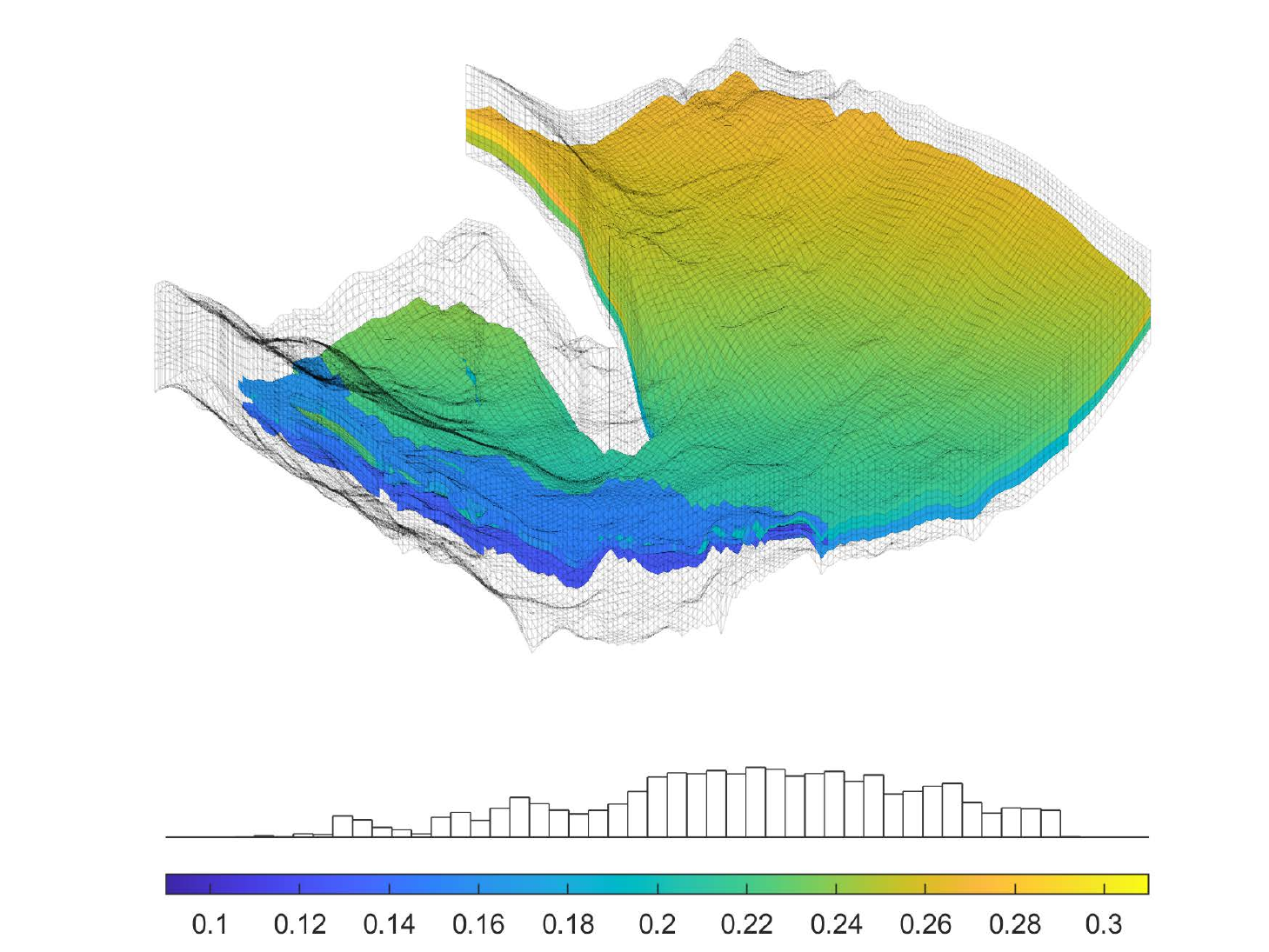}}
	\\
	\subcaptionbox{{\footnotesize Permeability of the entire formation}\label{Fig:pEDFM_CPG_TestCase4_Johansen_Permeability_Full}}
	{\includegraphics[width=0.23\textwidth]{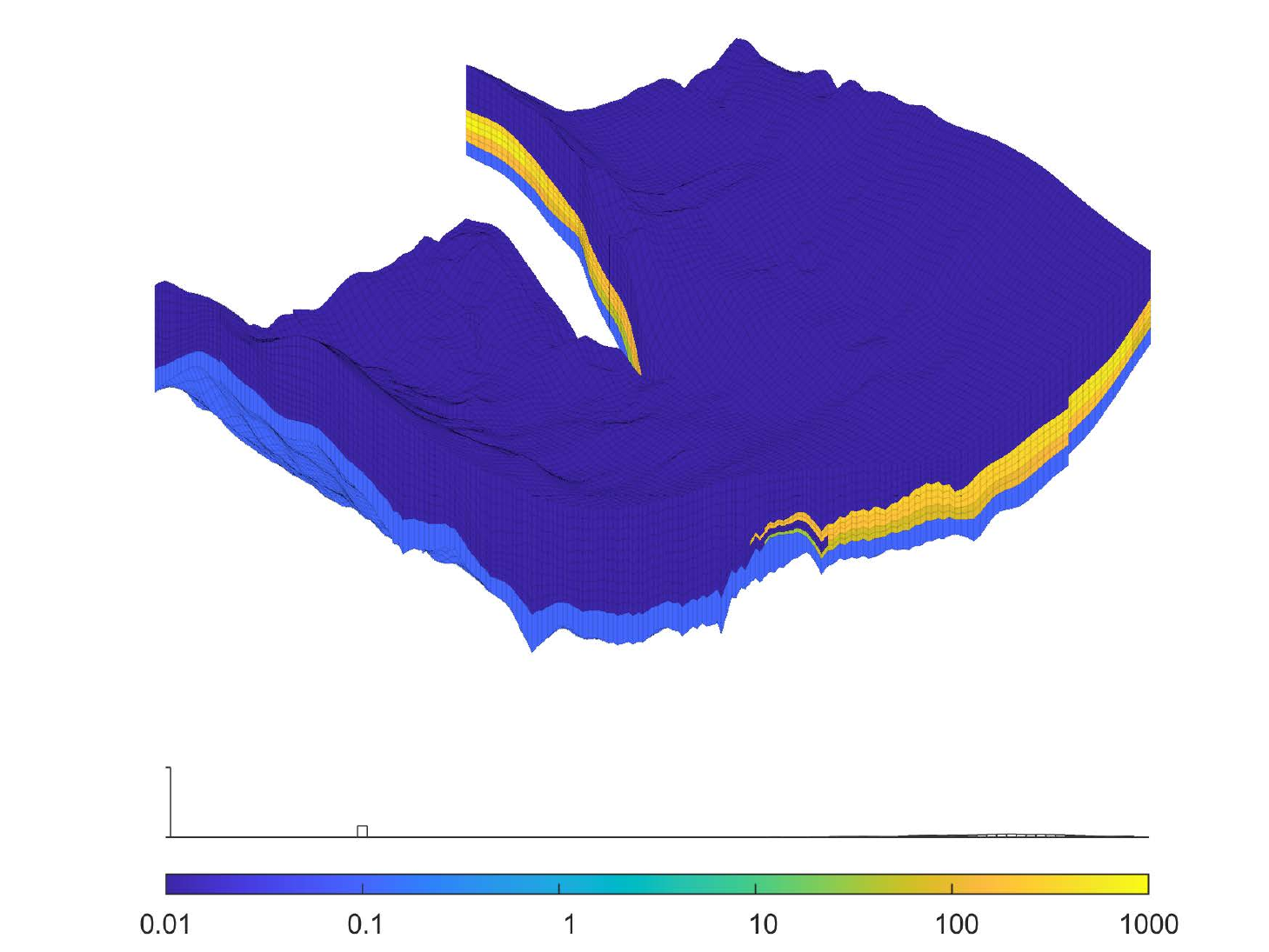}}
	\subcaptionbox{{\footnotesize Permeability of the NPD5 sector}\label{Fig:pEDFM_CPG_TestCase4_Johansen_Permeability_NPD5}}
	{\includegraphics[width=0.23\textwidth]{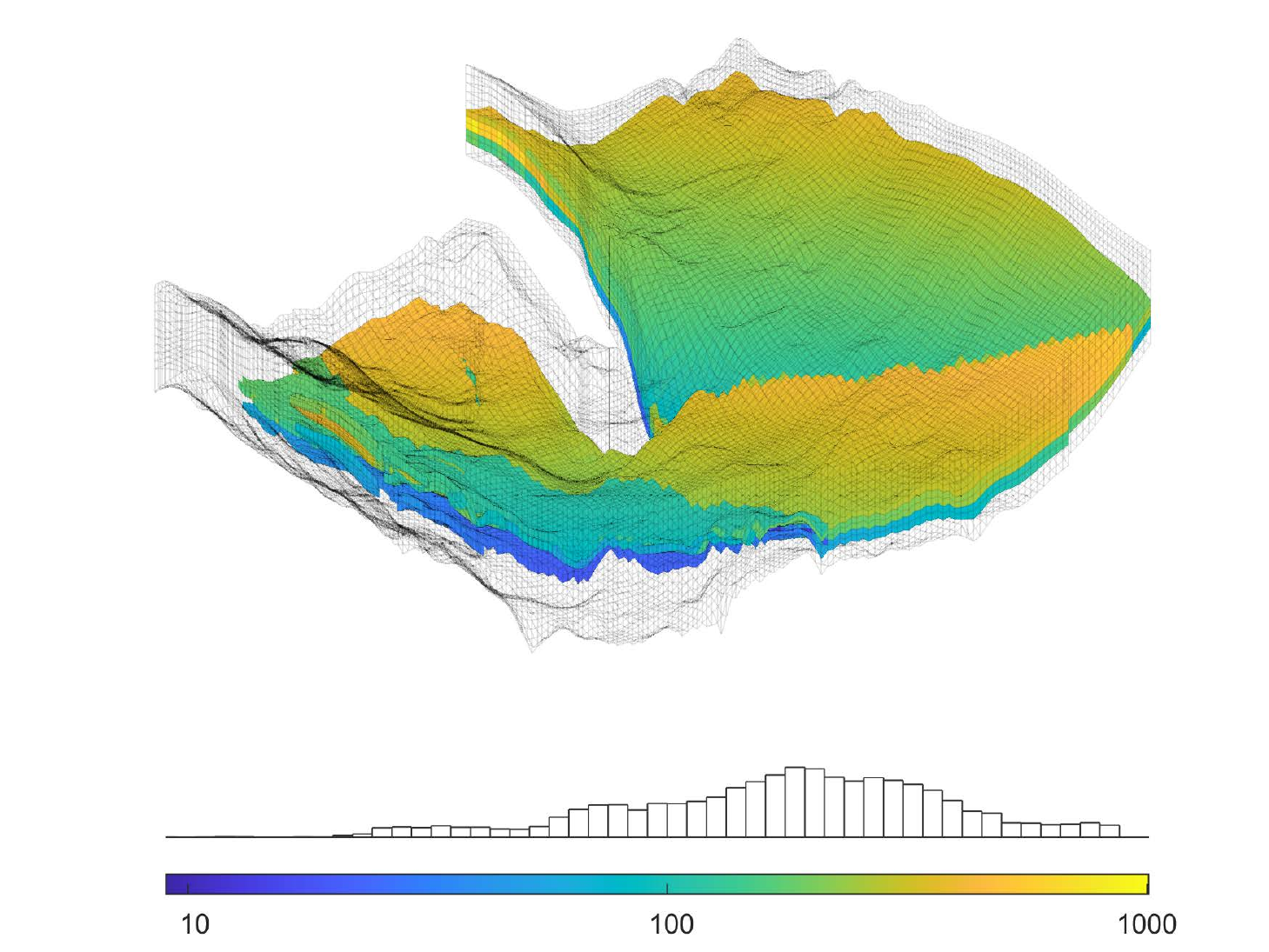}}
	\caption{Test Case 4: Porosity and permeability of the entire Johansen formation and the NPD5 sector). The left figures show porosity and permeability of the whole model, whereas the right figures show the porosity and permeability of the NPD5 sector (sandstone).}
	\label{Fig:pEDFM_CPG_TestCase4_Johansen_Porosity_Permeability}
\end{figure}

The Johansen sandstone is a structure with a wedge shape pinched out in the front part of the model and divided into two sections at the back. Figure \ref{Fig:pEDFM_CPG_TestCase4_Johansen_Porosity_Permeability} shows two different selections, i.e., the entire formation (figures \ref{Fig:pEDFM_CPG_TestCase4_Johansen_Porosity_Full} and \ref{Fig:pEDFM_CPG_TestCase4_Johansen_Permeability_Full}) and the NPD5 sector of the formation (figures \ref{Fig:pEDFM_CPG_TestCase4_Johansen_Porosity_NPD5} and \ref{Fig:pEDFM_CPG_TestCase4_Johansen_Permeability_NPD5}).

The porosity map of the entire model is visible in the figure \ref{Fig:pEDFM_CPG_TestCase4_Johansen_Porosity_Full}, and figure \ref{Fig:pEDFM_CPG_TestCase4_Johansen_Porosity_NPD5} shows the cells with porosity values larger than $0.1$ that belongs to Johansen formation. Similarly, the permeability map of the entire formation is shown in figures \ref{Fig:pEDFM_CPG_TestCase4_Johansen_Permeability_Full}, and figure \ref{Fig:pEDFM_CPG_TestCase4_Johansen_Permeability_NPD5} illustrates the permeability of the NPD5 sector where the Dunlin shale above the Johansen and the Amundsen shale below the Johansen formation are excluded. The permeability tensor is diagonal, with the vertical permeability equivalent to one-tenth of the horizontal permeability. In both graphs, the permeability is represented in a logarithmic color scale.

\subsubsection{Simulation results}
\begin{figure}[!htbp]
	\centering
	\subcaptionbox{{\footnotesize Fractures permeability (scenario 1)}\label{Fig:pEDFM_CPG_TestCase4_Johansen_Kf_1}}
	{\includegraphics[width=0.23\textwidth]{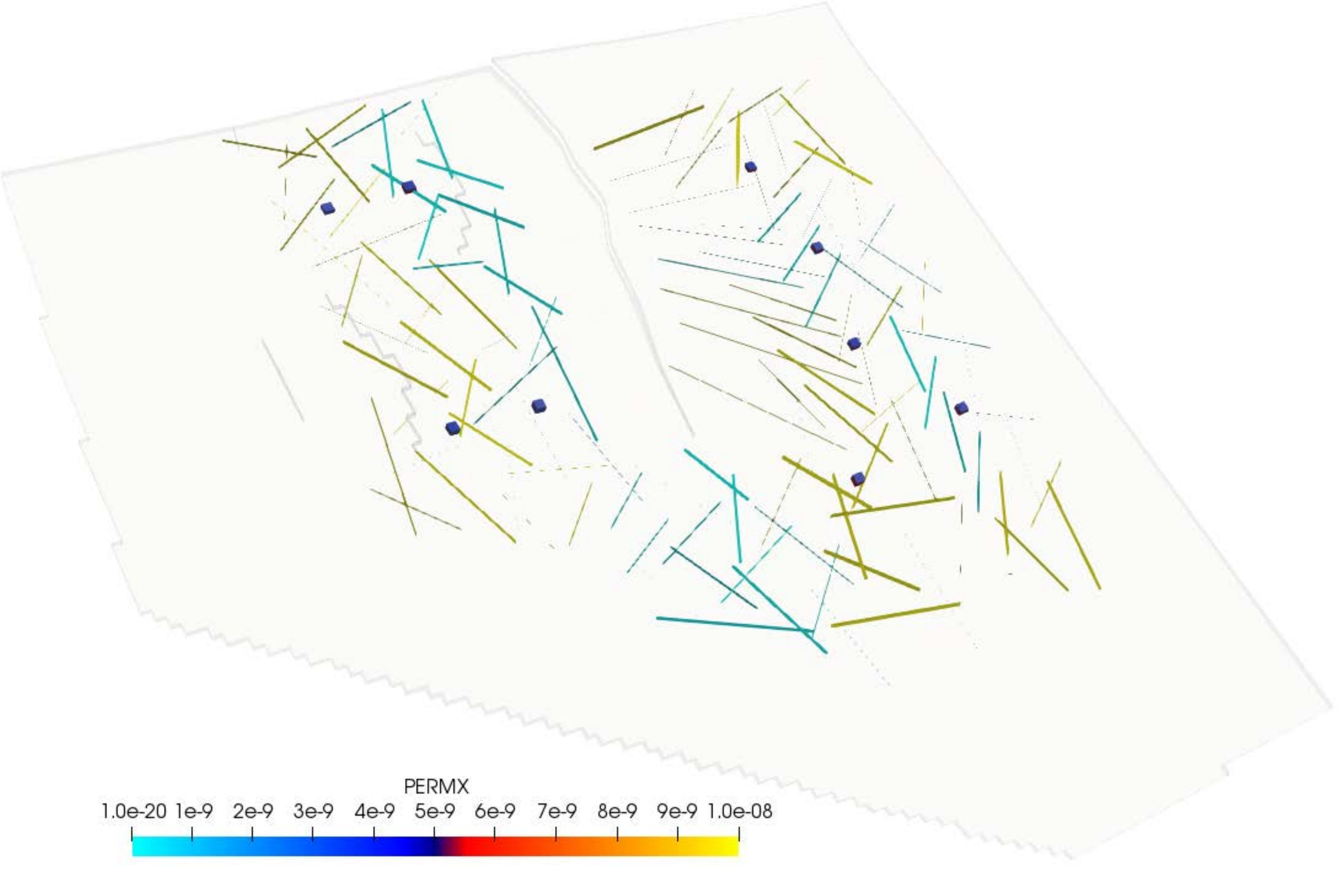}}
	\subcaptionbox{{\footnotesize Fractures permeability (scenario 2)}\label{Fig:pEDFM_CPG_TestCase4_Johansen_Kf_2}}
	{\includegraphics[width=0.23\textwidth]{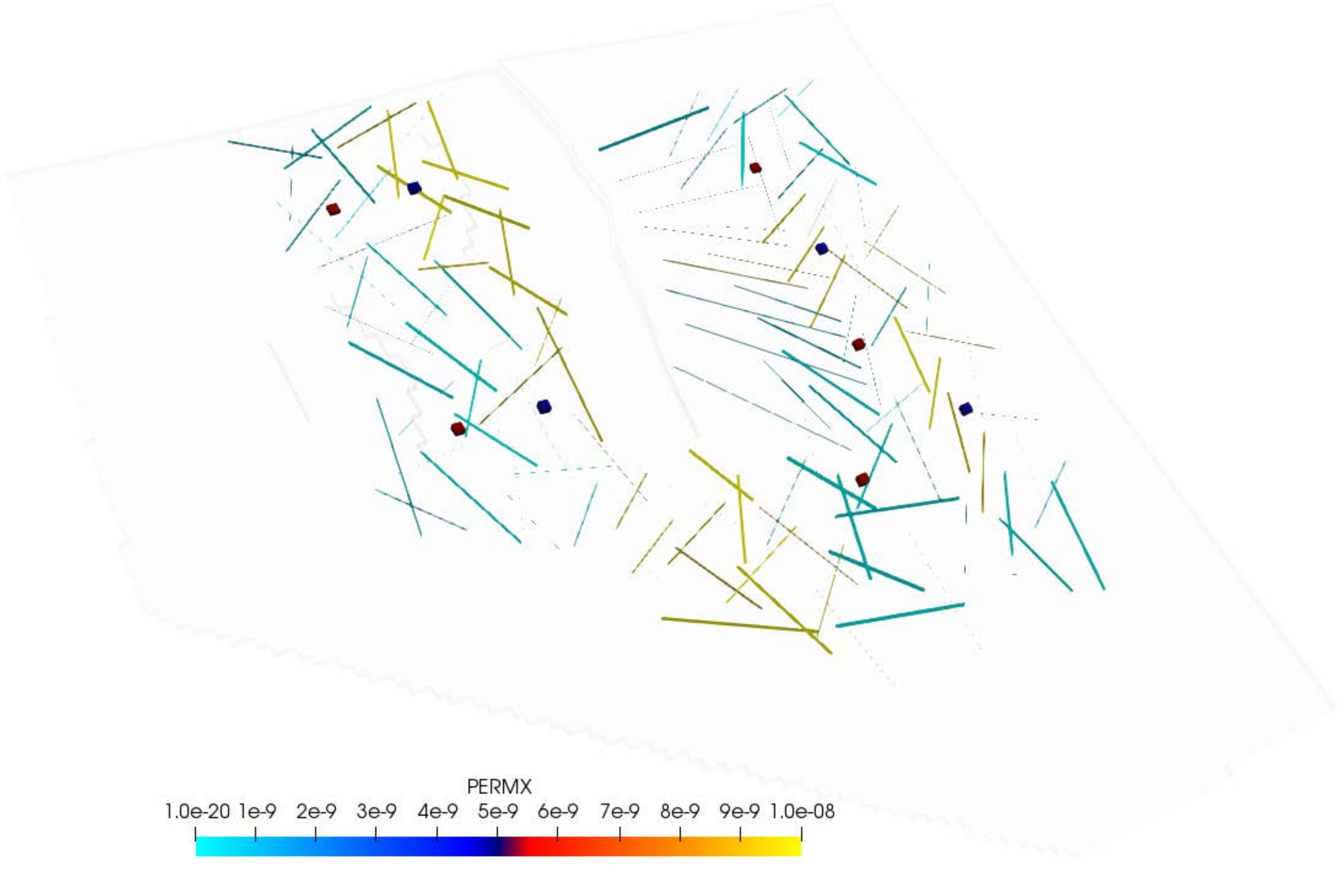}}
	\\
	\subcaptionbox{{\footnotesize Matrix cells overlapped by fractures}\label{Fig:pEDFM_CPG_TestCase4_Johansen_Fractured_Cells}}
	{\includegraphics[width=0.23\textwidth]{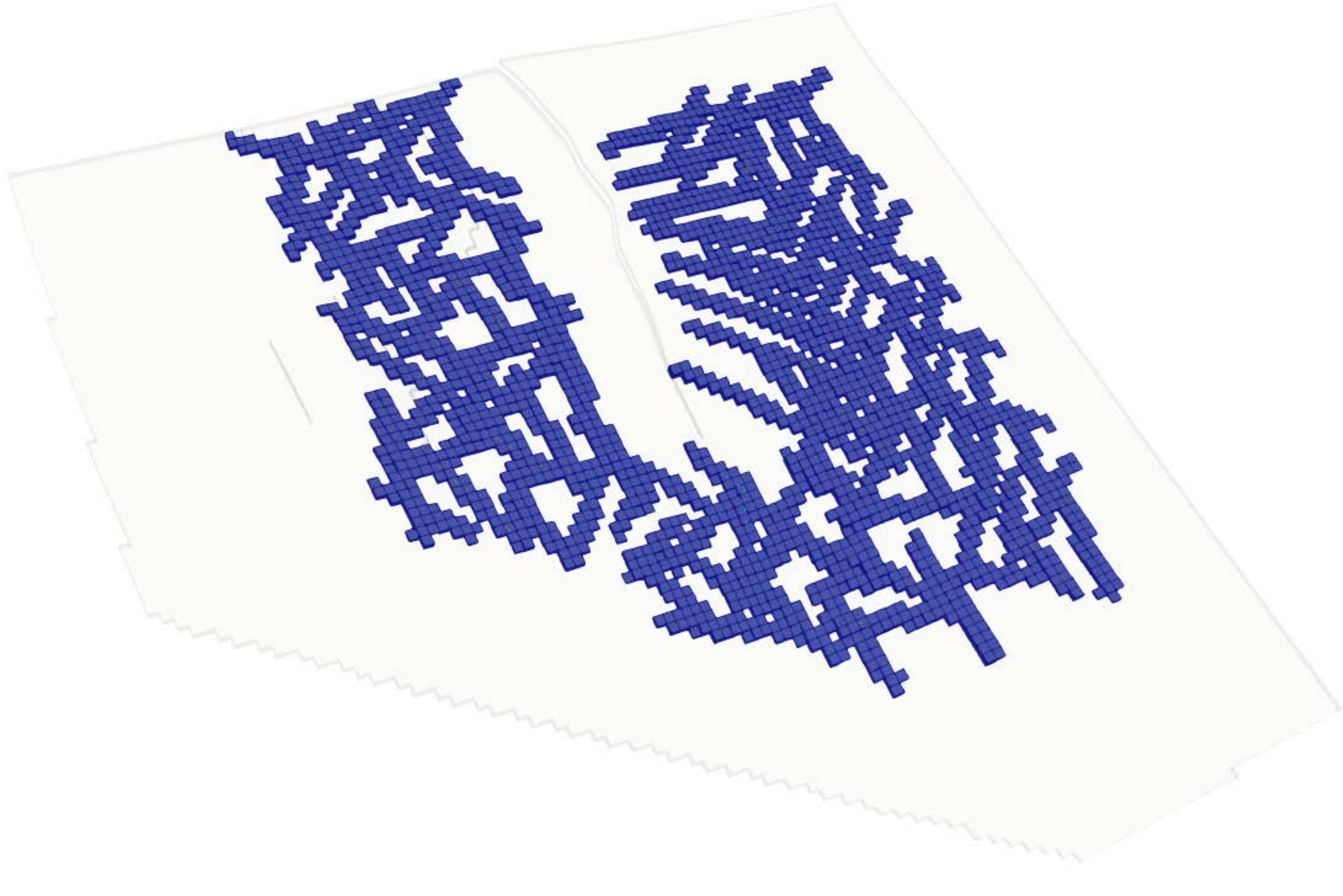}}
	\subcaptionbox{{\footnotesize Location of injection and production wells}\label{Fig:pEDFM_CPG_TestCase4_Johansen_Wells}}
	{\includegraphics[width=0.23\textwidth]{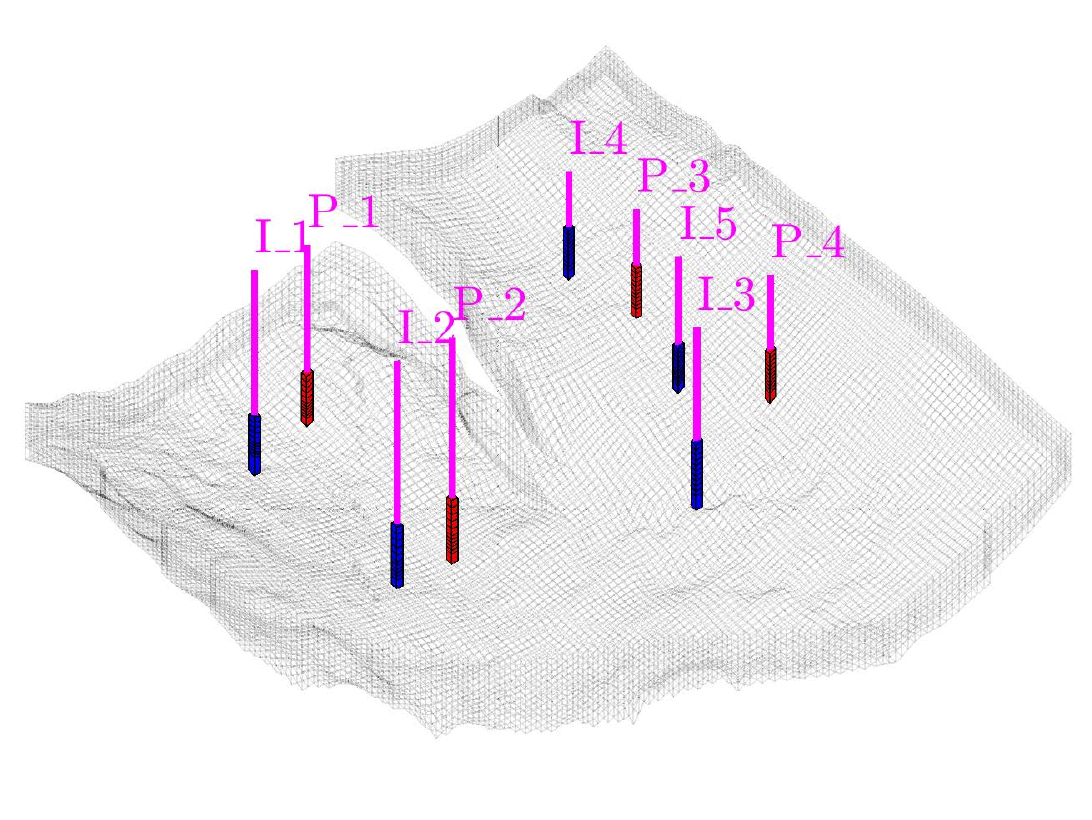}}
	\caption{Test case 4: The Johansen formation with $9$ wells and a set of 150 synthetic fractures (with mixed conductivities). The figures on top show the fractures network with different permeabilities for scenario $1$ (top left) and scenario $2$ (top right). The figure at bottom left illustrates the highlighted matrix cells that are overlapped by the fractures network. And the figure at bottom right shows the schematics of the injection and production wells.}
	\label{Fig:pEDFM_CPG_TestCase4_Johansen_Wells_Fractures}
\end{figure}

The "NPD5" sector of the Johansen formation model \cite{lie2019introduction} is used in the following test case. It is a corner-point grid reservoir model that consists of $100 \times 100 \times 11$ grid cells from which $88775$ grid cells are active. The rock properties of the Johansen formation available on public data were given as input in the simulation. A network of 121 fractures is embedded in the reservoir geological data set that contains both highly conductive fractures and flow barriers with permeability of $K_{f_{max}} = 10^{-8} \, [\text{m}^2]$ and $K_{f_{min}} = 10^{-20} \, [\text{m}^2]$ respectively. The model is bounded by two shale formations. Therefore the fractures were placed inside the Johansen formation (layers $6$ to $10$). In total $150$ fractures are embedded in the model, and the fracture network consists of $3494$ grid cells (in total $92269$ grid cells for matrix and fractures). Five injection wells with pressure of $p_{\text{inj}} = 5 \times 10^7 \, [\text{Pa}]$ and four production wells with pressure of $p_{\text{inj}} = 1 \times 10^7 \, [\text{Pa}]$ were placed in the model. Wells are vertical and drilled through the entire thickness of the model. Figure \ref{Fig:pEDFM_CPG_TestCase4_Johansen_Wells} illustrates the location of the injection and production wells in this test case.

Two scenarios are considered with two different fracture networks of mixed conductivities. While the geometry of both fracture networks is identical, the permeability values of the fractures from scenario $1$ are inverted for the scenario $2$. This implies that the highly conductive fractures in the fractures network of scenario $1$ act as flow barriers in the 2nd scenario and the flow barriers of scenario $1$ are modified to be highly conductive fractures in the scenario $2$. Figures \ref{Fig:pEDFM_CPG_TestCase4_Johansen_Kf_1} and \ref{Fig:pEDFM_CPG_TestCase4_Johansen_Kf_2} display the fractures networks of scenario $1$ and scenario $2$ respectively. The matrix grid cells
overlapped by the fractures are visible in figure \ref{Fig:pEDFM_CPG_TestCase4_Johansen_Fractured_Cells}.

\begin{figure}[!htbp]
	\centering
	\subcaptionbox{{\footnotesize Pressure in the matrix}\label{Fig:pEDFM_CPG_TestCase4_Johansen_Pm_scenario1}}
	{\includegraphics[width=0.23\textwidth]{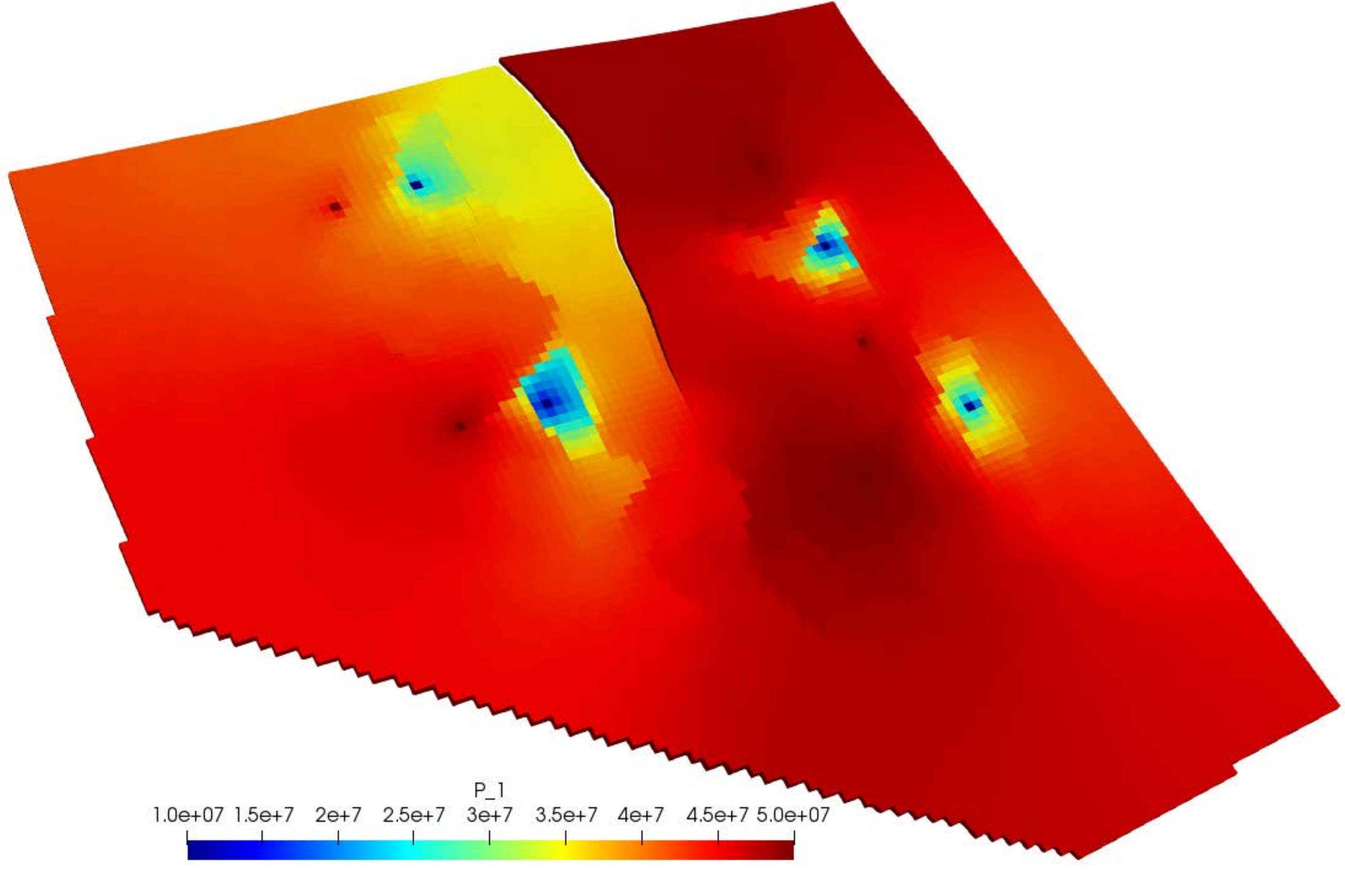}}
	\subcaptionbox{{\footnotesize Pressure in the fractures}\label{Fig:pEDFM_CPG_TestCase4_Johansen_Pf_scenario1}}
	{\includegraphics[width=0.23\textwidth]{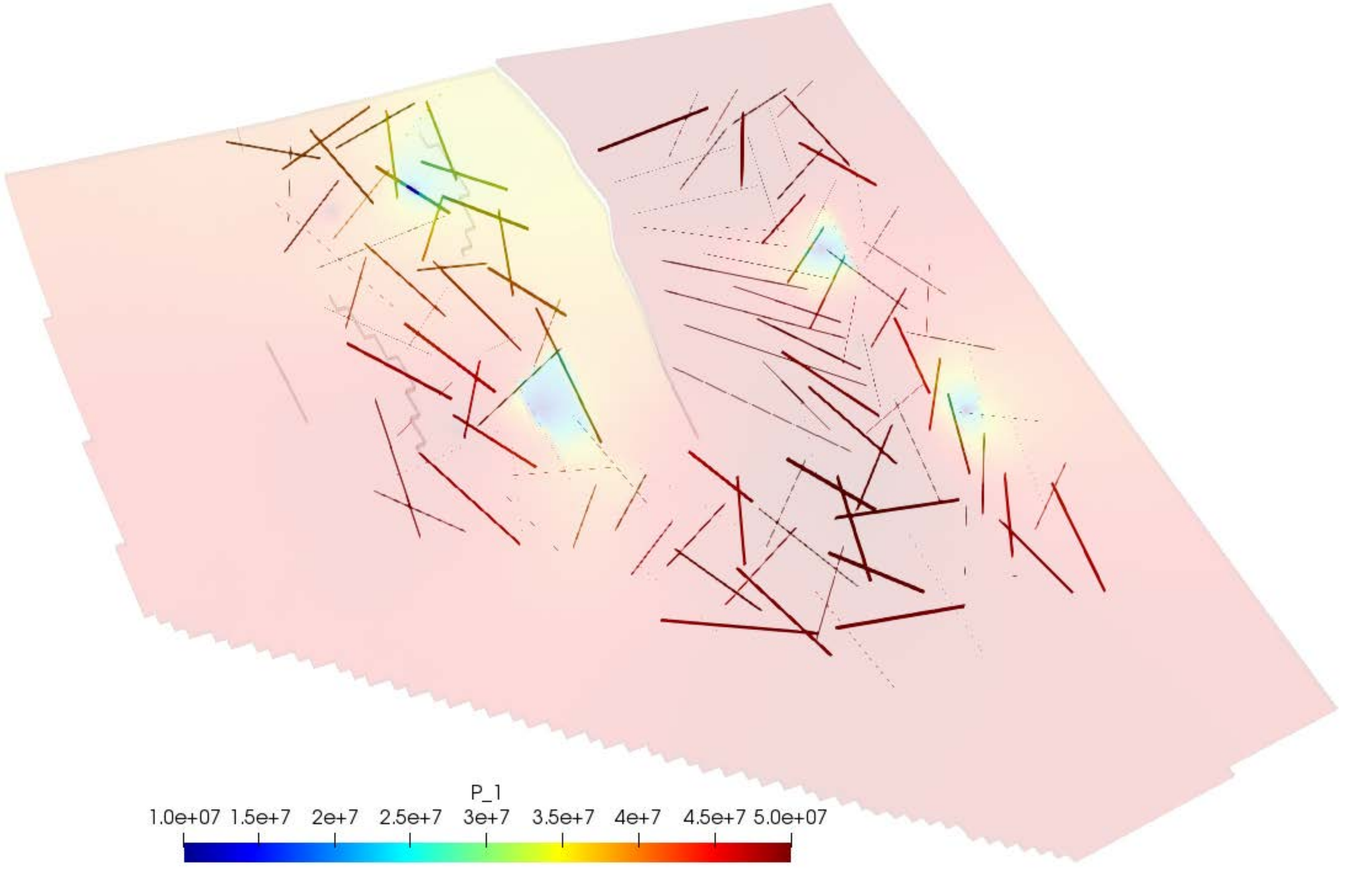}}
	\caption{Test case 4: The pressure profile of the Johansen formation for the simulation scenario $1$. The figure on the left illustrates the pressure distribution in the matrix grid cells. The transparency of this figure is increased to make the fractures visible and to display the pressure profile in the fractures in the figure on the right.}
	\label{Fig:pEDFM_CPG_TestCase4_Johansen_Pressure_Scenario1}
\end{figure}

\begin{figure}[!htbp]
	\centering
	\subcaptionbox{{\footnotesize Saturation in the matrix after $5000 [\text{days}]$ }\label{Fig:pEDFM_CPG_TestCase4_Johansen_Sm_scenario1_T10}}
	{\includegraphics[width=0.23\textwidth]{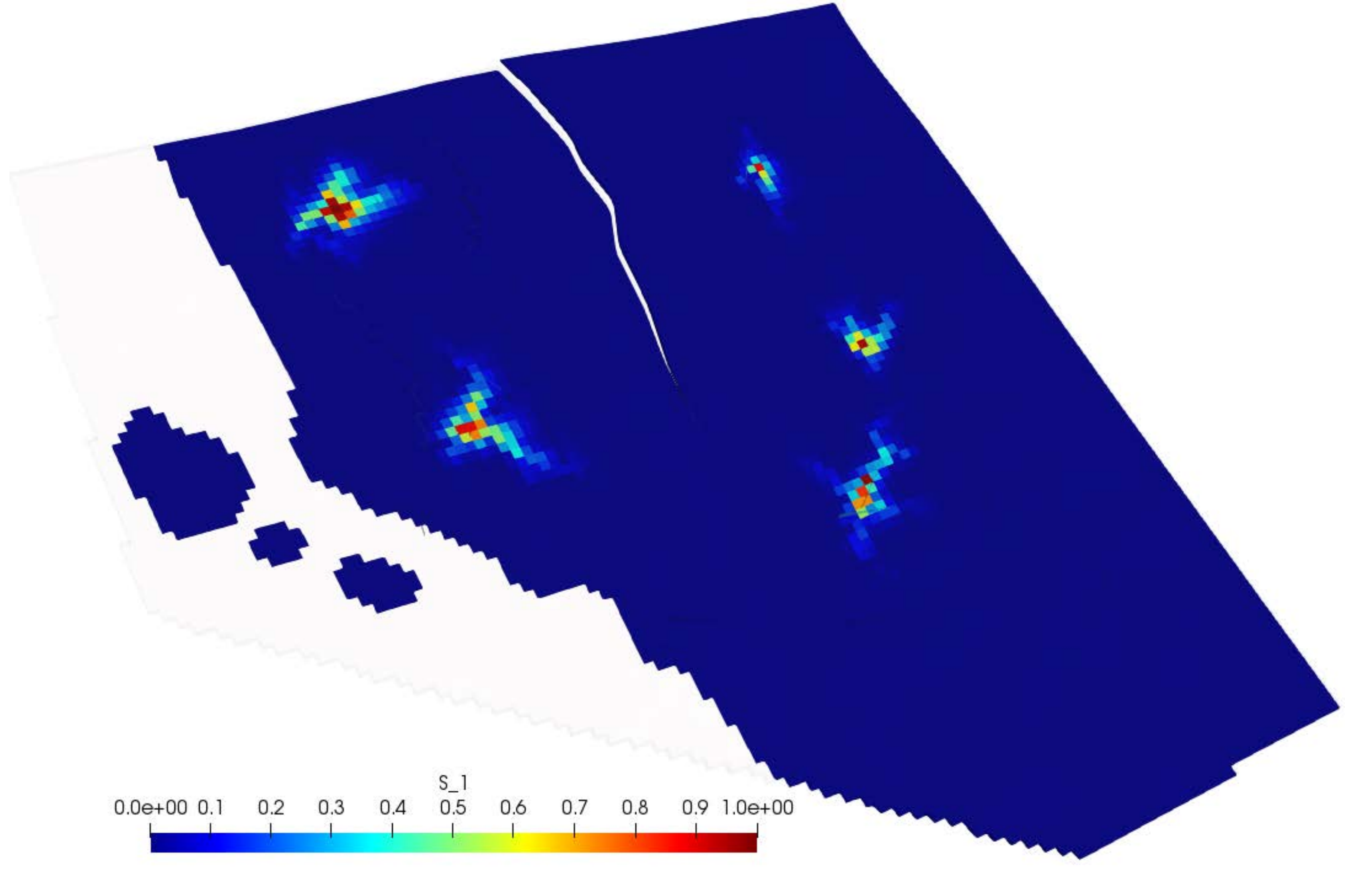}}
	\subcaptionbox{{\footnotesize Saturation in the fractures after $5000 [\text{days}]$ }\label{Fig:pEDFM_CPG_TestCase4_Johansen_Sf_scenario1_T10}}
	{\includegraphics[width=0.23\textwidth]{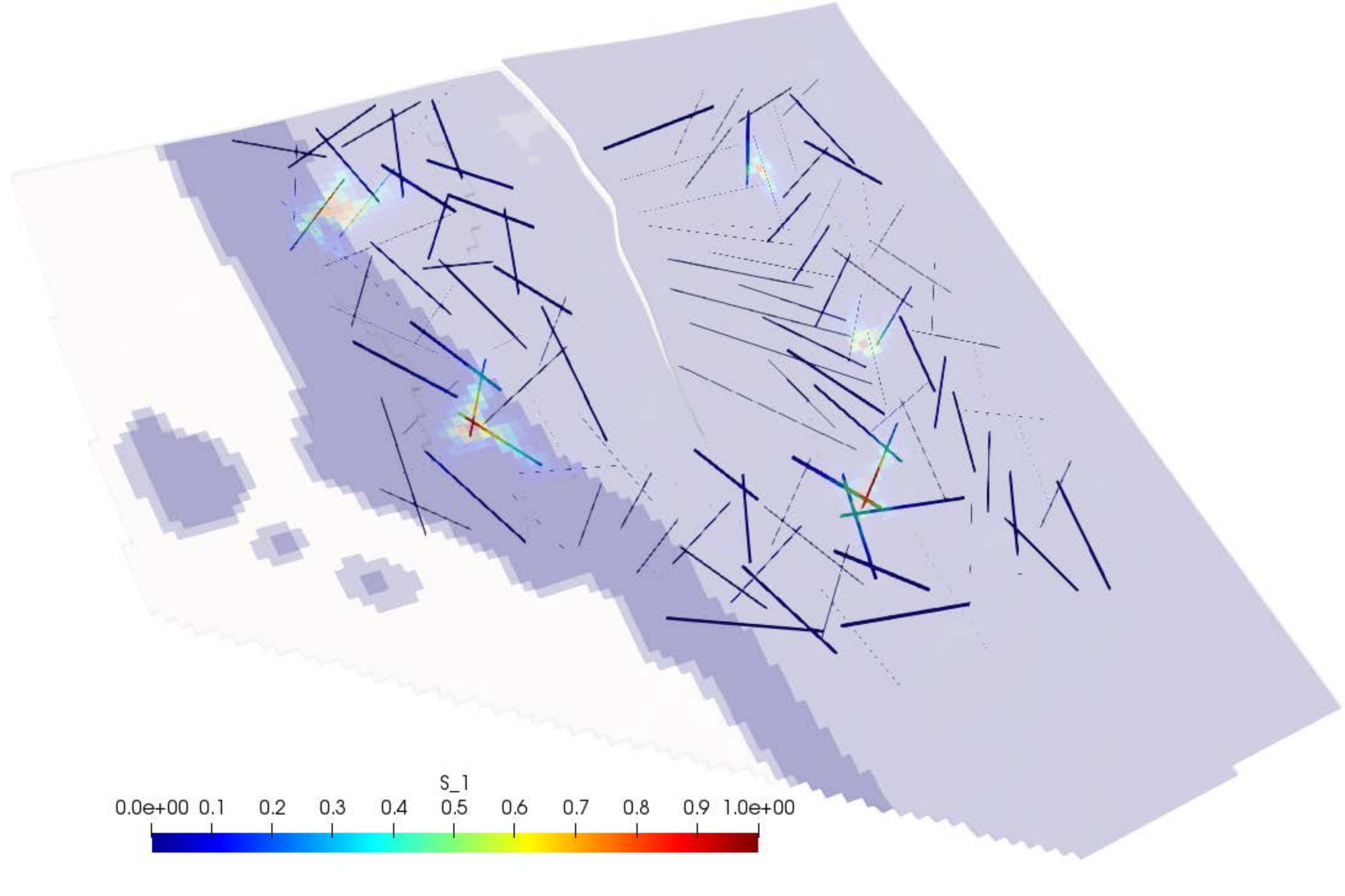}}
	\\
	\subcaptionbox{{\footnotesize Saturation in the matrix after $20000 [\text{days}]$ }\label{Fig:pEDFM_CPG_TestCase4_Johansen_Sm_scenario1_T40}}
	{\includegraphics[width=0.23\textwidth]{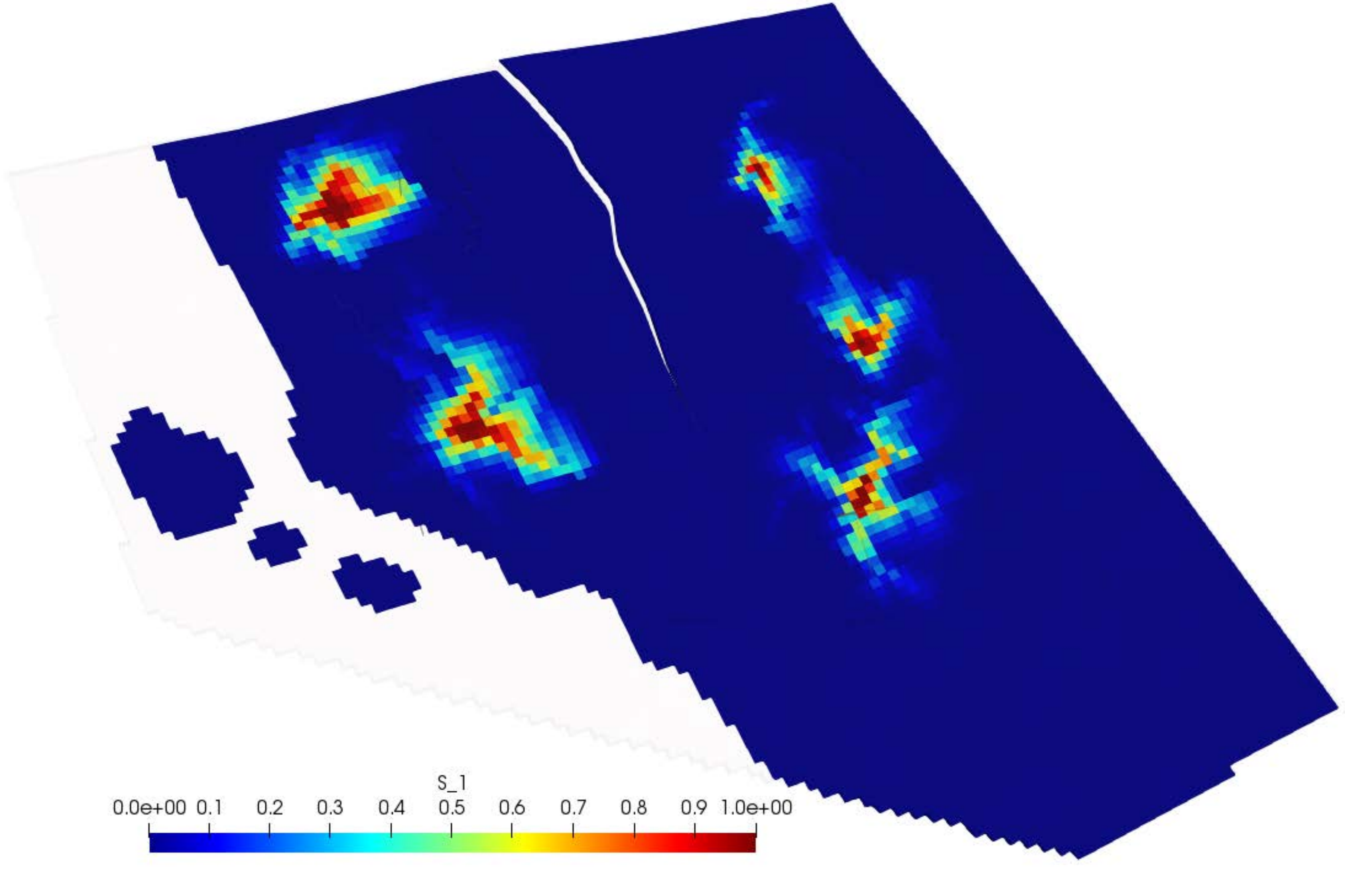}}
	\subcaptionbox{{\footnotesize Saturation in the fractures after $20000 [\text{days}]$ }\label{Fig:pEDFM_CPG_TestCase4_Johansen_Sf_scenario1_T40}}
	{\includegraphics[width=0.23\textwidth]{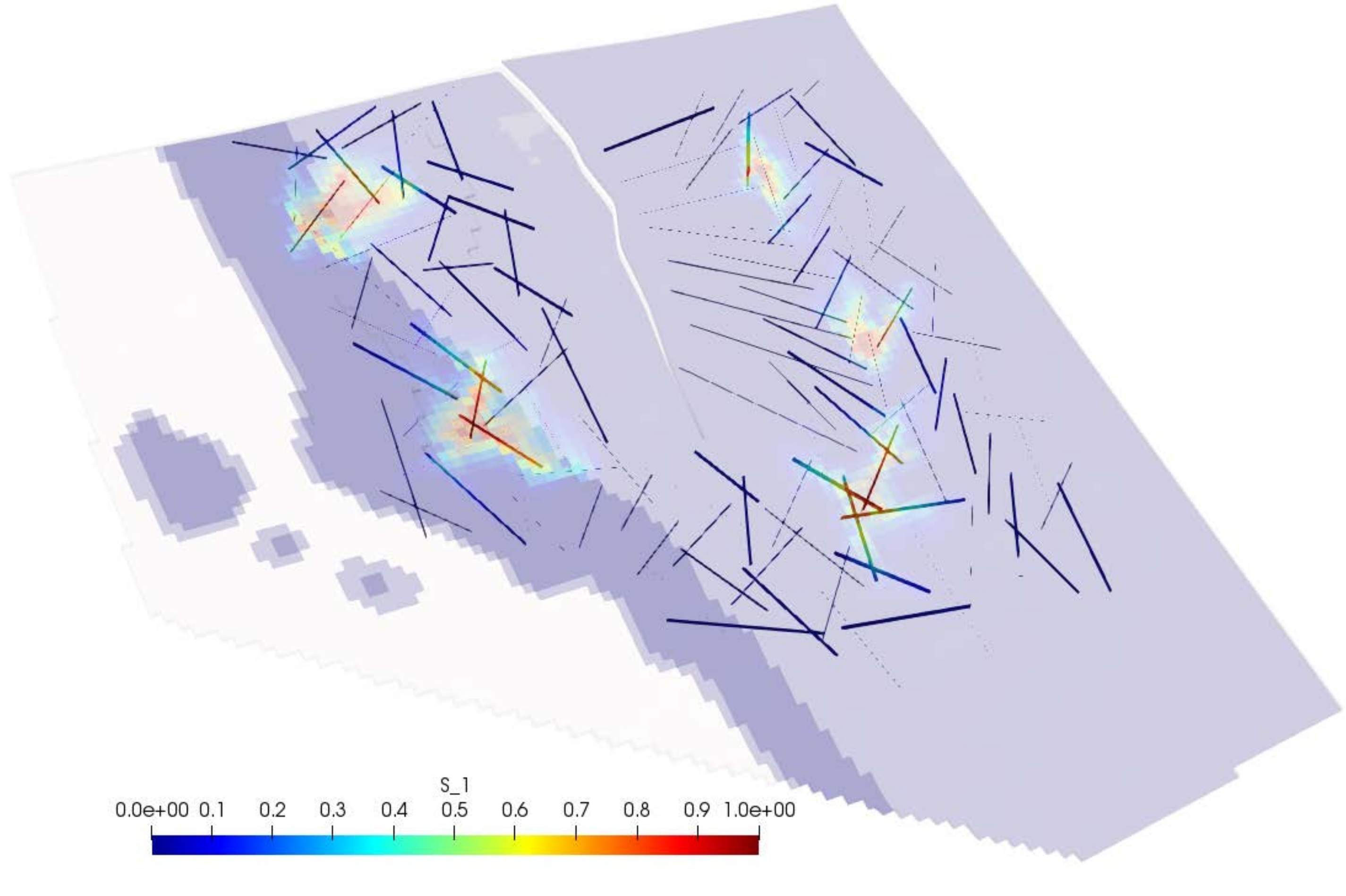}}
	\\
	\subcaptionbox{{\footnotesize Saturation in the matrix after $35000 [\text{days}]$ }\label{Fig:pEDFM_CPG_TestCase4_Johansen_Sm_scenario1_T70}}
	{\includegraphics[width=0.23\textwidth]{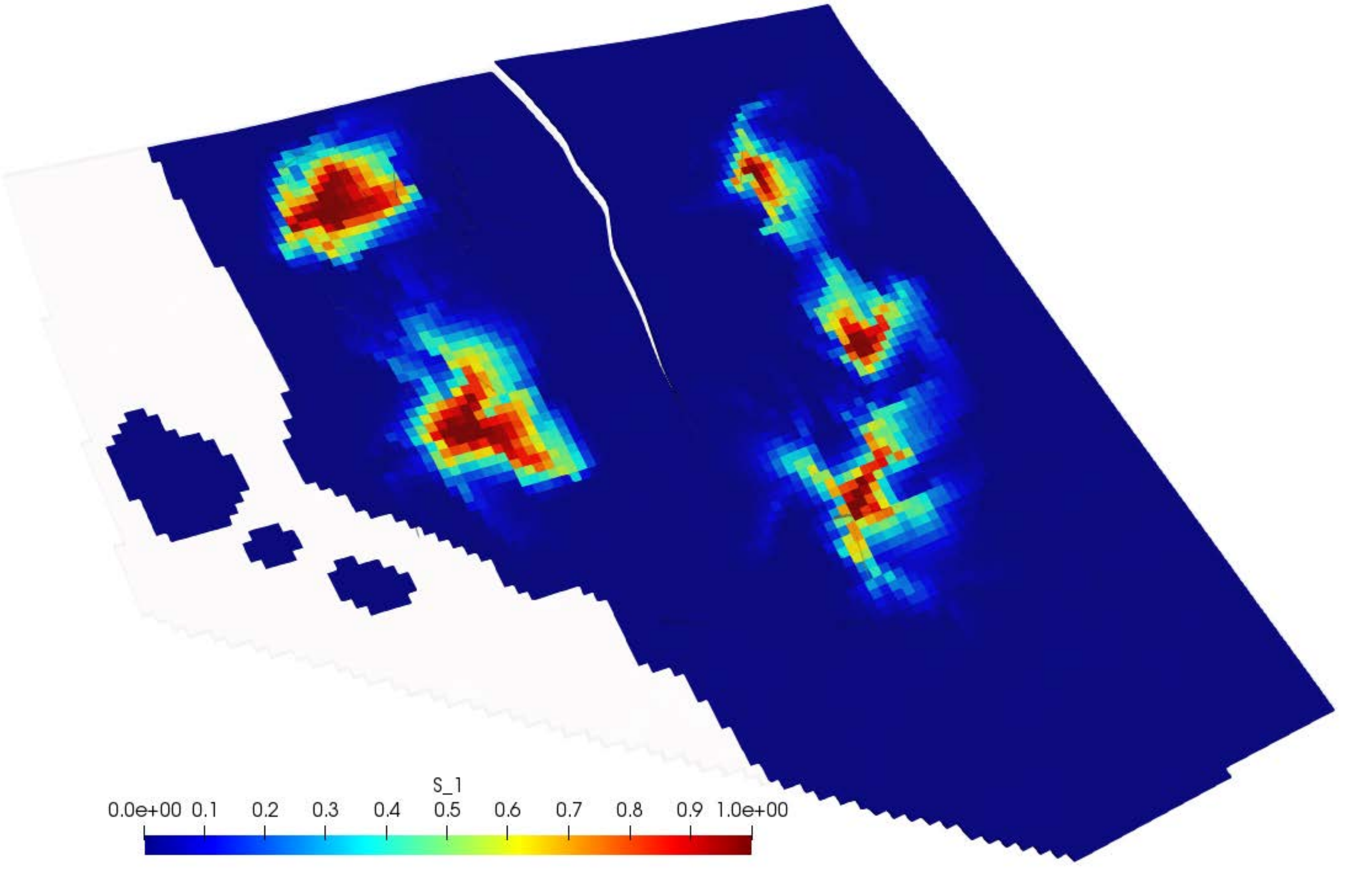}}
	\subcaptionbox{{\footnotesize Saturation in the fractures after $35000 [\text{days}]$ }\label{Fig:pEDFM_CPG_TestCase4_Johansen_Sf_scenario1_710}}
	{\includegraphics[width=0.23\textwidth]{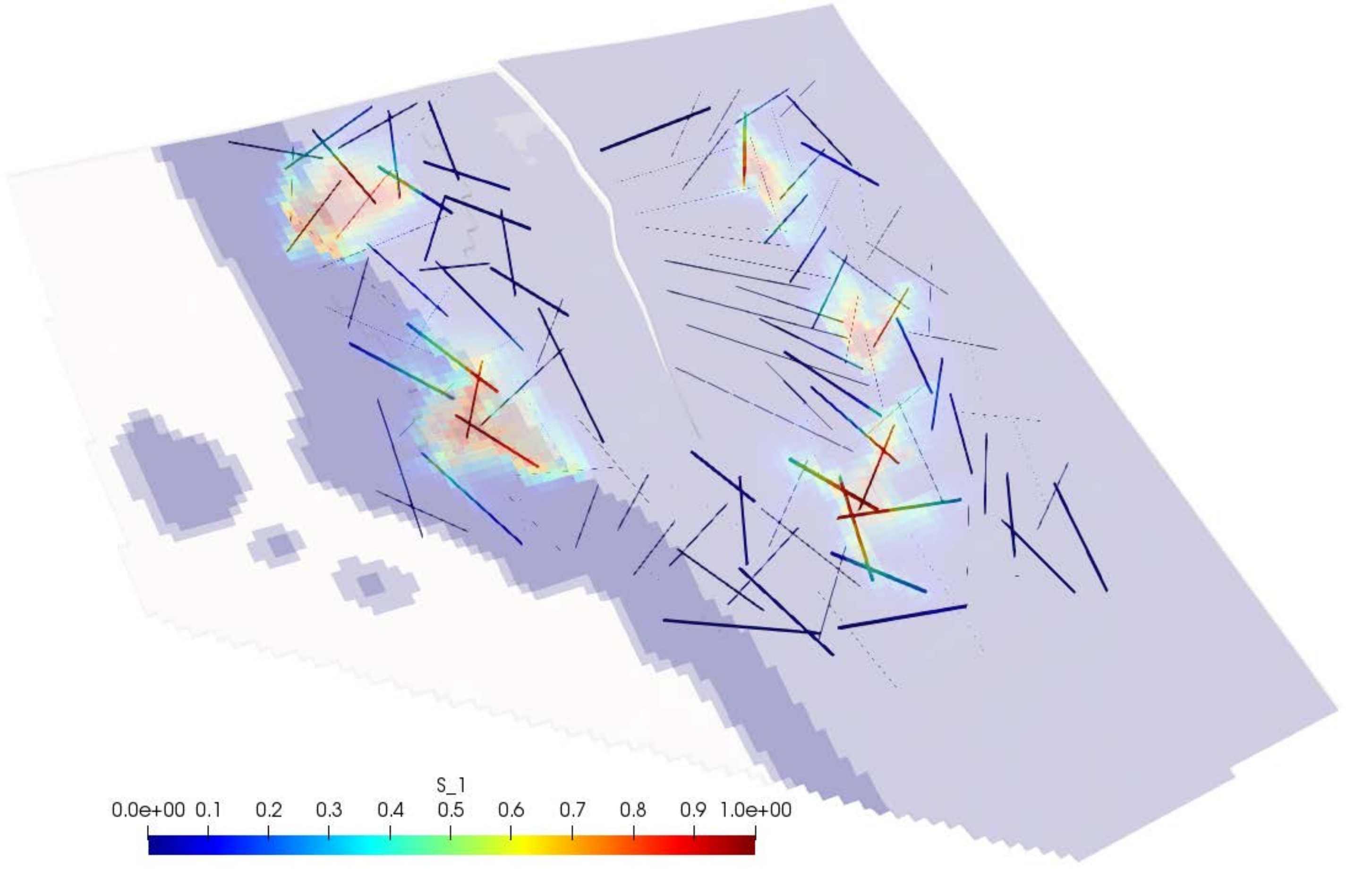}}
	\caption{Test case 4: The saturation profile of the Johansen formation for the simulation scenario $1$. The figures on the left illustrate the saturation profile in the matrix grid cells and the figures on the right side show the saturation maps in the fractures. From the top row towards the bottom row, the saturation profiles are displayed for simulations times $5000$, $20000$ and $35000 [\text{days}]$ respectively.}
	\label{Fig:pEDFM_CPG_TestCase4_Johansen_Saturation_Scenario1}
\end{figure}

The simulation results of the first scenario is presented in the figures \ref{Fig:pEDFM_CPG_TestCase4_Johansen_Pressure_Scenario1} and \ref{Fig:pEDFM_CPG_TestCase4_Johansen_Saturation_Scenario1}. The injection wells are surrounded by highly conductive fractures that facilitate the flow since the model's dimensions are considerably large (approximately $50 [\text{km}] \times 50 [\text{km}]$). The pressure distribution in the reservoir is shown in the figure \ref{Fig:pEDFM_CPG_TestCase4_Johansen_Pressure_Scenario1}. High pressure values is observed in a large section of the reservoir as there is no restriction for flow from the wells, and two shale formations bound the Johansen sandstone. One can interpret that the high pressure drops observed in some regions is caused by presence of low permeable fractures (or flow barriers) in those regions. The saturation displacement is considerably enhanced by the highly conductive fractures (figure \ref{Fig:pEDFM_CPG_TestCase4_Johansen_Saturation_Scenario1}) located near the injection wells.

\begin{figure}[!htbp]
	\centering
	\subcaptionbox{{\footnotesize Pressure in the matrix}\label{Fig:pEDFM_CPG_TestCase4_Johansen_Pm_scenario2}}
	{\includegraphics[width=0.23\textwidth]{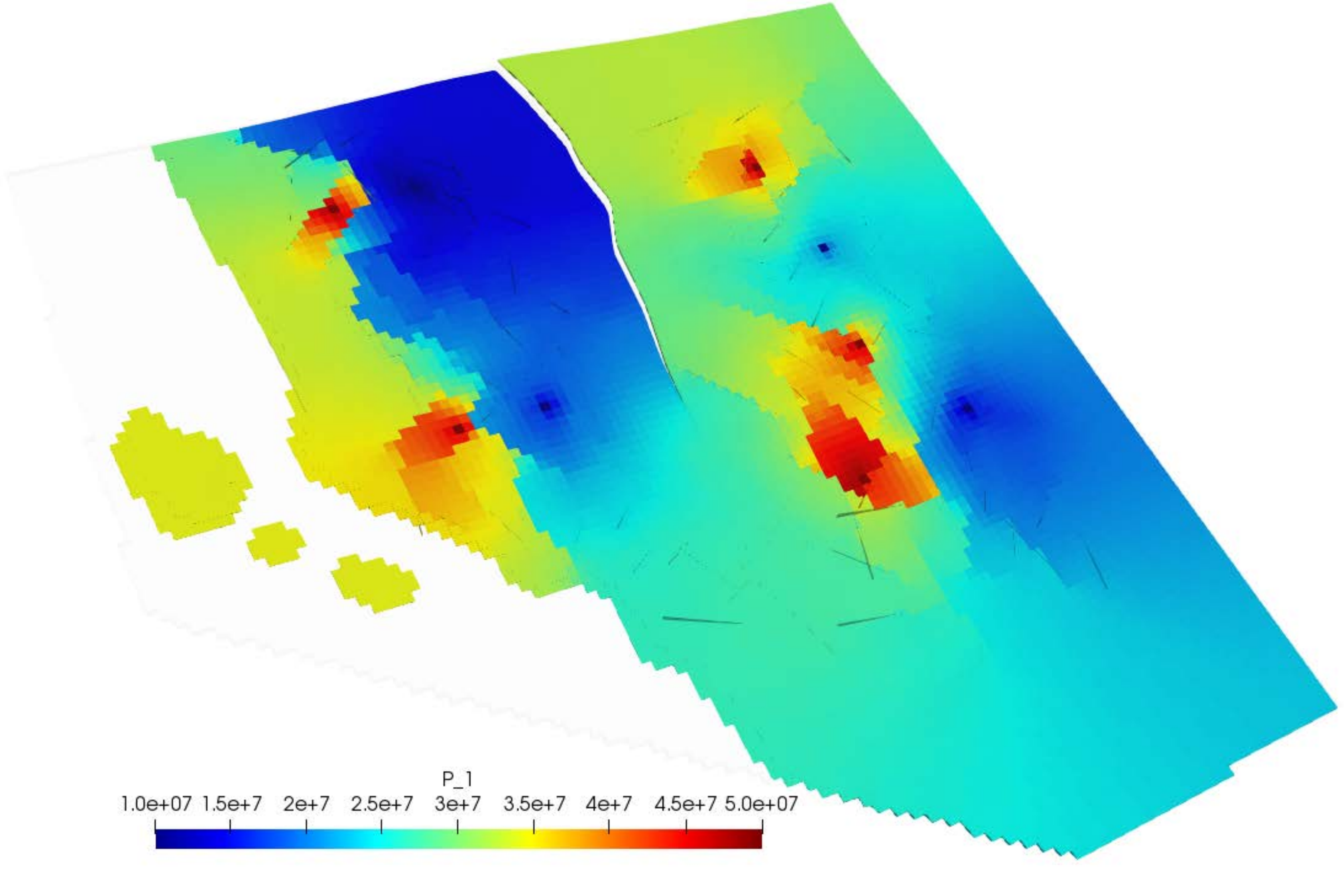}}
	\subcaptionbox{{\footnotesize Pressure in the fractures}\label{Fig:pEDFM_CPG_TestCase4_Johansen_Pf_scenario2}}
	{\includegraphics[width=0.23\textwidth]{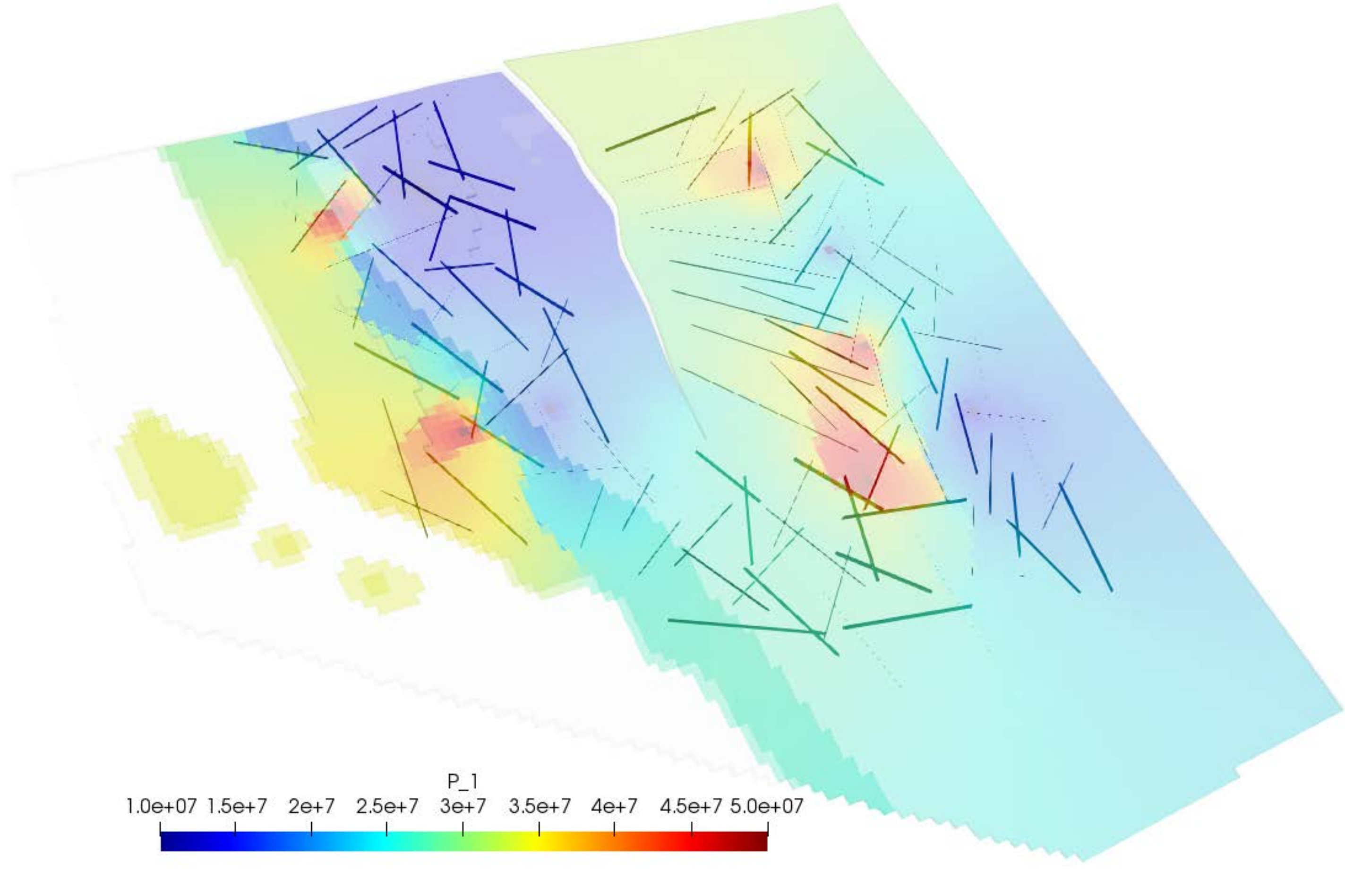}}
	\caption{Test case 4: The pressure profile of the Johansen formation for the simulation scenario $2$. The figure on the left illustrates the pressure distribution in the matrix grid cells. To make the pressure profile of the fractures visible, the transparency of the left side figure is increased and it is presented in the right side figure.}
	\label{Fig:pEDFM_CPG_TestCase4_Johansen_Pressure_Scenario2}
\end{figure}

\begin{figure}[!htbp]
	\centering
	\subcaptionbox{{\footnotesize Saturation in the matrix after $5000 [\text{days}]$ }\label{Fig:pEDFM_CPG_TestCase4_Johansen_Sm_scenario2_T10}}
	{\includegraphics[width=0.23\textwidth]{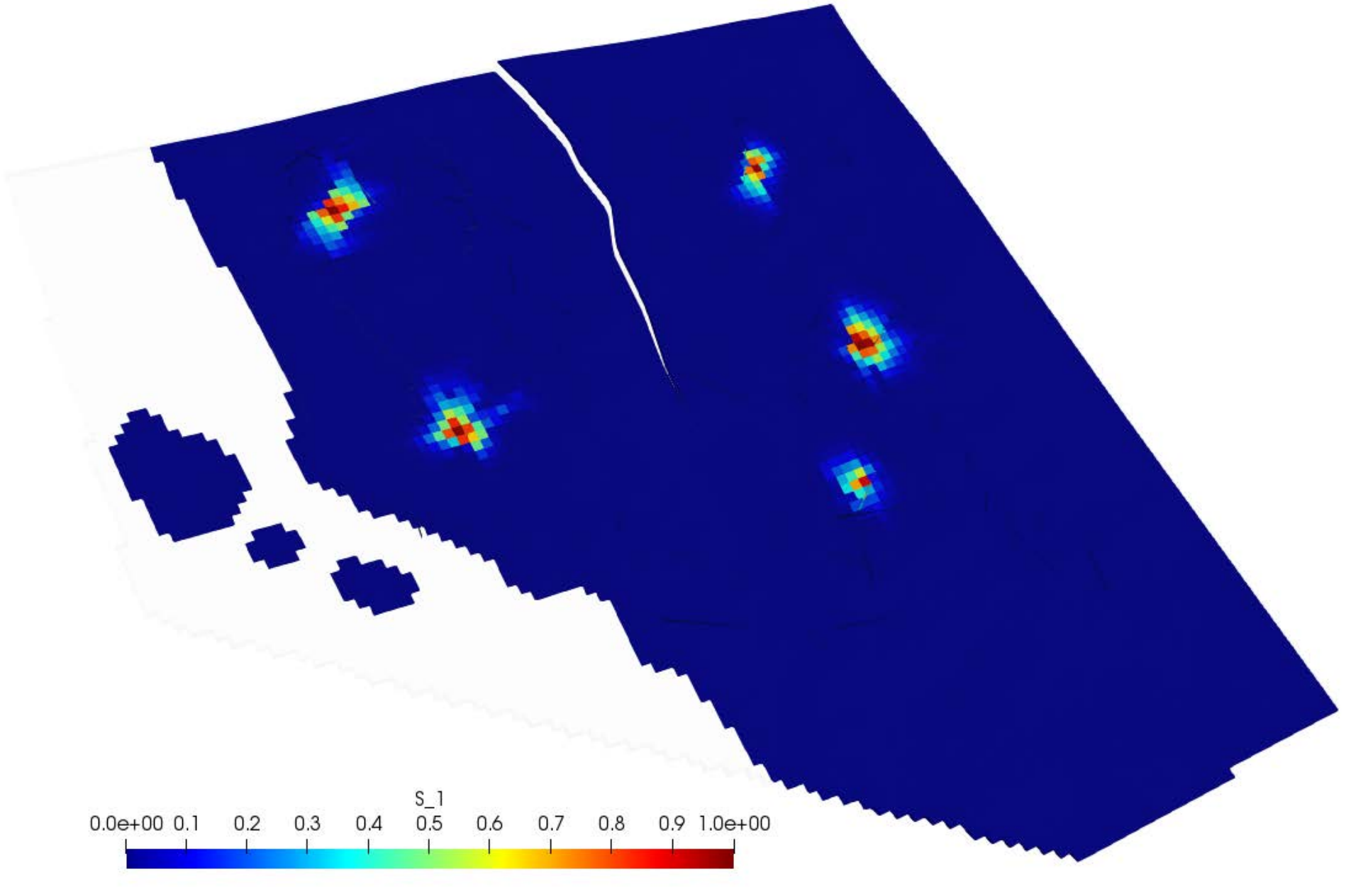}}
	\subcaptionbox{{\footnotesize Saturation in the fractures after $5000 [\text{days}]$ }\label{Fig:pEDFM_CPG_TestCase4_Johansen_Sf_scenario2_T10}}
	{\includegraphics[width=0.23\textwidth]{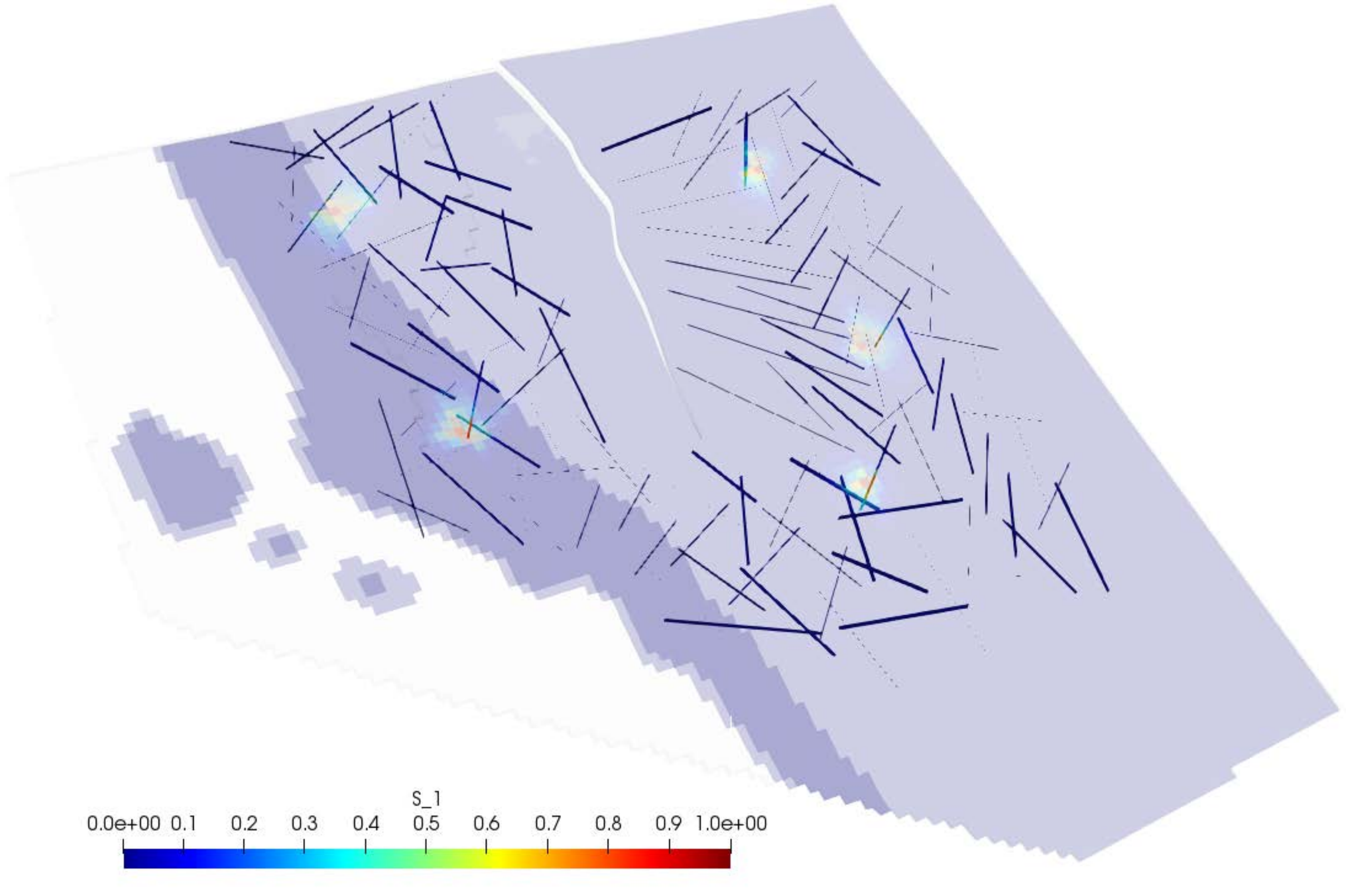}}
	\\
	\subcaptionbox{{\footnotesize Saturation in the matrix after $20000 [\text{days}]$ }\label{Fig:pEDFM_CPG_TestCase4_Johansen_Sm_scenario2_T40}}
	{\includegraphics[width=0.23\textwidth]{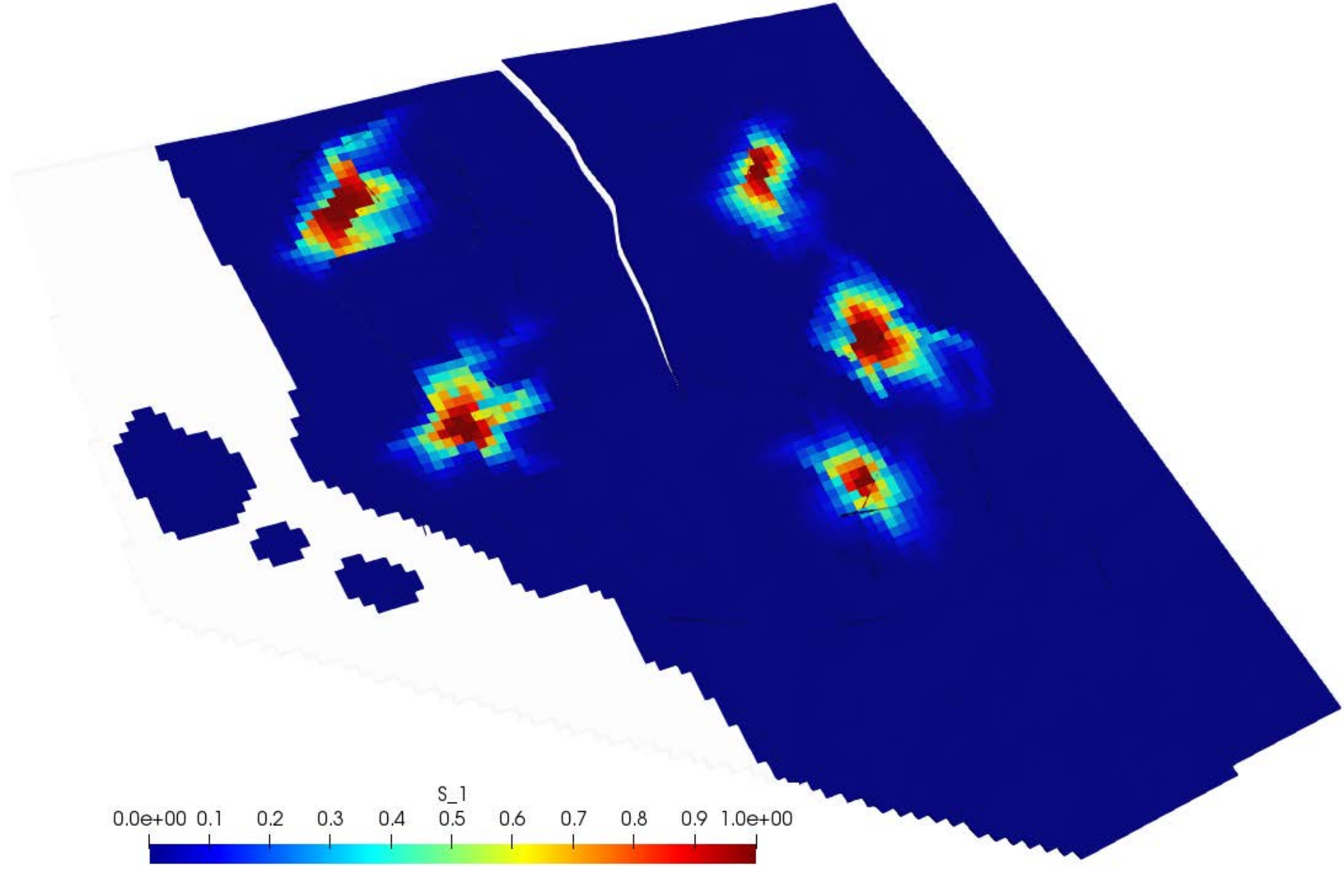}}
	\subcaptionbox{{\footnotesize Saturation in the fractures after $20000 [\text{days}]$ }\label{Fig:pEDFM_CPG_TestCase4_Johansen_Sf_scenario2_T40}}
	{\includegraphics[width=0.23\textwidth]{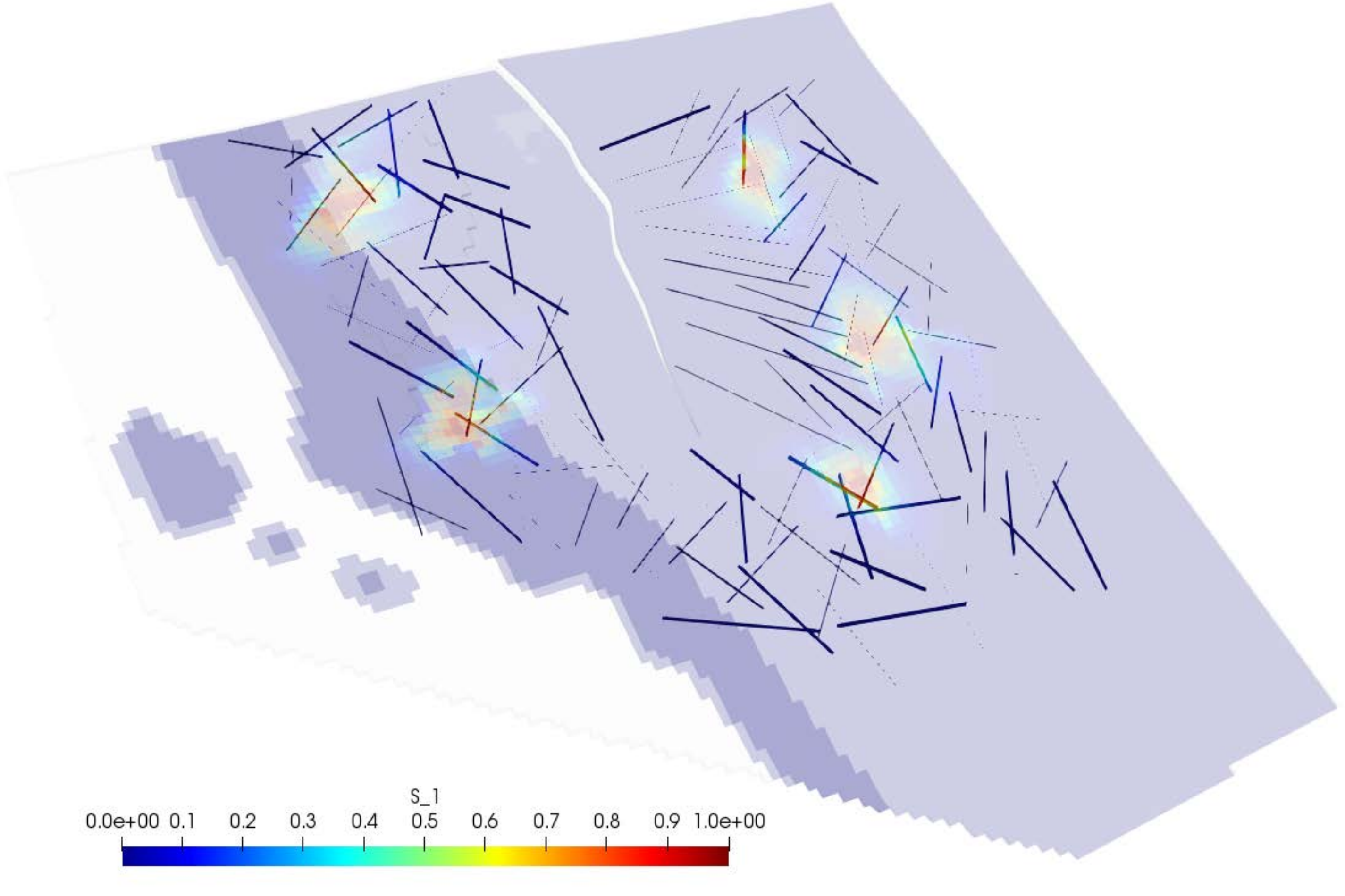}}
	\\
	\subcaptionbox{{\footnotesize Saturation in the matrix after $35000 [\text{days}]$ }\label{Fig:pEDFM_CPG_TestCase4_Johansen_Sm_scenario2_T70}}
	{\includegraphics[width=0.23\textwidth]{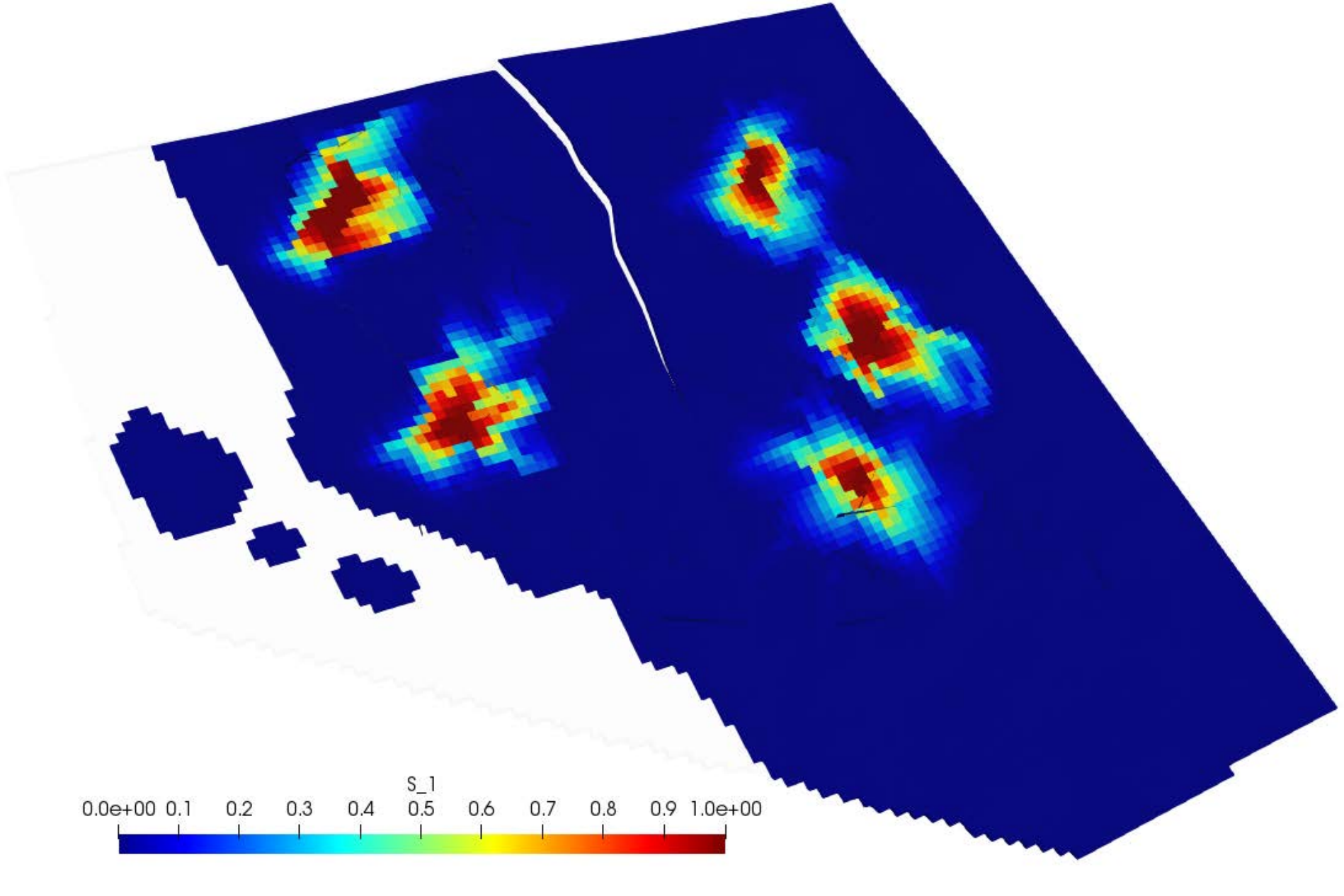}}
	\subcaptionbox{{\footnotesize Saturation in the fractures after $35000 [\text{days}]$ }\label{Fig:pEDFM_CPG_TestCase4_Johansen_Sf_scenario2_710}}
	{\includegraphics[width=0.23\textwidth]{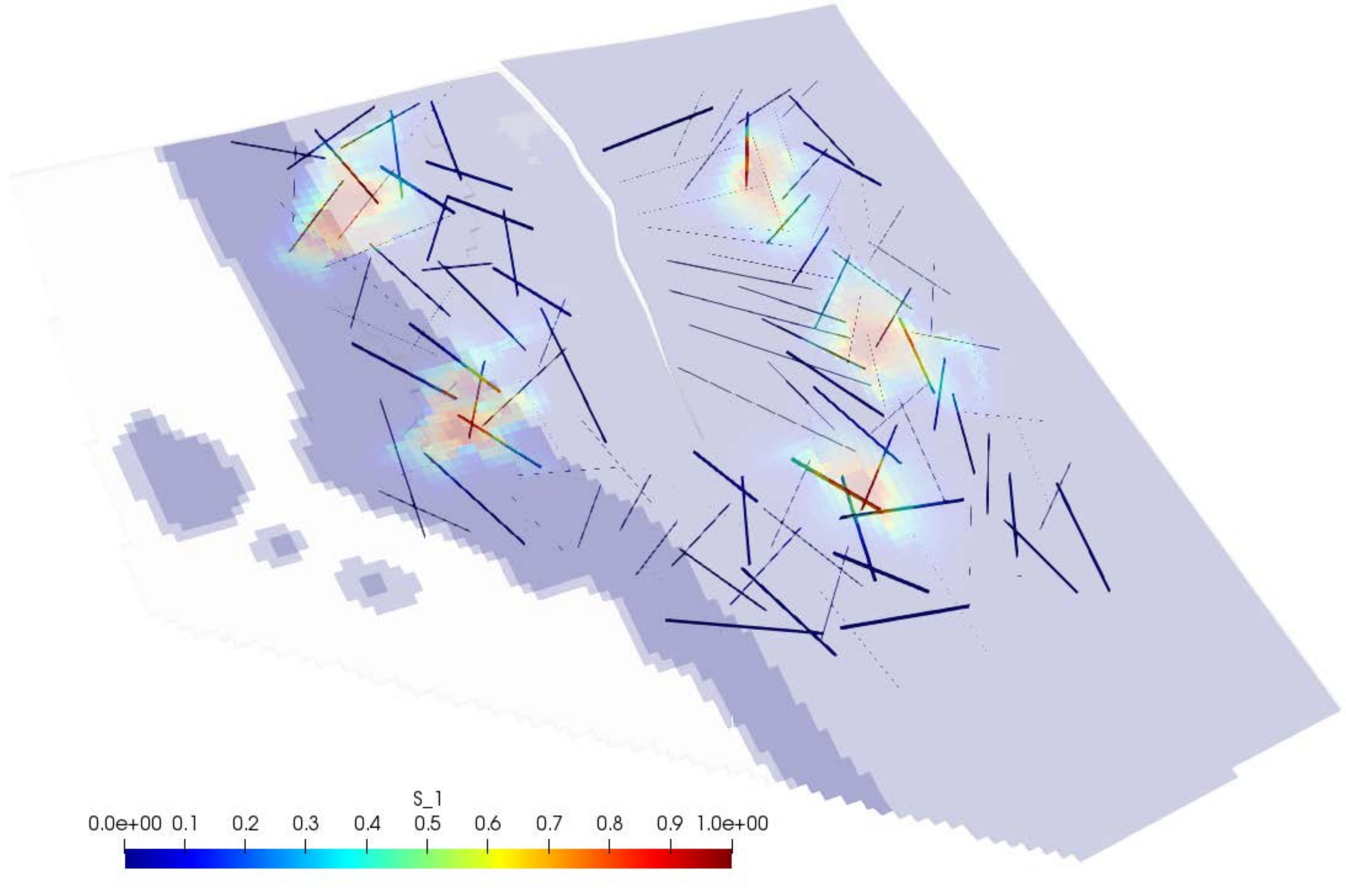}}
	\caption{Test case 4: The saturation profile of the Johansen formation for the simulation scenario $2$. The figures on the left side show the saturation profile in the matrix grid cells and the figures on the left side display the saturation maps in the fractures. From the top row towards the bottom row, the saturation profiles are displayed for simulations times $5000$, $20000$ and $35000 [\text{days}]$ respectively.}
	\label{Fig:pEDFM_CPG_TestCase4_Johansen_Saturation_Scenario2}
\end{figure}

The simulation results of the second scenario is presented in the figures \ref{Fig:pEDFM_CPG_TestCase4_Johansen_Pressure_Scenario2} and \ref{Fig:pEDFM_CPG_TestCase4_Johansen_Saturation_Scenario2}. The injection wells are surrounded by low conductive fractures which restrict the flow from the injection wells towards the production wells. The pressure distribution differs considerably when compared to the first scenario. The flow barriers near the wells result in high pressure drops in the vicinity of the injection wells. The saturation displacement (figure \ref{Fig:pEDFM_CPG_TestCase4_Johansen_Saturation_Scenario2}) is lower than that of scenario $1$ due to presence of low conductive fractures near the injection wells. 

\subsection{Test Case 5: The Brugge Model}\label{Sec:TestCase5_Brugge}
The Brugge model is an SPE benchmark study conceived as a reference platform to assess different closed-loop reservoir management methods \cite{Peters2010}. It is the largest and most complex test case on closed-loop optimization to represent real field management scenarios. The active Brugge field model has $44550$ corner-point grid cells, and the main geological features present in the model are a boundary fault and an internal fault. Seven different rock regions with their particular petrophysical properties are distributed in the whole model. Thirty wells are included in the field model's well production pattern: $20$ producers and $10$ injectors.

\begin{figure}[!htbp]
	\centering
	\begin{subfigure}[]{0.23\textwidth}
		\includegraphics[width=\textwidth]{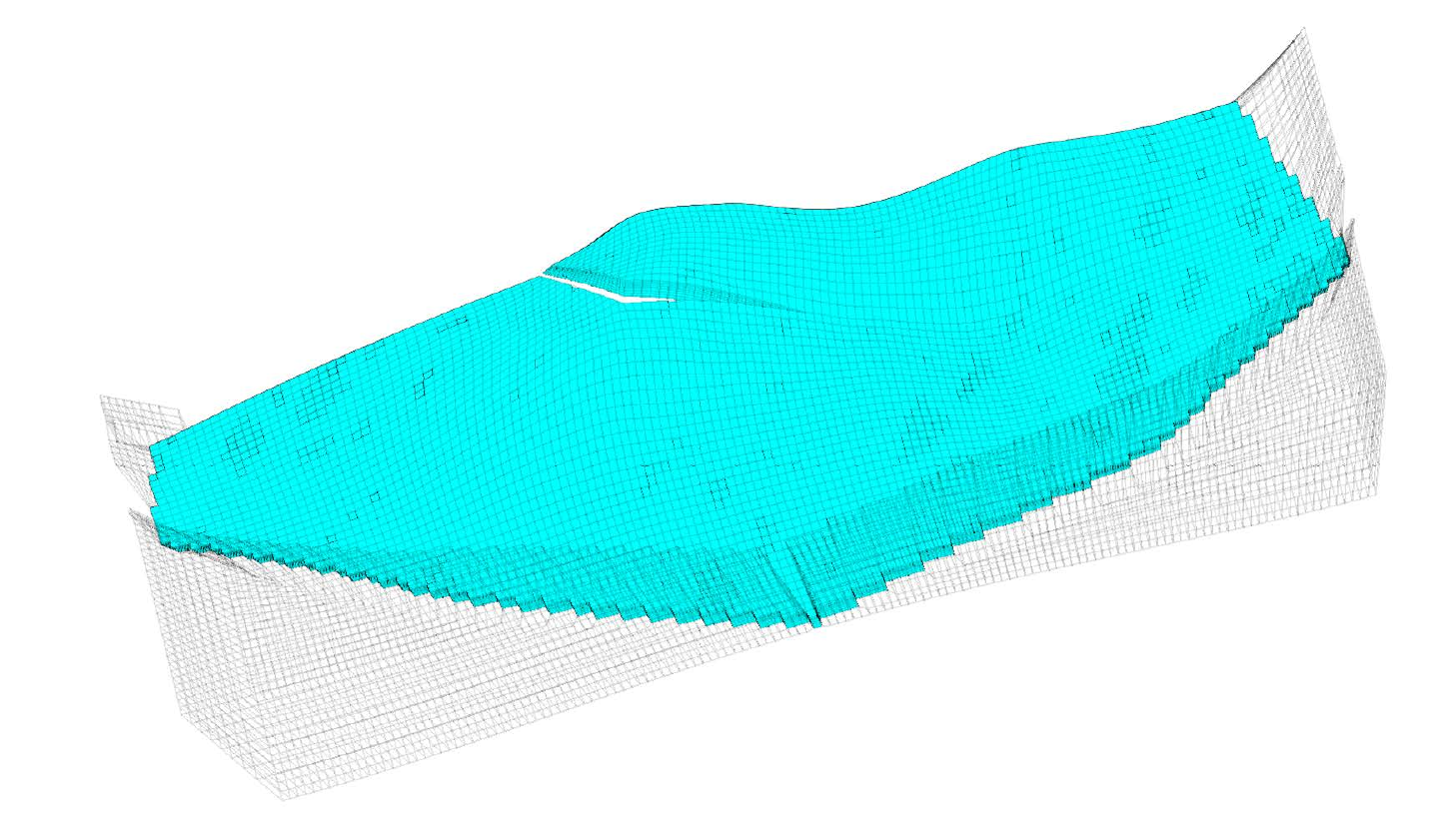}
		\caption{The whole model with active cells}
	\end{subfigure}
	\begin{subfigure}[]{0.23\textwidth}
		\includegraphics[width=\textwidth]{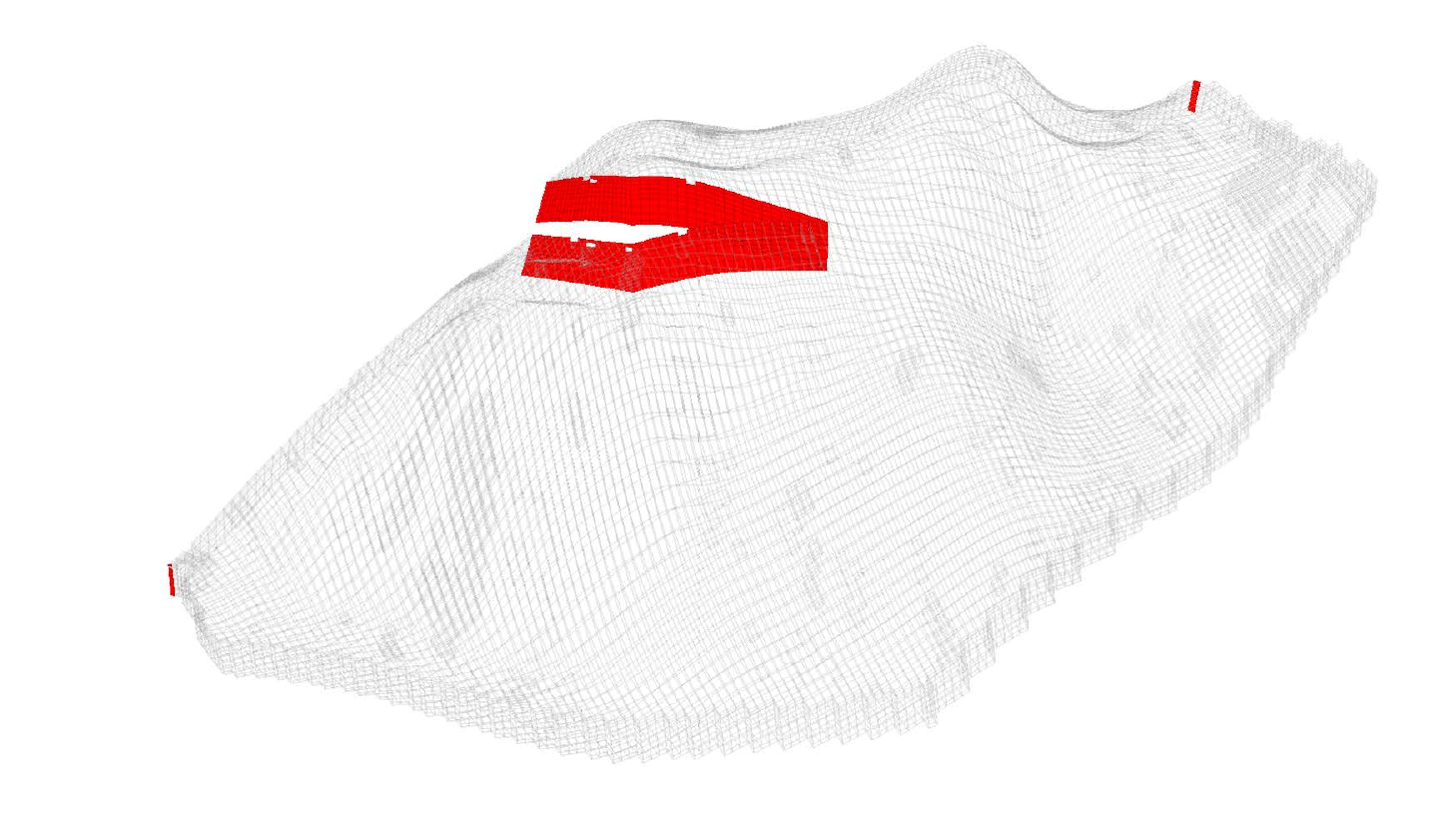}
		\caption{Faults present in the model}
	\end{subfigure}
	\caption{Test case 5: Illustration of the Brugge model. The left graph represents the active section (colored in blue) of the model, and the right figure shows the faults marked with red color.}
	\label{fig_Brugge1}
\end{figure}

\subsubsection{Geological model}
The geological structure of the Brugge field contains an east/west elongated half-dome with a boundary fault at its northern edge and an internal fault with a throw at an angle of nearly $20$ degrees to the northern fault edge. The dimensions of the field are approximately $10 [\text{km}] \times 3 [\text{km}]$.
The original high-resolution model consists of $20$ million grid cells, with average cell dimensions of $50 [\text{m}] \times 50 [\text{m}] \times 0.25 [\text{m}]$. In addition to the essential petrophysical properties for reservoir simulation (sedimentary facies, porosity, permeability, net-to-gross, and water saturation), the grid model includes properties measured in real fields (gamma-ray, sonic, bulk density, and neutron porosity).  The data were generated at a detailed scale to produce reliable well log data in the thirty wells drilled in the field. 

The original high-resolution model was upscaled to a $450000$ grid cells model, which established the foundation for all additional reservoir simulations of the reference case. A set of $104$ realizations, each containing $60000$ grid cells, was created from the data extracted from the reference case.

All the realizations used the same geological structure of the field. The North Sea Brent-type field was the reference to generate the reservoir zones' rock properties and thicknesses. An alteration of the formations' vertical sequence for the general Brent stratigraphy column (comprising the Broom-Rannoch-Etive-Ness-Tarbert Formations) was made and resulted in that the highly permeable reservoir zone switched locations with the underlying area (less permeable and heterogeneous).

\subsubsection{Rock properties}
A reservoir model with $60000$ grid cells was the reference to create $104$ upscaled realizations for the reservoir properties. The properties that contain the realizations are facies, porosity, a diagonal permeability tensor, net-to-gross ratio, and water saturation.

\begin{figure}[!htbp]
	\centering
	\begin{subfigure}[]{0.23\textwidth}
		\includegraphics[width=\textwidth]{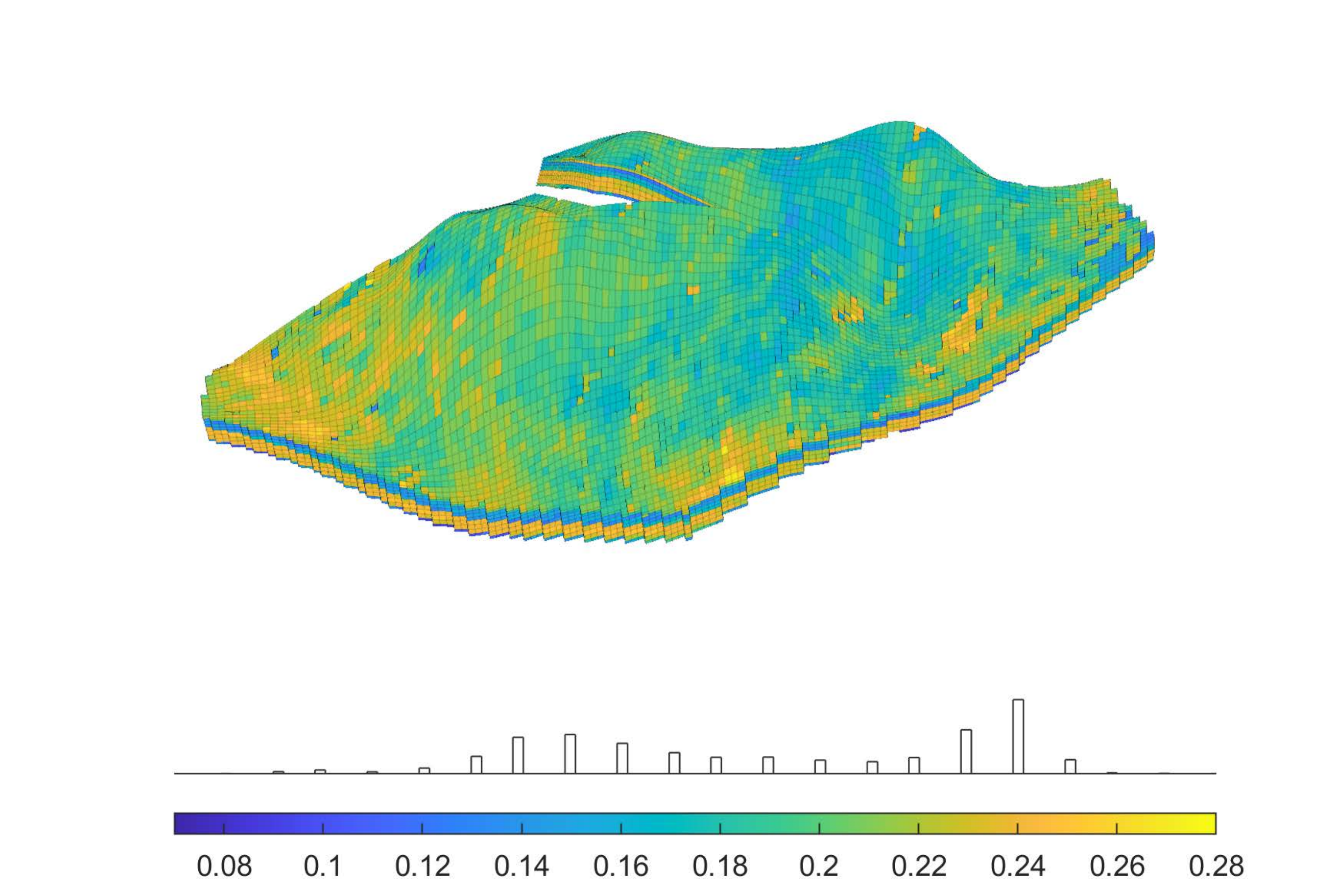}
	\end{subfigure}
	\begin{subfigure}[]{0.23\textwidth}
		\includegraphics[width=\textwidth]{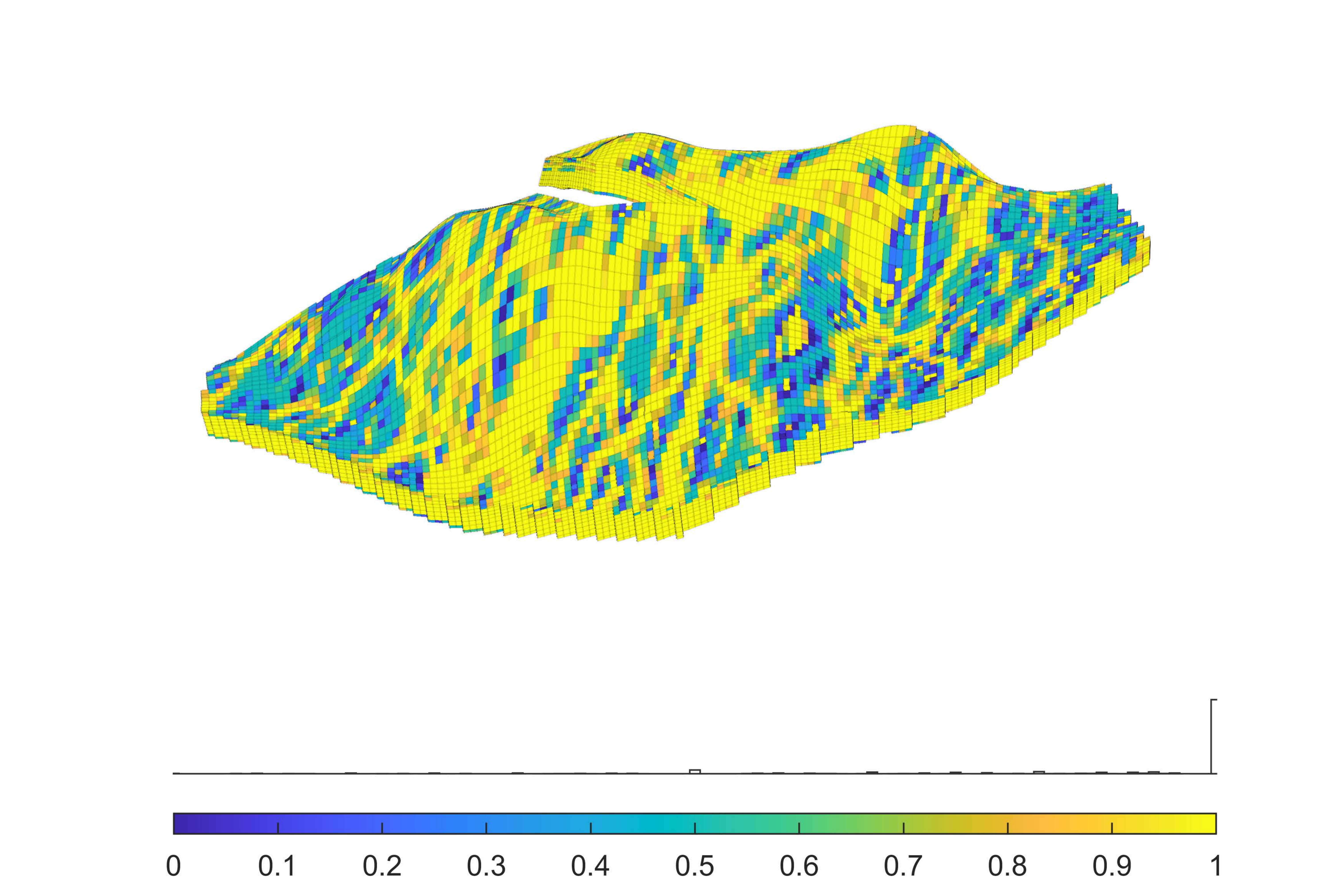}
	\end{subfigure}
	\caption{Test case 5: Porosity and Net-to-gross ratio for the Brugge model. The left graph shows the porosity of the model, and the right one shows the net-to-gross ratio map in the structural model.}
	\label{fig_Brugge2}
\end{figure}

\begin{figure}[!htbp]
	\centering
	\begin{subfigure}[]{0.23\textwidth}
		\includegraphics[width=\textwidth]{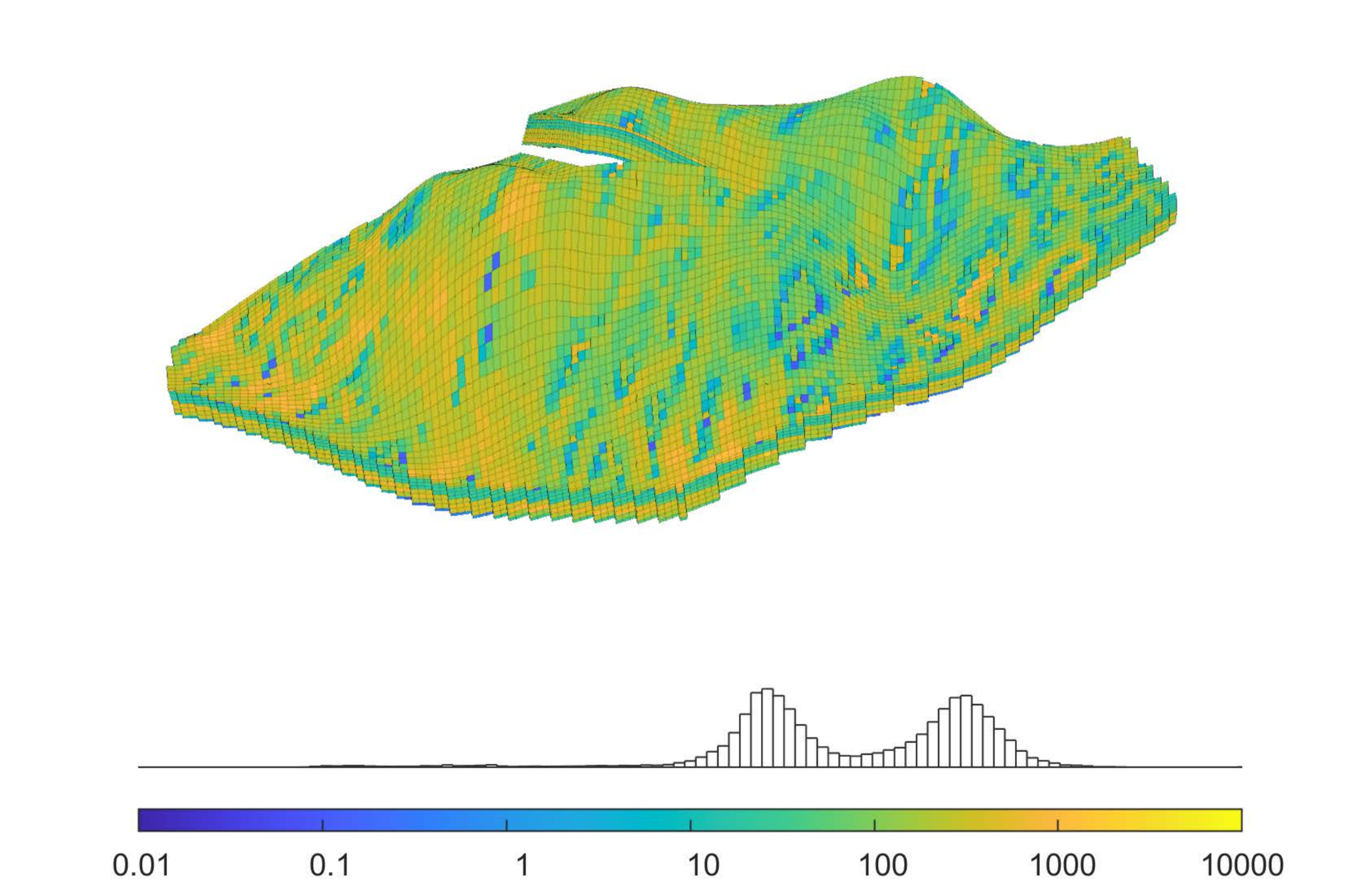}
	\end{subfigure}
	\begin{subfigure}[]{0.23\textwidth}
		\includegraphics[width=\textwidth]{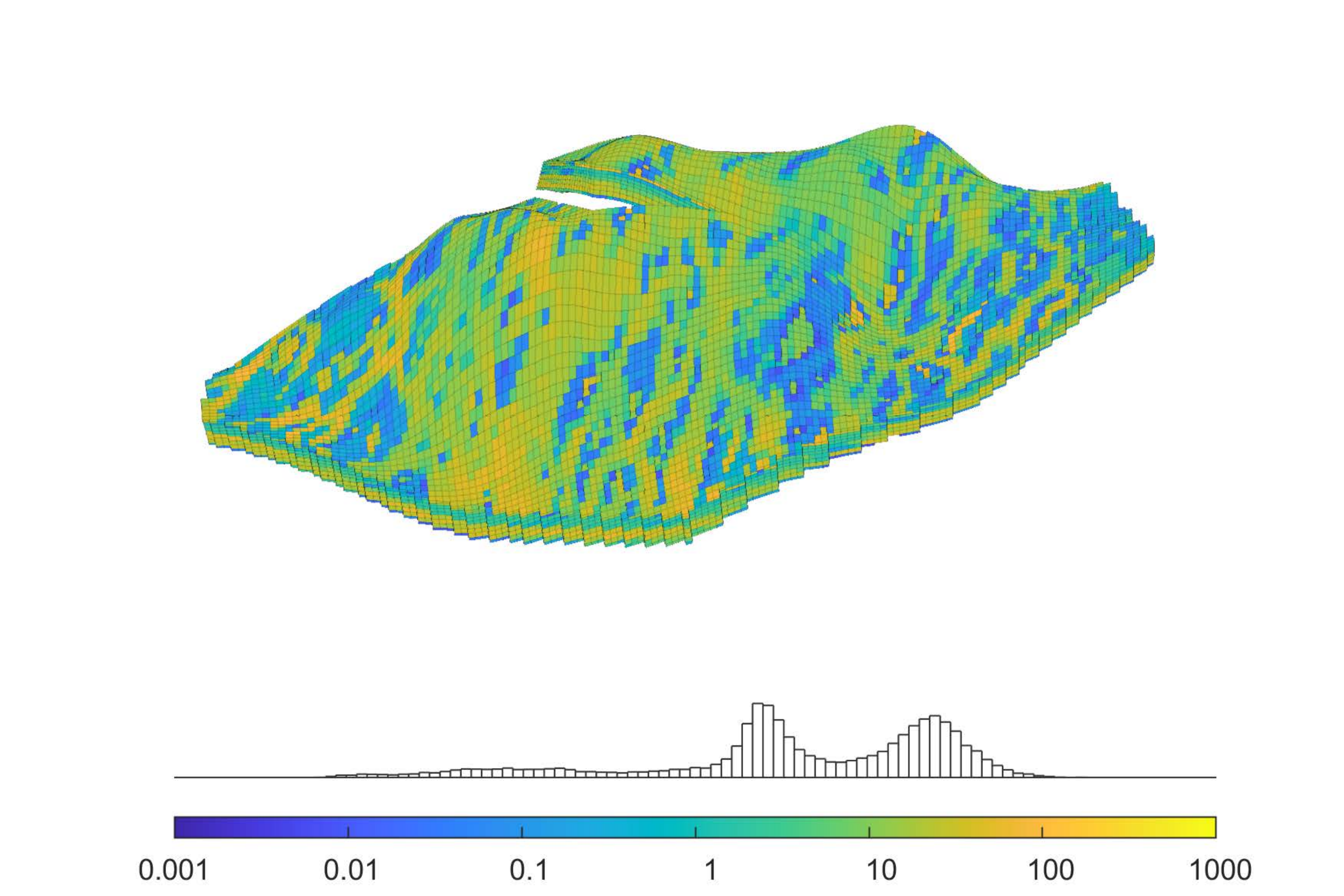}
	\end{subfigure}
	\caption{Test case 5: Permeability map of the Brugge model. The left figure shows the horizontal permeability, and the right figure illustrates the vertical permeability; both are plotted using a logarithmic color scale.}
	\label{fig_Brugge3}
\end{figure}

\begin{figure}[!htbp]
	\centering
	\begin{subfigure}[]{0.23\textwidth}
		\includegraphics[width=\textwidth]{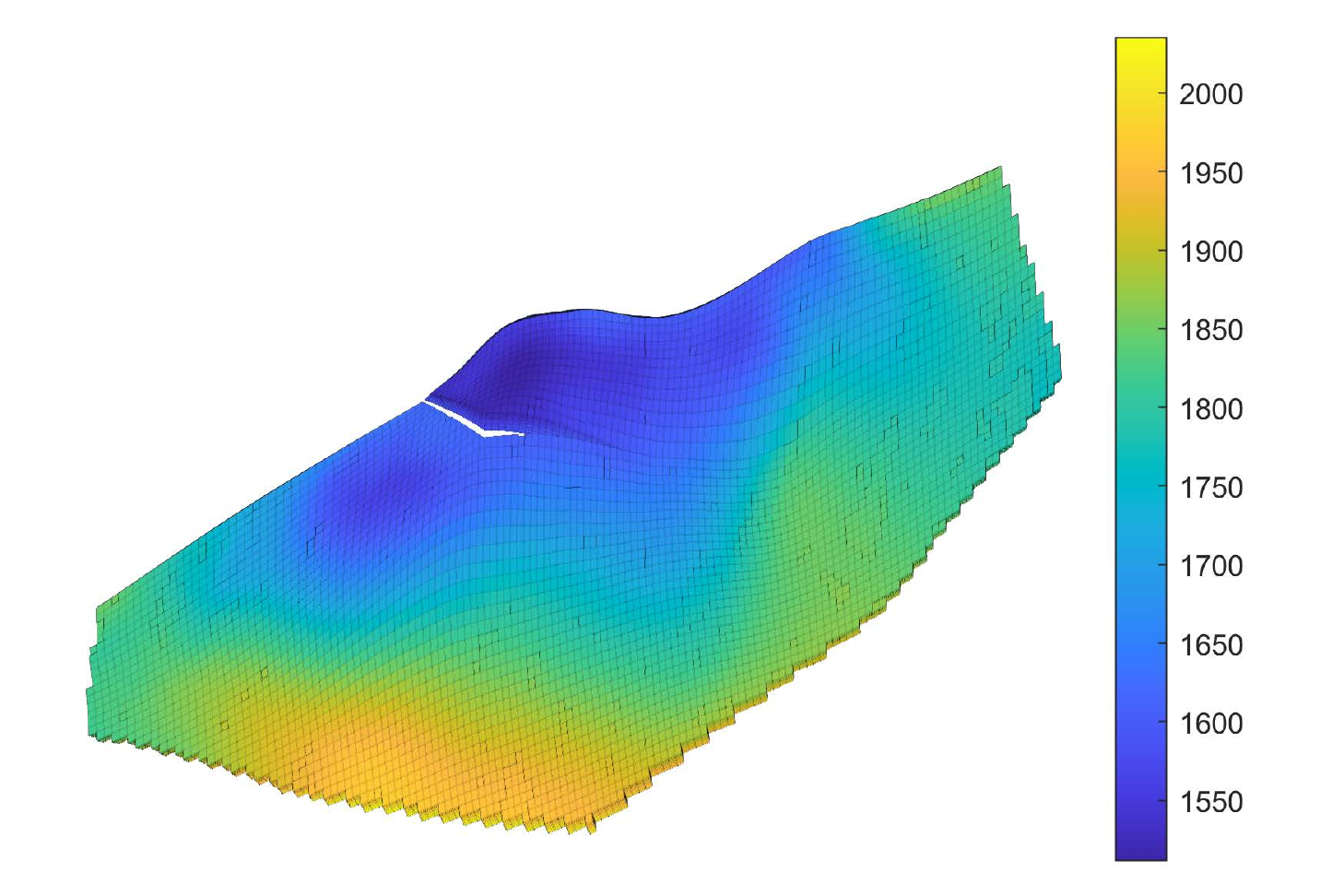}
	\end{subfigure}
	\begin{subfigure}[]{0.23\textwidth}
		\includegraphics[width=\textwidth]{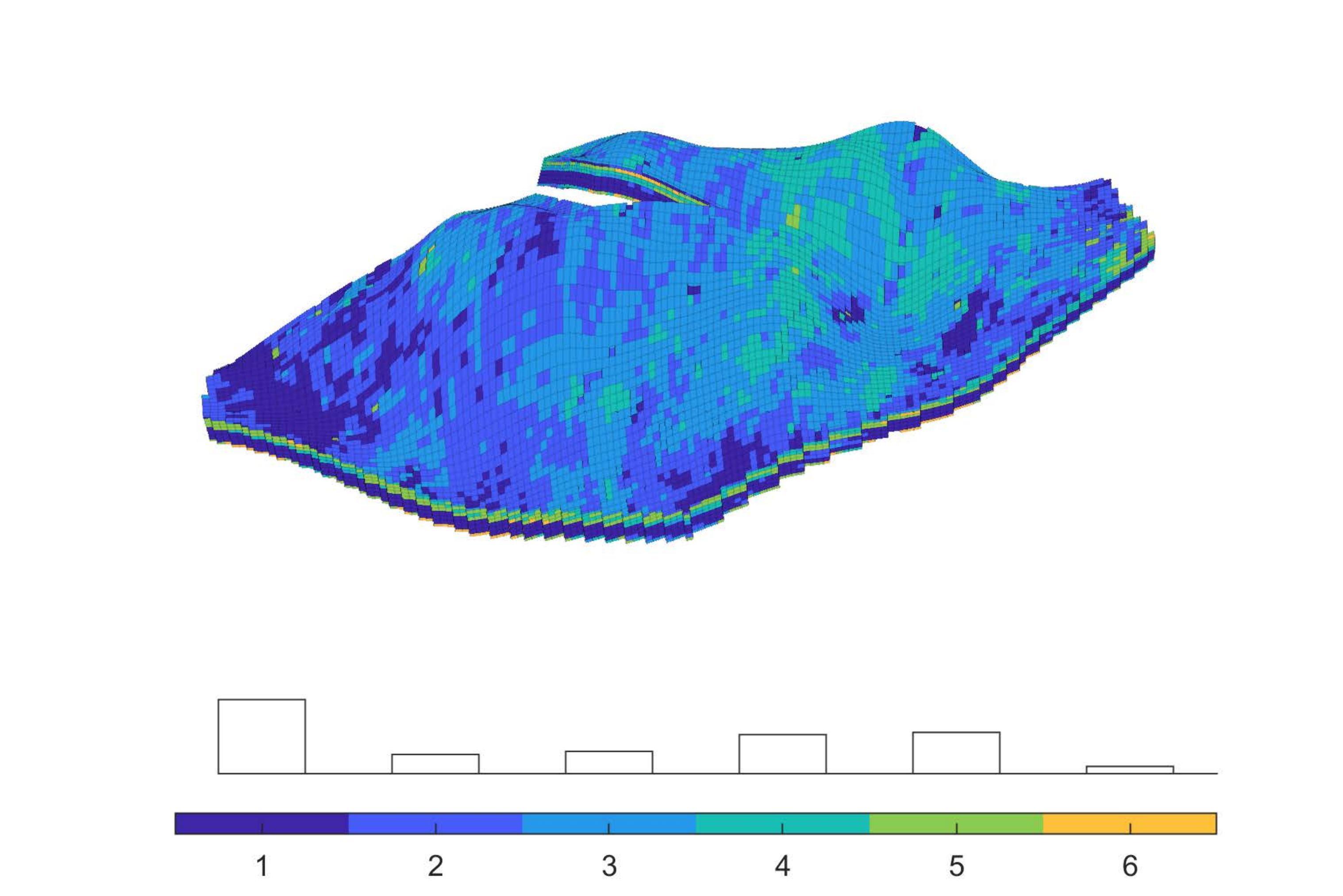}
	\end{subfigure}
	\caption{Test case 5: The depth map and saturation regions of the Brugge field. The left graph displays the depth map, and the right graph represents the rock type distribution displayed in the color bar.}
	\label{fig_Brugge4}
\end{figure}

\subsubsection{Simulation results}
The following test case from the Brugge model is used to show the pEDFM model's capability on fracture modeling in a synthetic geologically relevant model with corner-point grid geometry. The reservoir model consists of $138 \times 48 \times 9$ grid cells from which $43474$ grid cells are active.  Rock properties of the realization available on public data were used in the simulation. A network of $60$ fractures is defined in the reservoir domain containing both highly conductive fractures and flow barriers with permeability of $K_{f_{max}} = 10^{-8} \, [\text{m}^2]$ and $K_{f_{min}} = 10^{-20} \, [\text{m}^2]$ respectively. The fracture network consists of $5384$ grid cells (in total $48858$ grid cells). The well pattern used in this test case was a modified version of the original well pattern (with $30$ wells) \cite{Peters2010}. Four injection wells with $p_{\text{inj}} = 5 \times 10^7 \, [\text{Pa}]$ and three production wells with pressure of $p_{\text{prod}} = 1 \times 10^7 \, [\text{Pa}]$ were defined in the model. Wells are drilled vertical and through the entire thickness of the reservoir.

\begin{figure}[!htbp]
	\centering
	\subcaptionbox{{\footnotesize Fractures permeability (scenario 1)}\label{Fig:pEDFM_CPG_TestCase5_Brugge_Kf_1}}
	{\includegraphics[width=0.23\textwidth]{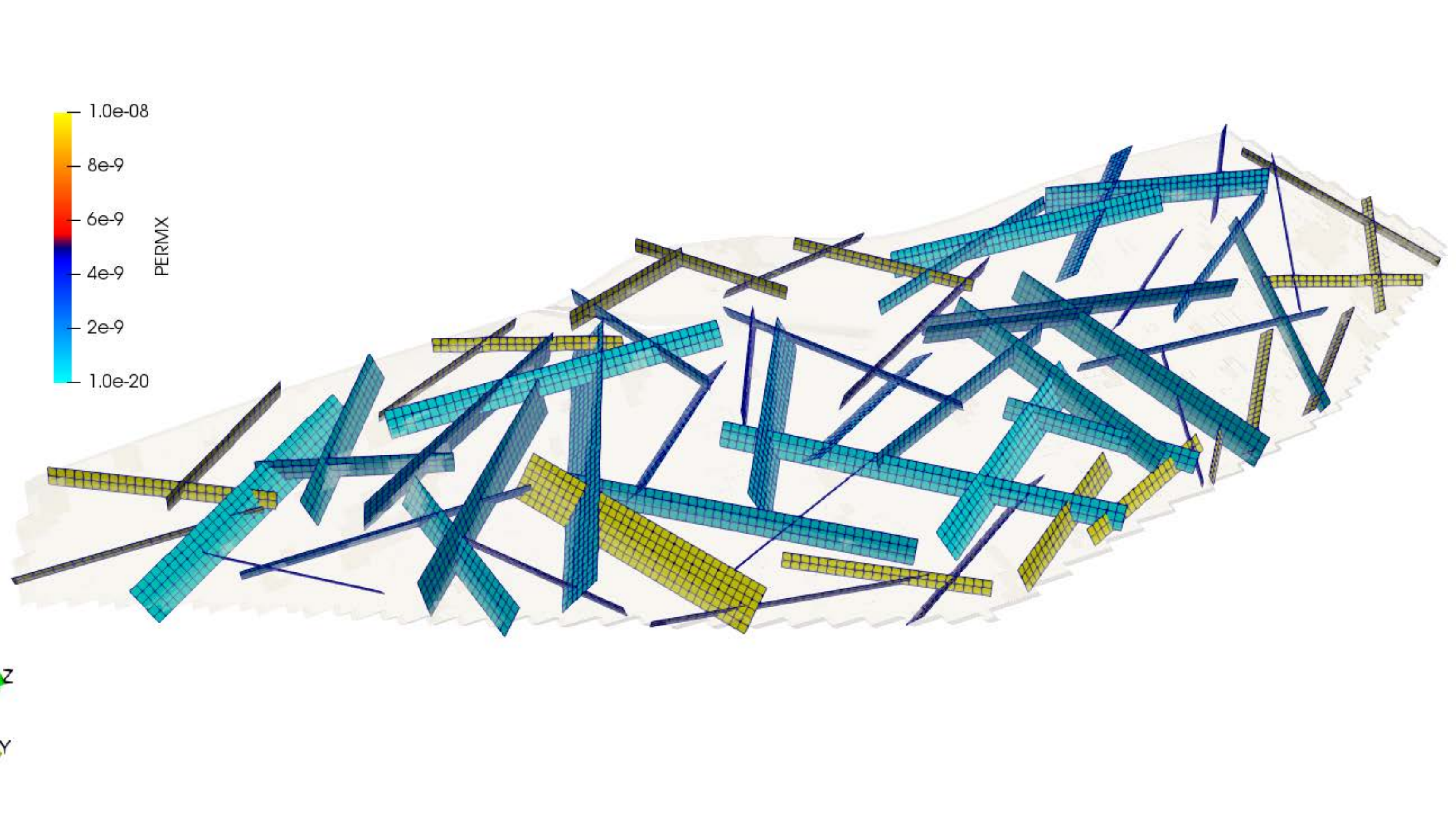}}
	\subcaptionbox{{\footnotesize Fractures permeability (scenario 2)}\label{Fig:pEDFM_CPG_TestCase5_Brugge_Kf_2}}
	{\includegraphics[width=0.23\textwidth]{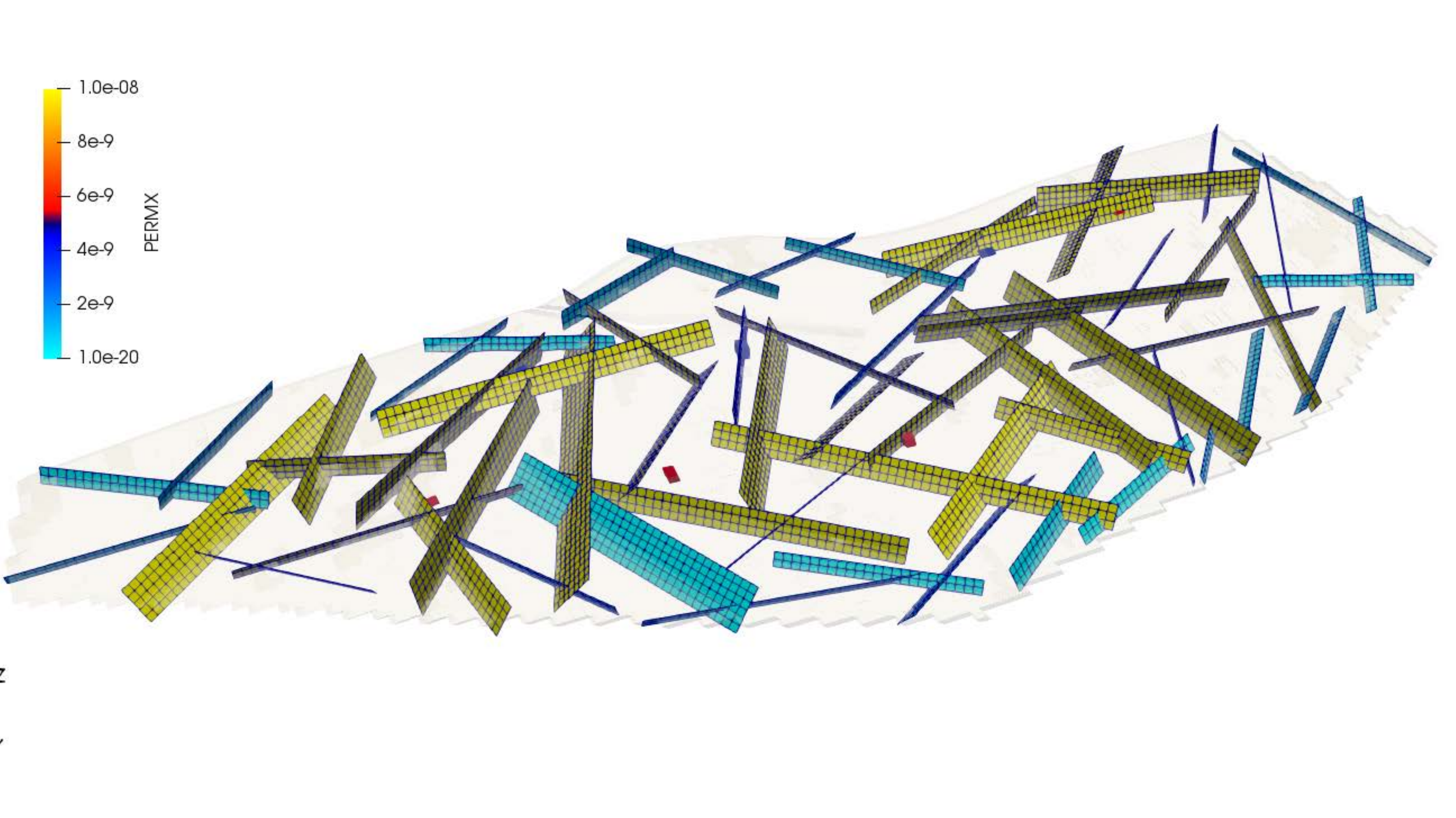}}
	\\
	\subcaptionbox{{\footnotesize Matrix cells overlapped by fractures}\label{Fig:pEDFM_CPG_TestCase5_Brugge_Fractured_Cells}}
	{\includegraphics[width=0.23\textwidth]{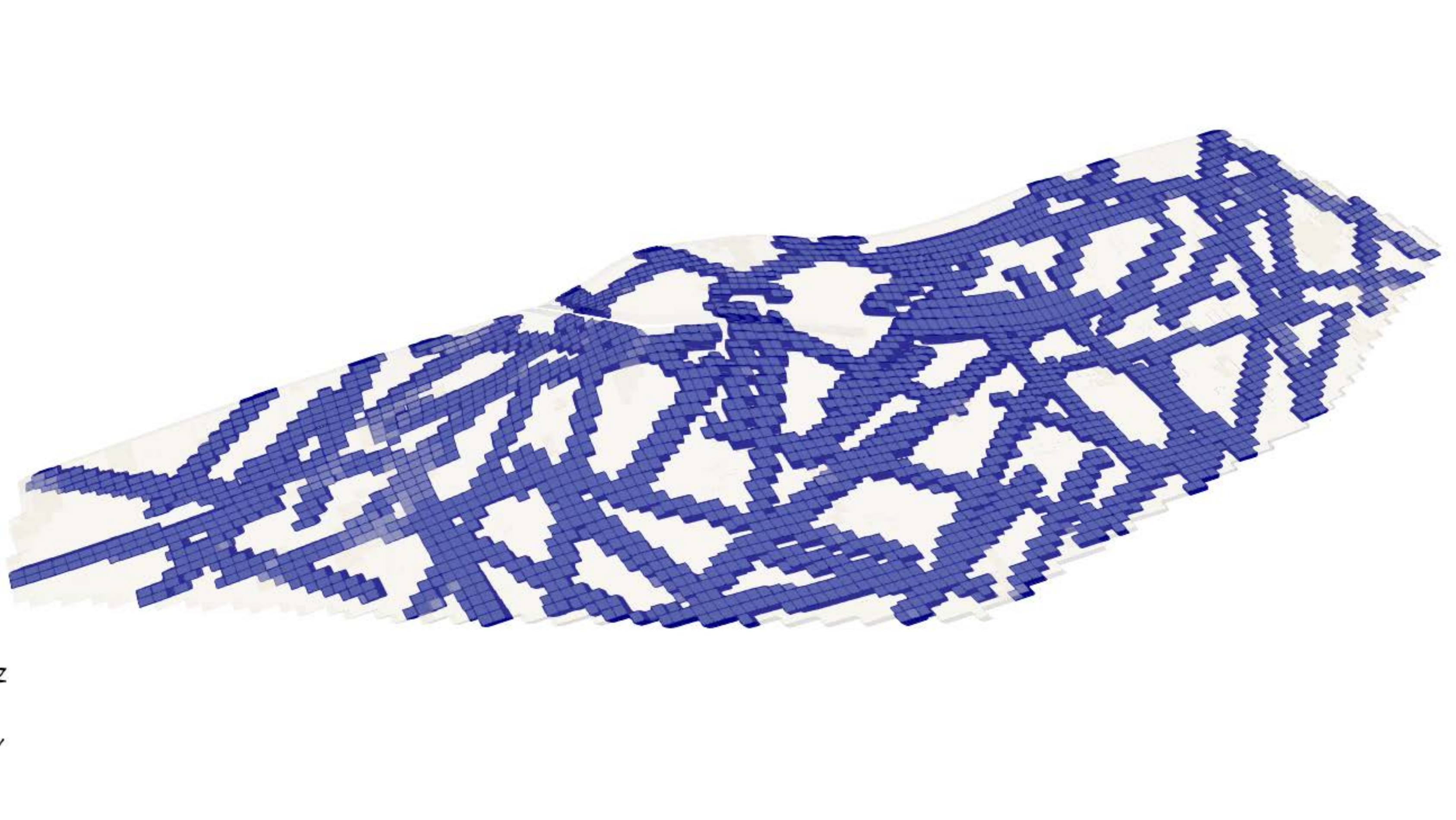}}
	\subcaptionbox{{\footnotesize Location of injection and production wells}\label{Fig:pEDFM_CPG_TestCase5_Brugge_Wells}}
	{\includegraphics[width=0.23\textwidth]{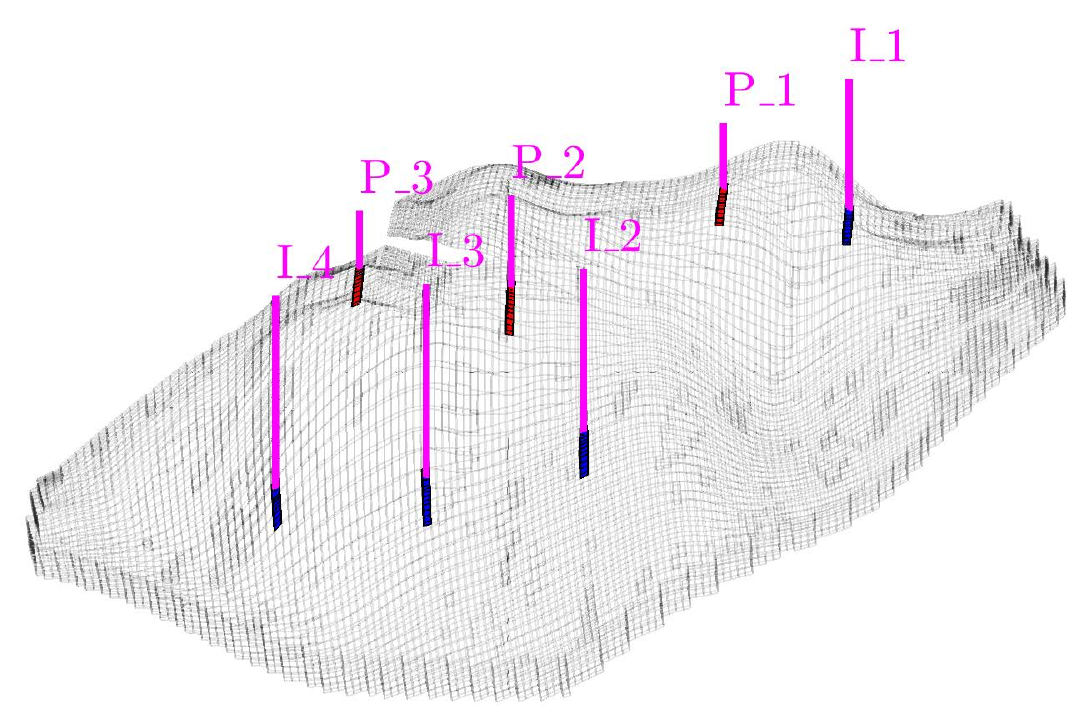}}
	\caption{Test case 5: The Brugge model with $7$ wells ($4$ injectors and $3$ producers) and a set of 60 synthetic fractures (with mixed conductivities). The figures on top show the fractures network with different permeabilities for scenario $1$ (top left) and scenario $2$ (top right). The figure at bottom left illustrates the highlighted matrix cells that are overlapped by the fractures network. And the figure at the bottom right shows the schematics of the injection and production wells.}
	\label{Fig:pEDFM_CPG_TestCase5_Brugge_Wells_Fractures}
\end{figure}

Two scenarios are created with two different fracture networks including mixed conductivities. The geometry of both fracture networks is identical but the permeability values of the fractures from scenario $1$ are inverted for the scenario $2$, namely, the highly conductive fractures in the fractures network of scenario $1$ act as flow barriers in the 2nd scenario and the flow barriers of scenario $1$ are modified to be highly conductive fractures in the scenario $2$. Figures \ref{Fig:pEDFM_CPG_TestCase5_Brugge_Kf_1} and \ref{Fig:pEDFM_CPG_TestCase5_Brugge_Kf_2} show the fracture networks of scenario $1$ and scenario $2$ respectively. The matrix grid cells
overlapped by the fractures are visible in figure \ref{Fig:pEDFM_CPG_TestCase4_Johansen_Fractured_Cells}.

\begin{figure}[!htbp]
	\centering
	\subcaptionbox{{\footnotesize Pressure in the matrix}\label{Fig:pEDFM_CPG_TestCase5_Brugge_Pm_scenario1}}
	{\includegraphics[width=0.23\textwidth]{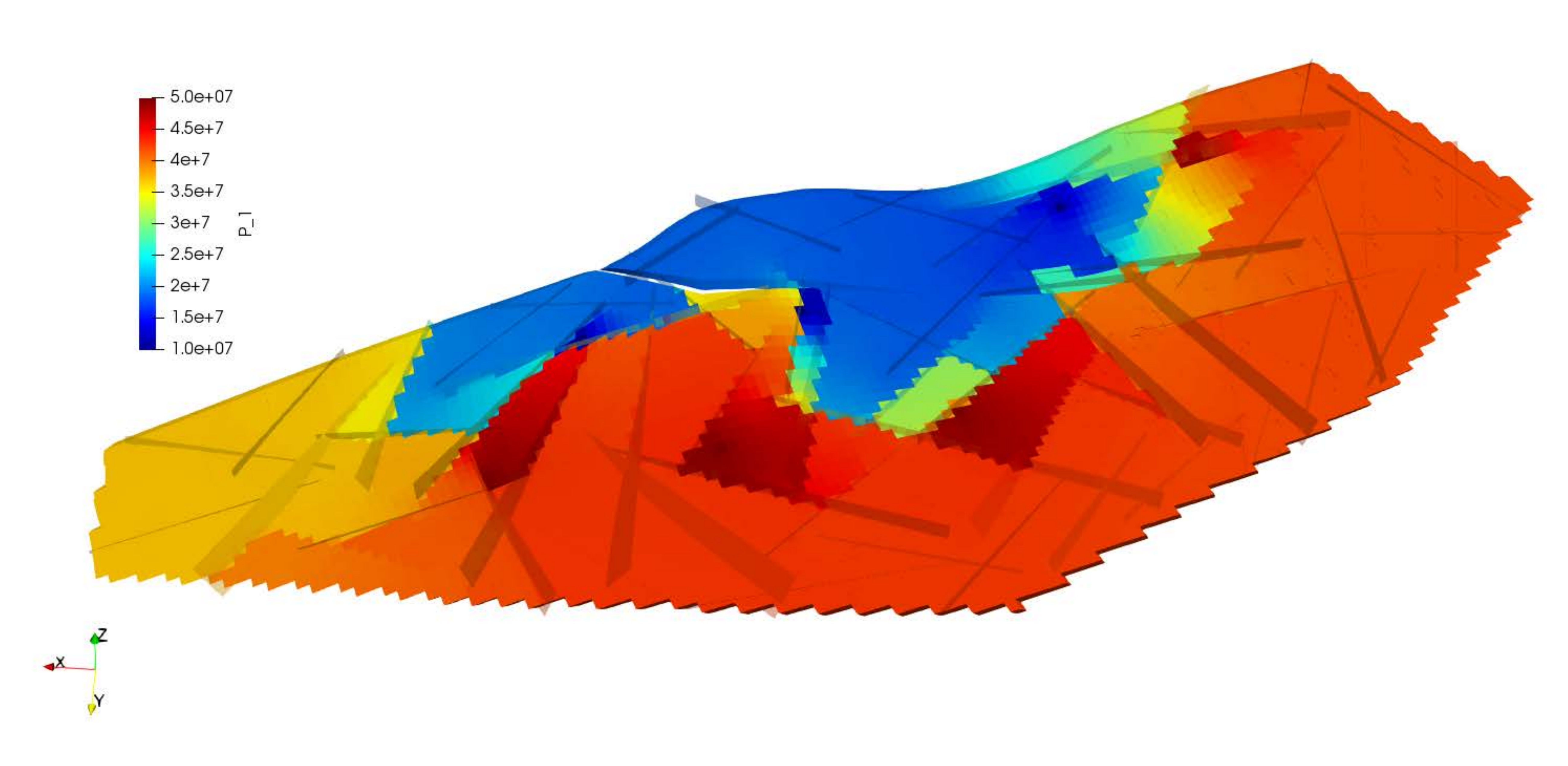}}
	\subcaptionbox{{\footnotesize Pressure in the fractures}\label{Fig:pEDFM_CPG_TestCase5_Brugge_Pf_scenario1}}
	{\includegraphics[width=0.23\textwidth]{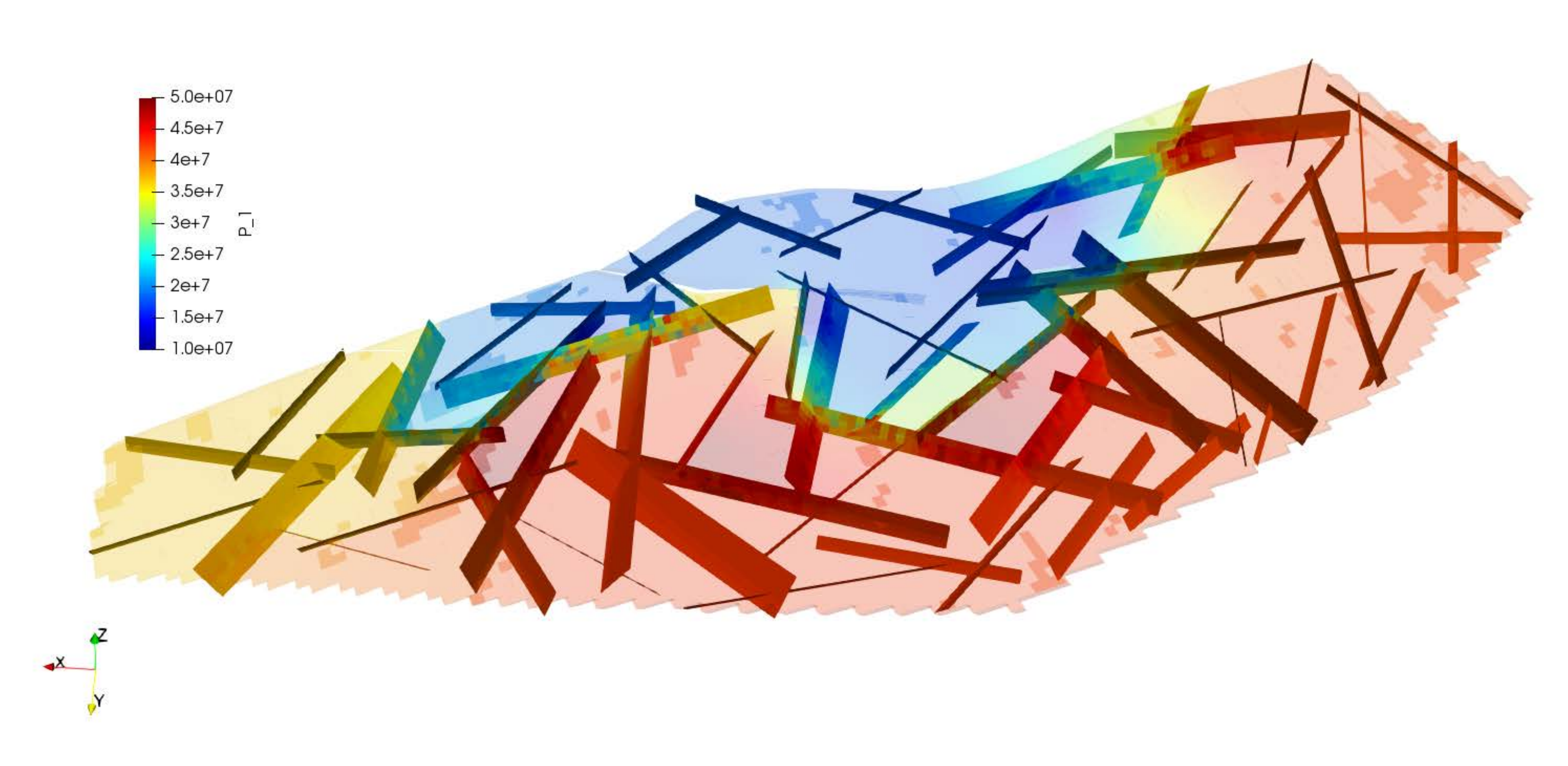}}
	\caption{Test case 5: The pressure profile of the Brugge model for the simulation scenario $1$. The figure on the left shows the pressure distribution in the matrix grid cells. The transparency of this figure is increased to make the pressure map in the fractures visible. This map is displayed on the right figure. The results are shown for the simulation time $5000 [\text{days}]$}
	\label{Fig:pEDFM_CPG_TestCase5_Brugge_Pressure_Scenario1}
\end{figure}

\begin{figure}[!htbp]
	\centering
	\subcaptionbox{{\footnotesize Saturation in the matrix after $2000 [\text{days}]$ }\label{Fig:pEDFM_CPG_TestCase5_Brugge_Sm_scenario1_T04}}
	{\includegraphics[width=0.23\textwidth]{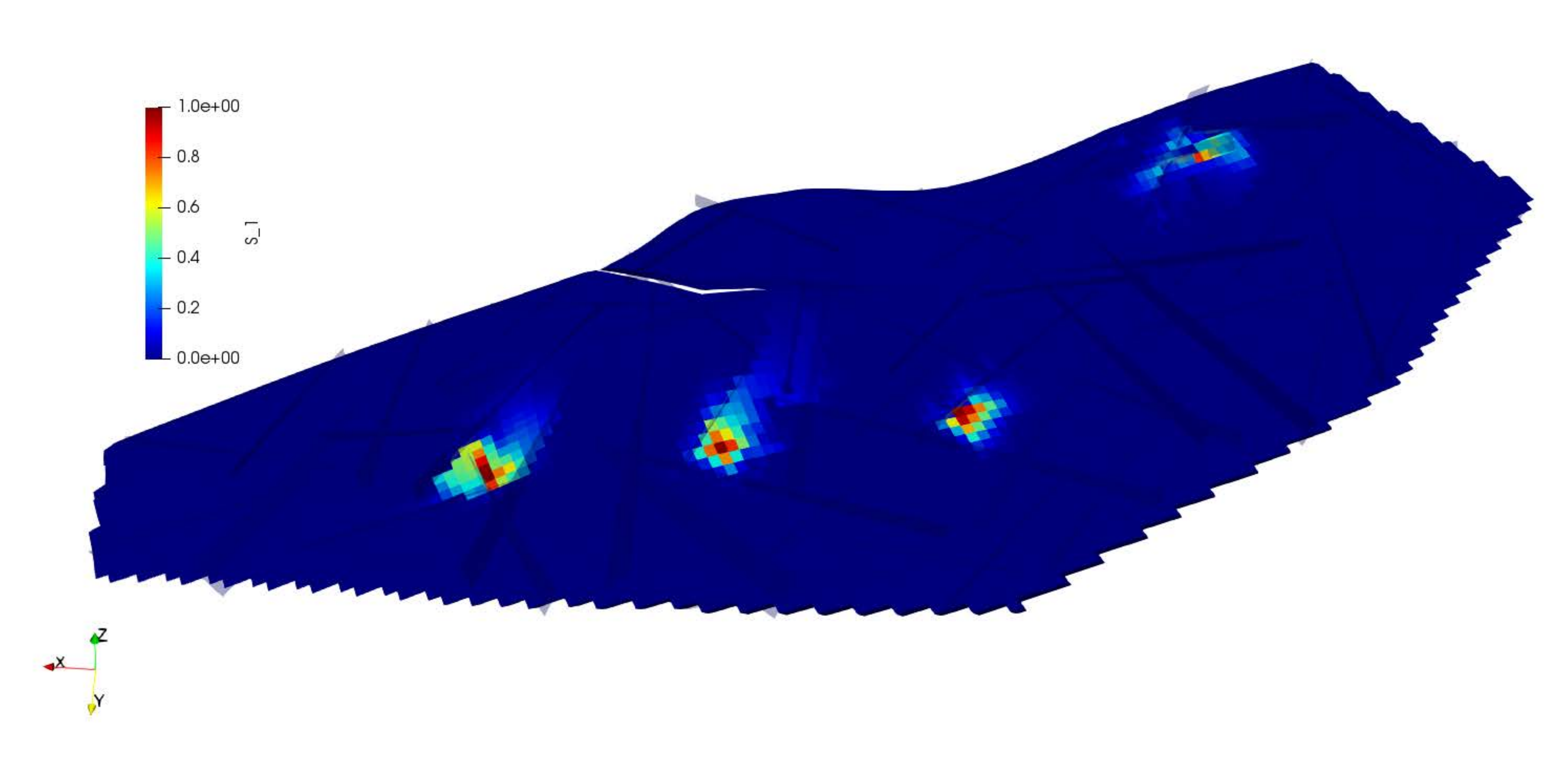}}
	\subcaptionbox{{\footnotesize Saturation in the fractures after $2000 [\text{days}]$ }\label{Fig:pEDFM_CPG_TestCase5_Brugge_Sf_scenario1_T04}}
	{\includegraphics[width=0.23\textwidth]{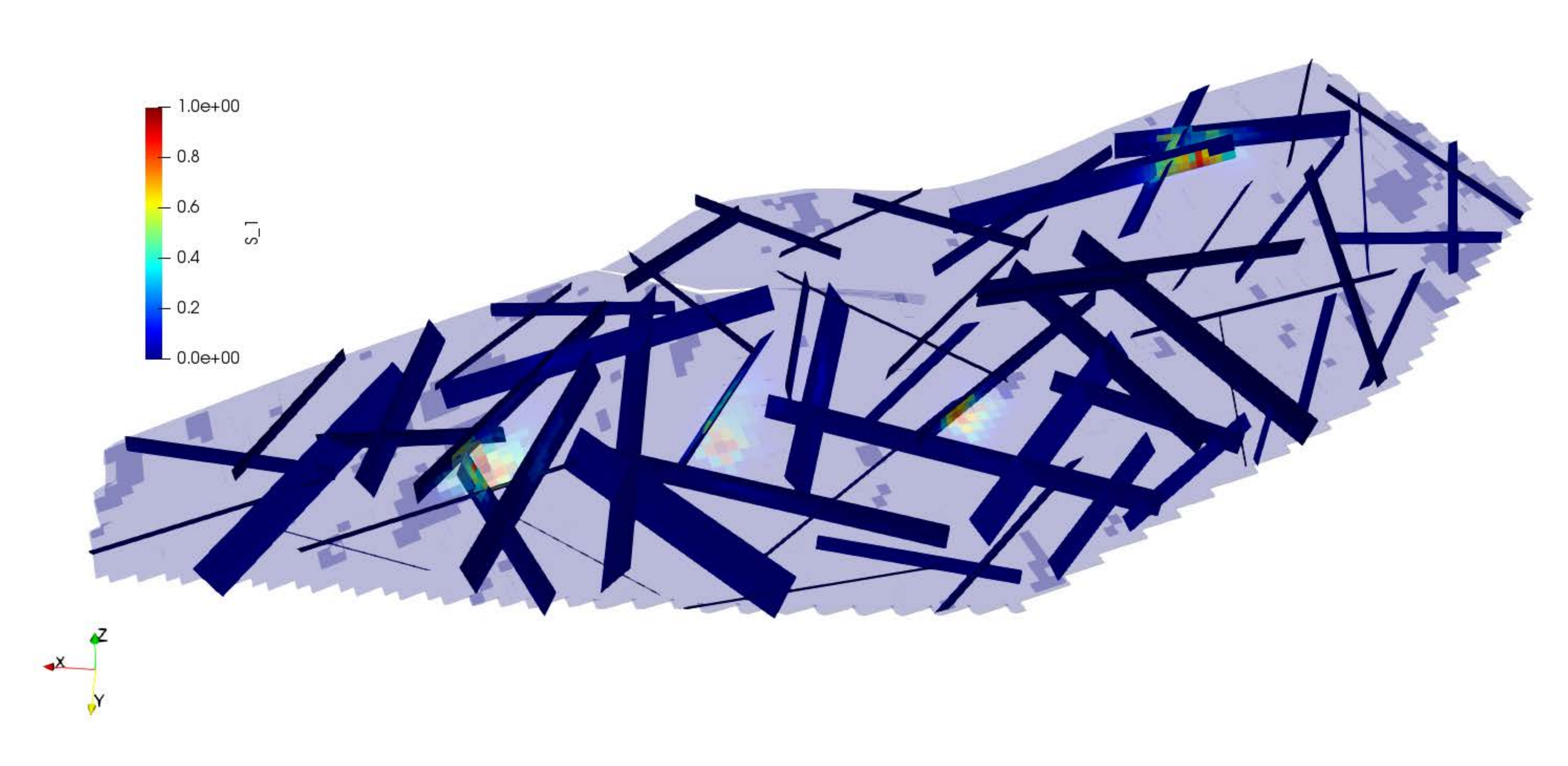}}
	\\
	\subcaptionbox{{\footnotesize Saturation in the matrix after $5000 [\text{days}]$ }\label{Fig:pEDFM_CPG_TestCase5_Brugge_Sm_scenario1_T10}}
	{\includegraphics[width=0.23\textwidth]{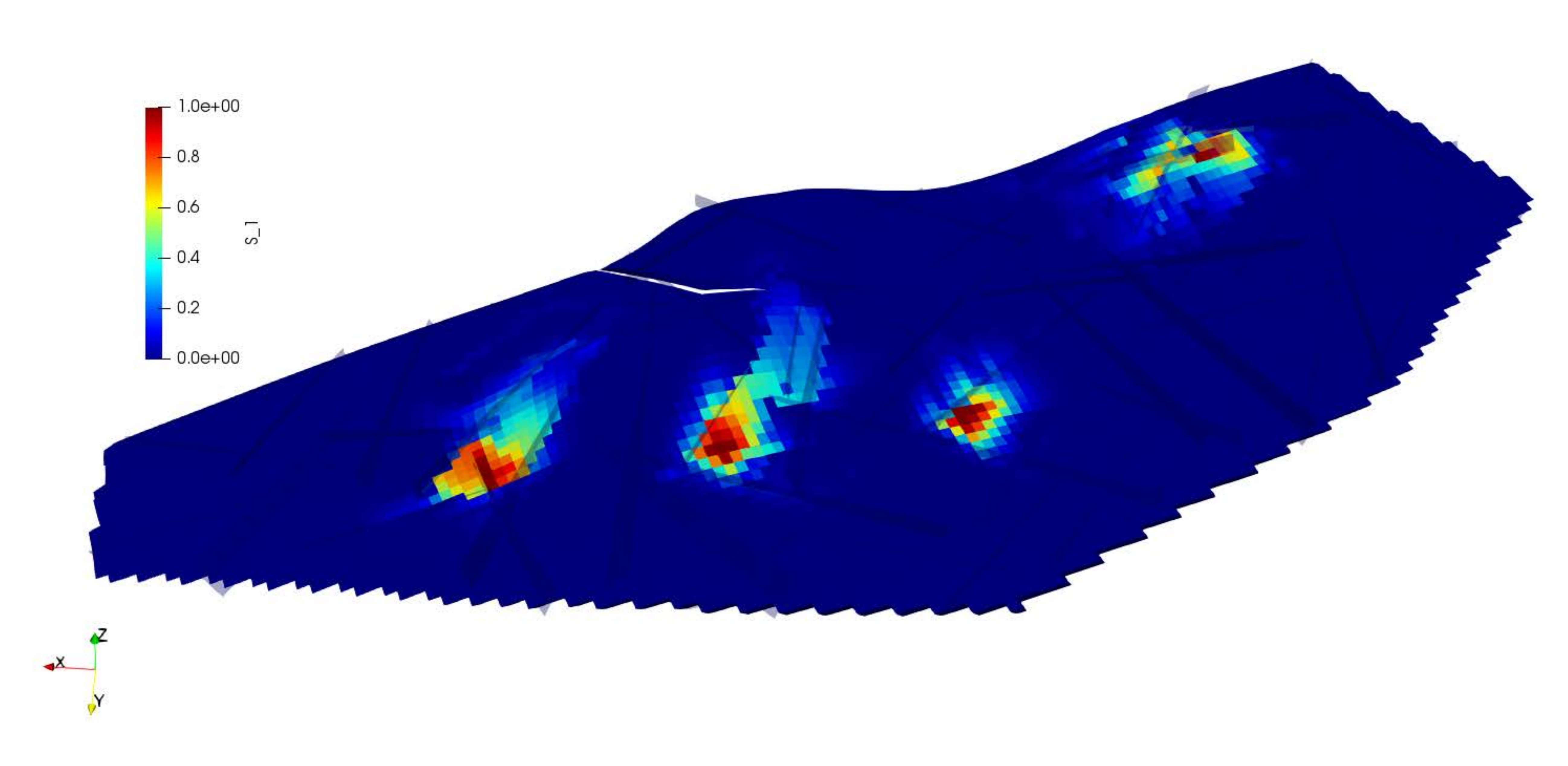}}
	\subcaptionbox{{\footnotesize Saturation in the fractures after $5000 [\text{days}]$ }\label{Fig:pEDFM_CPG_TestCase5_Brugge_Sf_scenario1_T10}}
	{\includegraphics[width=0.23\textwidth]{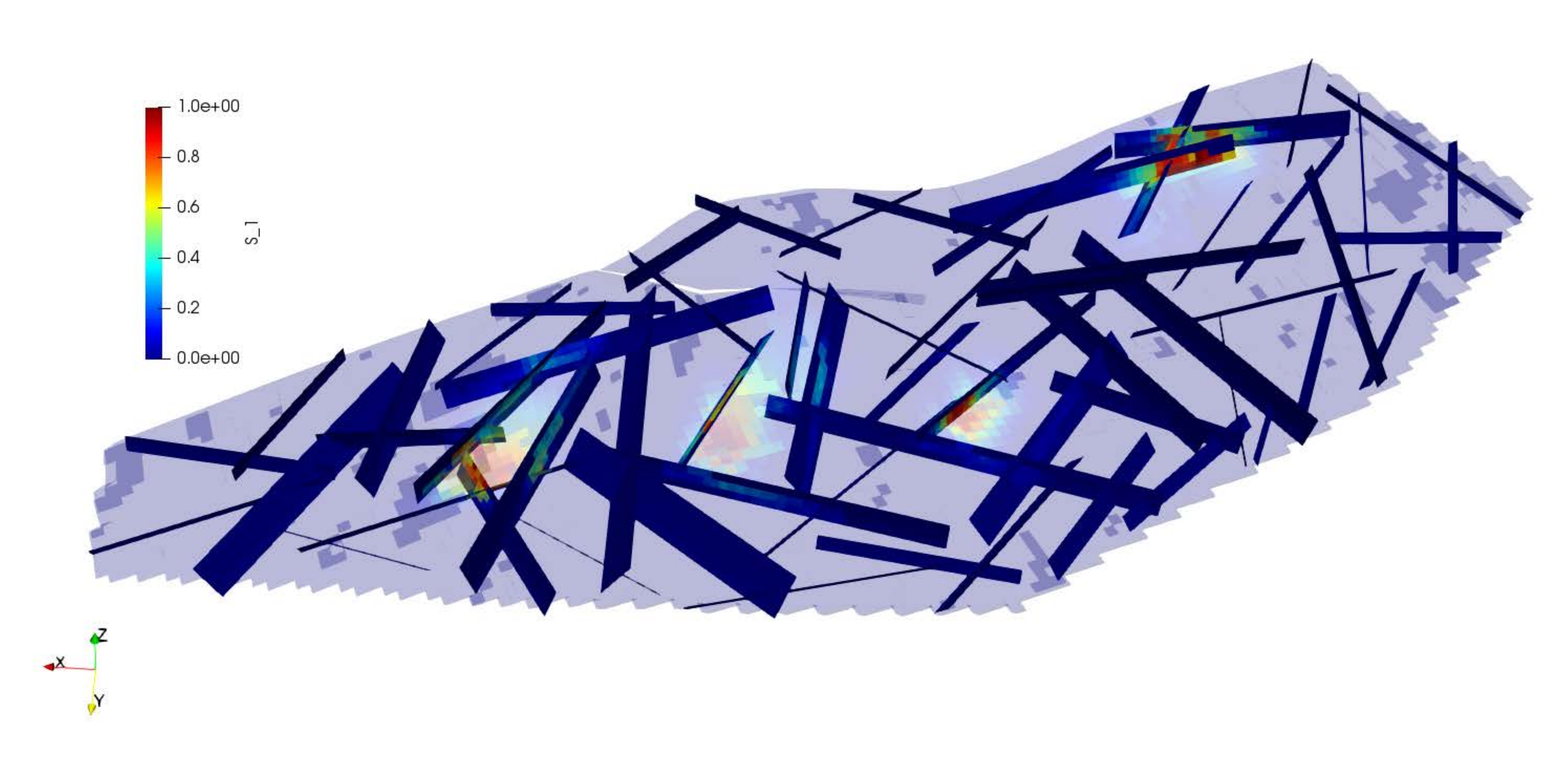}}
	\\
	\subcaptionbox{{\footnotesize Saturation in the matrix after $10000 [\text{days}]$ }\label{Fig:pEDFM_CPG_TestCase5_Brugge_Sm_scenario1_T40}}
	{\includegraphics[width=0.23\textwidth]{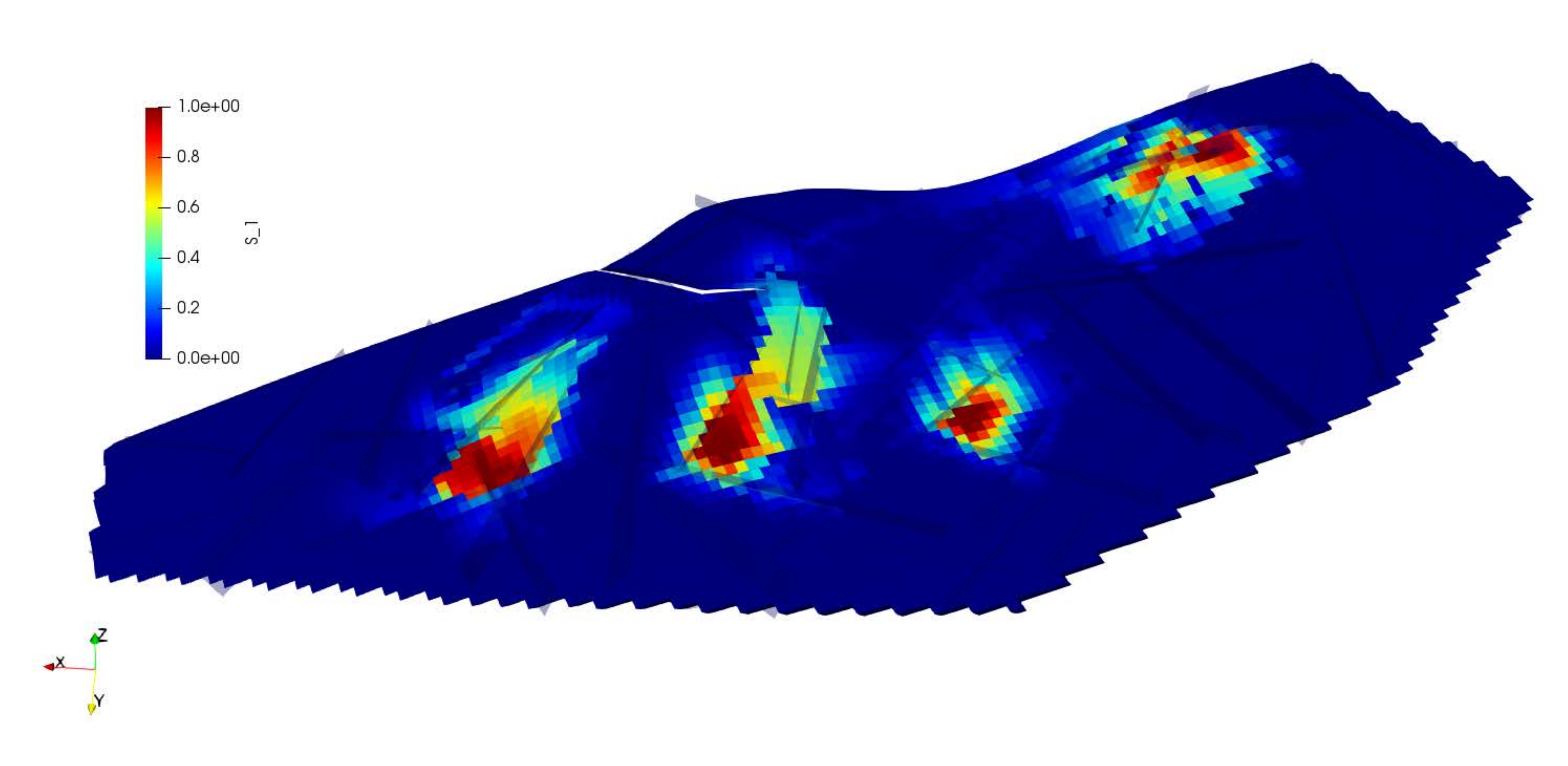}}
	\subcaptionbox{{\footnotesize Saturation in the fractures after $10000 [\text{days}]$ }\label{Fig:pEDFM_CPG_TestCase5_Brugge_Sf_scenario1_40}}
	{\includegraphics[width=0.23\textwidth]{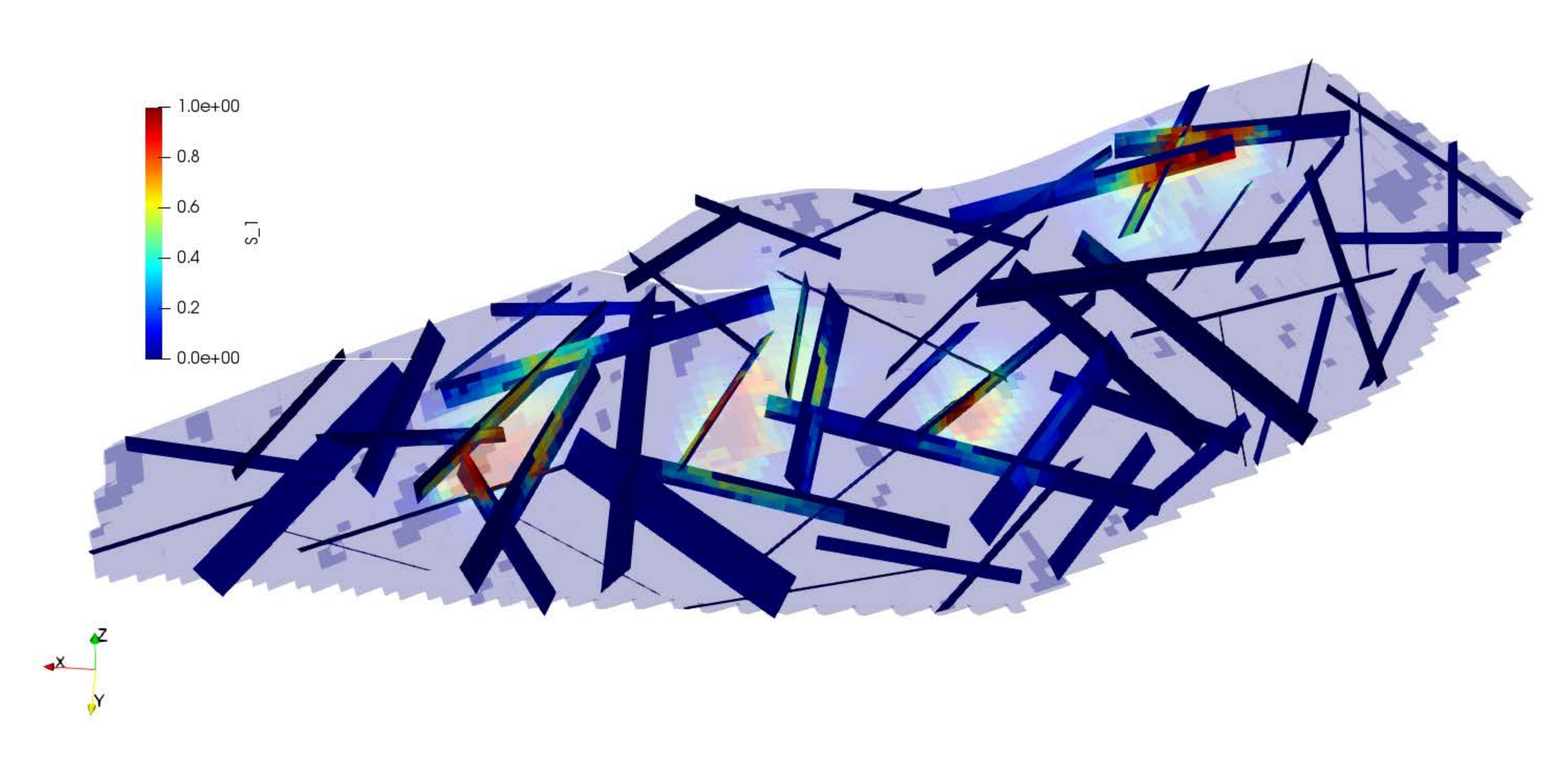}}
	\caption{Test case 5: The saturation profile of the Brugge model for the simulation scenario $1$. The figures on the left illustrate the saturation profile in the matrix grid cells and the figures on the right side show the saturation maps in the fractures. From the top row towards the bottom row, the saturation profiles are displayed for simulations times $2000$, $5000$ and $10000 [\text{days}]$ respectively.}
	\label{Fig:pEDFM_CPG_TestCase5_Brugge_Saturation_Scenario1}
\end{figure}

The pressure and saturation results of the scenario $1$ are showed in the figures \ref{Fig:pEDFM_CPG_TestCase5_Brugge_Pressure_Scenario1} and \ref{Fig:pEDFM_CPG_TestCase5_Brugge_Saturation_Scenario1} respectively. The pressure results are only shown for the simulation time $5000 [\text{days}]$, but the saturation profiles are presented for three time intervals of $2000$, $5000$ and $10000 [\text{days}]$. The injection wells are surrounded by highly conductive fractures that act as flow channels. As a result, the saturation of the injecting phase is considerably increased in larger distances from the injection phases and the pressure drop around the injection wells is not high.

\begin{figure}[!htbp]
	\centering
	\subcaptionbox{{\footnotesize Pressure in the matrix}\label{Fig:pEDFM_CPG_TestCase5_Brugge_Pm_scenario2}}
	{\includegraphics[width=0.23\textwidth]{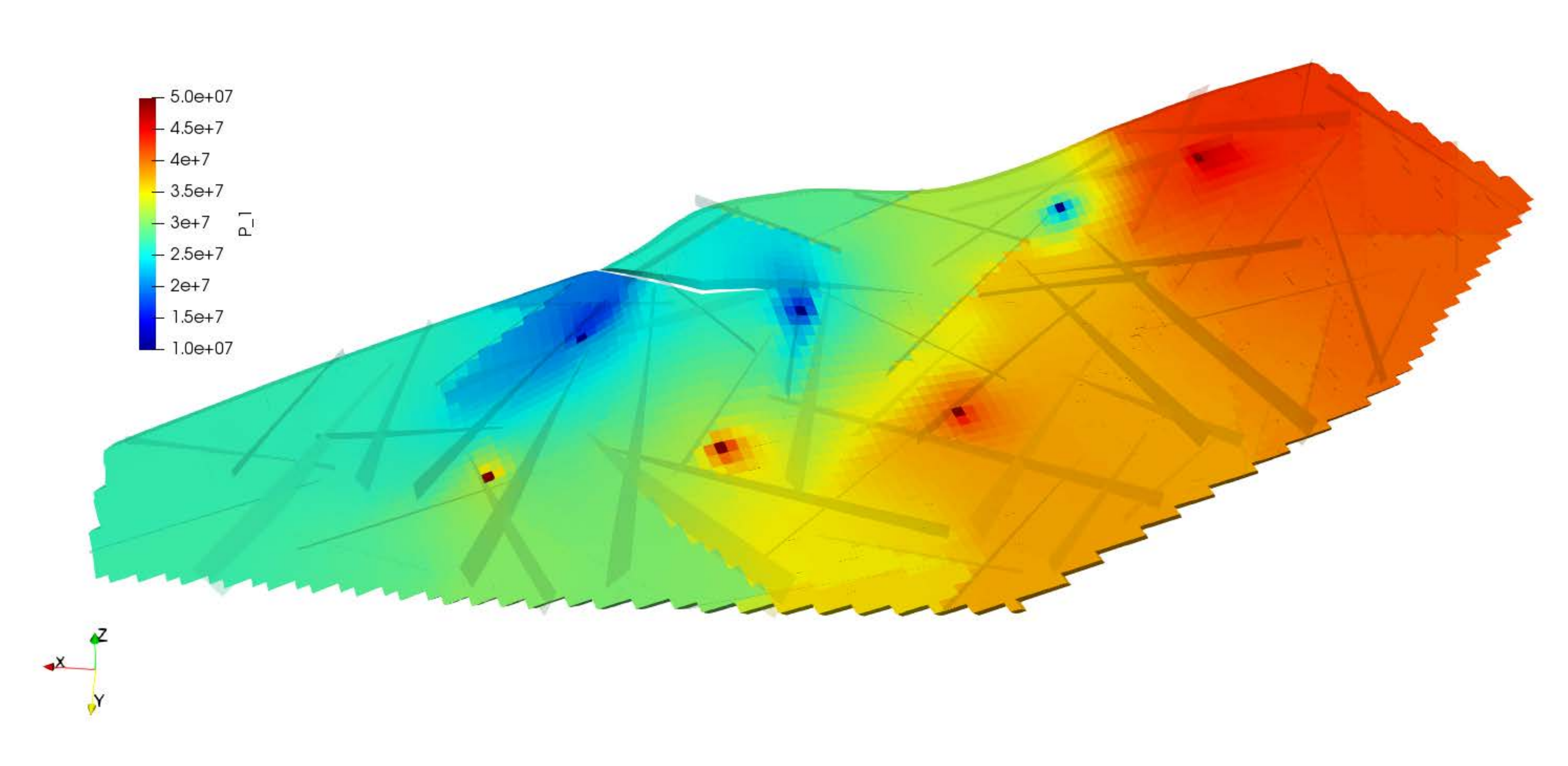}}
	\subcaptionbox{{\footnotesize Pressure in the fractures}\label{Fig:pEDFM_CPG_TestCase5_Brugge_Pf_scenario2}}
	{\includegraphics[width=0.23\textwidth]{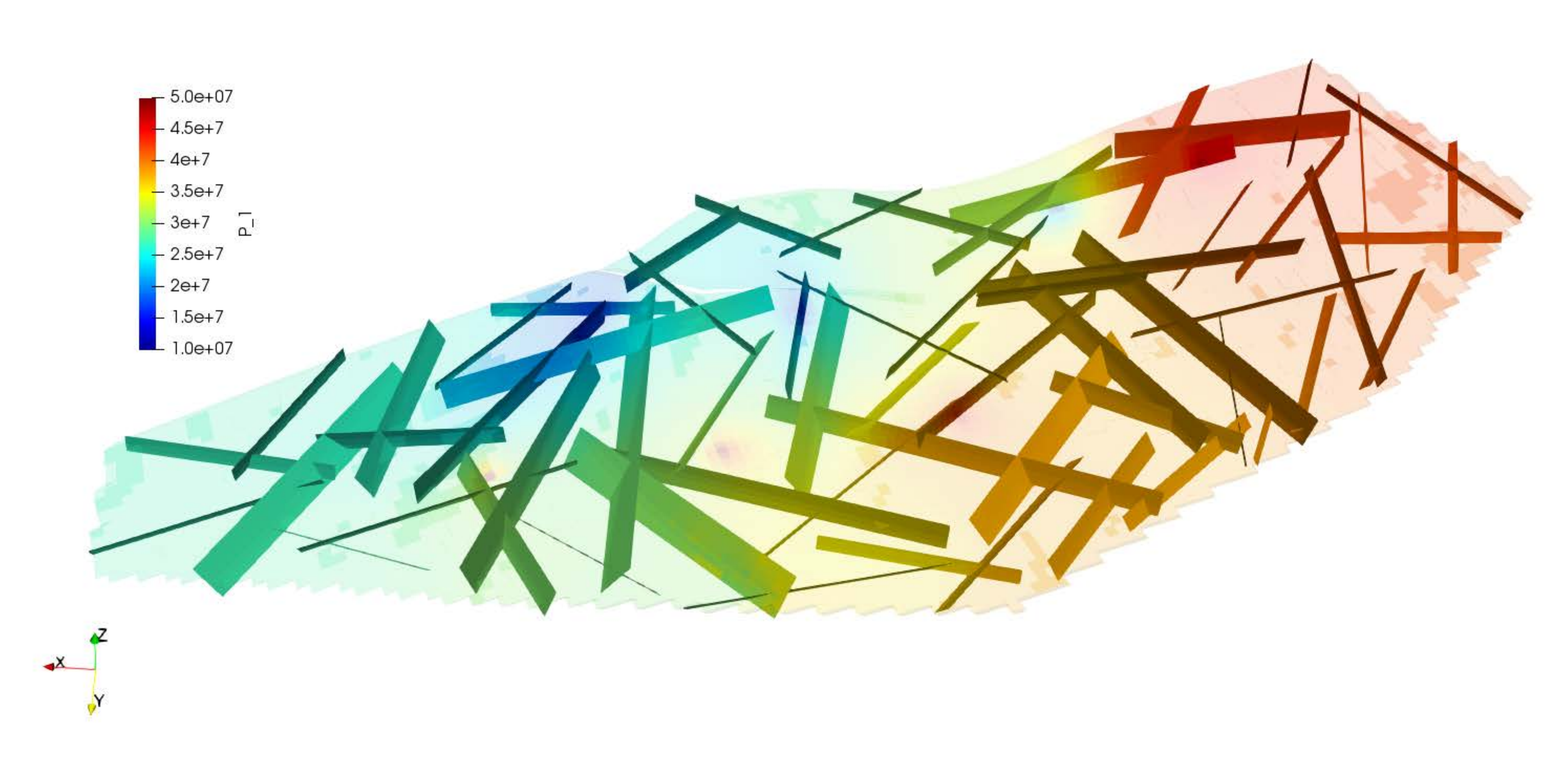}}
	\caption{Test case 5: The pressure profile of the Brugge model for the simulation scenario $2$. The figure on the left shows the pressure distribution in the matrix grid cells. The transparency of this figure is increased to make the pressure map in the fractures visible. This map is displayed on the right figure. The results are shown for the simulation time $5000 [\text{days}]$}
	\label{Fig:pEDFM_CPG_TestCase5_Brugge_Pressure_Scenario2}
\end{figure}

\begin{figure}[!htbp]
	\centering
	\subcaptionbox{{\footnotesize Saturation in the matrix after $2000 [\text{days}]$ }\label{Fig:pEDFM_CPG_TestCase5_Brugge_Sm_scenario2_T04}}
	{\includegraphics[width=0.23\textwidth]{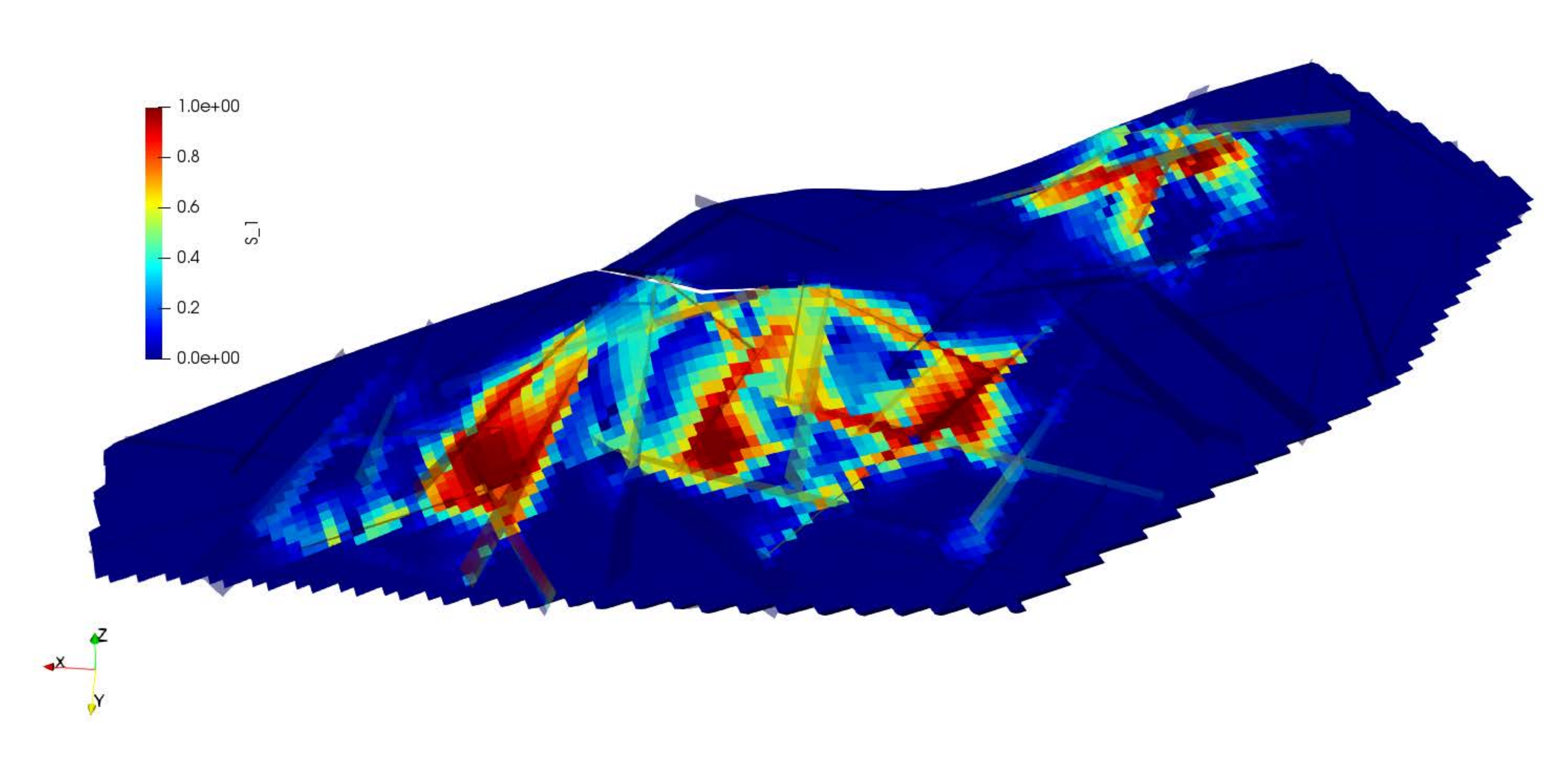}}
	\subcaptionbox{{\footnotesize Saturation in the fractures after $2000 [\text{days}]$ }\label{Fig:pEDFM_CPG_TestCase5_Brugge_Sf_scenario2_T04}}
	{\includegraphics[width=0.23\textwidth]{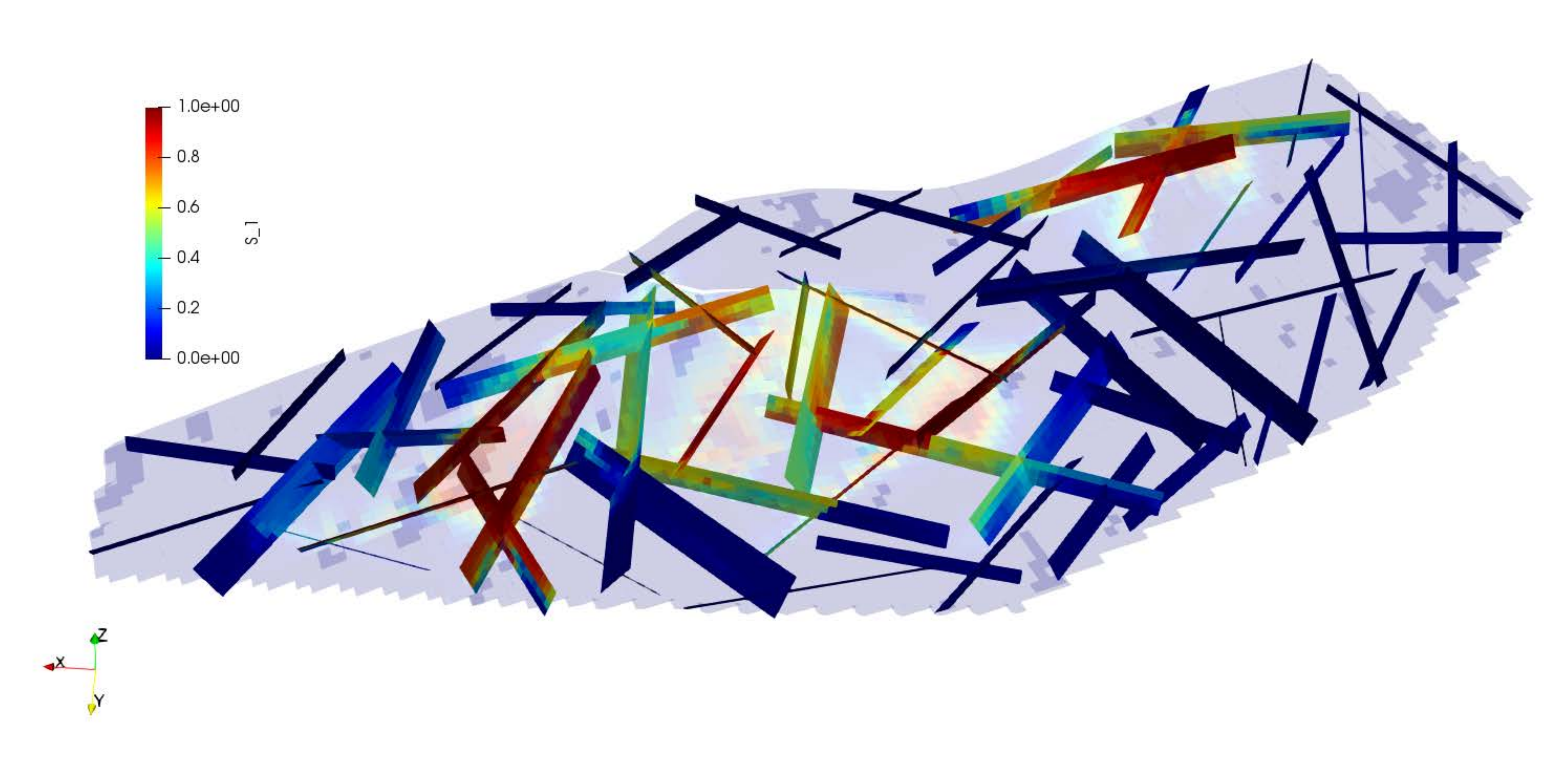}}
	\\
	\subcaptionbox{{\footnotesize Saturation in the matrix after $5000 [\text{days}]$ }\label{Fig:pEDFM_CPG_TestCase5_Brugge_Sm_scenario2_T10}}
	{\includegraphics[width=0.23\textwidth]{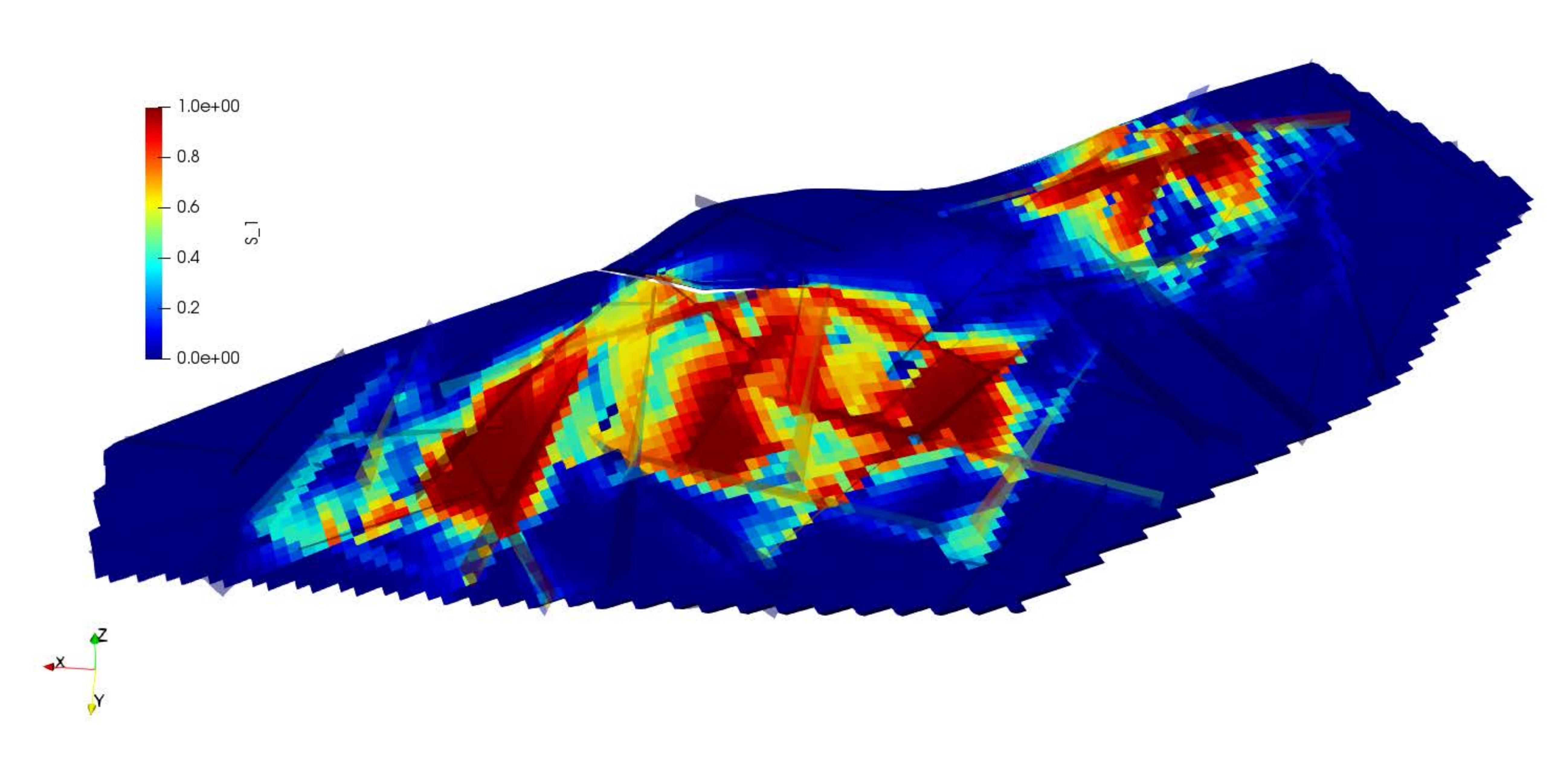}}
	\subcaptionbox{{\footnotesize Saturation in the fractures after $5000 [\text{days}]$ }\label{Fig:pEDFM_CPG_TestCase5_Brugge_Sf_scenario2_T10}}
	{\includegraphics[width=0.23\textwidth]{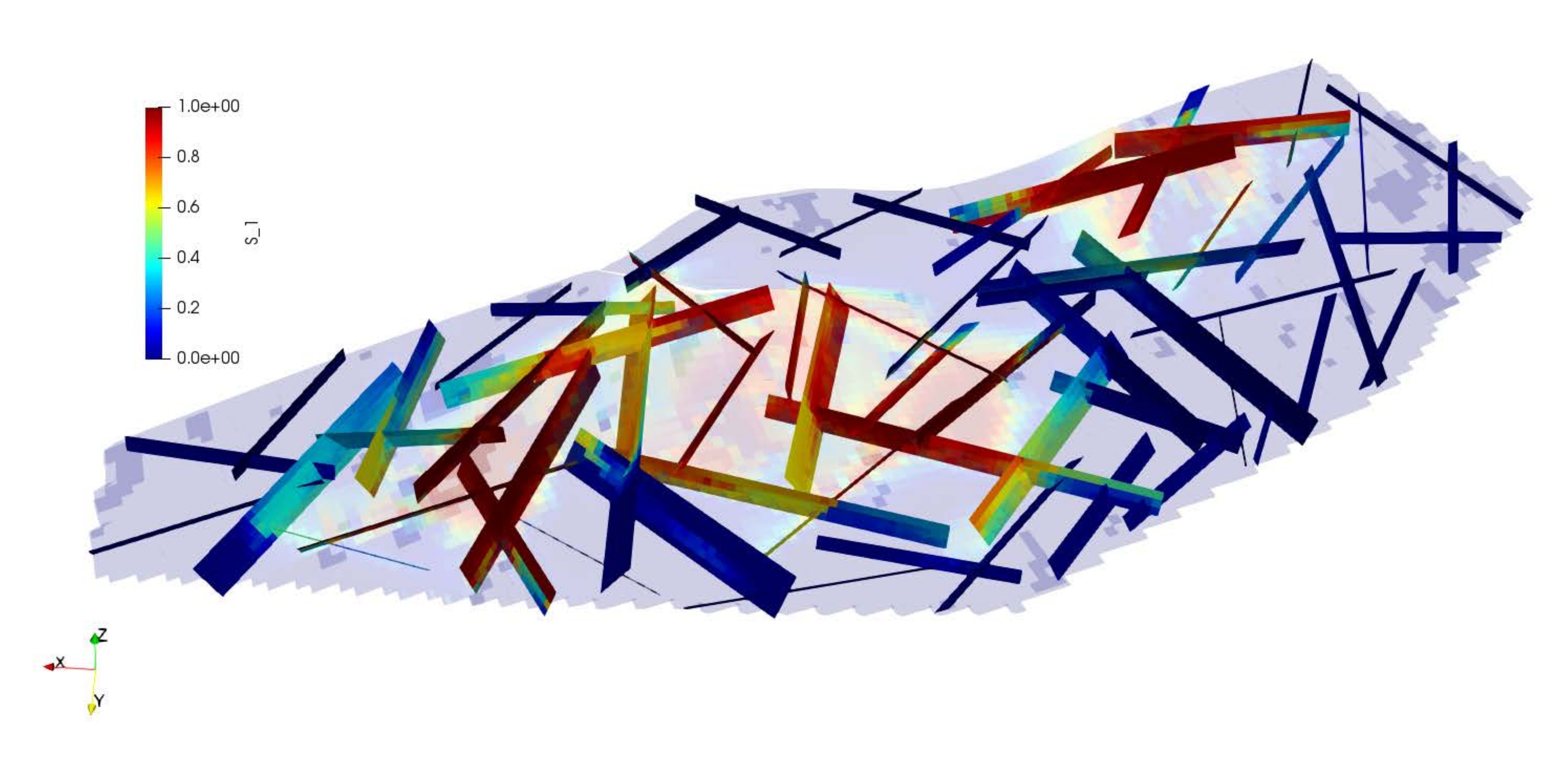}}
	\\
	\subcaptionbox{{\footnotesize Saturation in the matrix after $10000 [\text{days}]$ }\label{Fig:pEDFM_CPG_TestCase5_Brugge_Sm_scenario2_T40}}
	{\includegraphics[width=0.23\textwidth]{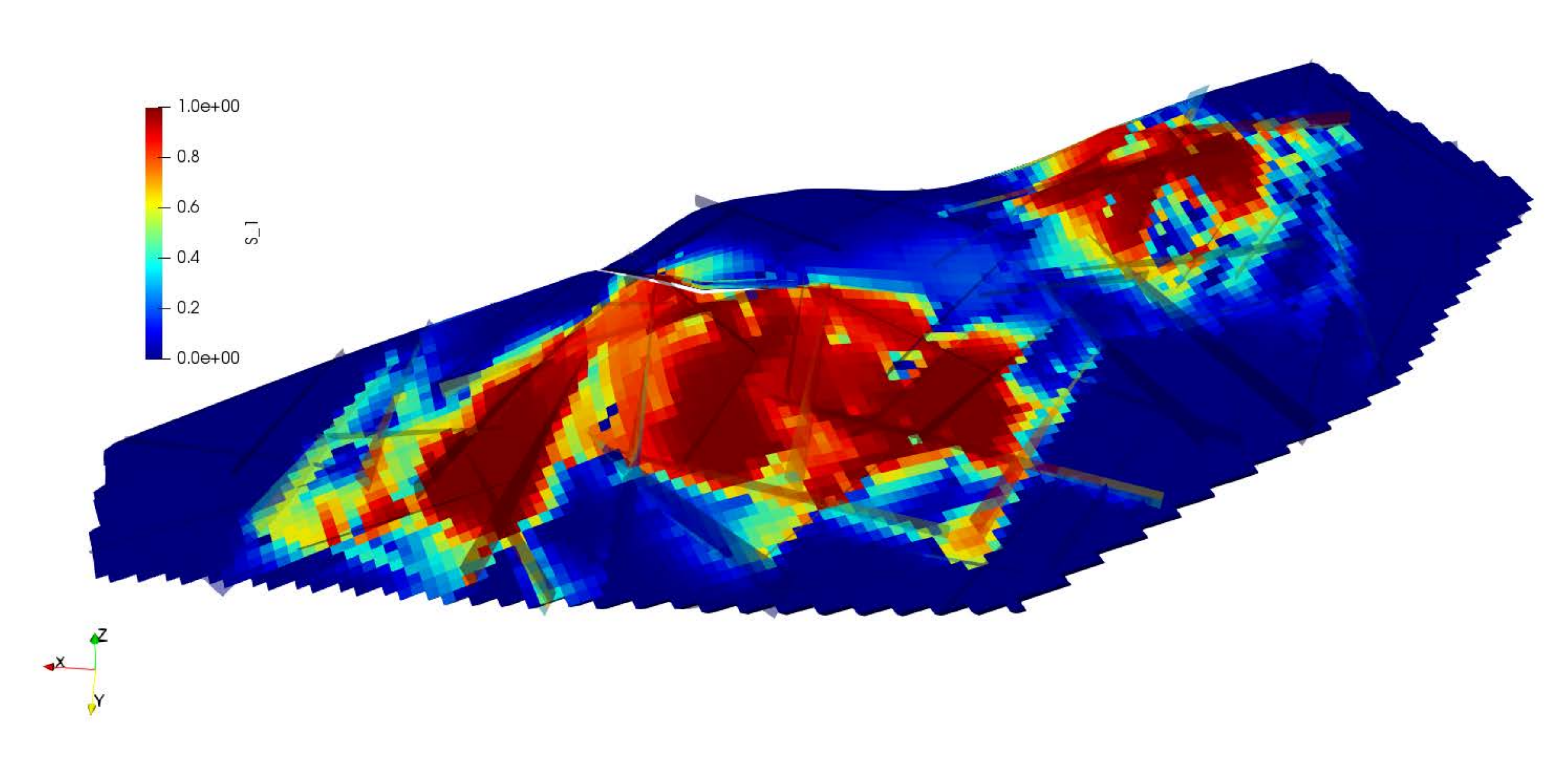}}
	\subcaptionbox{{\footnotesize Saturation in the fractures after $10000 [\text{days}]$ }\label{Fig:pEDFM_CPG_TestCase5_Brugge_Sf_scenario2_40}}
	{\includegraphics[width=0.23\textwidth]{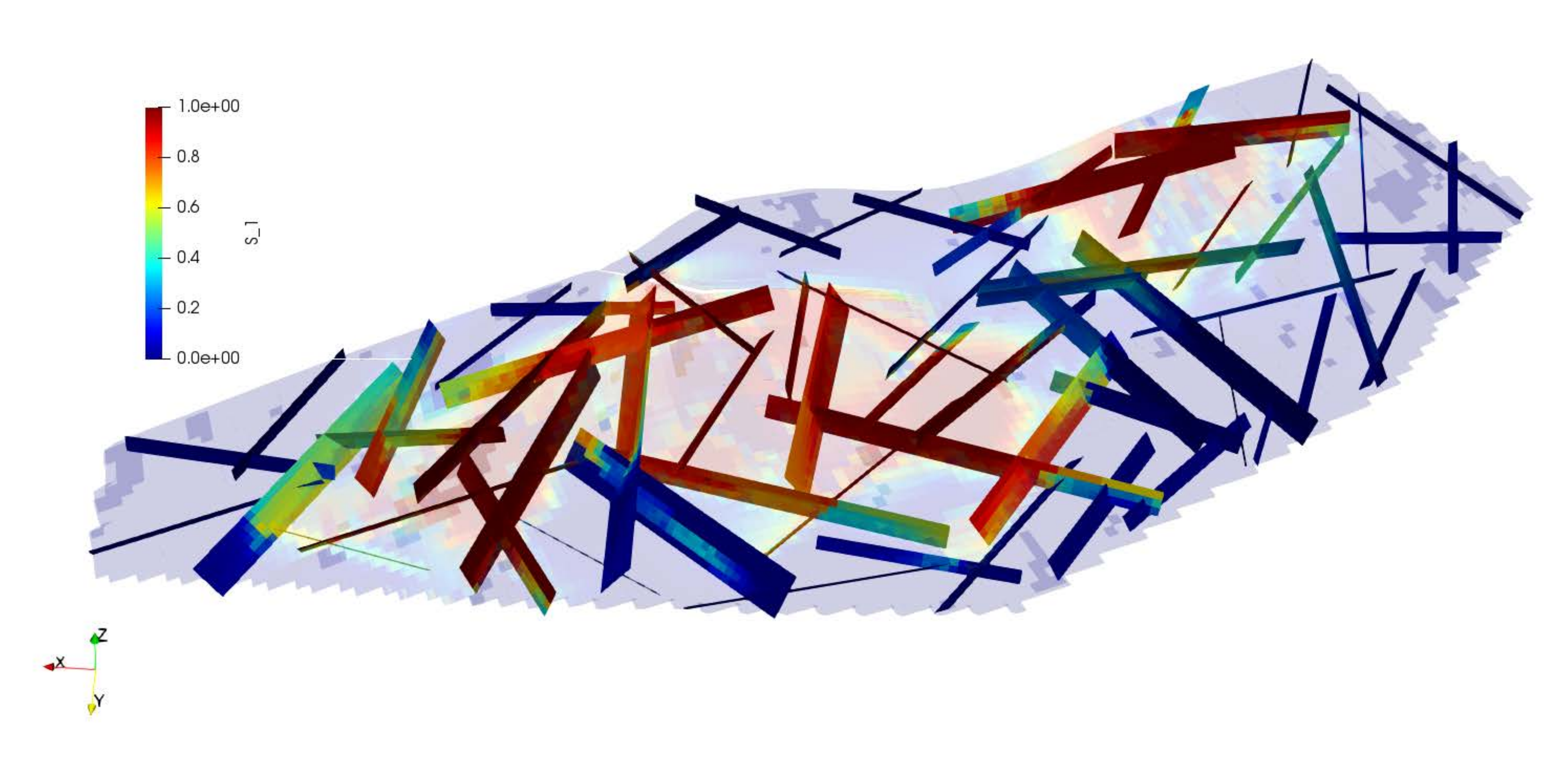}}
	\caption{Test case 5: The saturation profile of the Brugge model for the simulation scenario $2$. The figures on the left illustrate the saturation profile in the matrix grid cells and the figures on the right side show the saturation maps in the fractures. From the top row towards the bottom row, the saturation profiles are displayed for simulations times $2000$, $5000$ and $10000 [\text{days}]$ respectively.}
	\label{Fig:pEDFM_CPG_TestCase5_Brugge_Saturation_Scenario2}
\end{figure}

The pressure and saturation results of the scenario $2$ are showed in the figures \ref{Fig:pEDFM_CPG_TestCase5_Brugge_Pressure_Scenario2} and \ref{Fig:pEDFM_CPG_TestCase5_Brugge_Saturation_Scenario2} respectively. The pressure results are only shown for the simulation time $5000 [\text{days}]$, but the saturation profiles are presented for three time intervals of $2000$, $5000$ and $10000 [\text{days}]$. The injection wells are surrounded by flow barriers that restrict the flow. As a result, a high-pressure zone is formed near the wells since the central area of the reservoir is isolated with low permeability fractures. This is followed by a sharp pressure gradient. The saturation displacement is small due to the reservoir's low permeability values and the absence of highly conductive fractures near the wells. The saturation displacement is restricted to the area near the injection wells.

\subsection{Test Case 6 and 7: Norne field}\label{Sec:pEDFM_CPG_Norne_Field}
\begin{figure}[!htbp]
	\centering
	\begin{subfigure}[]{0.23\textwidth}
		\includegraphics[width=\textwidth]{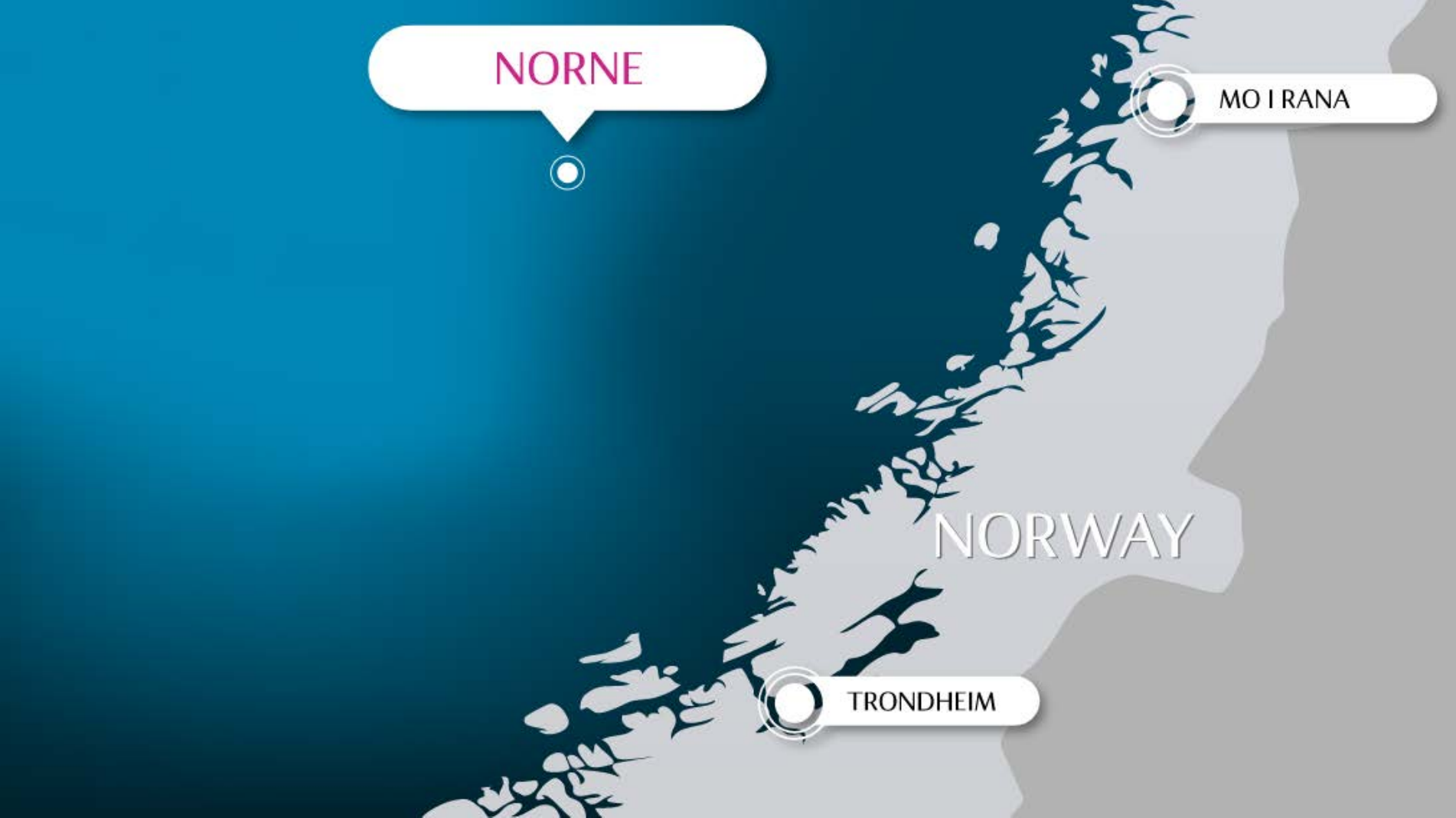}
	\end{subfigure}
	\begin{subfigure}[]{0.23\textwidth}
		\includegraphics[width=\textwidth]{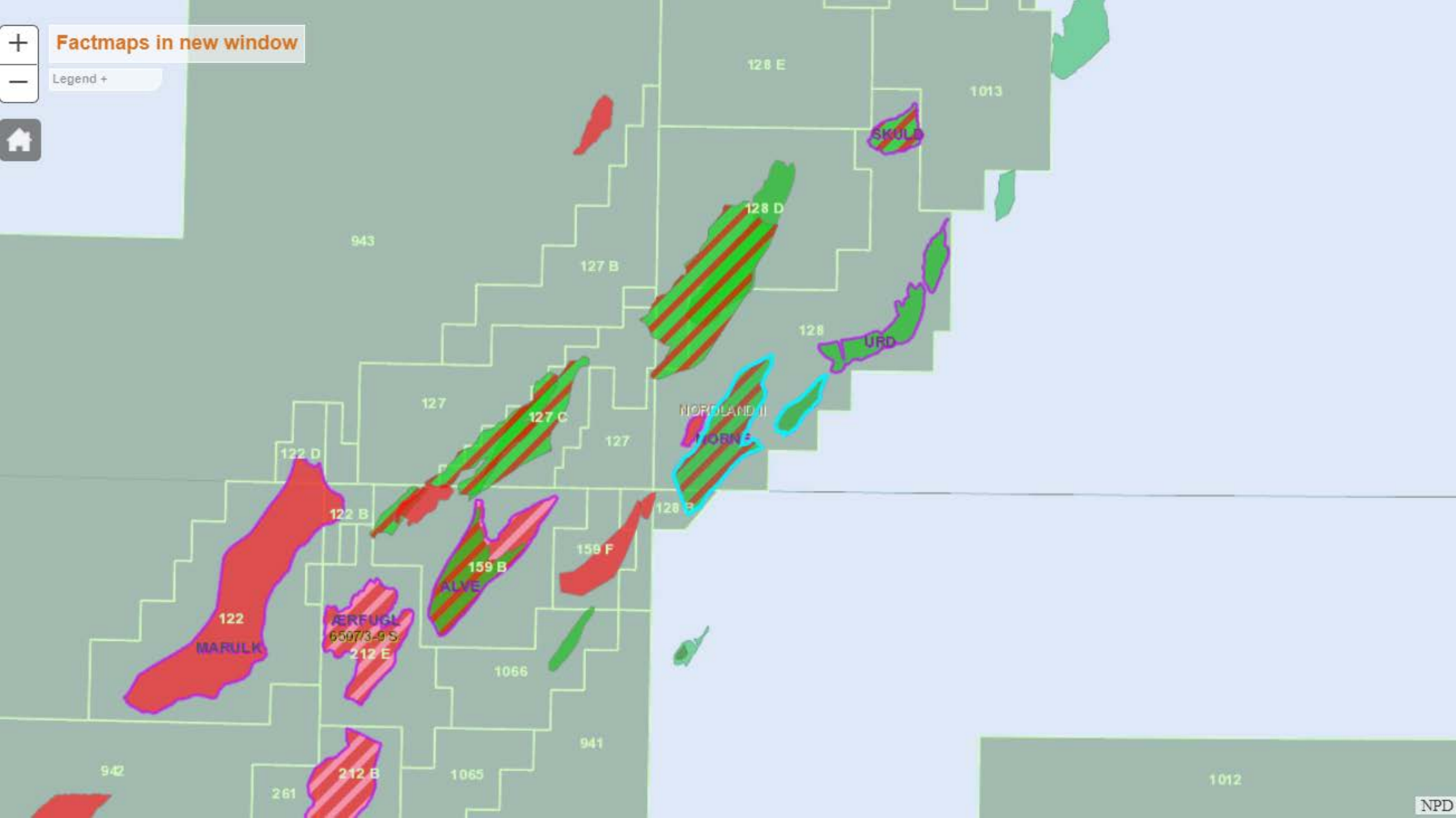}
	\end{subfigure}
	\caption{Test Case 6\&7: Location of the Norne Field. The left plot shows the field located in the Norwegian sea (source: Equinor), and the right picture shows the location of the licensed blocks (source: Norwegian Petroleum Directorate).}
	\label{Fig:pEDFM_CPG_TestCase6_Norne_Location}
\end{figure}

Norne \cite{Verlo2008} is an oil and gas field situated in the Norwegian Sea around 80 kilometers north of the Heidrun oil field. The field dimensions are approximately $9 [\text{km}] \times 3 [\text{km}]$ and the seawater depth in the area is $9 [\text{m}]$. The field is placed in a license awarded region in $1986$ and incorporates blocks $6608/10$ and $6608/11$ (see figure \ref{Fig:pEDFM_CPG_TestCase6_Norne_Location}).Equinor is the current field operator. The expected oil recovery factor is more than $60\%$, which is very high for an offshore sub-sea oil reservoir.

Subsurface data from the Norne field have been published for research and education purposes thanks to NTNU, Equinor, and partners' initiative. The full simulation model can be obtained through the Open Porous Media (OPM) project (opm-project.org) \cite{OPM}. The Norne field simulation model was the first benchmark case based on real field data available to the public. The model is based on the $2004$ geological model and consists of $46 \times 112 \times 22$ corner-point grid cells.

\subsubsection{Reservoir}
The Oil and gas production of Norne is obtained from a Jurassic sandstone, which lies at a depth of $2500$ meters below sea level. The original estimation of recoverable resources was $95.2$ million cubic meters for oil, mainly in the Ile and Tofte formations, and $13.01$ billion cubic meters for gas in the Garn formation.

\subsubsection{Field development}
The Alve field finding preceded the Norne field's discovery in $1992$. The plan for development and operation (PDO) was approved in $1995$, and the production started in $1997$. The field development infrastructure consists of production, storage, and offloading vessel (FPSO) attached to sub-sea templates. Water injection is the drive mechanism to produce from the field. Since $2001$, the gas has been exported from Norne, but in $2005$ the gas injection stopped as planned to be exported.

In $2019$, Norne FPSO was granted a lifetime extension to increase value creation from the Norne field and its satellite fields, and also, the blow-down of the gas cap in the Not Formation started. In $2020$ two production wells were planned to be drilled in the Ile Formation.

\begin{figure}[!htbp]
	\centering
	\begin{subfigure}[]{0.23\textwidth}
		\includegraphics[width=\textwidth]{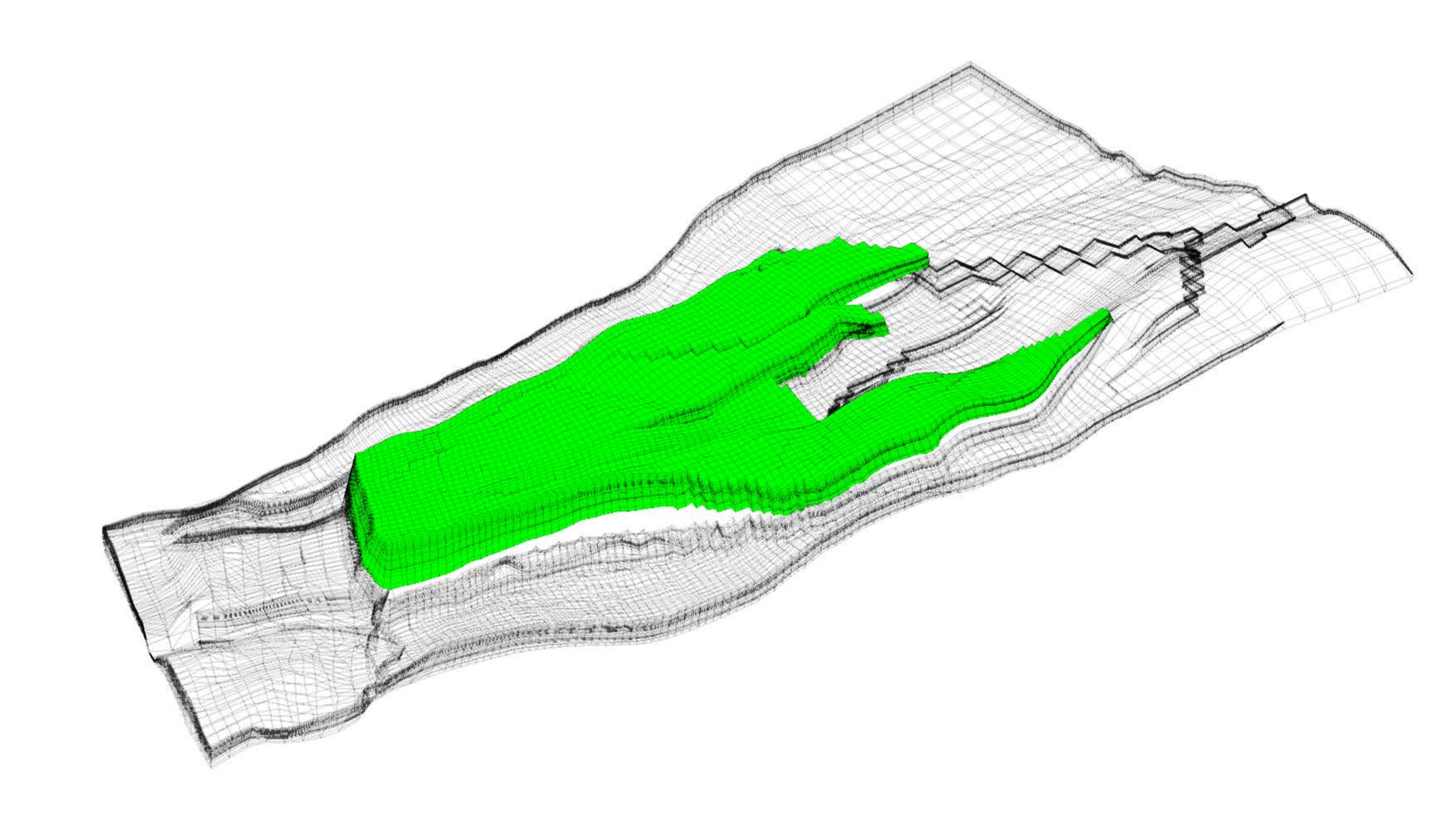}
	\end{subfigure}
	\begin{subfigure}[]{0.23\textwidth}
		\includegraphics[width=\textwidth]{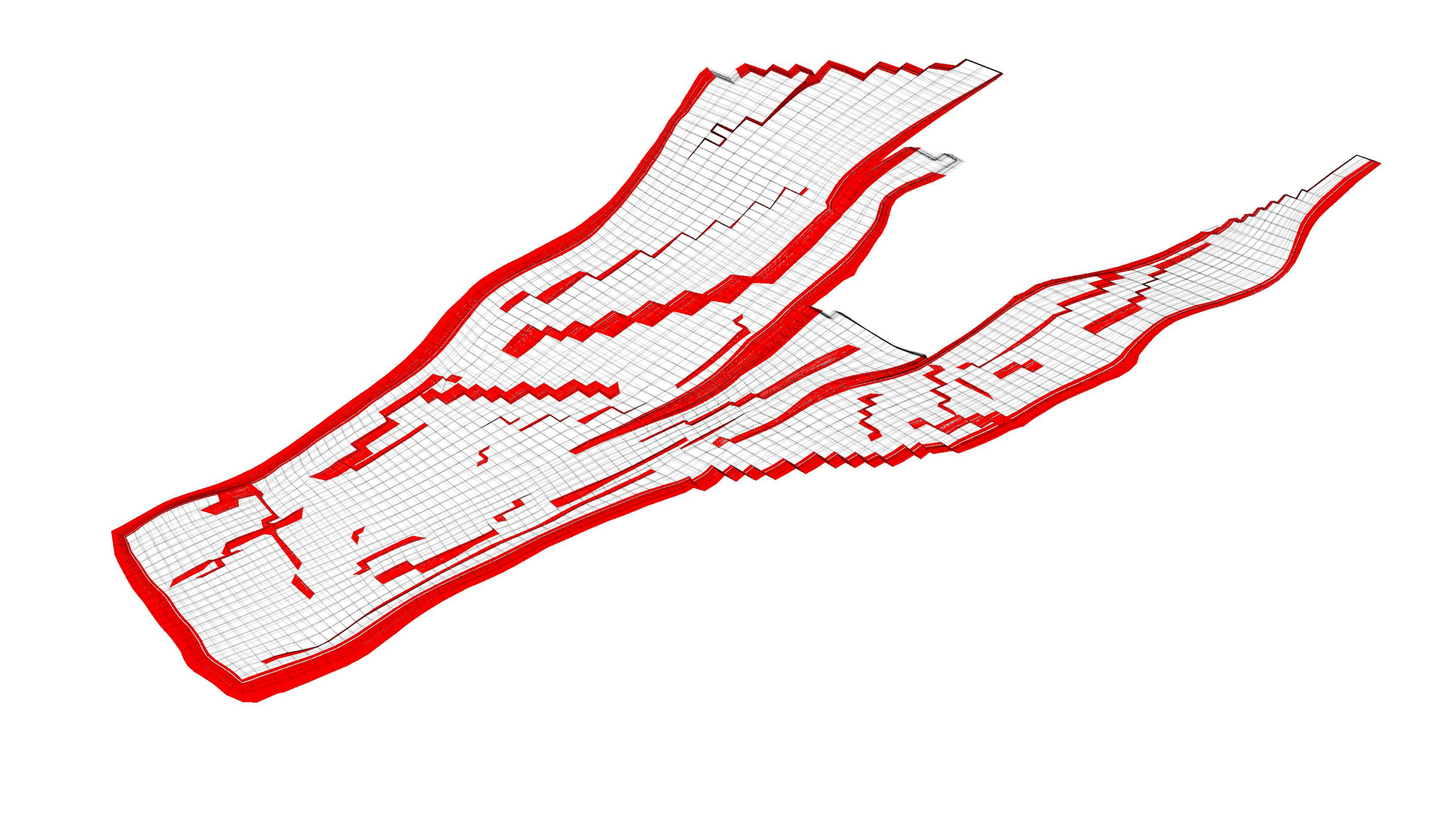}
	\end{subfigure}
	\caption{Test Case 6\&7: Illustration of the Norne field model. The left graph represents the active section (colored in green) of the model, and the right figure shows the faults marked with red color.}
	\label{Fig:pEDFM_CPG_TestCase6_Norne_Active_Cells_Faults}
\end{figure}

\subsubsection{Petrophysical data}
\begin{figure}[!htbp]
	\centering
	\begin{subfigure}[]{0.23\textwidth}
		\includegraphics[width=\textwidth]{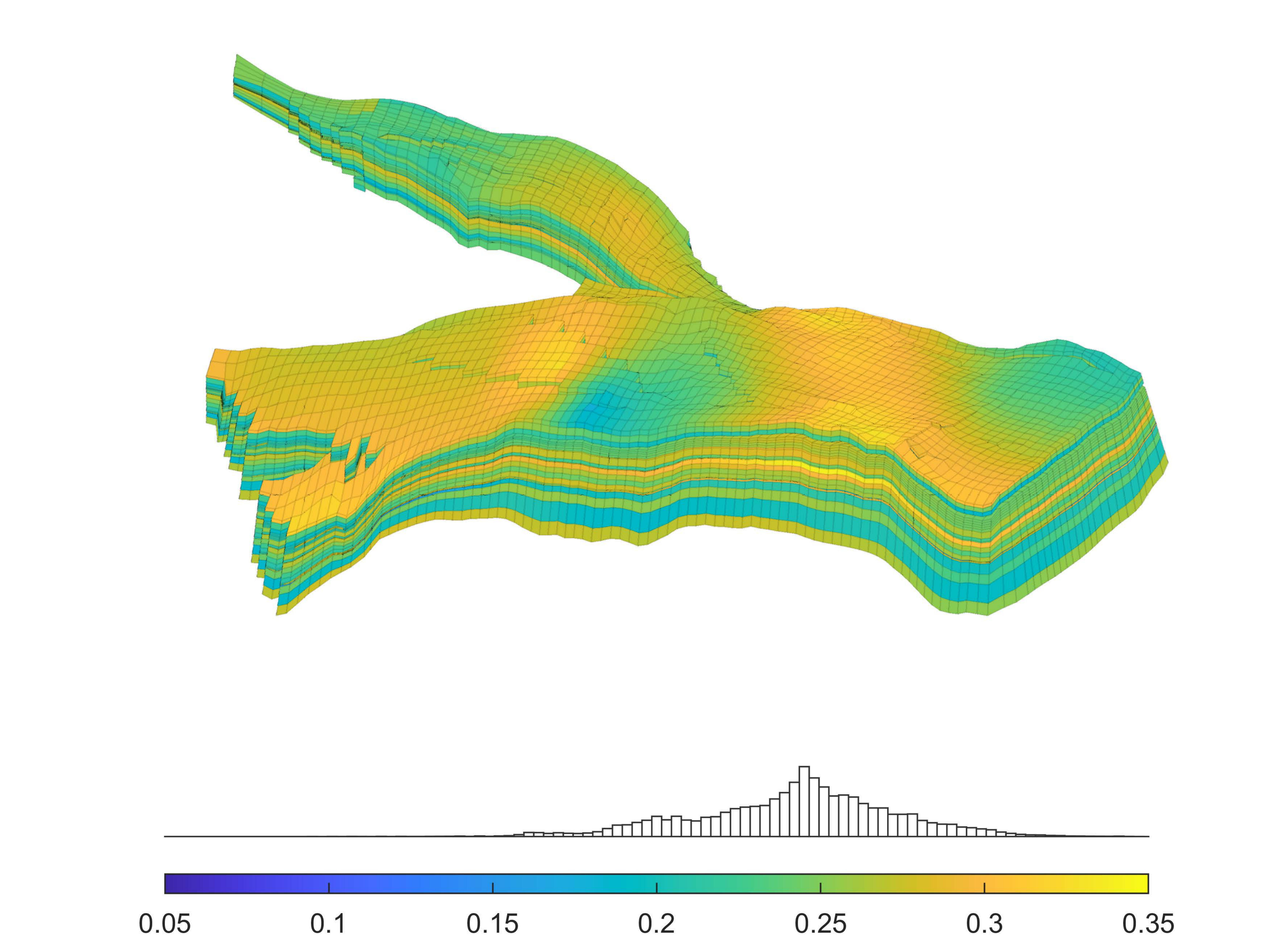}
	\end{subfigure}
	\begin{subfigure}[]{0.23\textwidth}
		\includegraphics[width=\textwidth]{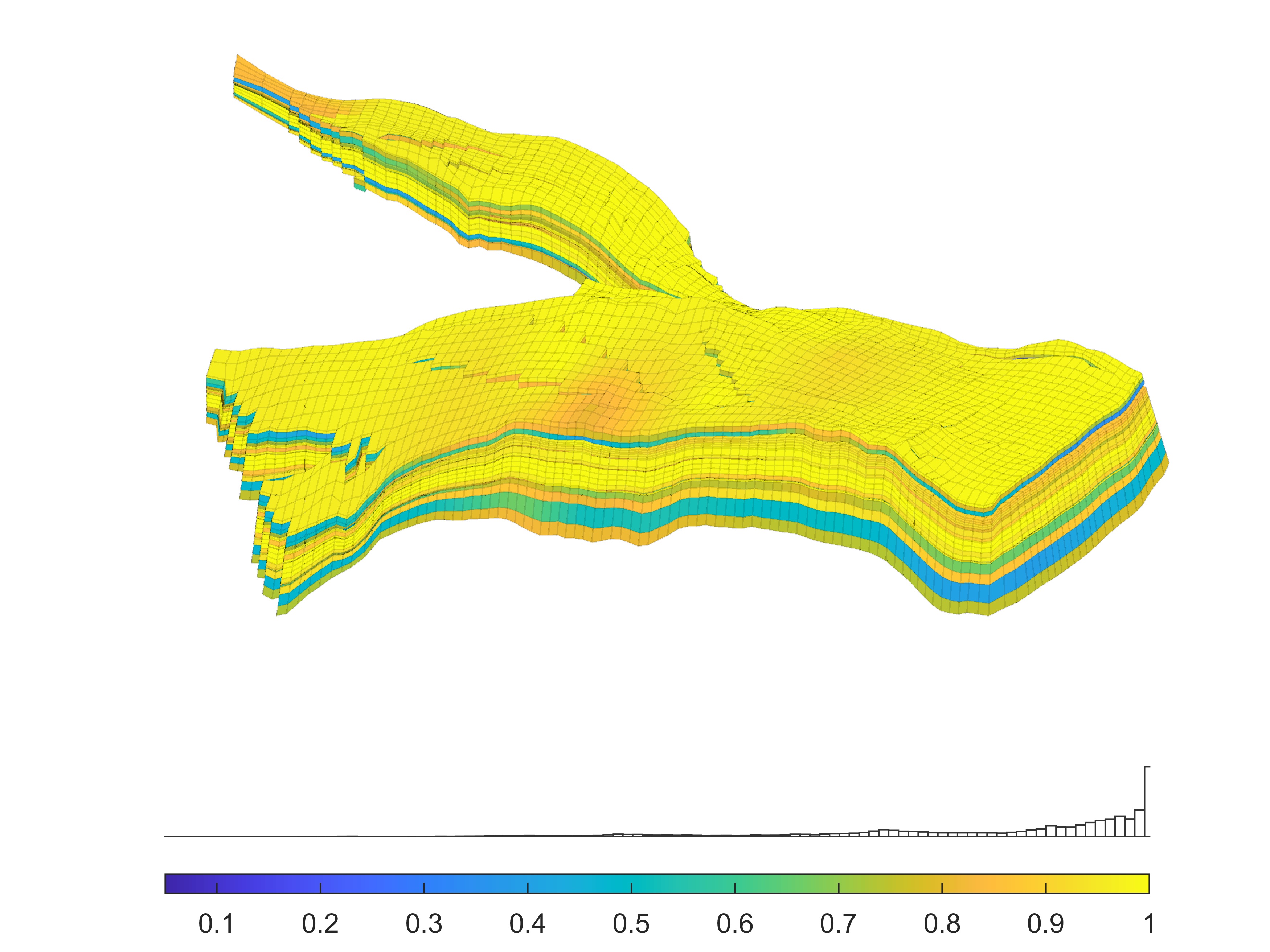}
	\end{subfigure}
	\caption{Test Case 6\&7: Porosity and net-to-gross ratio for the Norne Field. The left graph shows the model's porosity, and the right one shows net-to-gross ratio map in the structure of model.}
	\label{Fig:pEDFM_CPG_TestCase6_Norne_Porosity_NTG}
\end{figure}

The field simulation model's petrophysical data consist of porosity, permeability, net-to-gross, and transmissibility multiplier data. Permeability is anisotropic and heterogeneous, with a clear layered structure as expected for a real reservoir field model. The vertical communication is decreased in significant regions of the model by the transmissibility multiplier data that is available, resulting in intermediate layers of the reservoir with permeability values close to zero. The porosity values of the field are in the interval between $0.094$ and $0.347$ (see figure \ref{Fig:pEDFM_CPG_TestCase6_Norne_Porosity_NTG} on the left), and a reduction in effective porosity is expected since the net-to-gross data is available to use (see figure \ref{Fig:pEDFM_CPG_TestCase6_Norne_Porosity_NTG} on the right). A considerable percentage of impermeable shale is present in some regions in the model.

\begin{figure}[!htbp]
	\centering
	\begin{subfigure}[]{0.23\textwidth}
		\includegraphics[width=\textwidth]{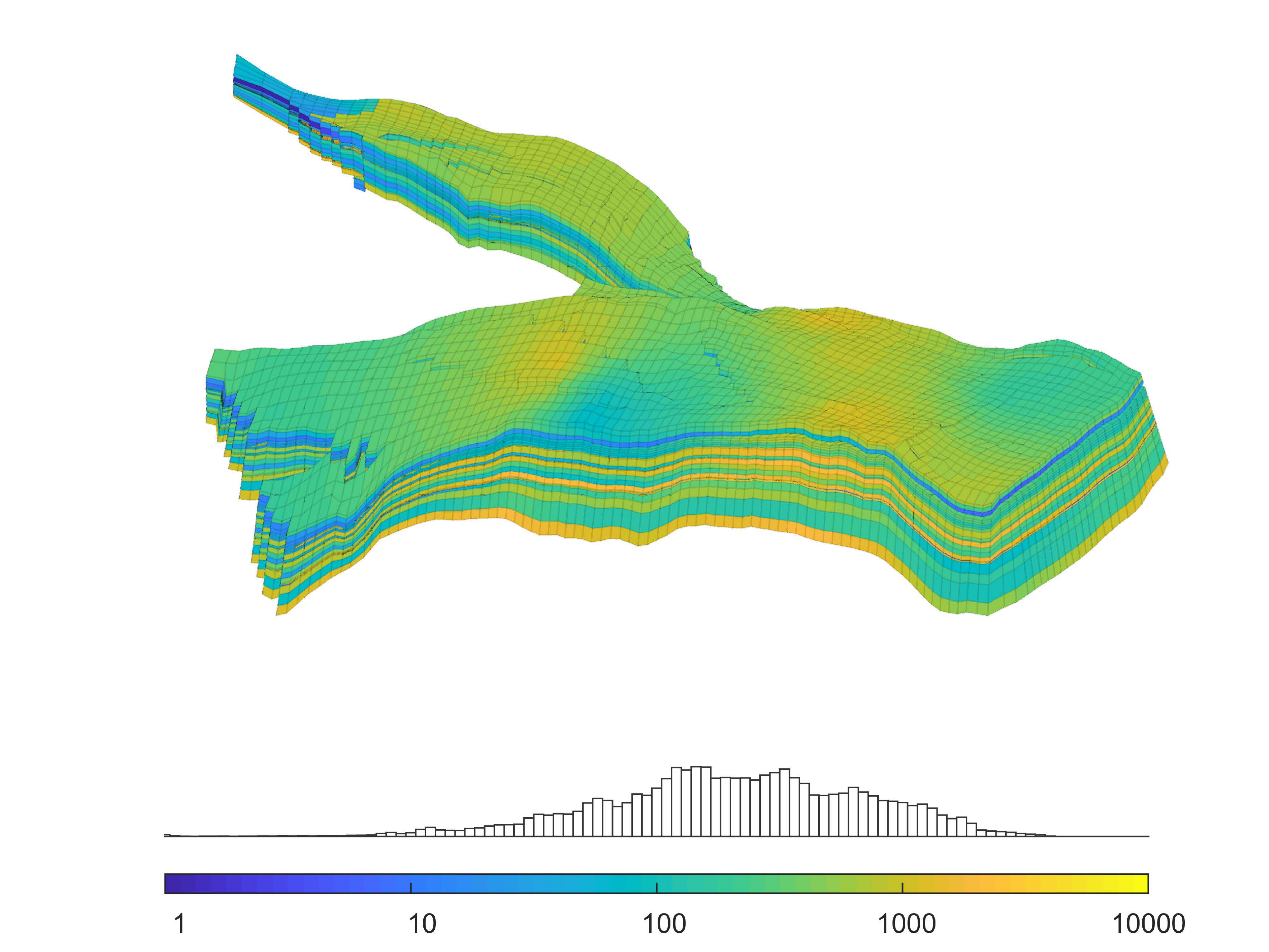}
	\end{subfigure}
	\begin{subfigure}[]{0.23\textwidth}
		\includegraphics[width=\textwidth]{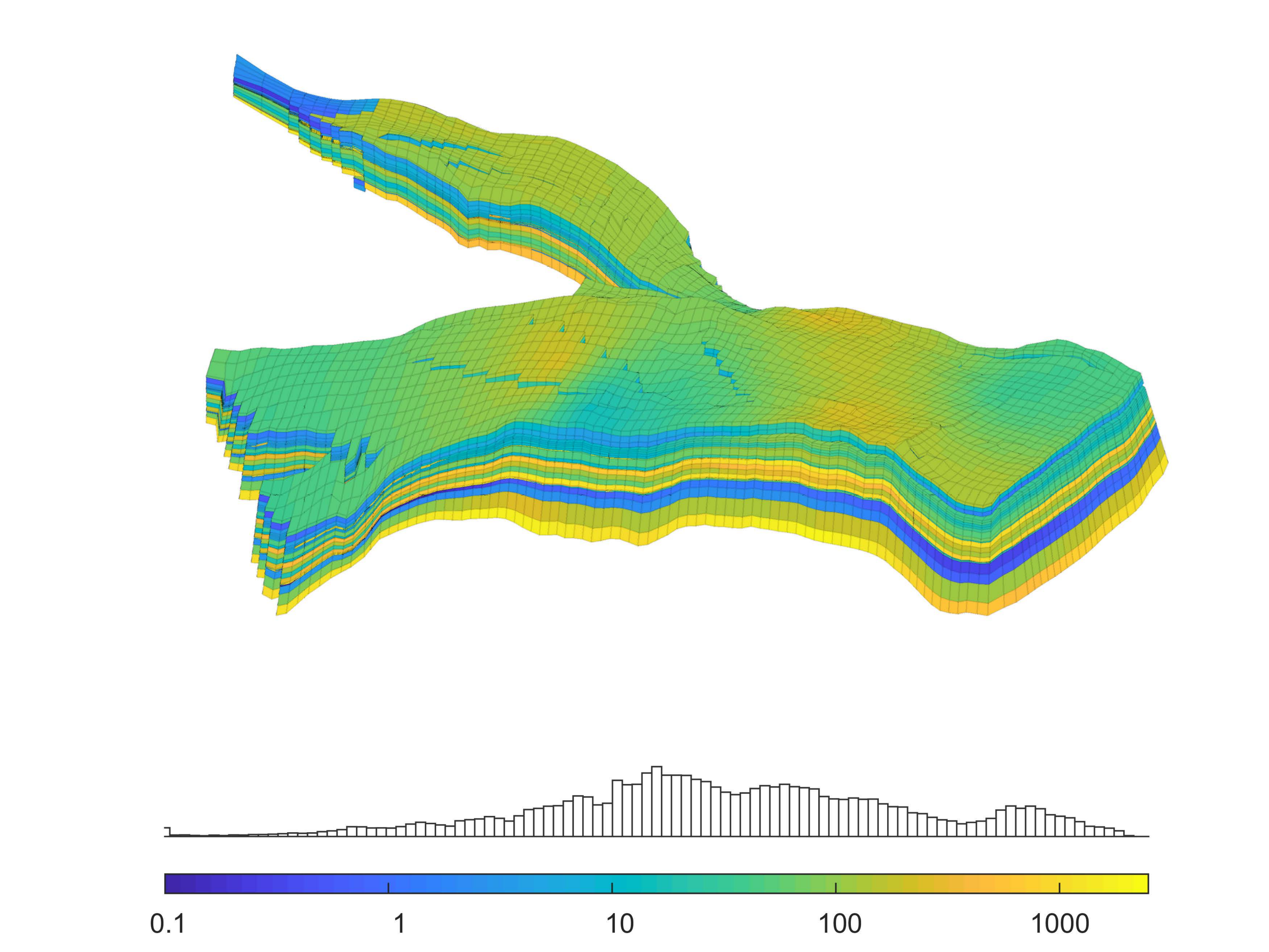}
	\end{subfigure}
	\caption{Test Case 6\&7: Permeability map of the Norne field. The left figure shows the horizontal permeability, and the right figure displays the vertical permeability; both are plotted using a logarithmic color scale.}
	\label{Fig:pEDFM_CPG_TestCase6_Norne_Permeability}
\end{figure}

\begin{figure}[!htbp]
	\centering
	\begin{subfigure}[]{0.23\textwidth}
		\includegraphics[width=\textwidth]{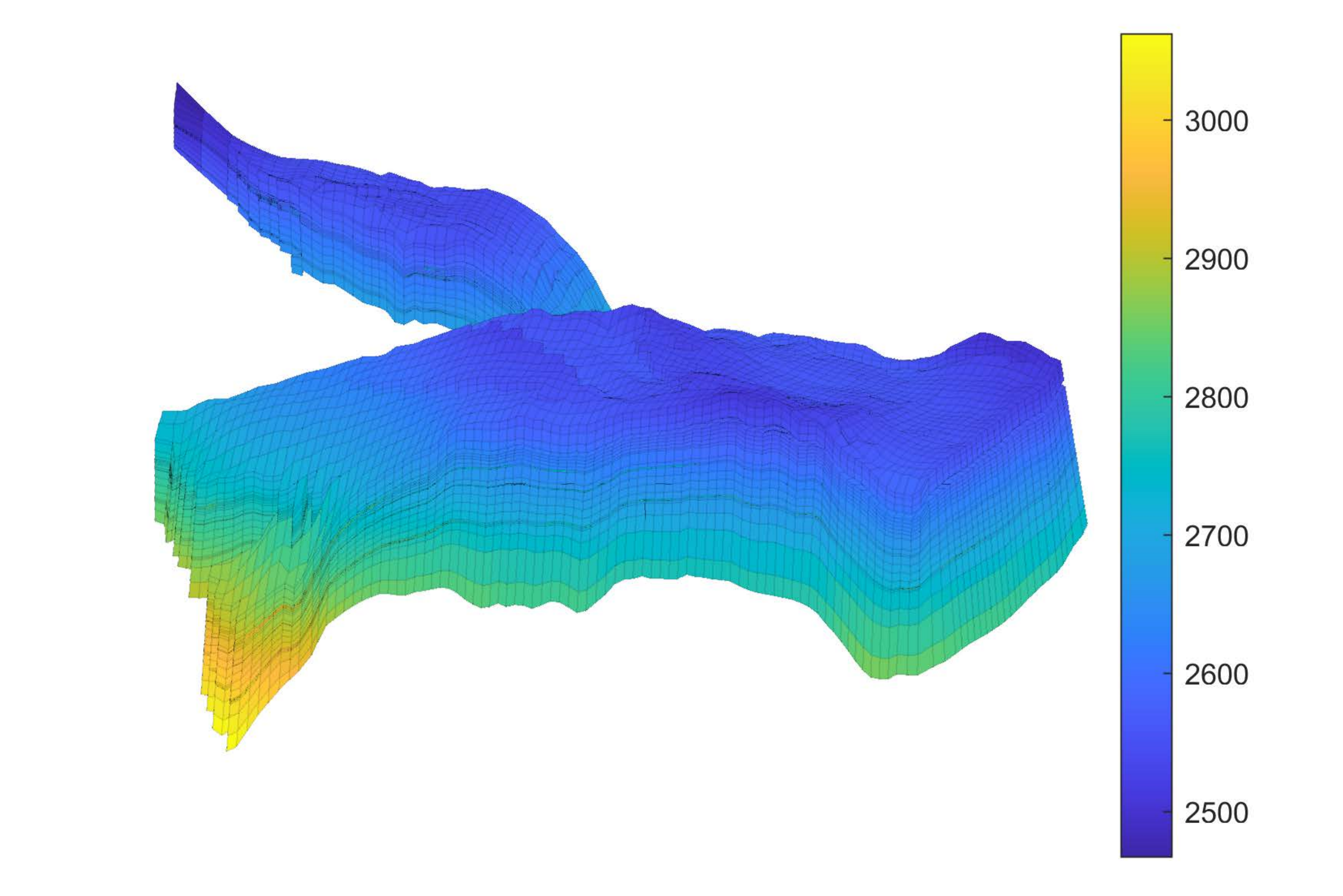}
		\caption{The depth map}
	\end{subfigure}
	\begin{subfigure}[]{0.23\textwidth}
		\includegraphics[width=\textwidth]{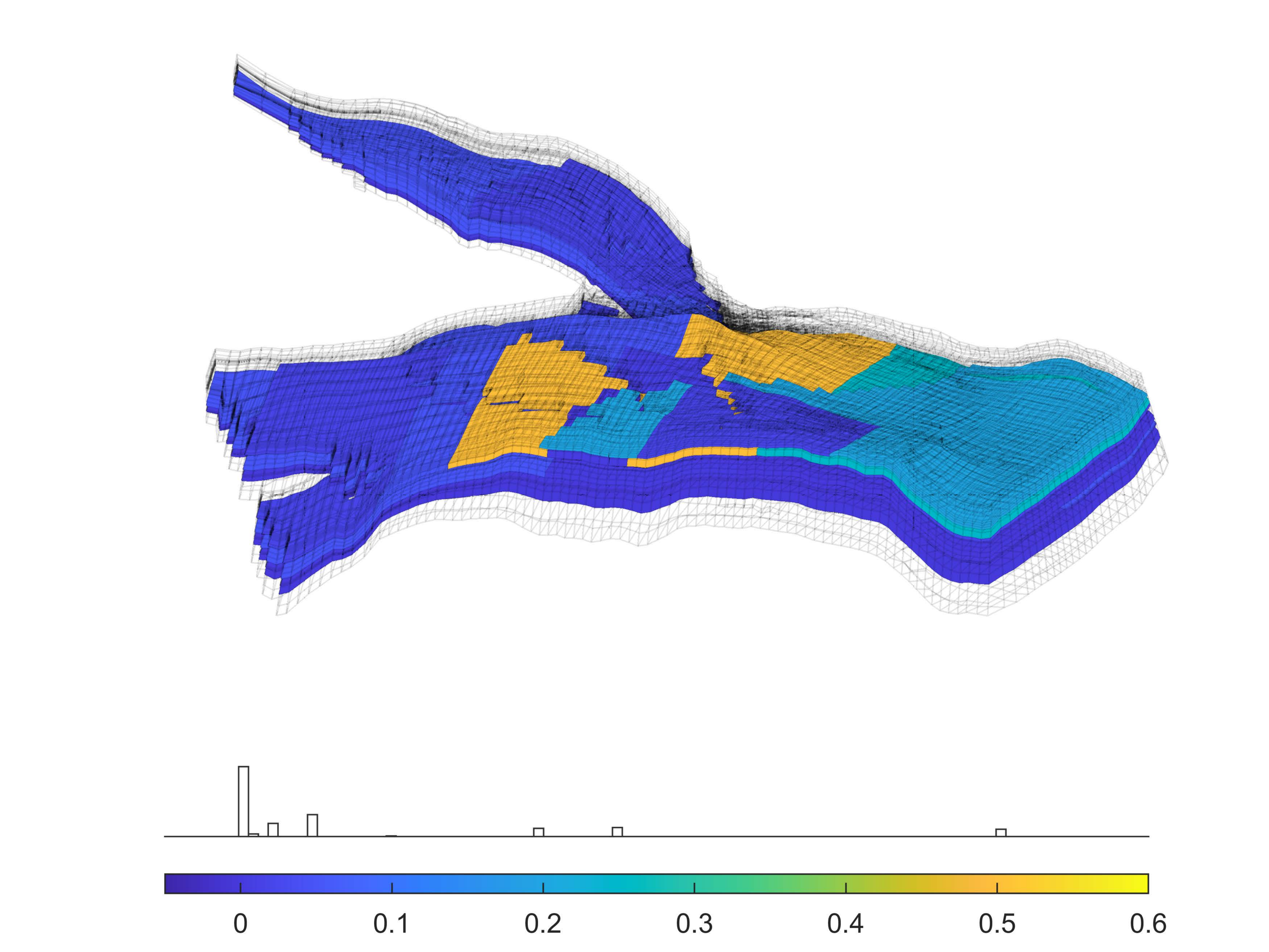}
		\caption{The map of the transmissibility multipliers}
	\end{subfigure}
	\caption{Test Case 6\&7: Depth map of the Norne field and the vertical transmissibility multipliers that reduce the vertical communication between the grid cells.}
	\label{Fig:pEDFM_CPG_TestCase6_Norne_Depth_Map} 
\end{figure}

\subsubsection{Simulation results of test case 6: Norne with highly conductive fractures}
This test case demonstrates the performance of the pEDFM model on Norne field. The corner-point grid data for this and the following test cases was extracted from the input files of MATLAB Reservoir Simulation Toolbox (MRST) \citep{lie2019introduction}. 

\begin{figure}[!htbp]
	\centering
	\subcaptionbox{{\footnotesize Pressure - Matrix}\label{Fig:pEDFM_CPG_TestCase6_Norne_Pm}}
	{\includegraphics[width=0.49\textwidth]{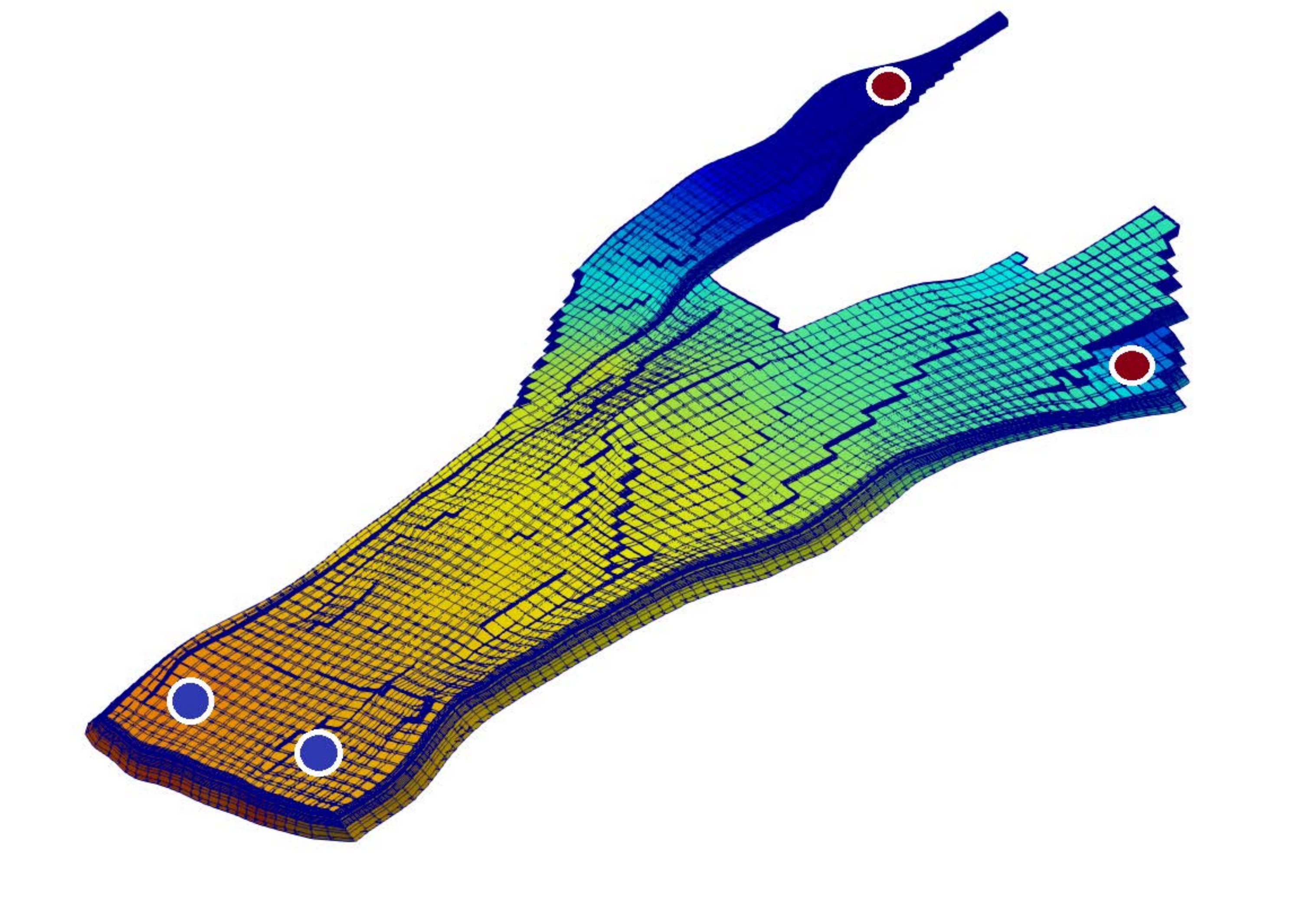}}
	\subcaptionbox{{\footnotesize Pressure - Fractures}\label{Fig:pEDFM_CPG_TestCase6_Norne_Pf}}
	{\includegraphics[width=0.49\textwidth]{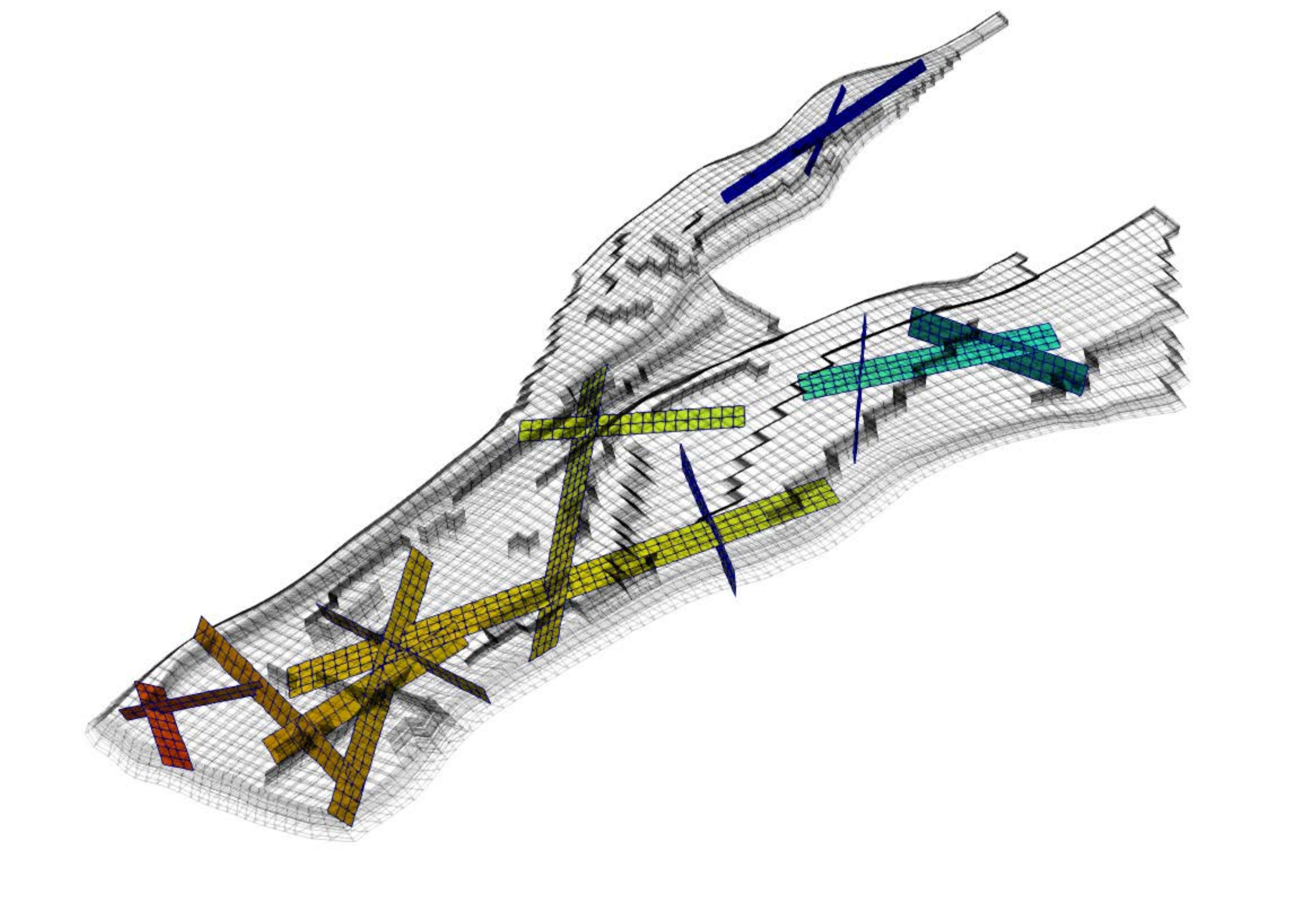}}
	\\
	\subcaptionbox{{\footnotesize Temperature}\label{Fig:pEDFM_CPG_TestCase6_Norne_T}}
	{\includegraphics[width=0.49\textwidth]{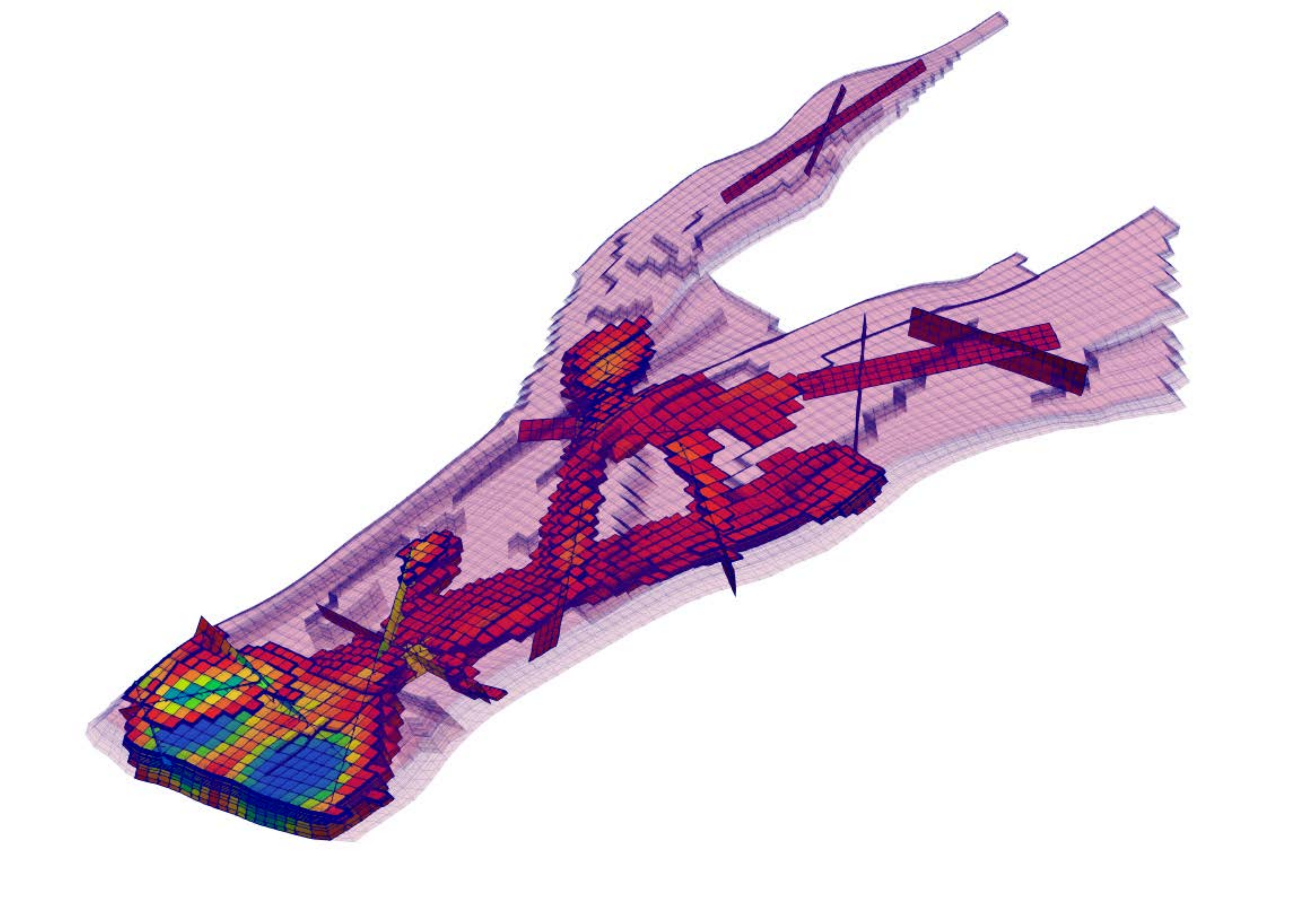}}
	\caption{Test case 6: Norne oil field. The figures \ref{Fig:pEDFM_CPG_TestCase6_Norne_Pm} and \ref{Fig:pEDFM_CPG_TestCase6_Norne_Pf} show the pressure solutions of the matrix and the embedded fractures. The figure on the bottom row (\ref{Fig:pEDFM_CPG_TestCase6_Norne_T}) visualizes the temperature solution on the same time-step.}
	\label{Fig:pEDFM_CPG_TestCase6_Norne_Pressure_Temperature}
\end{figure}

As explained above, Norne is an oil field located around $80$ kilometers north of the Heidrun oil field in the Norwegian Sea \citep{lie2019introduction}. As described in the MRST \citep{lie2019introduction}, the extent of this oil field is $10 \, [\text{Km}] \times 2 \, [\text{Km}] \times 100 \, [\text{m}]$. The corner-point grid skeleton consists of $46 \times 112 \times 22$ grid cells from which $44915$ grid cells are active forming the complex geometrical shape of this oil field. A synthetic network of $15$ fractures (designed by the author as a realization) is considered inside this domain. The permeability of the Norne rock matrix in this test case is assumed to be constant at $K_m = 10^{-14}\, [\text{m}^2]$ and the permeability data from the field was not used in this test case. All fractures are highly conductive with permeability of $K_f = 10^{-8}\, [\text{m}^2]$. Two injection wells with pressure of $p_{\text{inj}} = 5 \times 10^7 \, [\text{Pa}]$ and two production wells with pressure of $p_{\text{prod}} = 1 \times 10^7 \, [\text{Pa}]$ are located in the outer skirts of the reservoir as it can be seen on Fig. \ref{Fig:pEDFM_CPG_TestCase6_Norne_Pm}. All wells are vertical and perforate the entire thickness of the reservoir. For this test case, the low-enthalpy single-phase geothermal fluid model was used. The input parameters used in this test case are listed in table \ref{Tab:pEDFM_CPG_Input_Parameters_Geothermal}.

\subsubsection{Simulation results of test case 7: Norne with mix-conductive fractures}
In this test case, the Norne field model with skeleton of $46 \times 112 \times 22$ grid cells and a total of $44915$ grid cells active matrix grid cells is considered. Unlike in test case $6$, the real rock properties of the Norne field were used in this test case. A set of $56$ synthetic fractures are created and embedded in the reservoir domain which comprises highly conductive fractures and flow barriers with permeability of  $K_{f_{max}} = 10^{-8} \, [\text{m}^2]$ and $K_{f_{min}} = 10^{-20} \, [\text{m}^2]$ respectively. The fracture network consists of $2165$ grid cells. In total there are $48705$ grid cells in this test case. Four injection wells with a $p_{\text{inj}} = 5 \times 10^7 \, [\text{Pa}]$ and three production wells with a $p_{\text{prod}} = 1 \times 10^7 \, [\text{Pa}]$ were placed in the model. Wells are vertical and drilled through the entire thickness of the model.

Like the test cases in Johansen (\ref{Sec:TestCase4_Johansen}) and Brugge (\ref{Sec:TestCase5_Brugge}) models, two scenarios are considered for the fracture network used in this test case. In both scenarios, the gemoetrical properties of the fracture networks are identical. However, the permeability values of the highly conductive fractures and flow barriers from the scenario $1$ are inverted in the scenario $2$.

\begin{figure}[!htbp]
	\centering
	\subcaptionbox{{\footnotesize Fractures permeability (scenario 1)}\label{Fig:pEDFM_CPG_TestCase7_Norne_Kf_1}}
	{\includegraphics[width=0.23\textwidth]{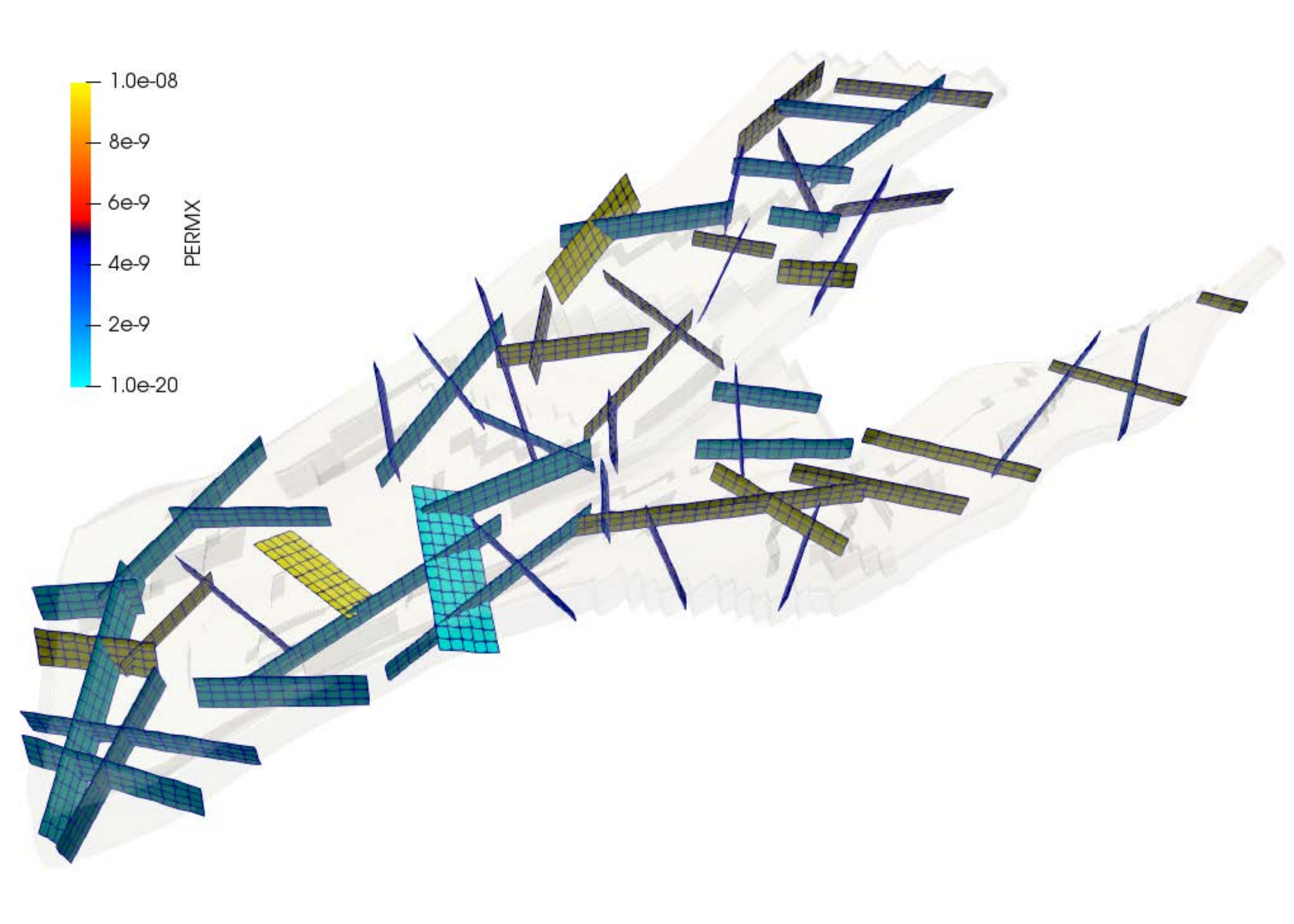}}
	\subcaptionbox{{\footnotesize Fractures permeability (scenario 2)}\label{Fig:pEDFM_CPG_TestCase7_Norne_Kf_2}}
	{\includegraphics[width=0.23\textwidth]{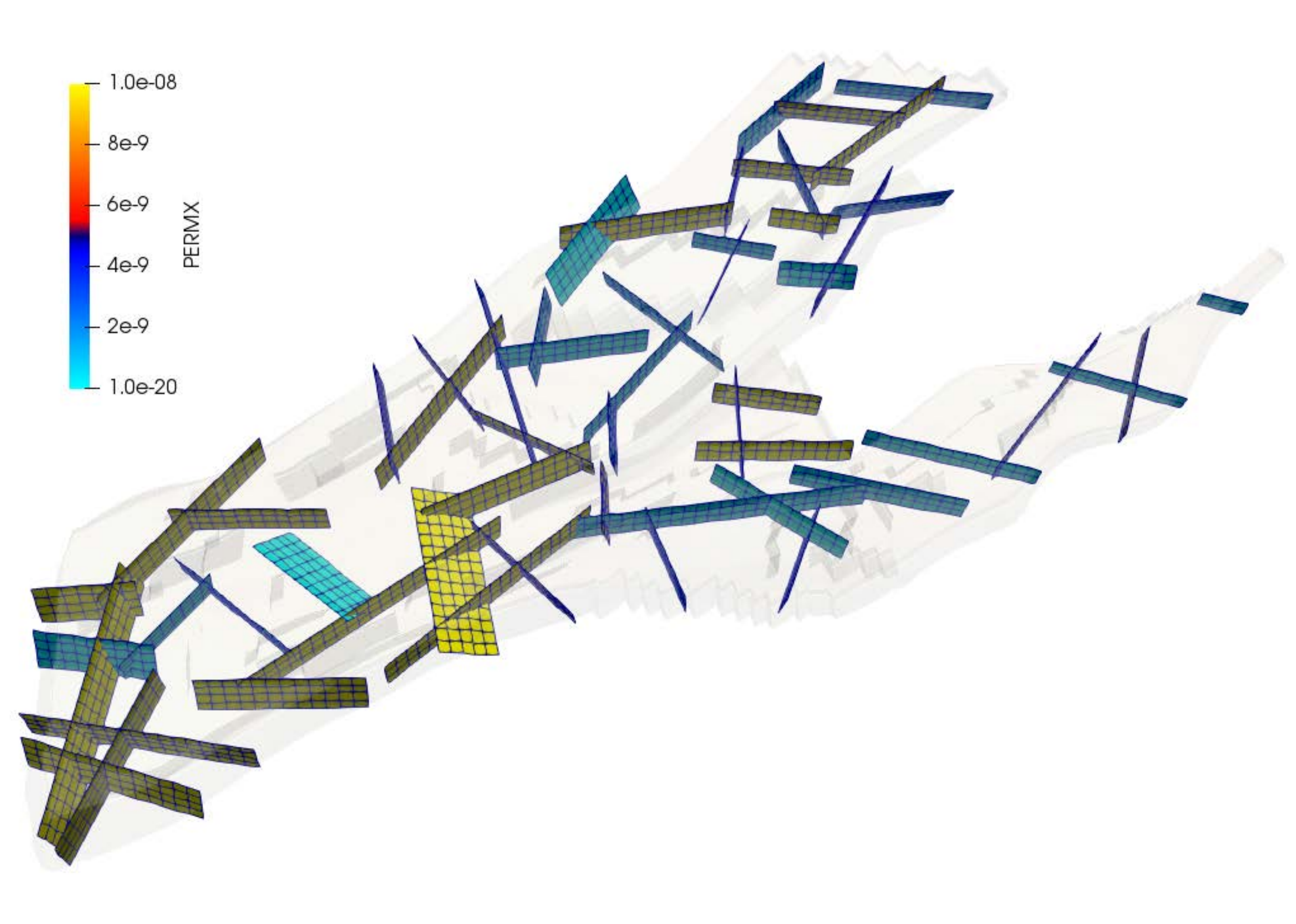}}
	\\
	\subcaptionbox{{\footnotesize Matrix cells overlapped by fractures}\label{Fig:pEDFM_CPG_TestCase7_Norne_Fractured_Cells}}
	{\includegraphics[width=0.23\textwidth]{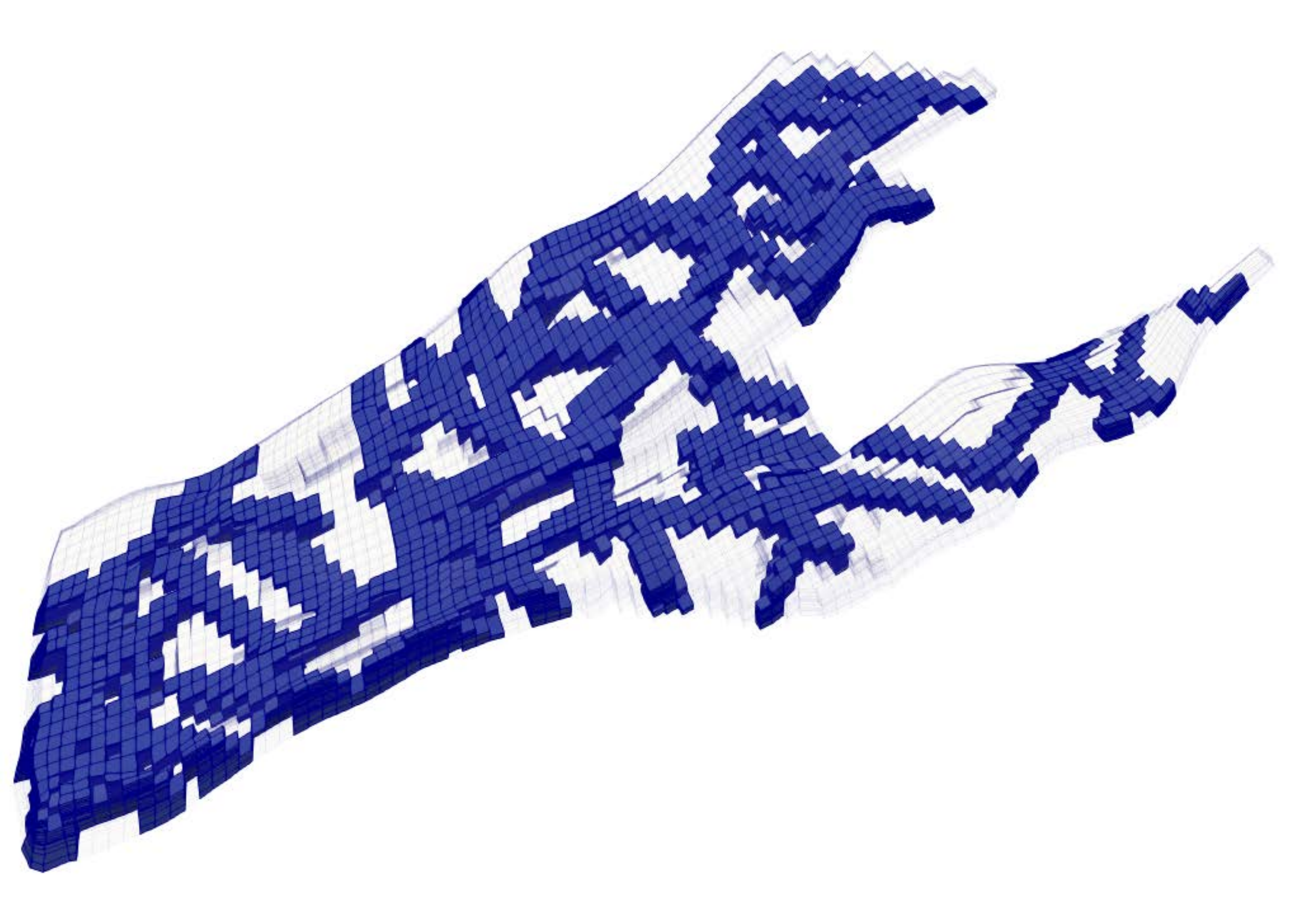}}
	\subcaptionbox{{\footnotesize Location of injection and production wells}\label{Fig:pEDFM_CPG_TestCase7_Norne_Wells}}
	{\includegraphics[width=0.23\textwidth]{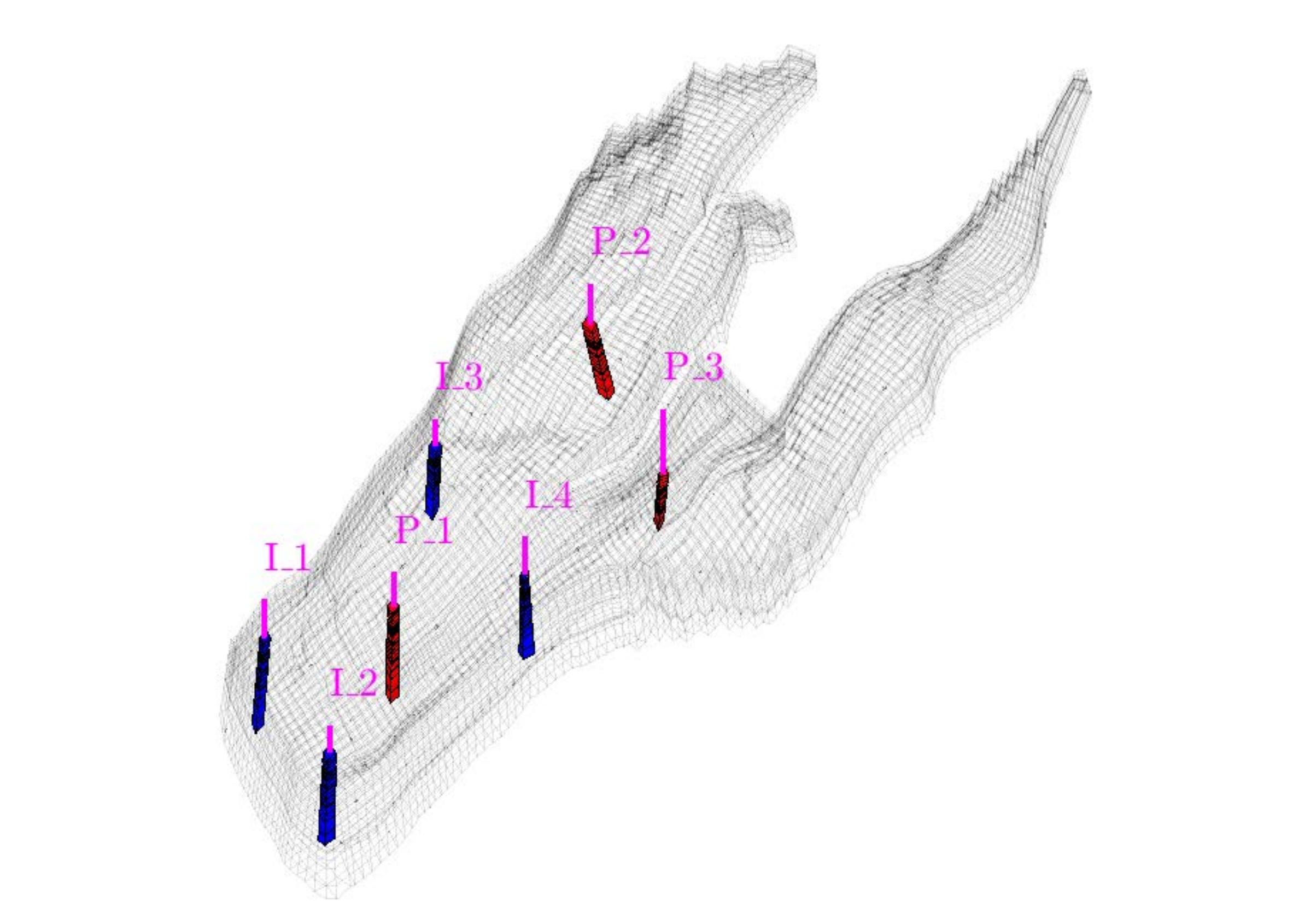}}
	\caption{Test case 7: The Norne model with $7$ wells ($4$ injectors and $3$ producers) and a set of $56$ synthetic fractures (with mixed conductivities). The figures on top show the fractures network with different permeabilities for scenario $1$ (top left) and scenario $2$ (top right). The figure at bottom left illustrates the highlighted matrix cells that are overlapped by the fractures network. And the figure at the bottom right shows the schematics of the injection and production wells.}
	\label{Fig:pEDFM_CPG_TestCase7_Norne_Wells_Fractures}
\end{figure}

\begin{figure}[!htbp]
	\centering
	\subcaptionbox{{\footnotesize Pressure in the matrix}\label{Fig:pEDFM_CPG_TestCase7_Norne_Pm_scenario1}}
	{\includegraphics[width=0.23\textwidth]{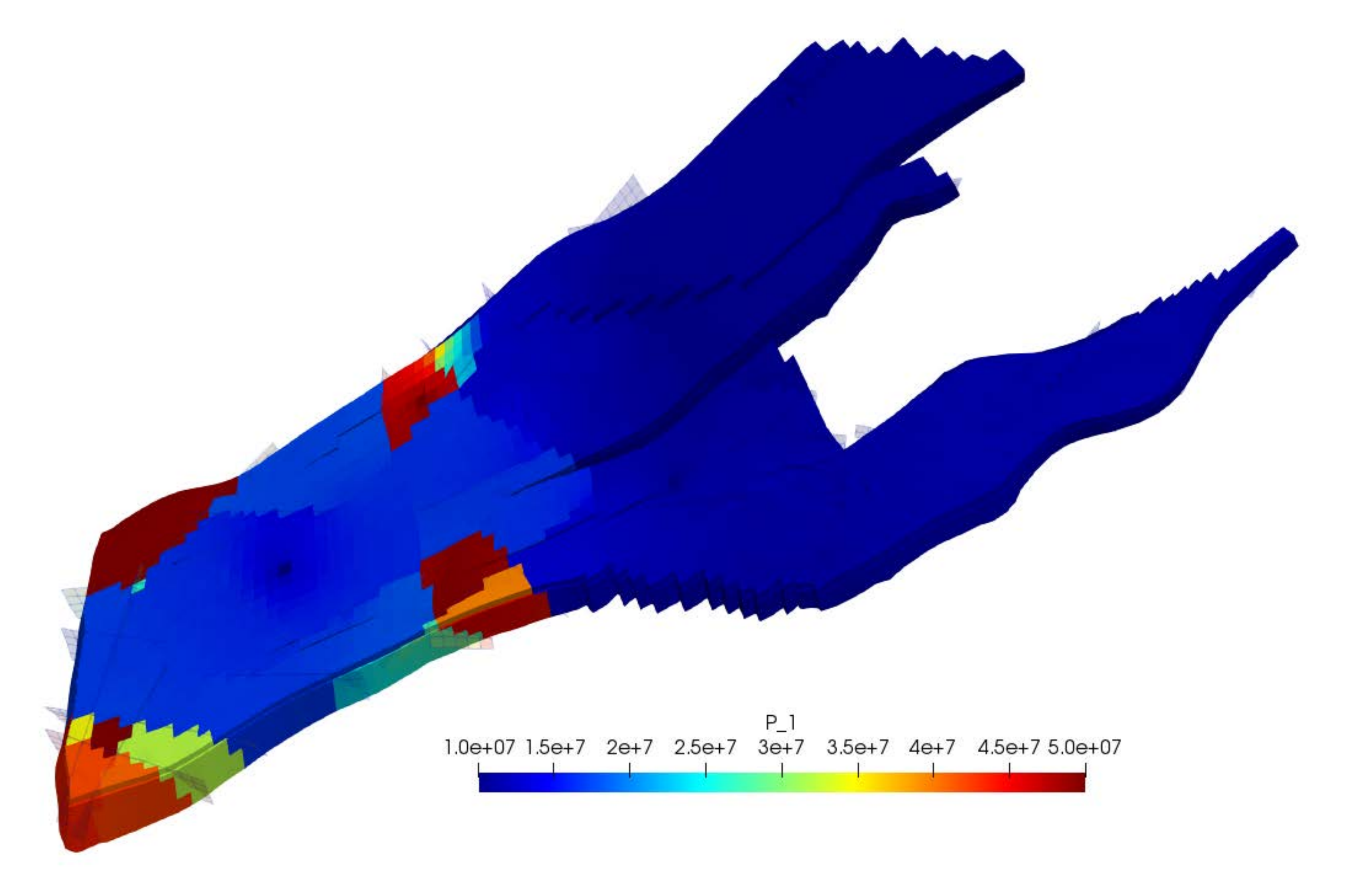}}
	\subcaptionbox{{\footnotesize Pressure in the fractures}\label{Fig:pEDFM_CPG_TestCase7_Norne_Pf_scenario1}}
	{\includegraphics[width=0.23\textwidth]{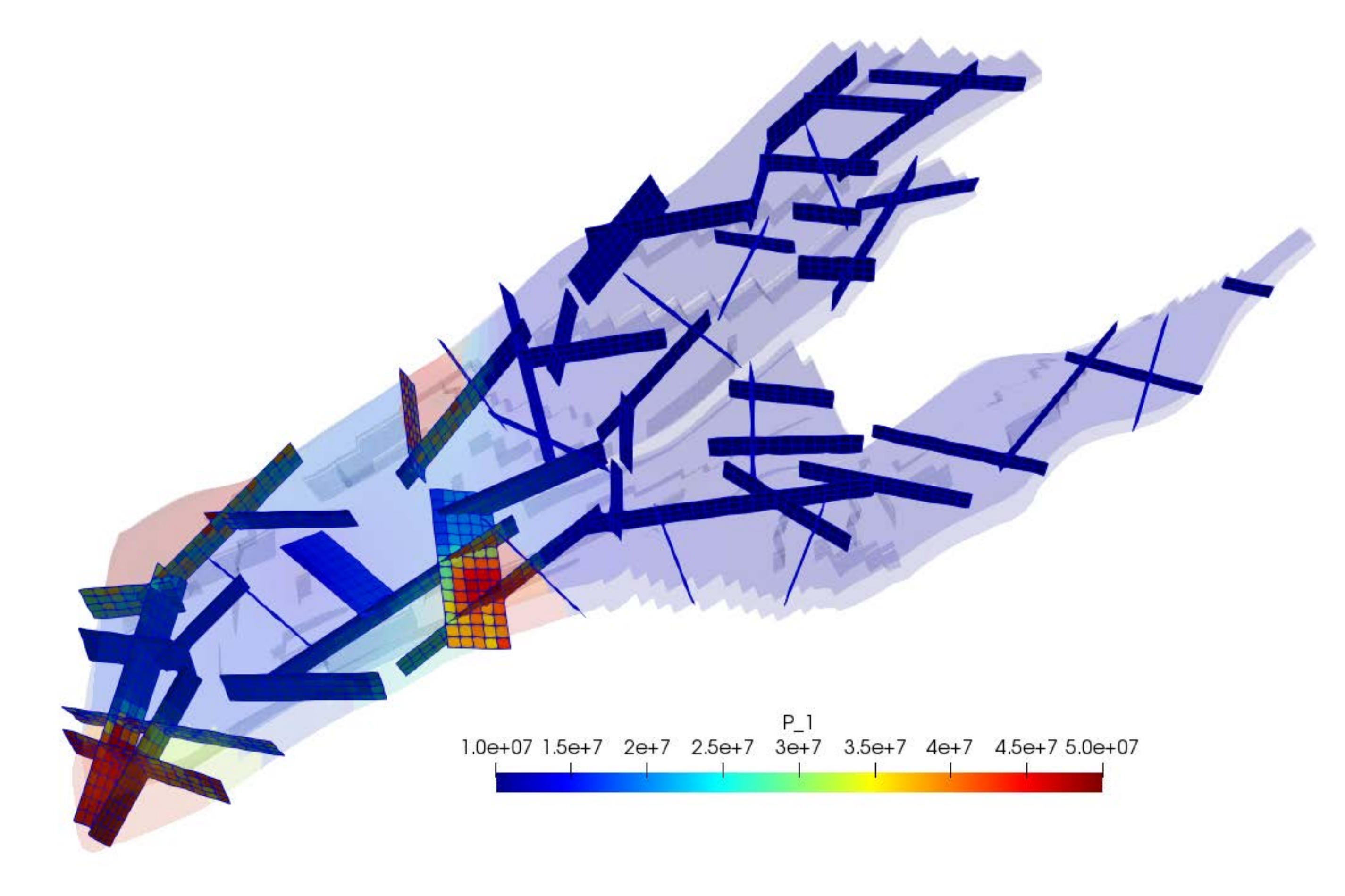}}
	\caption{Test case 7: The pressure profile of the Norne model for the simulation scenario $1$. The figure on the left shows the pressure distribution in the matrix grid cells. The transparency of this figure is increased to make the pressure map in the fractures visible. This map is displayed on the right figure. The results are shown for the simulation time $5000 [\text{days}]$}
	\label{Fig:pEDFM_CPG_TestCase7_Norne_Pressure_Scenario1}
\end{figure}

\begin{figure}[!htbp]
	\centering
	\subcaptionbox{{\footnotesize Saturation in the matrix after $2000 [\text{days}]$ }\label{Fig:pEDFM_CPG_TestCase7_Norne_Sm_scenario1_T04}}
	{\includegraphics[width=0.23\textwidth]{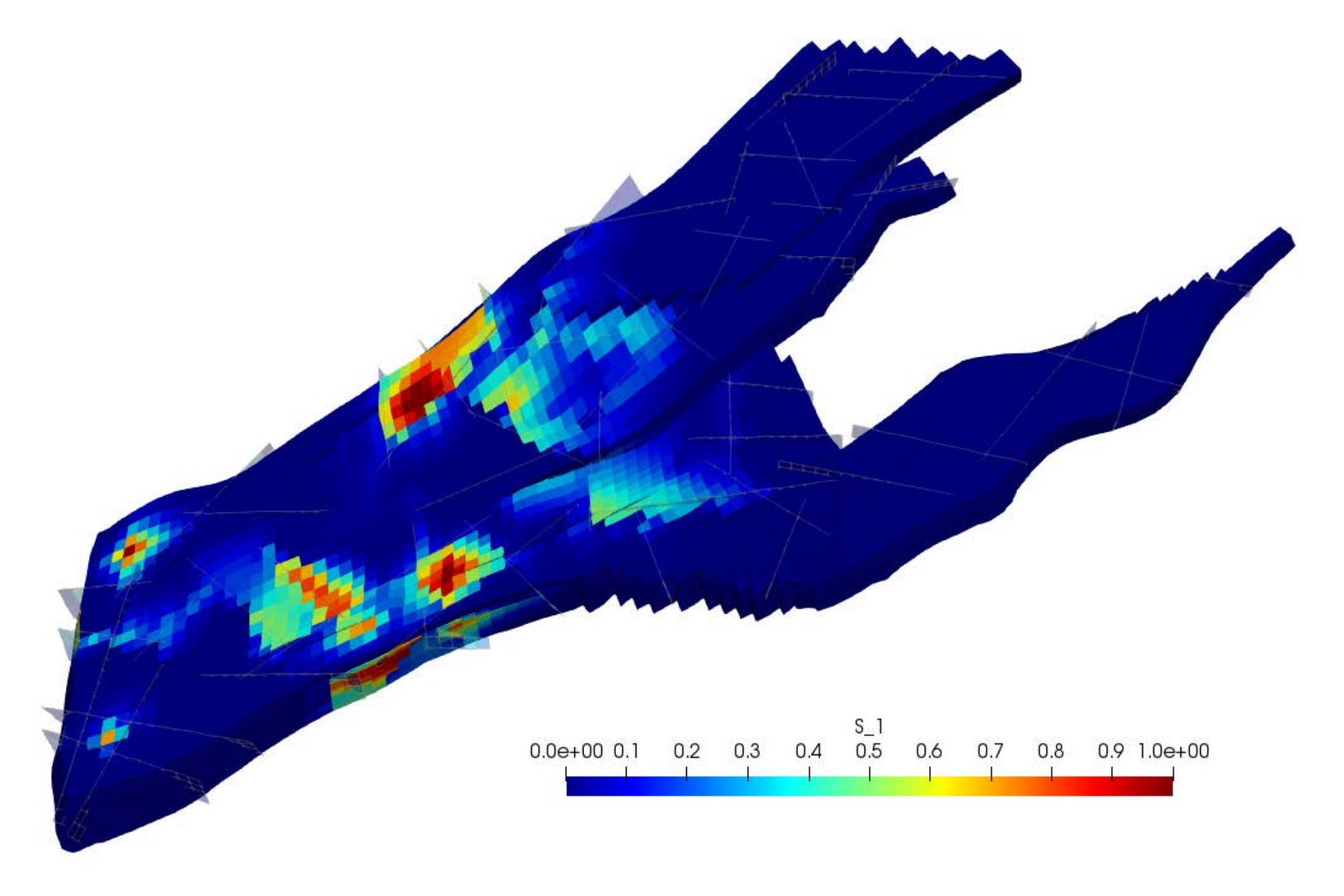}}
	\subcaptionbox{{\footnotesize Saturation in the fractures after $2000 [\text{days}]$ }\label{Fig:pEDFM_CPG_TestCase7_Norne_Sf_scenario1_T04}}
	{\includegraphics[width=0.23\textwidth]{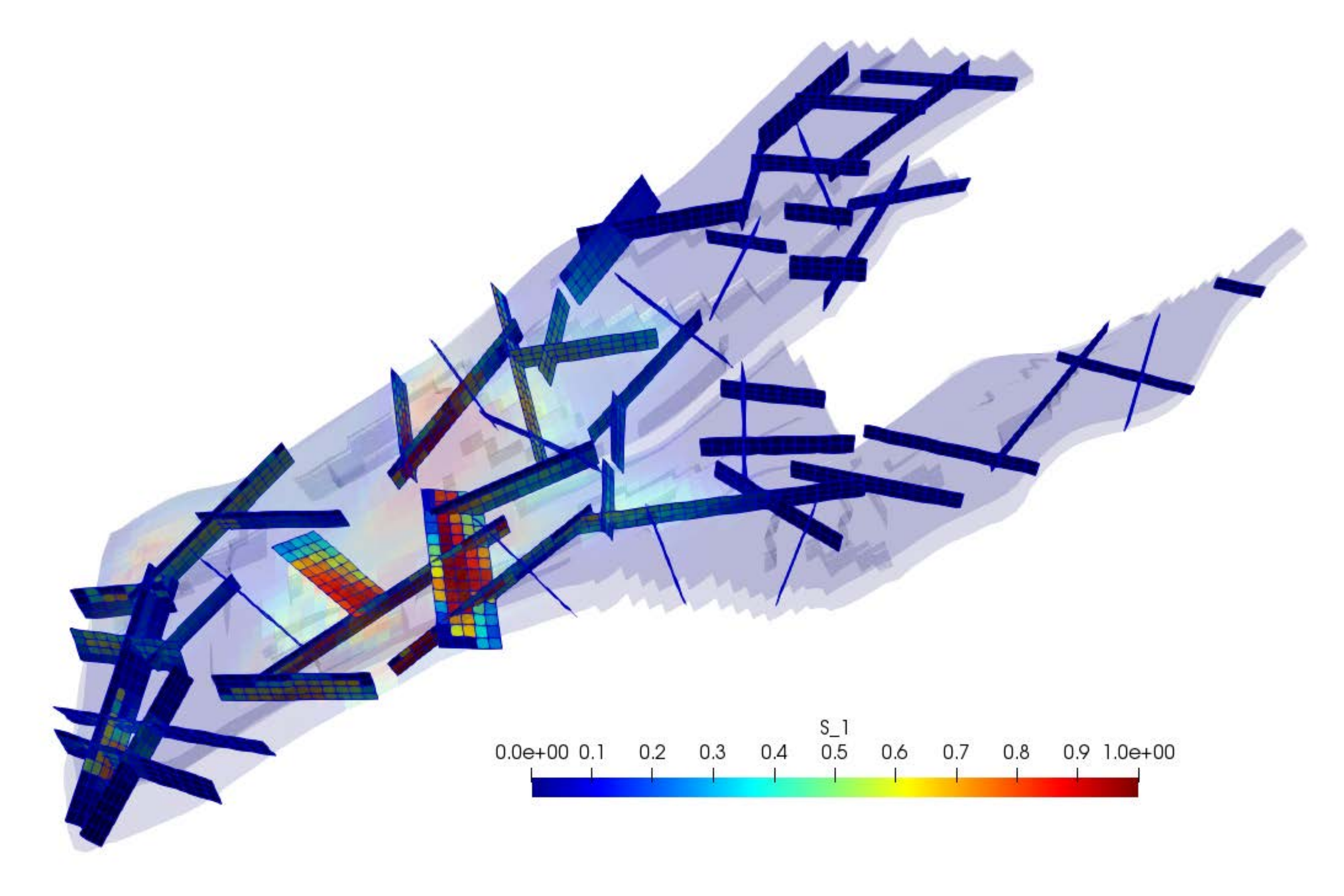}}
	\\
	\subcaptionbox{{\footnotesize Saturation in the matrix after $5000 [\text{days}]$ }\label{Fig:pEDFM_CPG_TestCase7_Norne_Sm_scenario1_T10}}
	{\includegraphics[width=0.23\textwidth]{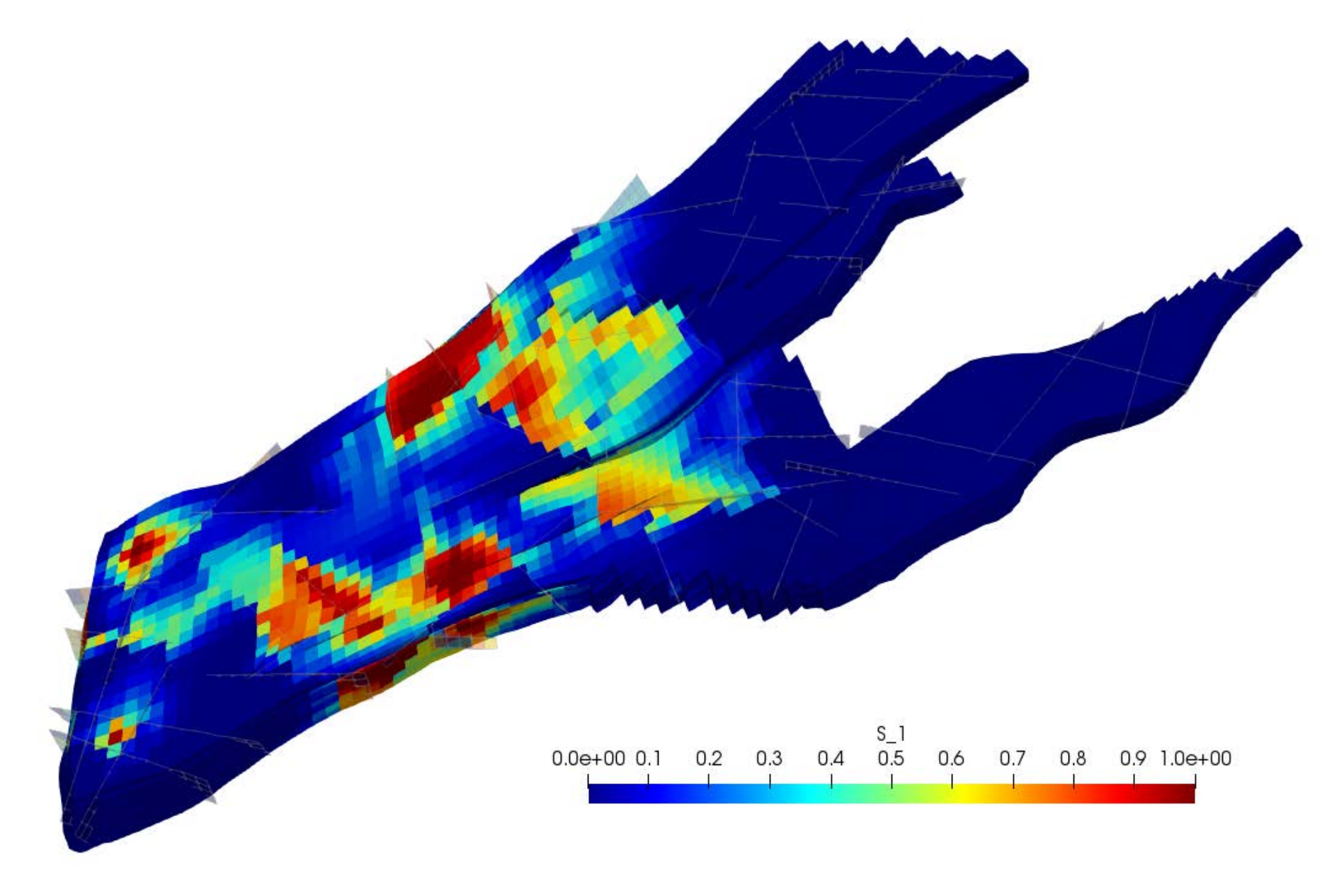}}
	\subcaptionbox{{\footnotesize Saturation in the fractures after $5000 [\text{days}]$ }\label{Fig:pEDFM_CPG_TestCase7_Norne_Sf_scenario1_T10}}
	{\includegraphics[width=0.23\textwidth]{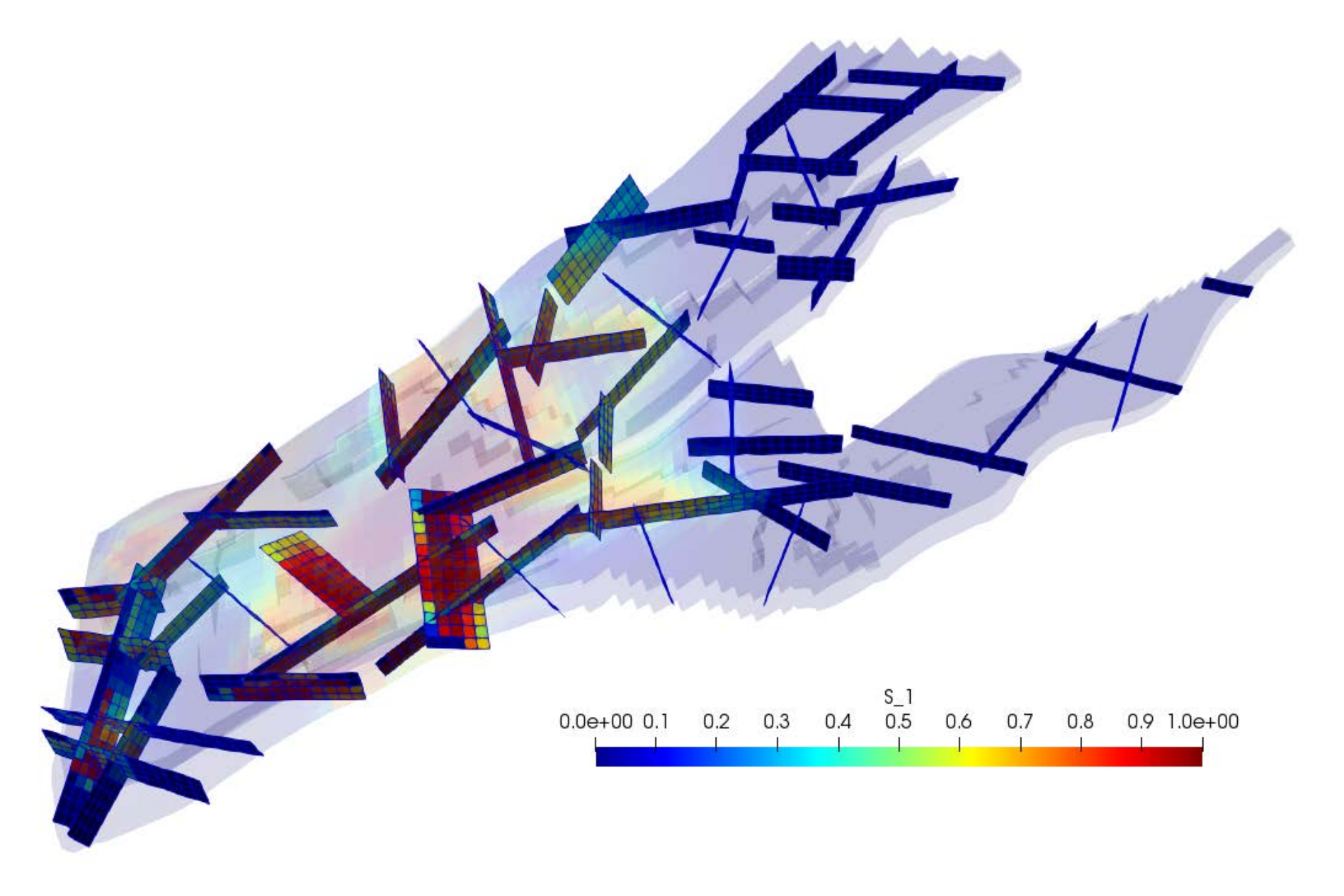}}
	\\
	\subcaptionbox{{\footnotesize Saturation in the matrix after $10000 [\text{days}]$ }\label{Fig:pEDFM_CPG_TestCase7_Norne_Sm_scenario1_T40}}
	{\includegraphics[width=0.23\textwidth]{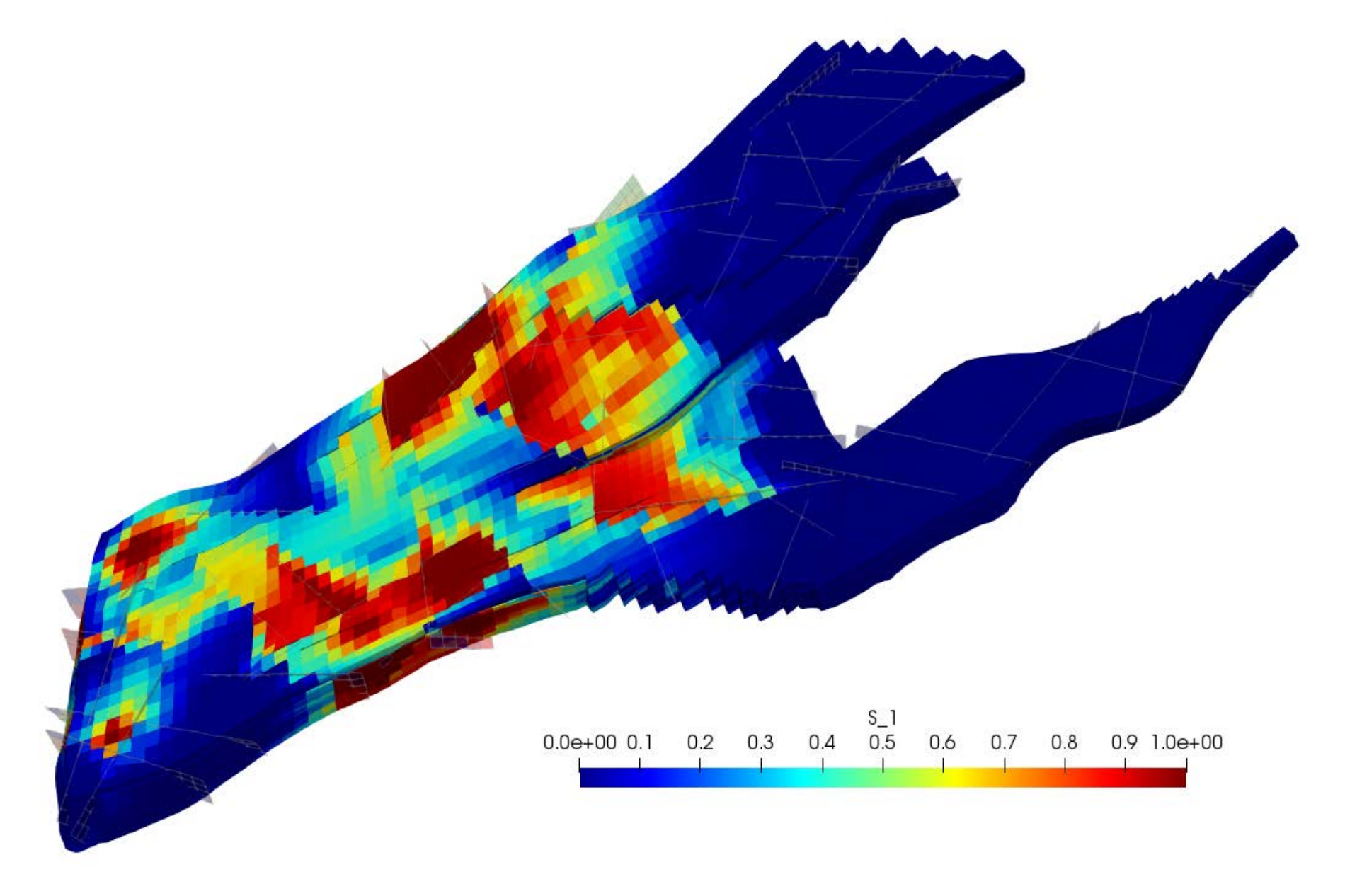}}
	\subcaptionbox{{\footnotesize Saturation in the fractures after $10000 [\text{days}]$ }\label{Fig:pEDFM_CPG_TestCase7_Norne_Sf_scenario1_40}}
	{\includegraphics[width=0.23\textwidth]{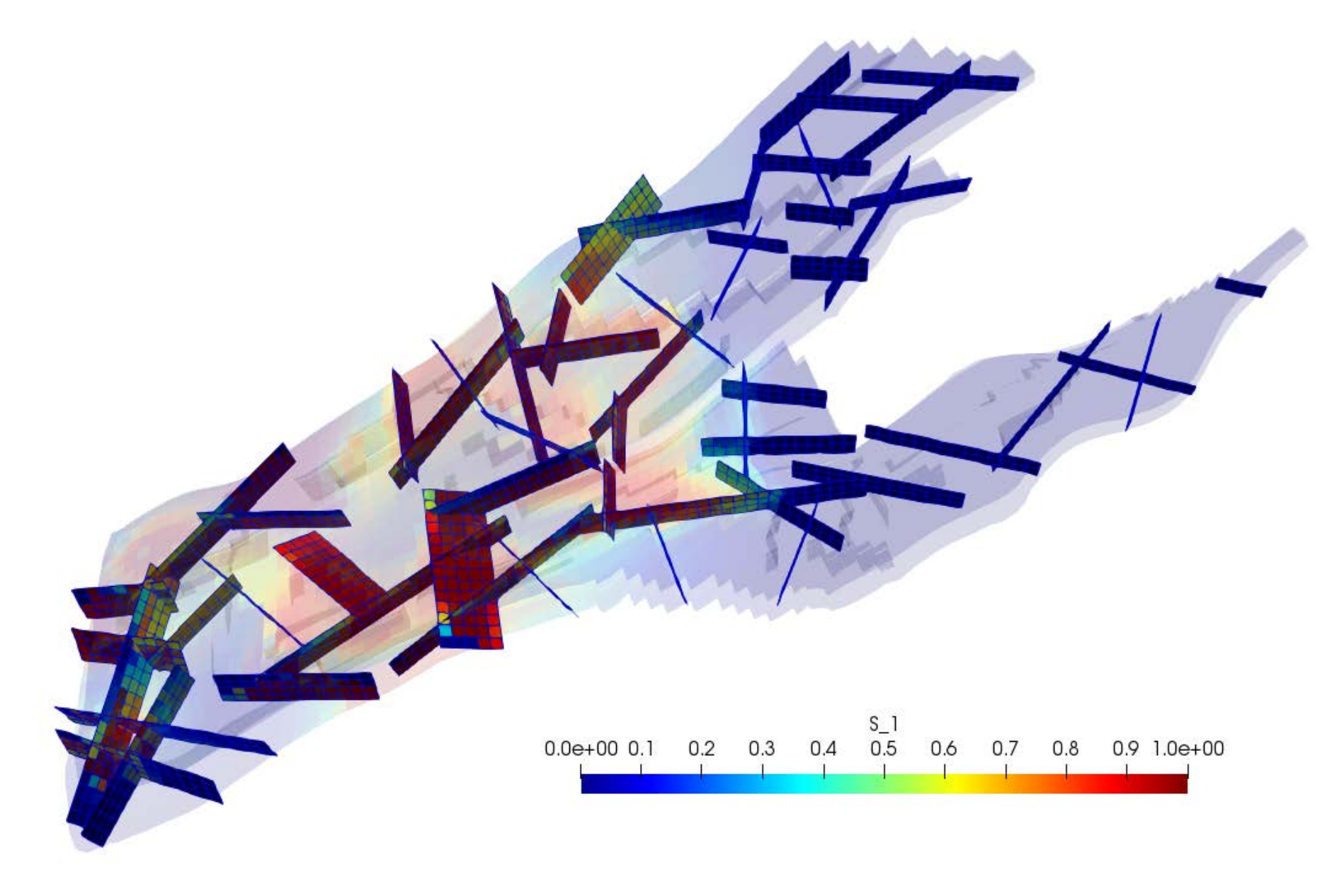}}
	\caption{Test case 7: The saturation profile of the Norne model for the simulation scenario $1$. The figures on the left illustrate the saturation profile in the matrix grid cells and the figures on the right side show the saturation maps in the fractures. From the top row towards the bottom row, the saturation profiles are displayed for simulations times $2000$, $5000$ and $10000 [\text{days}]$ respectively.}
	\label{Fig:pEDFM_CPG_TestCase7_Norne_Saturation_Scenario1}
\end{figure}

The pressure and saturation results of the scenario $1$ simulation are presented in the figures \ref{Fig:pEDFM_CPG_TestCase7_Norne_Pressure_Scenario1} and \ref{Fig:pEDFM_CPG_TestCase7_Norne_Saturation_Scenario1} respectively. The pressure results are only shown for the simulation time $5000 [\text{days}]$, but the saturation profiles are presented for three time intervals of $2000$, $5000$ and $10000 [\text{days}]$. The injection wells are surrounded by flow barriers that restrict the saturation displacement in the reservoir. The pressure is considerably high in the areas near the wells. These high-pressure areas are an indication that the pEDFM implementation in corner-point grid geometry is successful in the modeling of the fractures with low conductivities. High pressure drops can be seen at the location of the flow barriers. The increase in saturation is mainly carried out in two parts of the model. These two areas are not isolated from the rest of the model which allows a distribution of the injecting phase through the flow paths.

\begin{figure}[!htbp]
	\centering
	\subcaptionbox{{\footnotesize Pressure in the matrix}\label{Fig:pEDFM_CPG_TestCase7_Norne_Pm_scenario2}}
	{\includegraphics[width=0.23\textwidth]{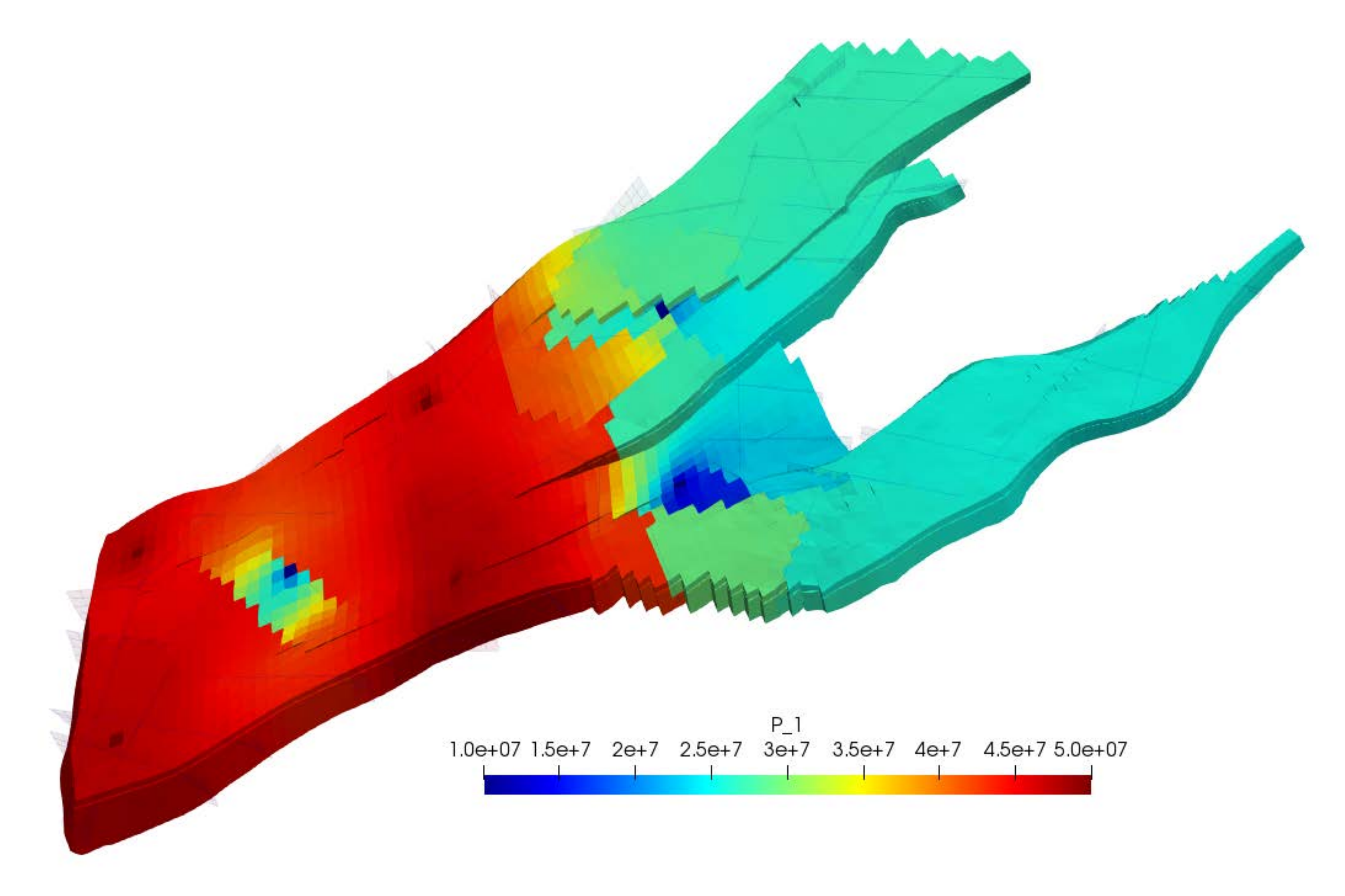}}
	\subcaptionbox{{\footnotesize Pressure in the fractures}\label{Fig:pEDFM_CPG_TestCase7_Norne_Pf_scenario2}}
	{\includegraphics[width=0.23\textwidth]{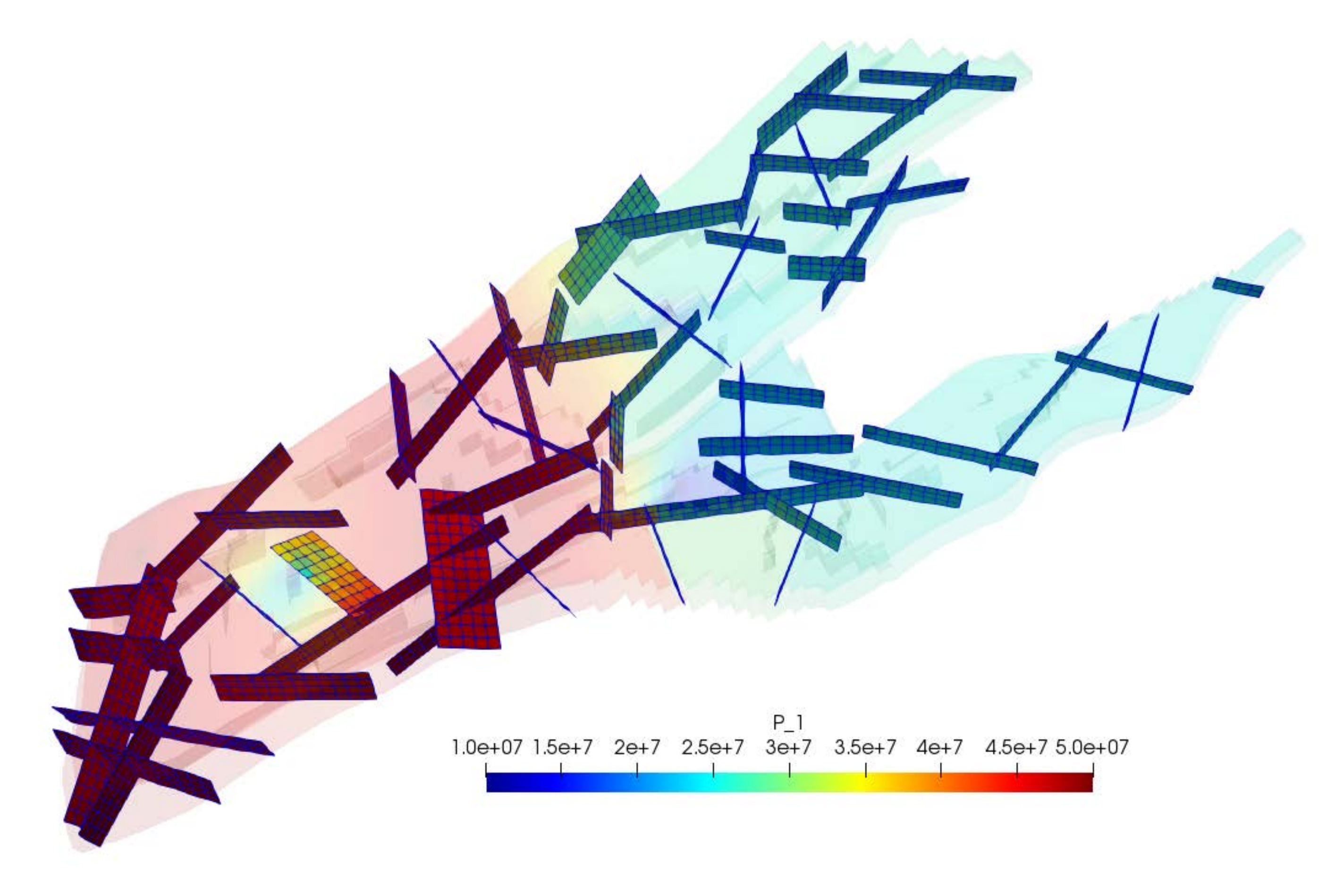}}
	\caption{Test case 7: The pressure profile of the Norne model for the simulation scenario $2$. The figure on the left shows the pressure distribution in the matrix grid cells. The transparency of this figure is increased to make the pressure map in the fractures visible. This map is displayed on the right figure. The results are shown for the simulation time $5000 [\text{days}]$}
	\label{Fig:pEDFM_CPG_TestCase7_Norne_Pressure_Scenario2}
\end{figure}

\begin{figure}[!htbp]
	\centering
	\subcaptionbox{{\footnotesize Saturation in the matrix after $2000 [\text{days}]$ }\label{Fig:pEDFM_CPG_TestCase7_Norne_Sm_scenario2_T04}}
	{\includegraphics[width=0.23\textwidth]{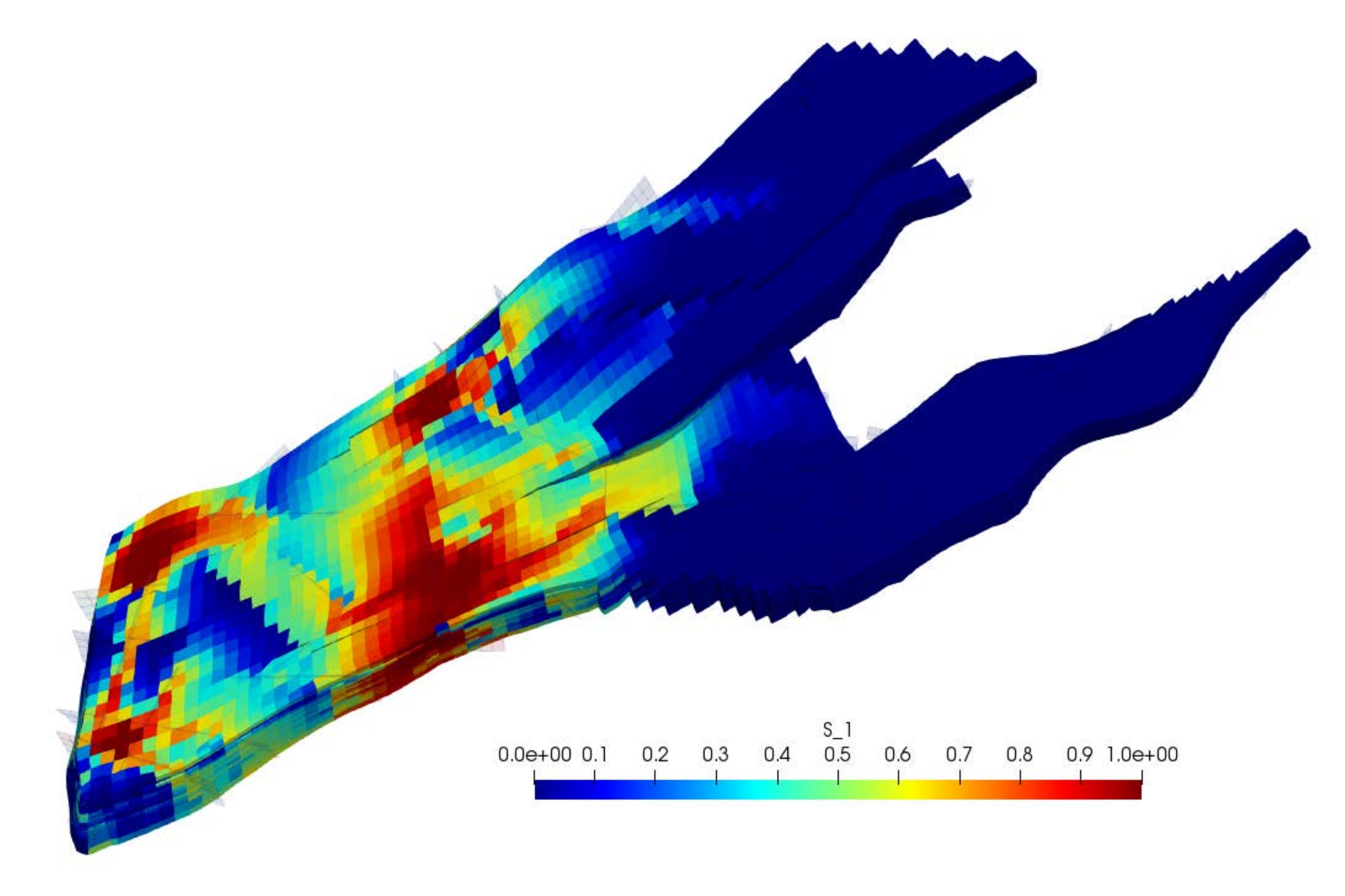}}
	\subcaptionbox{{\footnotesize Saturation in the fractures after $2000 [\text{days}]$ }\label{Fig:pEDFM_CPG_TestCase7_Norne_Sf_scenario2_T04}}
	{\includegraphics[width=0.23\textwidth]{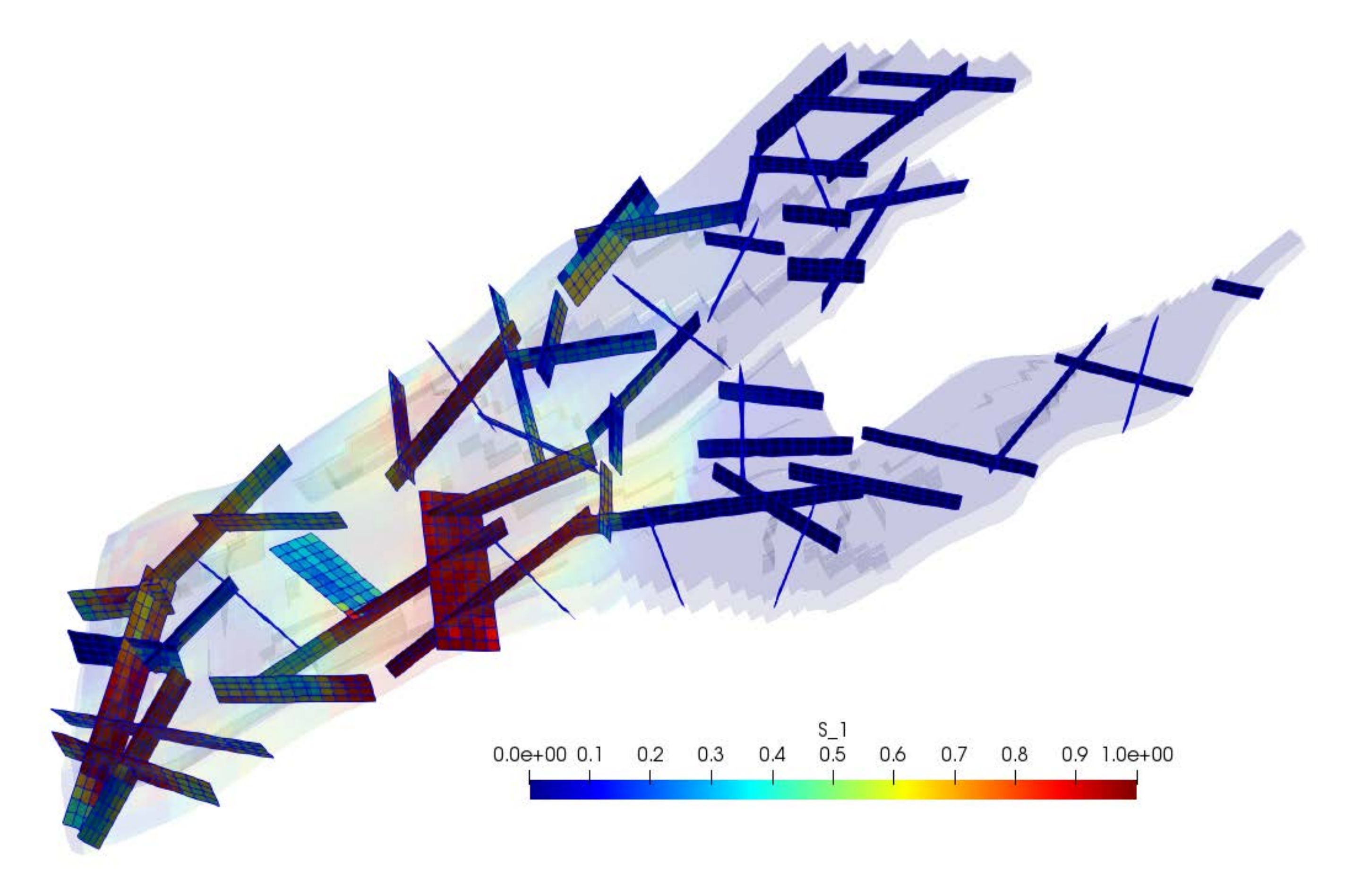}}
	\\
	\subcaptionbox{{\footnotesize Saturation in the matrix after $5000 [\text{days}]$ }\label{Fig:pEDFM_CPG_TestCase7_Norne_Sm_scenario2_T10}}
	{\includegraphics[width=0.23\textwidth]{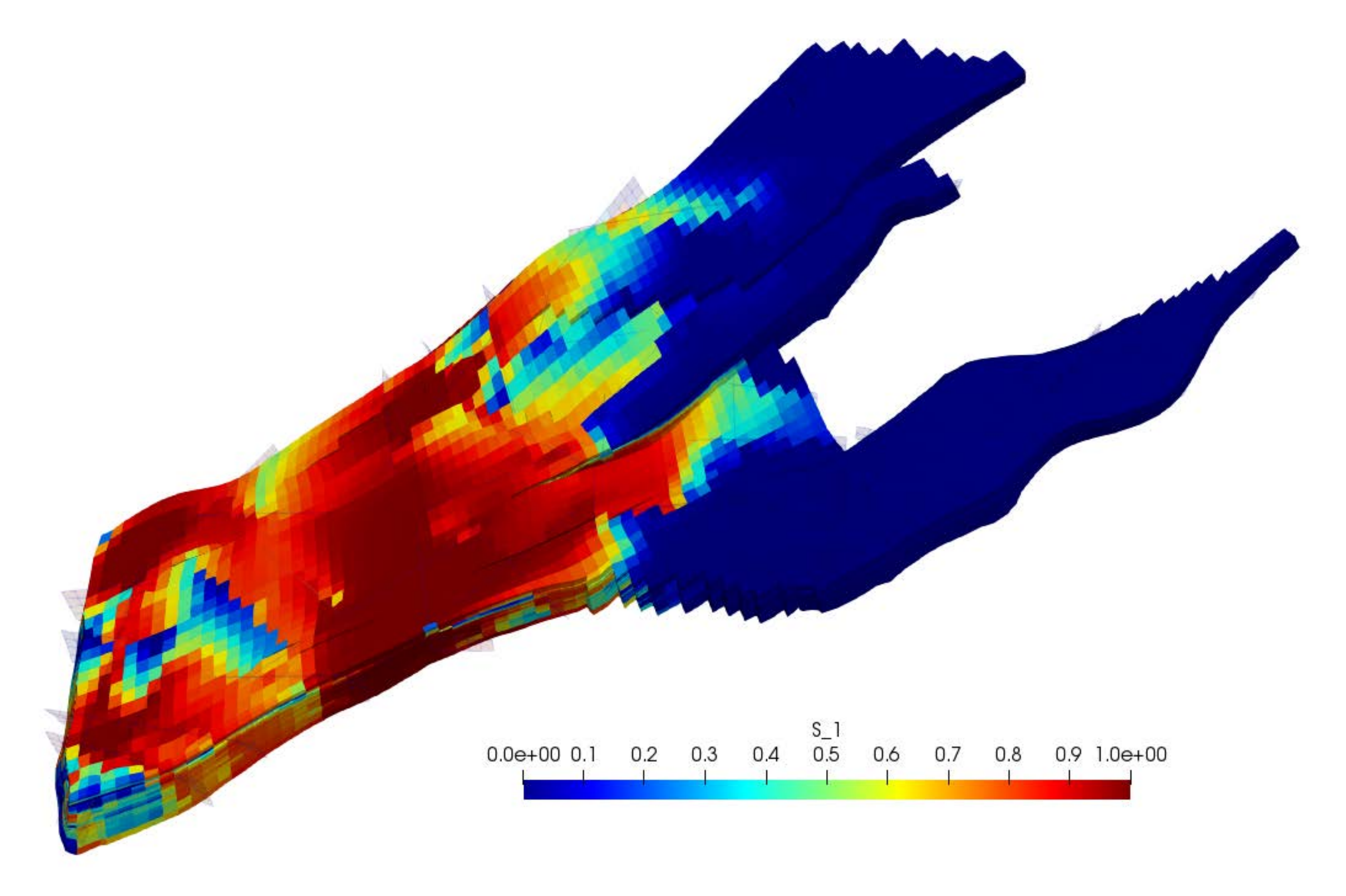}}
	\subcaptionbox{{\footnotesize Saturation in the fractures after $5000 [\text{days}]$ }\label{Fig:pEDFM_CPG_TestCase7_Norne_Sf_scenario2_T10}}
	{\includegraphics[width=0.23\textwidth]{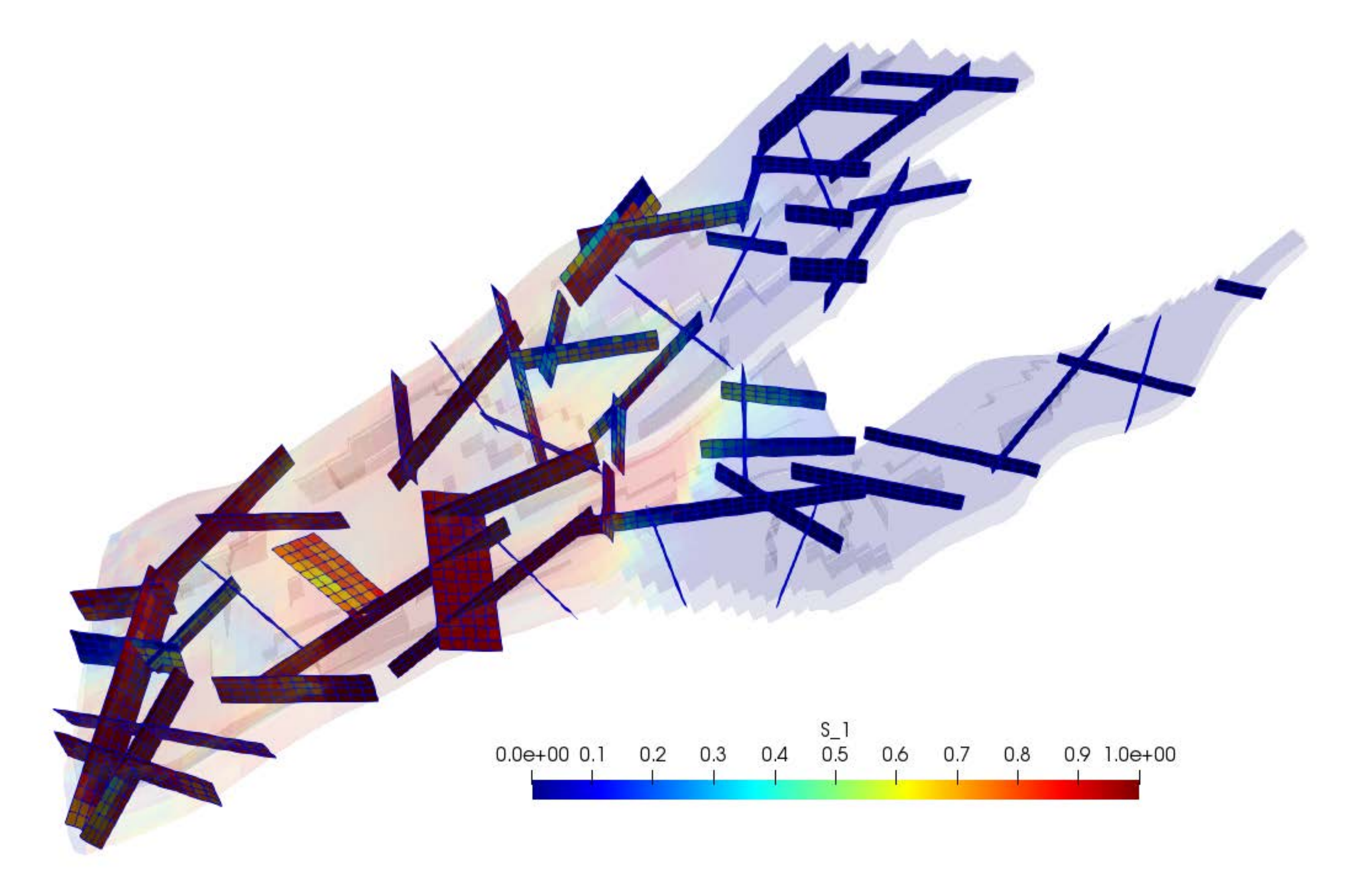}}
	\\
	\subcaptionbox{{\footnotesize Saturation in the matrix after $10000 [\text{days}]$ }\label{Fig:pEDFM_CPG_TestCase7_Norne_Sm_scenario2_T40}}
	{\includegraphics[width=0.23\textwidth]{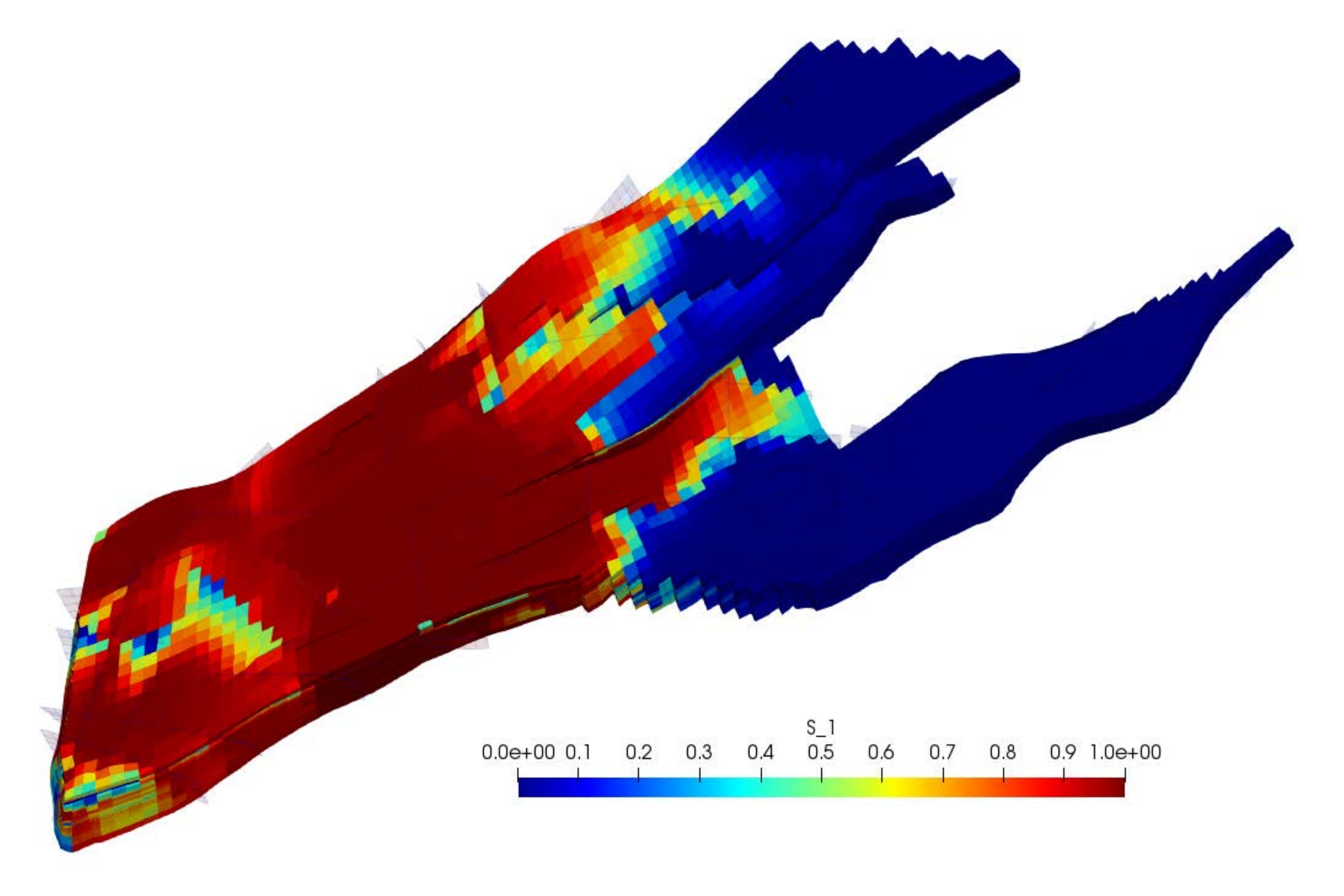}}
	\subcaptionbox{{\footnotesize Saturation in the fractures after $10000 [\text{days}]$ }\label{Fig:pEDFM_CPG_TestCase7_Norne_Sf_scenario2_40}}
	{\includegraphics[width=0.23\textwidth]{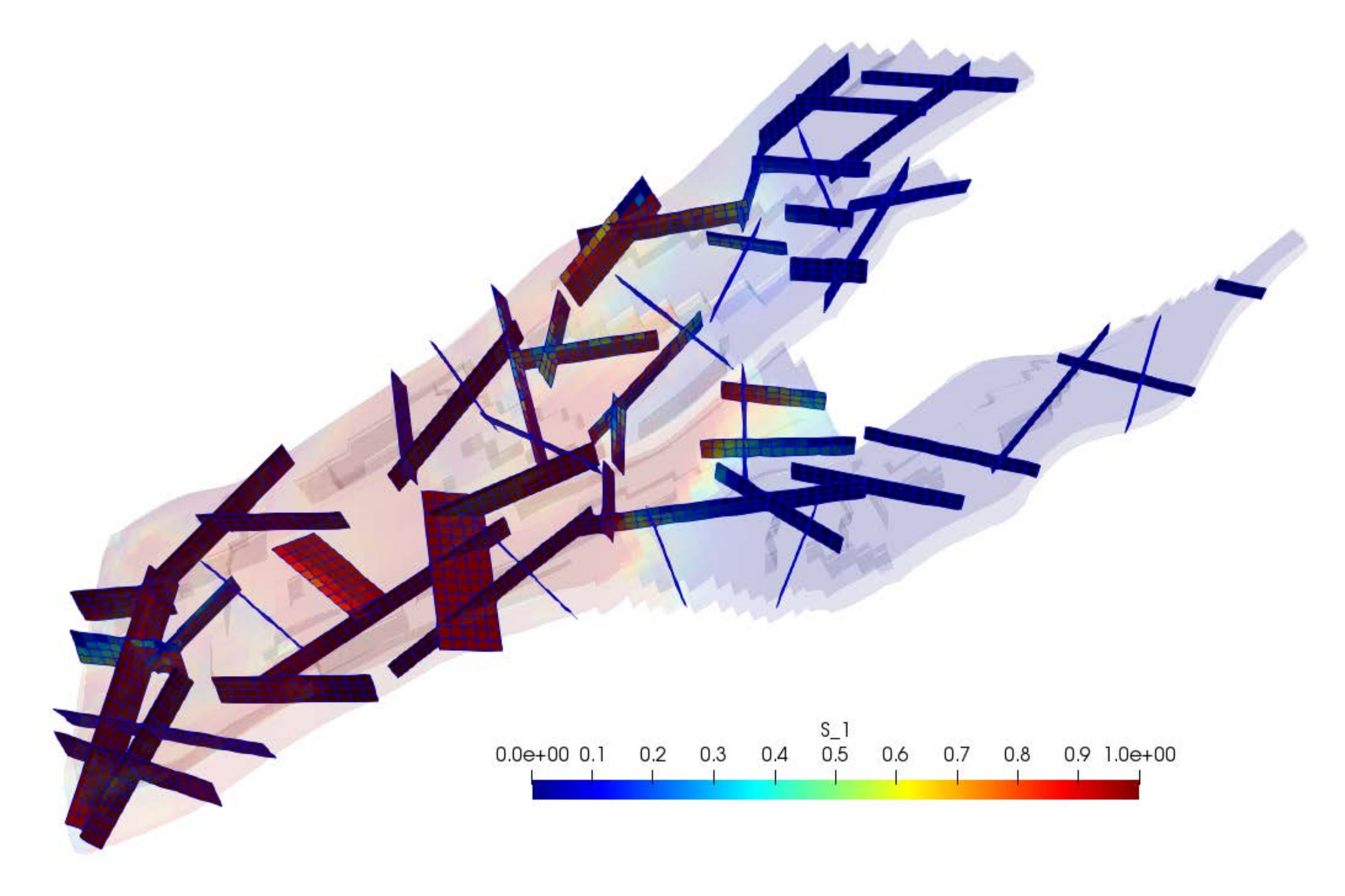}}
	\caption{Test case 7: The saturation profile of the Norne model for the simulation scenario $2$. The figures on the left illustrate the saturation profile in the matrix grid cells and the figures on the right side show the saturation maps in the fractures. From the top row towards the bottom row, the saturation profiles are displayed for simulations times $2000$, $5000$ and $10000 [\text{days}]$ respectively.}
	\label{Fig:pEDFM_CPG_TestCase7_Norne_Saturation_Scenario2}
\end{figure}

The results of the scenario $2$ are showed in the figures \ref{Fig:pEDFM_CPG_TestCase7_Norne_Pressure_Scenario2} and \ref{Fig:pEDFM_CPG_TestCase7_Norne_Saturation_Scenario2} respectively. Same as previous scenario, the pressure results are only shown for the simulation time $5000 [\text{days}]$, while the saturation profiles are shown for time intervals of $2000$, $5000$ and $10000 [\text{days}]$. The injection wells are surrounded by highly conductive fractures that act as flow channels. The pressure is more uniformly distributed. As a result, the effect of high permeable fractures near the injector wells has increased the saturation displacement across larger distances in the domain.

\section{Conclusions and future work}\label{Sec:Conclusion}
In this paper, a projection-based embedded discrete fracture model (pEDFM) for corner-point grid (CPG) geometry was developed and presented. This method was used with different fluid models, i.e., for fully-implicit simulation of isothermal multiphase fluid flow and low-enthalpy single-phase coupled mass-heat flow in fractured heterogeneous porous media. First, the corner-point grid geometry and its discretization approach were briefly described. Afterwards, the pEDFM model \cite{Tene2017,Hosseinimehr2020_Geothermal} was extended to account for fully 3D fracture geometries on generic corner-point grid discrete system. Through a few box-shaped 2D and 3D homogeneous and heterogeneous test cases, the accuracy of pEDFM on corner-point grid geometry was briefly compared against the Cartesian grid geometry. The new method presented similar results of satisfactory accuracy on corner-point grid geometry when compared to Cartesian grid-geometry. The 3D box-shaped reservoir was then converted into non-orthogonal gridding system to asses the pEDFM method further.

Moreover, numerical results were obtained on a number of geologically-relevant test cases. Different scenarios with various synthetic fracture networks were considered for these test cases. These fine-scale simulations allowed for mix-conductivity fractures. It was shown that pEDFM can accurately capture the physical influence of both highly conductive fractures and flow barriers on the flow patterns. The performance of the pEDFM on corner-point grid geometry casts a promising solution for increasing the discretization flexibility and enhancing the computational performance while honoring the accuracy. Many geo-models (including ones used in the test cases above) contain millions of grid cells that have complex geometrical alignments to match the positioning of fractures and faults, causing significant computational complexity and lack of flexibility especially when taking geomechanical deformation into account. The pEDFM model can provide an appropriate opportunity to avoid the complexity of gridding in such models by explicitly representing the discontinuities such as fractures and faults. In presence of elastic and (more importantly) plastic deformations, one could only modify the gridding structure of the affected region in the rock matrix, fully independent of the fractures and faults. This advantage results in significant computational gains especially in the realm of poromechanics.

The developments of pEDFM on corner-point grid geometry and all the related software implementations of this work are made open source and accessible on  \href{https://gitlab.com/darsim}{https://gitlab.com/DARSim}.

\section*{Acknowledgments}
Many thanks are due towards the members of the DARSim (Delft Advanced Reservoir Simulation) research group of TU Delft, for the useful discussions during the development of the pEDFM on corner-point grid geometry. 


\bibliographystyle{elsarticle-num}
\bibliography{References}

\begin{thebibliography}{10}
\expandafter\ifx\csname url\endcsname\relax
  \def\url#1{\texttt{#1}}\fi
\expandafter\ifx\csname urlprefix\endcsname\relax\def\urlprefix{URL }\fi
\expandafter\ifx\csname href\endcsname\relax
  \def\href#1#2{#2} \def\path#1{#1}\fi

\bibitem{Jansen2005}
J.-D. Jansen, D.~Brouwer, G.~Naevdal, C.~Van~Kruijsdijk, Closed-loop reservoir
  management, First Break 23~(1) (2005) 43--48.

\bibitem{OSullivan2001}
M.~J. OSullivan, K.~Pruess, M.~J. Lippmann, State of the art of geothermal
  reservoir simulation, Geothermics 30~(4) (2001) 395--429.

\bibitem{Axelsson2003}
G.~Axelsson, V.~Stefansson, Y.~Xu, Sustainable management of geothermal
  resources, in: Proceedings of the International Geothermal Conference, 2003,
  pp. 40--48.

\bibitem{Burnell2012}
J.~Burnell, E.~Clearwater, C.~A., W.~Kissling, J.~OSullivan, M.~OSullivan,
  A.~Yeh, {Future directions in geothermal modelling}, in: 34rd New Zealand
  Geothermal Workshop, 2012, pp. 19--21.

\bibitem{Burnell2015}
J.~Burnell, M.~OSullivan, J.~OSullivan, W.~Kissling, A.~Croucher, J.~Pogacnik,
  S.~Pearson, G.~Caldwell, S.~Ellis, S.~Zarrouk, M.~Climo, Geothermal
  supermodels: the next generation of integrated geophysical, chemical and flow
  simulation modelling tools., in: World Geothermal Congress, 2015, pp. 19--21.

\bibitem{gan2016production}
Q.~Gan, D.~Elsworth, Production optimization in fractured geothermal reservoirs
  by coupled discrete fracture network modeling, Geothermics 62 (2016)
  131--142.

\bibitem{gholizadeh2016study}
N.~Gholizadeh~Doonechaly, R.~R. Abdel~Azim, S.~S. Rahman, A study of
  permeability changes due to cold fluid circulation in fractured geothermal
  reservoirs, Groundwater 54~(3) (2016) 325--335.

\bibitem{salimzadeh2019novel}
S.~Salimzadeh, M.~Grandahl, M.~Medetbekova, H.~Nick, A novel radial jet
  drilling stimulation technique for enhancing heat recovery from fractured
  geothermal reservoirs, Renewable energy 139 (2019) 395--409.

\bibitem{WONG2018236}
Z.~Y. Wong, R.~N. Horne, H.~A. Tchelepi, Sequential implicit nonlinear solver
  for geothermal simulation, Journal of Computational Physics 368 (2018) 236 --
  253.
\newblock \href {http://dx.doi.org/https://doi.org/10.1016/j.jcp.2018.04.043}
  {\path{doi:https://doi.org/10.1016/j.jcp.2018.04.043}}.

\bibitem{Rossi2018}
E.~Rossi, M.~A. Kant, C.~Madonna, M.~O. Saar, P.~R. von Rohr, The effects of
  high heating rate and high temperature on the rock strength: Feasibility
  study of a thermally assisted drilling method, Rock Mechanics and Rock
  Engineering.

\bibitem{Garipov2016}
T.~T. Garipov, M.~Karimi-Fard, H.~A. Tchelepi, Discrete fracture model for
  coupled flow and geomechanics, Computational Geosciences 20~(1) (2016)
  149--160.
\newblock \href {http://dx.doi.org/10.1007/s10596-015-9554-z}
  {\path{doi:10.1007/s10596-015-9554-z}}.

\bibitem{gholizadeh2016evaluation}
N.~Gholizadeh~Doonechaly, R.~Abdel~Azim, S.~Rahman, Evaluation of recoverable
  energy potential from enhanced geothermal systems: a sensitivity analysis in
  a poro-thermo-elastic framework, Geofluids 16~(3) (2016) 384--395.

\bibitem{geochem_1}
F.~Morel, J.~Morgan, Numerical method for computing equilibriums in aqueous
  chemical systems, Environmental Science \& Technology 6~(1) (1972) 58 -- 67.

\bibitem{Saar_2}
A.~M.~M. Leal, D.~A. Kulik, W.~R. Smith, M.~O. Saar, An overview of
  computational methods for chemical equilibrium and kinetic calculations for
  geochemical and reactive transport modeling, Pure and Applied Chemistry
  89~(5) (2017) 597 -- 643.

\bibitem{salimzadeh2019coupled}
S.~Salimzadeh, H.~Nick, A coupled model for reactive flow through deformable
  fractures in enhanced geothermal systems, Geothermics 81 (2019) 88--100.

\bibitem{Lie_unst}
K.-A. Lie, T.~S. Mykkeltvedt, O.~M{\o}yner,
  \href{https://doi.org/10.1007/s10596-019-9829-x}{A fully implicit weno scheme
  on stratigraphic and unstructured polyhedral grids}, Computational
  Geosciences 24~(2) (2020) 405--423.
\newblock \href {http://dx.doi.org/10.1007/s10596-019-9829-x}
  {\path{doi:10.1007/s10596-019-9829-x}}.
\newline\urlprefix\url{https://doi.org/10.1007/s10596-019-9829-x}

\bibitem{Rainer2006}
V.~Reichenberger, H.~Jakobs, P.~Bastian, R.~Helmig, A mixed-dimensional finite
  volume method for two-phase flow in fractured porous media, Advances in Water
  Resources 29 (2006) 1020--1036.

\bibitem{Ahmed2015_DFM}
R.~Ahmed, M.~G. Edwards, S.~Lamine, B.~A.~H. Huisman, M.~Pal, Control-volume
  distributed multi-point flux approximation coupled with a lower-dimensional
  fracture model, J. Comput. Phys. 284 (2015) 462--489.

\bibitem{KarimiFard2004}
Karimi-Fard, L.~Durlofsky, K.Aziz, An efficient discrete-fracture model
  applicable for general-purpose reservoir simulators, SPE Journal (2004)
  227--236.

\bibitem{rami_edfmDFM}
J.~Jiang, R.~M. Younis, \href{https://doi.org/10.2118/178430-PA}{Hybrid coupled
  discrete-fracture/matrix and multicontinuum models for
  unconventional-reservoir simulation}, SPE Journal 21~(03) (2016) 1009--1027.
\newblock \href {http://dx.doi.org/10.2118/178430-PA}
  {\path{doi:10.2118/178430-PA}}.
\newline\urlprefix\url{https://doi.org/10.2118/178430-PA}

\bibitem{ponting1989corner}
D.~K. Ponting, {Corner point geometry in reservoir simulation}, in: ECMOR I-1st
  European Conference on the Mathematics of Oil Recovery, European Association
  of Geoscientists {\&} Engineers, 1989, pp. cp----234.

\bibitem{ding1995use}
Y.~Ding, P.~Lemonnier, Others, {Use of corner point geometry in reservoir
  simulation}, in: International Meeting on Petroleum Engineering, Society of
  Petroleum Engineers, 1995.

\bibitem{geoquest2014eclipse}
S.~GeoQuest, {ECLIPSE reference manual}, Schlumberger, Houston, Texas.

\bibitem{lie2019introduction}
K.-A. Lie, An introduction to reservoir simulation using MATLAB/GNU Octave:
  User guide for the MATLAB Reservoir Simulation Toolbox (MRST), Cambridge
  University Press, 2019.

\bibitem{berkowitz2002characterizing}
B.~Berkowitz, Characterizing flow and transport in fractured geological media:
  A review, Advances in water resources 25~(8-12) (2002) 861--884.

\bibitem{KUMAR2020109138}
K.~Kumar, F.~List, I.~S. Pop, F.~A. Radu,
  \href{https://www.sciencedirect.com/science/article/pii/S0021999119308435}{Formal
  upscaling and numerical validation of unsaturated flow models in fractured
  porous media}, Journal of Computational Physics 407 (2020) 109138.
\newblock \href {http://dx.doi.org/https://doi.org/10.1016/j.jcp.2019.109138}
  {\path{doi:https://doi.org/10.1016/j.jcp.2019.109138}}.
\newline\urlprefix\url{https://www.sciencedirect.com/science/article/pii/S0021999119308435}

\bibitem{Warren1963}
J.~Warren, P.~Root, The behavior of naturally fractured reservoirs, SPE J.
  (1963) 245--255.

\bibitem{Barenblatt1983}
G.~Barenblatt, Y.~Zheltov, I.~Kochina, Basic concepts in the theory of seepage
  of homogeneous fluids in fissurized rocks, J. Appl. Math. Mech. 5~(24) (1983)
  1286--1303.

\bibitem{Kazemi1996}
H.~Kazemi, L.~Merrill, K.~Porterfield, P.~Zeman, Numerical simulation of
  water-oil flow in naturally fractured reservoirs, SPE Journal~(5719) (1996)
  317--326.

\bibitem{RainerBook2005}
P.~Dietrich, R.~Helmig, M.~Sauter, H.~Hotzl, J.~Kongeter, G.~Teutsch, Flow and
  Transport in Fractured Porous Media, Springer, 2005.

\bibitem{Lee_Jensen_1999}
C.~L.~J. S.~H.~Lee, M.~F. Lough, An efficient finite difference model for flow
  in a reservoir with multiple length-scale fractures, SPE ATCE.

\bibitem{Lee_Lough_2001}
S.~Lee, M.~Lough, C.~Jensen, Hierarchical modeling of flow in naturally
  fractured formations with multiple length scales, Water Resource Research
  37~(3) (2001) 443--455.

\bibitem{Hajibeygi2012_JCP}
H.~Hajibeygi, D.~Karvounis, P.~Jenny, An upstream finite element method for
  solution of transient transport equation in fractured porous media, Journal
  of Computational Physics 230 (2012) 8729--8743.

\bibitem{Li2008}
L.~Li, S.~H. Lee, Efficient field-scale simulation of black oil in naturally
  fractured reservoir through discrete fracture networks and homogenized media,
  SPE Reservoir Evaluation \& Engineering (2008) 750--758.

\bibitem{Dimitrios2013}
D.~C. Karvounis, Simulations of enhanced geothermal systems with an adaptive
  hierarchical fracture representation, Ph.D. thesis, ETH Zurich (2013).

\bibitem{Moinfar2014}
A.~Moinfar, A.~Varavei, K.~Sepehrnoori, R.~T. Johns, Development of an
  efficient embedded discrete fracture model for 3d compositional reservoir
  simulation in fractured reservoirs, SPE J. 19 (2014) 289--303.

\bibitem{FLEMISCH2018239}
B.~Flemisch, I.~Berre, W.~Boon, A.~Fumagalli, N.~Schwenck, A.~Scotti,
  I.~Stefansson, A.~Tatomir,
  \href{https://www.sciencedirect.com/science/article/pii/S0309170817300143}{Benchmarks
  for single-phase flow in fractured porous media}, Advances in Water Resources
  111 (2018) 239--258.
\newblock \href
  {http://dx.doi.org/https://doi.org/10.1016/j.advwatres.2017.10.036}
  {\path{doi:https://doi.org/10.1016/j.advwatres.2017.10.036}}.
\newline\urlprefix\url{https://www.sciencedirect.com/science/article/pii/S0309170817300143}

\bibitem{LI2021108657}
L.~Li, D.~Voskov,
  \href{https://www.sciencedirect.com/science/article/pii/S092041052100317X}{A
  novel hybrid model for multiphase flow in complex multi-scale fractured
  systems}, Journal of Petroleum Science and Engineering 203 (2021) 108657.
\newblock \href
  {http://dx.doi.org/https://doi.org/10.1016/j.petrol.2021.108657}
  {\path{doi:https://doi.org/10.1016/j.petrol.2021.108657}}.
\newline\urlprefix\url{https://www.sciencedirect.com/science/article/pii/S092041052100317X}

\bibitem{Ali_EDFM01}
A.~Moinfar, A.~Varavei, K.~Sepehrnoori, R.~T. Johns,
  \href{https://doi.org/10.2118/154246-PA}{{Development of an Efficient
  Embedded Discrete Fracture Model for 3D Compositional Reservoir Simulation in
  Fractured Reservoirs}}, SPE Journal 19~(02) (2013) 289--303.
\newblock \href
  {http://arxiv.org/abs/https://onepetro.org/SJ/article-pdf/19/02/289/2099537/spe-154246-pa.pdf}
  {\path{arXiv:https://onepetro.org/SJ/article-pdf/19/02/289/2099537/spe-154246-pa.pdf}},
  \href {http://dx.doi.org/10.2118/154246-PA} {\path{doi:10.2118/154246-PA}}.
\newline\urlprefix\url{https://doi.org/10.2118/154246-PA}

\bibitem{SHAH201636}
S.~Shah, O.~MÃžyner, M.~Tene, K.-A. Lie, H.~Hajibeygi,
  \href{https://www.sciencedirect.com/science/article/pii/S0021999116301267}{The
  multiscale restriction smoothed basis method for fractured porous media
  (f-msrsb)}, Journal of Computational Physics 318 (2016) 36--57.
\newblock \href {http://dx.doi.org/https://doi.org/10.1016/j.jcp.2016.05.001}
  {\path{doi:https://doi.org/10.1016/j.jcp.2016.05.001}}.
\newline\urlprefix\url{https://www.sciencedirect.com/science/article/pii/S0021999116301267}

\bibitem{SANDVE20123784}
T.~Sandve, I.~Berre, J.~Nordbotten,
  \href{https://www.sciencedirect.com/science/article/pii/S0021999112000447}{An
  efficient multi-point flux approximation method for discrete fracture-matrix
  simulations}, Journal of Computational Physics 231~(9) (2012) 3784--3800.
\newblock \href {http://dx.doi.org/https://doi.org/10.1016/j.jcp.2012.01.023}
  {\path{doi:https://doi.org/10.1016/j.jcp.2012.01.023}}.
\newline\urlprefix\url{https://www.sciencedirect.com/science/article/pii/S0021999112000447}

\bibitem{Hajibeygi2011c_Hierarchical}
H.~Hajibeygi, D.~Karvounis, P.~Jenny, A hierarchical fracture model for the
  iterative multiscale finite volume method, J. Comput. Phys. 230~(24) (2011)
  8729--8743.

\bibitem{Hosseinimehr2018}
M.~HosseiniMehr, M.~Cusini, C.~Vuik, H.~Hajibeygi, Algebraic dynamic multilevel
  method for embedded discrete fracture model (f-adm), Journal of Computational
  Physics 373 (2018) 324--345.

\bibitem{Ali_EDFM02}
Y.~Xu, K.~Sepehrnoori, \href{https://doi.org/10.2118/195572-PA}{{Development of
  an Embedded Discrete Fracture Model for Field-Scale Reservoir Simulation With
  Complex Corner-Point Grids}}, SPE Journal 24~(04) (2019) 1552--1575.
\newblock \href
  {http://arxiv.org/abs/https://onepetro.org/SJ/article-pdf/24/04/1552/2118111/spe-195572-pa.pdf}
  {\path{arXiv:https://onepetro.org/SJ/article-pdf/24/04/1552/2118111/spe-195572-pa.pdf}},
  \href {http://dx.doi.org/10.2118/195572-PA} {\path{doi:10.2118/195572-PA}}.
\newline\urlprefix\url{https://doi.org/10.2118/195572-PA}

\bibitem{Tene2017}
M.~Tene, S.~B.~M. Bosma, M.~S.~A. Kobaisi, H.~Hajibeygi,
  \href{http://www.sciencedirect.com/science/article/pii/S0309170817300994}{Projection-based
  embedded discrete fracture model (pedfm)}, Adv. Water Resour. 105 (2017) 205
  -- 216.
\newblock \href
  {http://dx.doi.org/https://doi.org/10.1016/j.advwatres.2017.05.009}
  {\path{doi:https://doi.org/10.1016/j.advwatres.2017.05.009}}.
\newline\urlprefix\url{http://www.sciencedirect.com/science/article/pii/S0309170817300994}

\bibitem{Hosseinimehr2020_Geothermal}
M.~HosseiniMehr, C.~Vuik, H.~Hajibeygi, Adaptive dynamic multilevel simulation
  of fractured geothermal reservoirs, Journal of Computational Physics: X 7
  (2020) 100061.

\bibitem{Rami_pedfm_copyCatz}
J.~Jiang, R.~M. Younis,
  \href{https://www.sciencedirect.com/science/article/pii/S0309170817304657}{An
  improved projection-based embedded discrete fracture model (pedfm) for
  multiphase flow in fractured reservoirs}, Advances in Water Resources 109
  (2017) 267--289.
\newblock \href
  {http://dx.doi.org/https://doi.org/10.1016/j.advwatres.2017.09.017}
  {\path{doi:https://doi.org/10.1016/j.advwatres.2017.09.017}}.
\newline\urlprefix\url{https://www.sciencedirect.com/science/article/pii/S0309170817304657}

\bibitem{Peaceman1978}
D.~W. Peaceman, Interpretation of well-block pressures in numerical reservoir
  simulation, SPE J. 18 (3) (1978) 183--194.

\bibitem{WANG2020114693}
Y.~Wang, D.~Voskov, M.~Khait, D.~Bruhn,
  \href{https://www.sciencedirect.com/science/article/pii/S0306261920302051}{An
  efficient numerical simulator for geothermal simulation: A benchmark study},
  Applied Energy 264 (2020) 114693.
\newblock \href
  {http://dx.doi.org/https://doi.org/10.1016/j.apenergy.2020.114693}
  {\path{doi:https://doi.org/10.1016/j.apenergy.2020.114693}}.
\newline\urlprefix\url{https://www.sciencedirect.com/science/article/pii/S0306261920302051}

\bibitem{Al_Shemmeri_Book}
T.~Al-Shemmeri, {Engineering Fluid Mechanics}, Bookboon, 2012, Ch.~1, p.~18.

\bibitem{KHCoats1977}
K.~H. Coats, {Geothermal Reservoir Modelling}, in: SPE Annual Fall Technical
  Conference and Exhibition, 1977.
\newblock \href {http://dx.doi.org/10.2118/6892-MS}
  {\path{doi:10.2118/6892-MS}}.

\bibitem{Timothy_GeoTMsFV}
T.~Praditia, R.~Helmig, H.~Hajibeygi, Multiscale formulation for coupled
  flow-heat equations arising from single-phase flow in fractured geothermal
  reservoirs, Computat. Geo. 22 (2018) 1305--1322.
\newblock \href {http://dx.doi.org/10.1007/s10596-018-9754-4}
  {\path{doi:10.1007/s10596-018-9754-4}}.

\bibitem{IAPWS}
W.~Wagner, H.~Kretzschmar, International Steam Tables - Properties of Water and
  Steam based on the Industrial Formulation IAPWS-IF97, 2nd Edition, Springer,
  2008.

\bibitem{Bosma2017145}
S.~B.~M. Bosma, H.~Hajibeygi, M.~Tene, H.~A. Tchelepi, Others, {Multiscale
  Finite Volume Method for Discrete Fracture Modeling with Unstructured Grids},
  SPE Reservoir Simulation Conference 351 (2017) 145--164.
\newblock \href {http://dx.doi.org/10.2118/182654-MS}
  {\path{doi:10.2118/182654-MS}}.

\bibitem{Eigestad2009}
G.~Eigestad, H.~Dahle, B.~Hellevang, F.~Riis, W.~Johansen, E.~Tian, Geological
  modeling and simulation of co2 injection in the johansen formation,
  Computational Geosciences 13(1) (2009) 435--450.
\newblock \href {http://dx.doi.org/10.1007/s10596-009-9153-y}
  {\path{doi:10.1007/s10596-009-9153-y}}.

\bibitem{Peters2010}
L.~Peters, R.~Arts, G.~Brouwer, Results of the brugge benchmark study for
  flooding optimization and history matching, SPE Reservoir Evaluation and
  Engineering 294--295 (2010) 391--405.
\newblock \href {http://dx.doi.org/10.2118/119094-PA}
  {\path{doi:10.2118/119094-PA}}.

\bibitem{Verlo2008}
S.~B. Verlo, M.~Hetland, Development of a field case with real production and
  4d data from the norne field as a benchmark case for future reservoir
  simulation model testing. msc thesis, Norwegian University of Science and
  Technology, Trondheim, Norway.

\bibitem{OPM}
Open porous media (opm), \url{http://opm-project.org}, accessed: 2020-10-30.

\end{thebibliography}

\end{document}